\documentclass[printer,longauth,usenatbib]{aa}

\usepackage{natbib}
\usepackage{times}
\usepackage[T1]{fontenc}
\usepackage{graphicx}
\usepackage{tabularx}
\usepackage[skip=1ex]{caption}

\usepackage[utf8]{inputenc}
\usepackage{tabularx}

\def\deg{\hbox{$^\circ$}}
\newcommand{\chisq}[0]{$\bar{\chi}^2_{\mathrm{min}}$}

\begin{document}

\title{Thermal properties of slowly rotating asteroids: \break 
  Results from a targeted survey}

\authorrunning{Marciniak et al.}
\titlerunning{Thermal properties of slowly rotating asteroids}

\author{A. Marciniak \inst{1} 
  \and V. Al{\'i}-Lagoa \inst{2}
  \and T. G. M{\"u}ller \inst{2}
  \and R. Szak{\'a}ts \inst{3}
  \and L. Moln{\'a}r \inst{3,4}
  \and A. P{\'a}l \inst{3,5} 
  \and E. Podlewska - Gaca \inst{1,6}
  \and N.~Parley \inst{7}
  \and P. Antonini \inst{8} 
  \and E. Barbotin \inst{9} %
  \and R. Behrend \inst{10}  
  \and L. Bernasconi \inst{11} 
  \and M. Butkiewicz - B\k{a}k \inst{1}
  \and R. Crippa \inst{12} 
  \and R. Duffard \inst{13}
  \and R.~Ditteon \inst{14}
  \and M. Feuerbach \inst{10} 
  \and S. Fauvaud \inst{15}
  \and J. Garlitz \inst{16} 
  \and S. Geier \inst{17,18}
  \and R. Goncalves \inst{19} 
  \and J. Grice \inst{20}
  \and I. Grze{\'s}kowiak \inst{1}
  \and R.~Hirsch \inst{1}
  \and J. Horbowicz \inst{1}
  \and K. Kami{\'n}ski \inst{1}
  \and M.~K.~Kami{\'n}ska \inst{1}
  \and D.-H. Kim \inst{21,22} 
  \and M.-J. Kim \inst{22}
  \and I. Konstanciak \inst{1}
  \and V.~Kudak \inst{23,24}
  \and P. Kulczak \inst{1}
  \and J. L.~Maestre \inst{25} 
  \and F. Manzini \inst{12} 
  \and S. Marks \inst{10} 
  \and F. Monteiro \inst{26} 
  \and W. Og{\l}oza \inst{27} 
  \and D. Oszkiewicz \inst{1}
  \and F.~Pilcher \inst{28}
  \and V. Perig \inst{24} 
  \and T. Polakis \inst{29}
  \and M. Poli{\'n}ska \inst{1}
  \and R. Roy \inst{30}  
  \and J. J. Sanabria \inst{17}
  \and T. Santana-Ros \inst{1}
  \and B. Skiff \inst{31}
  \and J.~Skrzypek \inst{1}
  \and K. Sobkowiak \inst{1}
  \and E. Sonbas \inst{32}
  \and O. Thizy \inst{10} 
  \and P. Trela \inst{1}
  \and S. Urakawa \inst{33}
  \and M. {\.Z}ejmo \inst{34}
  \and K. {\.Z}ukowski \inst{1}
}

\institute{Astronomical Observatory Institute, Faculty of Physics, A. Mickiewicz University,
  S{\l}oneczna 36, 60-286 Pozna{\'n}, Poland. E-mail: am@amu.edu.pl 
  \and Max-Planck-Institut f{\"u}r Extraterrestrische Physik, Giessenbachstrasse 1, 85748 Garching, Germany 
  \and Konkoly Observatory, Research Centre for Astronomy and Earth Sciences, Hungarian Academy of Sciences, 
  H-1121 Budapest, Konkoly Thege Mikl{\'o}s {\'u}t 15-17, Hungary 
  \and MTA CSFK Lend{\"u}let Near-Field Cosmology Research Group 
  \and Astronomy Department, E\"otv\"os Lor\'and University, P\'azm\'any P. s. 1/A, H-1171 Budapest, Hungary 
  \and Institute of Physics, University of Szczecin, Wielkopolska 15, 70-453 Szczecin, Poland 
  \and The IEA, University of Reading, Philip Lyle Building, Whiteknights Campus, Reading, RG6 6BX 
  \and Observatoire des Hauts Patys, F-84410 B\'edoin, France 
  \and Villefagnan Observatory, France 
  \and Geneva Observatory, CH-1290 Sauverny, Switzerland 
  \and Les Engarouines Observatory, F-84570 Mallemort-du-Comtat, France 
  \and Stazione Astronomica di Sozzago, I-28060 Sozzago, Italy 
  \and Departamento de Sistema Solar, Instituto de Astrof{\'i}sica de Andaluc{\'i}a (CSIC),
  Glorieta de la Astronom{\'i}a s/n, 18008 Granada, Spain 
  \and Rose-Hulman Institute of Technology, CM 171 5500 Wabash Ave., Terre Haute, IN 47803, USA 
  \and Observato{\'i}re du Bois de Bardon, 16110 Taponnat, France 
  \and 1155 Hartford St; Elgin, OR USA 
  \and Instituto de Astrof{\'i}sica de Canarias, C/ V{\'i}a Lactea, s/n, 38205 La Laguna, Tenerife, Spain 
  \and Gran Telescopio Canarias (GRANTECAN), Cuesta de San Jos{\'e} s/n, E-38712, Bre{\~n}a Baja, La Palma, Spain 
  \and Lincaheira Observatory, Instituto Polit{\'e}cnico de Tomar, 2300-313 Tomar Portugal 
  \and School of Physical Sciences, The Open University, MK7 6AA, UK 
  \and Chungbuk National University, 1, Chungdae-ro, Seowon-gu, Cheongju-si, Chungcheongbuk-do, Republic of Korea 
  \and Korea Astronomy and Space Science Institute, 776 Daedeokdae-ro, Yuseong-gu, 305-348 Daejeon, Korea 
  \and Institute of Physics, Faculty of Natural Sciences, University of P. J. \v{S}af{\'a}rik, Park Angelinum 9, 
  040 01 Ko\v{s}ice, Slovakia 
  \and Laboratory of Space Researches, Uzhhorod National University, Daleka st. 2a, 88000, Uzhhorod, Ukraine 
  \and Albox, Spain 
  \and Observat{\'o}rio Nacional Rua General Jos{\'e} Cristino, 77, 20921-400 Bairro Imperial de S{\~a}o Crist{\'o}v{\~a}o 
  Rio de Janeiro, RJ, Brasil 
  \and Mt. Suhora Observatory, Pedagogical University, Podchor\k{a}{\.z}ych 2, 30-084, Cracow, Poland 
  \and 4438 Organ Mesa Loop, Las Cruces, New Mexico 88011 USA 
  \and Command Module Observatory, 121 W. Alameda Dr., Tempe, AZ 85282 USA 
  \and Observatoire de Blauvac, 293 chemin de St Guillaume, F-84570 St-Est{\`e}ve, France 
  \and Lowell Observatory, 1400 West Mars Hill Road, Flagstaff, Arizona, 86001 USA 
  \and University of Adiyaman, Department of Physics, 02040 Adiyaman, Turkey 
  \and Bisei Spaceguard Center, Japan Spaceguard Association, 1716-3, Okura, Bisei, Ibara,
  Okayama 714-1411, Japan 
  \and Kepler Institute of Astronomy, University of Zielona G{\'o}ra, Lubuska 2, 65-265 Zielona G{\'o}ra, Poland 
}
\date{Received 25 January 2019 / Accepted 9 April 2019}

\abstract
{
    Earlier work suggests that slowly rotating asteroids should have higher
    thermal inertias than faster rotators because the heat wave penetrates
    deeper into the sub-surface. However, thermal inertias have been determined
    mainly for fast rotators due to selection effects in the available
    photometry used to obtain shape models required for thermophysical
    modelling (TPM). 
}
{
    Our aims are to mitigate these selection effects by producing shape models of
    slow rotators, to scale them and compute their thermal inertia with TPM,
    and to verify whether thermal inertia increases with the rotation period.
}
{
  To decrease the bias against slow rotators, we conducted a photometric
  observing campaign of main-belt asteroids with periods longer than 12 hours,
  from multiple stations worldwide, adding in some cases data from WISE and
  Kepler space telescopes. For spin and shape reconstruction we used the
  lightcurve inversion method, and to derive thermal inertias we applied
    a thermophysical model to fit available infrared data from IRAS, AKARI, and
    WISE.
}
{
    We present new models of 11 slow rotators that provide a good fit to
    the thermal data. In two cases, the TPM analysis showed a clear preference 
    for one of the two possible mirror solutions. We derived the diameters and
    albedos of our targets in addition to their thermal inertias, which ranged
    between 3$^{+33}_{-3}$ and 45$^{+60}_{-30}$ J\,m$^{-2}$\,s$^{-1/2}$\,K$^{-1}$.
}
{
    Together with our previous work, we have analysed 16 slow rotators from our
    dense survey with sizes between 30 and 150 km. The current sample thermal
    inertias vary widely, which does not confirm the earlier suggestion that slower
    rotators have higher thermal inertias. 
}


\maketitle

\section{Introduction}

  Thermal infrared flux from asteroids carries information on their surface regolith properties like 
  thermal inertia, surface roughness, regolith grain size, and their compactness. 
  Thermal data also allow us to precisely determine the size and albedo of these objects, when coupled with visual absolute magnitudes. 
  Asteroid sizes are otherwise hard to determine, unless multi-chord stellar occultations, adaptive optics images, or radar echoes 
  are available \citep{Durech2015}, yet they are essential characteristics for studies of for example 
  the collisional evolution based on currently observed size-frequency distribution \citep{Bottke2005}, 
  asteroid densities \citep{Carry2012}, 
  or asteroid family members dispersion under thermal recoil force \citep{Vokrouhlicky2015}.
  Physical properties of the regolith covering asteroid surfaces are connected to their age and composition. 
  Young, fresh surfaces display high thermal inertia, because of the small amount of fine-grained regolith on the surface, 
  while the surfaces of old targets are covered with thick layer of insulating regolith, resulting in small thermal inertia values. 
 Also, bodies larger than 100 km in diameter 
 are usually covered with very fine-grained regolith (thus displaying smaller thermal inertia), while smaller sized ones 
 display signatures of coarser grains in the surface, resulting in larger thermal inertia \citep{Gundlach2013}.

  These properties can provide valuable insights into the dynamical, collisional and thermal history of asteroids, some of which 
  are intact leftover planetesimals that created planets and shaped the solar system. It might also be possible to study deeper layers 
  under the immediate surface by using longer wavelengths that penetrate deeper, or by studying targets with long episodes of the Sun 
  heating the same surface area, like slow rotators or targets with low spin axis inclination to the orbit.

  For detailed studies of asteroid thermal properties, various thermophysical models (henceforth: TPM) are being used 
  \citep[for their overview see][]{Delbo2015}. 
  The essential input to apply these models is the knowledge of the spin and shape of the objects under study. 
  The rotation period and spin axis inclination dictate the duration and intensity of the alternating cycles 
  of surface heating and cooling. Such cycles, when especially intense, can result in thermal cracking \citep{Delbo2014}. 
  As for the shape, at first approximation it can be represented by a sphere, 
 although such a simple shape often fails to explain thermal lightcurves or separate thermal measurements taken at different viewing 
 aspects and phase angles.

 Today the availability of spin and shape modelled targets is the main limiting factor for asteroid studies by TPM. 
 Also, in spite of an abundance of thermal data available mainly from WISE (Wide-field Infrared Survey Explorer), 
 AKARI (meaning ``light'' in Japanese), IRAS (The Infrared Astronomical Satellite), or Herschel space observatories, 
 they also possess their limitations: for unique solutions from thermophysical modelling, thermal data have to come 
 from a range of aspect and phase angles, probing the target rotation with sufficient resolution at the same time. 
  As a result, so far detailed TPM have been applied to 
 less than 200 asteroids \citep[see the compilations by][]{Delbo2015, Hanus2018b}, and only a small fraction of them 
 rotate slowly, with periods exceeding 12 hours. Slow rotators are 
 especially interesting in this context, as a trend of increasing thermal inertia with period has been found by \cite{Harris2016}, 
 based on estimated beaming parameters ($\eta$) determined on data from the WISE space telescope.

 The distribution of thermal inertia amongst the members of asteroid families would be crucial for example in the search for evidence for asteroid 
 differentiation, separating iron rich from iron poor family members \citep{Matter2013}. 
 This quantity is also essential for estimates of the orbital drift caused by the Yarkovsky effect, and for studies 
 of regolith properties. However, the relatively small number of targets with known spin and shape parameters 
 prevents detailed thermophysical studies on large groups of asteroids, 
 limiting these to using simple thermal models with spherical shape approximation, often insufficient to explain thermal data 
 taken in various bands and viewing geometries. It is especially vital in the light of high precision thermal infrared data 
 from the WISE and Spitzer space telescopes now available, with uncertainties comparable to average uncertainties of the shape models 
 based on lightcurve inversion \citep{Delbo2015, Hanus2015}.

 Nowadays asteroid spin and shape models are created mainly from 
 sparse-in-time visual data (of the order of one to two points per day)  
 from large surveys like those compiled in the Lowell photometric database
 \citep{Hanus2011}. 
 The low photometric precision of such data (average $\sigma \sim$ 0.15 - 0.2 mag),  
 originally gathered for astrometric purposes, results in the favouring of some targets in the modelling, for example mainly those displaying 
 large amplitudes in each apparition \citep{Durech2016}. A partial solution to the problem is joining sparse data 
 with dense lightcurves \citep[like in e.g.][]{Hanus2013}, an approach limited though by the availability of the latter, 
 which are also strongly biased towards large amplitudes and short periods. 
 On the other hand, targeted surveys gathering dense lightcurves favour certain 
 types of objects, like near Earth asteroids (NEAs) 
 or members of specific groups or families like Flora, Eos, or Hungaria asteroids. 
   The resulting spin and shape determinations, even if taken altogether, do
   not provide unbiased information on for example the spin axis distributions of the
   whole asteroid population. Instead, they are biased by spin clusters within
   families \citep{Slivan2002, Kryszczynska2013} and by preferential retrograde
   rotation of NEAs \citep{Vokrouhlicky2015}.
 Taking all this together can distort the results for the whole population,
 missing certain specific cases, so that many elements essential for properly
 understanding asteroid dynamics, physics, and evolution are lacking 
 \citep{Warner2011}. For example, Jupiter Trojans on average rotate much more slowly than main belt asteroids, 
 so there might be a gradation of rotation periods with growing heliocentric distance \citep{Marzari2011}.
 Also, as has been stressed by \cite{Warner2011}, as much as 40\% of lightcurves considered in the Lightcurve Database 
 (LCDB)\footnote{http://www.MinorPlanet.info/lightcurvedatabase.html.}
 as reliable, have small maximum amplitude: a$_{max}$ $\leq$ 0.2 mag, posing additional challenges in period determination 
 due to potential ambiguities, and having profound effects on the results from any survey using asparse-in-time 
 observing cadence.

  Some light on the subject of biases in asteroid models was shed by recent 
  results based on data for asteroids from Gaia Data Release 2 (DR2), which facilitated unique determination of spin parameters 
  for around 200 targets \citep{Durech2018a}. This first, largely unbiased sample of targets 
  \citep[with some exceptions, see][]{Santana-Ros2015} modelled using very precise absolute brightnesses 
  revealed interesting results. Half of the asteroids in this sample turned out to rotate slowly, with periods longer than 12 hours, 
  unlike in the sets of models from the majority of previous surveys, based on ground-based data, where shorter periods dominated 
  \citep[e.g.][]{Hanus2011, Hanus2013}. Results from Gaia DR2 confirm the findings of \cite{Szabo2016} and \cite{Molnar2018} 
  based on data from the Kepler Space Telescope of the substantial contribution of slow rotators in the asteroid population.

  It should also be noted that models based on sparse data provide reliable spin parameters, 
  but only low-resolution, coarse shape representations \citep[flag '1' or '2' in the shape quality system proposed by][]{Hanus2018a}, 
  limiting their use in other applications \citep{Hanus2016}. 
  For example, detailed thermophysical modelling needs rather high-resolution (flag '3') shape models as input, only possible 
  with large datasets of dense lightcurves obtained at various viewing geometries. Such models  
  usually provide a much better fit to thermal data than shapes approximated by a sphere 
  with the same spin axis and period \citep{Marciniak2018}.

  In the next section we refer to our targeted survey, then in Sect. 3 we describe our modelling methods. 
  Results from both lightcurve inversion and thermophysical modelling are presented in Sect. 4, 
  and discussion and conclusions in Sect. 5.
  Appendix A contains details of the observing runs and composite lightcurves, while Appendix B presents 
  O-C plots from thermophysical analysis.

\section{Targeted survey}

 Taking all the above-mentioned facts into account, it is clear that there is a continuing need for targeted, dense-in-time observations 
 of objects omitted by previous and ongoing surveys. Such targets are mainly slow rotators, and also bodies that either 
 constantly or temporarily display lightcurves of small amplitudes. 
  To counteract both selection effects, our observing campaign 
 is targeted at main-belt slow rotators (P $>$ 12 h), which at the same time have small maximum amplitudes (a$_{max}\leq$0.25 mag). 
 Neither sparse data from large surveys nor available datasets of dense lightcurves allow for their precise spin and shape 
 modelling, so they require a dedicated observing campaign. It is worth noting that it is exactly these features -- 
 small and slow flux variations -- that make them challenging for spin and shape reconstructions,   
 which make these targets perfect calibration sources for a range of infrared observatories 
 \citep[like ALMA -- Atacama Large Millimeter Array, APEX -- Atacama Pathfinder Experiment, 
 or IRAM -- Institut de Radioastronomie Millim{\'e}trique,
][]{MullerLagerros2002}, on the condition that these variations can be exactly predicted. 
 To this purpose their reliable spin and shape model are necessary. 
 
 The details of our campaign conducted at over 20 stations from 12 countries worldwide, the target selection procedure, and first results 
 are described in \citet{Marciniak2015} and \citet{Marciniak2016}. 
 In a nutshell, our survey revealed a substantial number of slow rotators which, in spite of having good data from dense 
 observations, had wrongly determined rotation periods, summarised in LCDB \citep{Warner2009}, a database 
 used in a variety of further studies on asteroid spins and dynamics. We corrected those determinations with the new values 
 for the periods, and confirmed them in consecutive apparitions. 
 We also constructed spin and shape models for the first sample our targets, using dense lightcurves from multiple apparitions, 
 and scaled these models by stellar occultations fitting and thermophysical modelling, with consistent size determinations
 (Marciniak et al. 2018). 
 In the latter work we also presented simultaneous fits to data from 
 three different infrared missions (IRAS, AKARI, and WISE), and in spite of potential calibration problems of each 
 of the missions alone and cross-calibrations between them, we obtained good fits. This gave us confidence in our models 
 and methods used.

 The most important of our previous results presented in \citet{Marciniak2018} were large thermal inertia values obtained 
 for most of our slow rotators, 
 which followed the trend found by \citet{Harris2016}. This seemed to support the idea that for slow rotators we observe thermal 
 emission from deeper, more compact layers of the surface, a fact that might open new paths for studies of asteroid regolith. 
 However, due to the small number of models for slowly rotating asteroids, until recently thermal inertia values from detailed TPM 
 have been determined for only two such asteroids (227 Elvira and 956 Elisa, Delbo \& Tanga 2009, Lim et al. 2011,  
 see fig 5. in \citet{Harris2016}, based on compilation by \citet{Delbo2015}, reproduced here in Fig. \ref{HD})
 The above-mentioned trend was found on the values derived from estimated beaming parameters using a simplified approach. 
 Recent results from TPM by \citet{Hanus2018b} showed a rather large diversity of thermal inertia values for 
 slow rotators within the size range of 10-100 km. Here we further investigate this issue by studying more 
 slow rotators in both visible light and thermal infrared radiation.

 Our photometric campaign is ongoing, and new observing stations have joined. 
 Our data sources now also include observations from the Kepler Space Telescope in its extended mission, K2 \citep{Howell2014}.
 Kepler could not track moving targets during a campaign, so solar system objects could be observed either by using 
 previously allocated, curved, or boomerang-shaped masks in accordance with the expected positions of the target minor bodies 
 \citep[see e.g.][]{Pal2015} or larger continuous pixel masks (so-called supermasks) where such objects also appeared for several days, 
 but not on purpose \citep[see e.g.][]{Molnar2018}. Data processing has been safely established for these observations, 
 regardless of whether these were targeted or not. This processing is based on the registration  of the images 
 (to correct for spacecraft jitter) followed by a differential-image photometry by involving oblong-shaped apertures 
 (see P{\'a}l et al. 2015, Szab{\'o} et al. 2017, and Moln{\'a}r et al. 2018 for further details). 
 The model of (100) Hekate presented here has been partially based on Kepler data.
 The asteroid (100) Hekate was observed with a dedicated set of masks for 5.5 and 4.5 days continuously in Campaigns 16 and 18, respectively. 
  Table \ref{obs} in the Appendix gives details of each observing run obtained within the present work. 
 Scarce literature data had to be complemented with from $\sim$100 up to $\sim$500 hours of new observations for each target before 
 unique models were feasible.

\section{Lightcurve inversion and thermophysical modelling}\label{sec:model}

For spin and shape reconstructions we use lightcurve inversion by \citet{Kaasalainen2001}, which represents 
the shape model by a convex polyhedron. 
All the lightcurves were treated as relative, and not absolutely calibrated, as from our campaign 
we obtain mostly relative brightness measurements.
Only the models with a unique solution, clearly the best in terms of 
$\chi^2$ fit of observed to modelled lightcurves, have been accepted. Each of the acceptable shape model solutions 
has been visually inspected to fulfil the criterion of rotation around the axis of greatest inertia. 
From the range of accepted solutions the uncertainty range of spin axis position and period has been evaluated. 
As shown by \citet{Kaasalainen_et_al2001} this is the only practical way of obtaining realistic uncertainty 
for spin axis position because formal errors like those obtained from the covariance matrix tend to be strongly underestimated. 
The effects of random noise are much smaller than systematic errors not uncommon in photometric data, or the combined effects 
of model shape and spin axis uncertainties. The uncertainty of period is also determined from the range of best-fit solutions, 
and is dictated by the time span of the whole photometric dataset and the period duration itself. If the minimum in the periodogram 
is substantially lower than the others, the period uncertainty can be assumed to be one hundreth of this minimum width, 
which is usually of the same order as the one determined from the best-fit solutions. 

As is usually the case when models are based exclusively on dense lightcurves, the uncertainty of the 
spin axis position (pole) was of the order of a few degrees (see Table \ref{results}), and the shape models had a
smooth appearance. Formally large uncertainty in some values of pole longitude ($\lambda_p$) is due to high 
inclination of the pole ($\beta_p$), translating actually to a relatively small distance on the celestial sphere. 
\begin{table*}[t!]
  \centering
\begin{small}
\caption{Spin parameters of asteroid models obtained in this work, with their uncertainty values. 
The first column gives the sidereal period of rotation, next there are two sets of pole J2000.0 longitude and latitude. 
The sixth column gives the rms deviations of the model lightcurves from the data, and next follow the photometric dataset 
parameters (observing span, number of apparitions, and number of individual lightcurve fragments).
Pole solutions preferred by TPM are marked in bold. }
\label{results}
\begin{tabular}{rrrrccccc}
\hline
Sidereal      & \multicolumn{2}{c}{Pole 1} & \multicolumn{2}{c}{Pole 2} & rmsd  & Observing span & $N_{app}$ & $N_{lc}$ \\
period [hours]& $\lambda_p$ & $\beta_p$    & $\lambda_p$ & $\beta_p$    & [mag] &   (years)      &           &          \\
\hline
&&&&&&&&\\
{\bf (100) Hekate}  & & & & & & & & \\
$27.07027$   & $104\deg$     & $+51\deg$   & $306\deg$    & $+52\deg$   & 0.012 & 1977--2018 & 8 & 62 \\  
$\pm 0.00006$& $\pm 7\deg$  & $\pm 2\deg$ & $\pm 8\deg$ & $\pm 3\deg$ &  &  &  &  \\
&&&&&&&&\\
{\bf (109) Felicitas} & & & & & & & &  \\
$13.190550$   & ${\bf 77\deg}$     & ${\bf -26\deg}$   & $252\deg$    & $-49\deg$   & 0.015 & 1980--2018 & 8 & 62 \\  
$\pm 0.000004$& $\pm 1\deg$   & $\pm 5\deg$ & $\pm 4\deg$  & $\pm 7\deg$ &  &  &  &  \\
&&&&&&&&\\
{\bf (195) Eurykleia} & & & & & & &  & \\
$16.52178$   & $101\deg$     & $+71\deg$   & $352\deg$    & $+83\deg$    & 0.015 & 2001--2017 & 7 & 51 \\  
$\pm 0.00001$& $\pm 60\deg$  & $\pm 15\deg$ & $\pm 60\deg$ & $\pm 15\deg$ &  &  &  &  \\
&&&&&&&&\\
{\bf (301) Bavaria} & & & & & & &  & \\
$12.24090$   & $ 46\deg$     & $+61\deg$   & $226\deg$    & $+70\deg$    & 0.014 & 2004--2018 & 5 & 30 \\  
$\pm 0.00001$& $\pm 5\deg$   & $\pm 6\deg$ & $\pm 14\deg$ & $\pm 6\deg$  &  &  &  &  \\
&&&&&&&&\\
{\bf (335) Roberta} & & & & & &  & & \\
$12.02713$   & $ 105\deg$     & $ +48\deg$  & $297\deg$    & $+54\deg$    & 0.012  & 1981--2017 & 9 & 52 \\  
$\pm 0.00003$& $\pm 7\deg$   & $\pm 2\deg$ & $\pm 9\deg$ & $\pm 6\deg$  &  &  &  &  \\
&&&&&&&&\\
{\bf (380) Fiducia} & & & & & &  & & \\
$13.71723$   & $ 21\deg$   &   $+34\deg$   &  $ 202\deg$   &   $+44\deg$    & 0.016  & 2008--2018 & 6 & 37 \\  
$\pm 0.00002$& $\pm 2\deg$  & $\pm 3\deg$  &  $\pm 1\deg$  & $\pm 2\deg$    &  &  &  &  \\
&&&&&&&&\\
{\bf (468) Lina} & & & & & & &  & \\
$16.47838$   & $ 74\deg$     & $+68\deg$   & $255\deg$     & $+68\deg$    & 0.013  & 1977--2018 & 7 & 40 \\  
$\pm 0.00003$& $\pm 18\deg$  & $\pm 6\deg$ & $\pm 17\deg$  & $\pm 8\deg$  &  &  &  &  \\
&&&&&&&&\\
{\bf (538) Friederike} & & & &  & & & & \\
$46.739$   & $168\deg$     & $-58\deg$   & $328\deg$     & $-59\deg$     & 0.015  & 2003--2018 & 8 & 98 \\  
$\pm 0.001$& $\pm 30\deg$  & $\pm 5\deg$ & $\pm 25\deg$  & $\pm 10\deg$  &  &  &  &  \\
&&&&&&&&\\
{\bf (653) Berenike} & & & & &  & & & \\
$12.48357$   & ${\bf 147\deg}$     & ${\bf +18\deg}$   & $335\deg$     & $+8\deg$     & 0.009  & 1984--2018 & 6 & 39 \\  
$\pm 0.00003$& $\pm 3\deg$   & $\pm 2\deg$ & $\pm 2\deg$   & $\pm 1\deg$  &  &  &  &  \\
&&&&&&&&\\
{\bf (673) Edda} & & & & & & &  & \\
$22.33411$   & $ 66\deg$    & $+64\deg$    & $236\deg$     & $+63\deg$     & 0.022  & 2005--2017 & 7 & 49 \\  
$\pm 0.00004$& $\pm 15\deg$  & $\pm 6\deg$ & $\pm 13\deg$  & $\pm 6\deg$   &  &  &  &  \\
&&&&&&&&\\
{\bf (834) Burnhamia} & & & &  & & & & \\
$13.87594$   & $ 77\deg$    & $+60\deg$    & $256\deg$     & $+69\deg$     & 0.018  & 2005--2017 & 6 & 32 \\  
$\pm 0.00002$& $\pm 10\deg$  & $\pm 6\deg$ & $\pm 8\deg$  & $\pm 7\deg$   &  &  &  &  \\
&&&&&&&&\\
\hline
\end{tabular}
\end{small}
\end{table*}

In thermophysical modelling, the diameter ($D$) and the thermal inertia
($\Gamma$) are fitted and different surface roughness are tried by
varying the opening angle of hemispherical craters covering 0.6 of the
area of the facets \citep[following ][]{Lagerros1996I}, covering rms
values between 0.2 and $\sim$1. Heat diffusion is 1-D and we use the
Lagerros approximation \citep{Lagerros1996I,Lagerros1998,Mueller1998,Mueller2002}.
The TPM implementation is the one used in \citet{Ali-Lagoa2014}, based
on that of \citet{Delbo2002}, however the colour correction is treated 
differently: instead of colour correcting each facet's flux based on its
effective temperature, we now use the $H$ and $G$ values tabulated in
the JPL (Jet Propulsion Laboratory) Horizons database to compute an effective temperature based on
the heliocentric distance at which each observation was taken
\citep[following the approach in][]{Usui2011}. Another simplifying
assumption is that spectral emissivity is constant and equal to 0.9
\citep[see e.g.][]{Delbo2015}, which does not seem unreasonable given
the small spectral contrast of 1--3\% found by \citet{Licandro2012} in
the 10-$\mu$m features of 50--60~km Themis family members (the 20-$\mu$
emission plateau is expected to be flatter). Once we obtain $D$,
  we compute the visible geometric albedo using the $H$-$G_{12}$ values
  from \citet{Oszkiewicz2011}. The Bond albedo that would be derived
  from these values is sometimes different from the Bond albedo value
  used as input for the TPM ($A_{i}$). This value was obtained by
  first averaging the tabulated diameter from IRAS, AKARI, and WISE 
  (\citealt{Tedesco2005,Usui2011,Ali-Lagoa2018,Mainzer2016}) and
  occultations \citep[compiled by][]{Dunham2016}. Then, we used this
  size to compute the Bond albedo from all available $H$-$G$, $H$-$G_{12}$,
  and/or $H$-$G_1$-$G_2$ values from the Minor Planet Center
  (\citet{Oszkiewicz2011} or \citet{Veres2015}), and took again the
  average value. This approach is somewhat arbitrary but the TPM results
  are not very sensitive to the value of the Bond albedo. Justification
  for this, further details, and discussion, including information on
  how we estimate our error bars, are given in Appendix B.

\section{Results}

In this section for each target separately we refer to previous works containing lightcurves, briefly describe our data, 
the character of brightness variations, and its implications for the spin and shape. Later, the thermal data 
availability and range are described, followed by the results of applying these models in TPM analysis. 
In plots like Fig. \ref{fig:ThLC_100}, the fit of model infrared flux variations to thermal lightcurves is presented, 
and plots like Fig. \ref{fig:100_OMR} in the Appendix show the observation-to-model ratios versus wavelength, 
helicentric distance, rotational angle, and phase angle.

Table \ref{results} presents spin parameters obtained within this work (sidereal period, spin axis position, and used 
lightcurve data). 
Results from thermophysical modelling with our spin and shape models applied are summarised in Table \ref{TPMresults}, 
presenting the diameter, albedo, and thermal inertia for the best-fitting model solution. 
The values obtained here for diameter are ''scaling values'' of the given spin and shape
solutions listed in Table 1 that would otherwise be scale-free. 
For comparison we also refer there to the diameters obtained previously from the AKARI \citep{Usui2011}, IRAS \citep{Tedesco2005}, 
and WISE \citep{Mainzer2011, Masiero2011} surveys. We also added their taxonomic class following \citet{Bus2002a, Bus2002b}
and \citet{Tholen1989}, so that the albedos could be verified for agreement with taxonomy. 
\begin{table*}[t!]
\centering
  \caption{Asteroid diameters from AKARI, IRAS, and WISE ($D_A, D_I, D_W$)
      \citep{Tedesco2005,Usui2011,Mainzer2016},  
      compared to our values from TPM (sixth column). We also provide the
      corresponding visible geometric albedo ($p_V$) as described in
      Sect.~\ref{sec:model}. Finally, we tabulate the nominal thermal inertia ($\Gamma$)
      and its value normalised at 1 AU using the mid-value (Rh) between the
      maximum and minimum heliocentric distances spanned by each object's IR
      data set. 
      To convert $\Gamma$ values to 1 AU we assumed that $\Gamma$ is proportional to Rh$^{-3/4}$ 
      (see the Discussion section). Error bars are 3-$\sigma$.
  } 
\label{TPMresults}
\begin{tabular}{cccccllccc}
\hline
                 &           &          &          &           &  \multicolumn{3}{c}{Radiometric solution for combined data} &      &Thermal inertia\\
Target           &$D_{A}$&$D_{I}$&$D_{W}$& Taxonomic & Diameter & $p_V$ & Thermal inertia                          &  Rh  &at 1 AU \\
                 &   [km]    &   [km]   &  [km]    &   type    &  [km]    &        & [SI units]                               & [AU] &[SI units]  \\
\hline                                                       
&&&&&&&&&\\
100 Hekate       &   88.52   &   88.66  &  91.421  &    S      & 87$^{+5}_{-4}$  & 0.22 $^{+0.03}_{-0.03}$  &  4$^{+66}_{-2}$  & 3.10 &  9$^{+154}_{-5}$\\
&&&&&&&&&\\
109 Felicitas    &   80.81   &   89.44  &  89.000  &    Ch     & 85$^{+7}_{-5}$  & 0.065$^{+0.008}_{-0.01}$ & 40$^{+100}_{-40}$& 3.15 & 95$^{+236}_{-95}$\\
&&&&&&&&&\\
195 Eurykleia    &   89.38   &   85.71  &  80.330  &    Ch     & 87$^{+11}_{-9}$ & 0.06 ${\pm 0.02}$        & 15$^{+55}_{-15}$ & 2.85 & 33$^{+121}_{-33}$\\                                
&&&&&&&&&\\
301 Bavaria      &   51.90   &   54.32  &  55.490  &    C      & 55$^{+2}_{-2}$  & 0.047$^{+0.004}_{-0.003}$& 45$^{+60}_{-30}$ & 2.65 & 94$^{+125}_{-62}$\\
&&&&&&&&&\\
335 Roberta      &   92.12   &   89.07  &  89.703  &    B      & 98$^{+10}_{-11}$& 0.046$^{+0.014}_{-0.008}$& unconstrained    & 2.80 & unc.\\                             
&&&&&&&&&\\
380 Fiducia      &   75.72   &   73.19  &  69.249  &    C      & 72$^{+9}_{-5}$  & 0.057$^{+0.009}_{-0.012}$& 10$^{+140}_{-10}$& 2.50 & 20$^{+278}_{-20}$\\                                 
&&&&&&&&&\\
468 Lina         &   59.80   &   69.34  &  64.592  &   CPF     & 69$^{+11}_{-4}$ & 0.052$^{+0.006}_{-0.014}$& 20$^{+280}_{-20}$& 3.00 & 46$^{+638}_{-46}$\\
&&&&&&&&&\\
538 Friederike   &   72.86   &   72.49  &  79.469  &    C      & 76$^{+5}_{-2}$  & 0.06 ${\pm 0.01}$  & 20$^{+25}_{-20}$ & 3.50 & 50$^{+115}_{-50}$\\
&&&&&&&&&\\
653 Berenike     &   46.91   &   39.22  &  56.894  &    K      & 46$^{+4}_{-2}$  & 0.18 $^{+0.02}_{-0.03}$  & 40$^{+120}_{-40}$& 3.00 & 91$^{+273}_{-91}$\\
&&&&&&&&&\\
673 Edda         &   39.38   &   37.53  &  41.676  &    S      & 38$^{+6}_{-2}$  & 0.13 $^{+0.03}_{-0.05}$  &  3$^{+33}_{-3}$  & 2.82 &  7$^{+72}_{-7}$\\
&&&&&&&&&\\
834 Burnhamia    &   61.44   &   66.65  &  66.151  &   GS:     & 67$^{+8}_{-6}$  & 0.074$^{+0.014}_{-0.016}$& 22$^{+30}_{-20}$ & 2.75 & 47$^{+64}_{-43}$\\
&&&&&&&&&\\
\hline
\end{tabular}
\end{table*}

For practical reasons, the projections of the shape models and the fit to all the visible lightcurves are not presented here, 
but will be available from the DAMIT (Database of Astroid Models from Inversion Techniques)\footnote{http://astro.troja.mff.cuni.cz/projects/asteroids3D.} 
created by \citet{Durech2010}, for models viewing and download, to be used in other applications.  
In Appendix A (Figs. from \ref{100composit2006} to \ref{834composit2017}), 
we present all the previously unpublished photometric data 
in the form of composite lightcurves, as they testify the good quality of our spin and shape solutions: 
photometric accuracy here is mostly at the level of a few millimagnitudes, which is one to two  orders of magnitude better 
than in sparse data standardly used for asteroid modelling. Also, with such slow rotation, 
separate lightcurves would not show the whole character of the brightness variability, but only a small fraction 
of the full lightcurve, which is visible only in the folded plots. Moreover, such plots enable us to see 
lightcurve evolution caused by phase angle effect, which due to various levels of shadowing 
highlight various shape features. Since we applied no photometric phase correction, 
their signatures are visible on the overlapping fragments 
that cover the same rotational phase, but are spaced by more than a month in time
like for example in Fig. \ref{301composit2018}. 
Apparitions with only one lightcurve fragment or those with data covering only a small fraction of full rotation, 
are not presented here, because creating composite lightcurves would be either impossible 
or would not present much information on the period or the lightcurve appearance. 
However, such data have been used in the modelling, helping to constrain spin and shape. 
The full list of observing runs is summarised in Table \ref{obs} in Appendix A.

\subsection{(100) Hekate}

We compiled archival lightcurves from four apparitions, published by \citet{Tedesco1979}, \citet{Gil-Hutton1990}, 
\citet{Hainaut-Rouelle1995}, and \citet{Galad2009}, and complemented them with data from the SuperWASP
survey (Wide Angle Search for Planets, Grice et al. 2017), 
and our own data from three more apparitions 
(see Figs. \ref{100composit2006} to \ref{100composit2018} in the Appendix). 
SuperWASP cameras are known to suffer from the detector having a temperature dependence, so the absolute lightcurves can be 
systematically shifted in magnitude depending on the temperature of the detector. However, as we treat them as relative lightcurves
this should not be an issue.
In the last apparition, Hekate was observed for our project 
by the Kepler Space Telescope, resulting in two continuous, four-day-long lightcurves of great quality. 
The K2 data have been reduced with the fitsh package, using the same methods that were already applied to targeted observations 
of Trojan asteroids and chance observations of main-belt asteroids (MBAs) 
in the mission \citep{Pal2012, Szabo2016, Szabo2017, Molnar2018}.
Overall, Hekate displayed interesting, asymmetric lightcurves of amplitudes ranging from $0.11$ to $0.23$ magnitude 
and a long, 27.07 hours period.

The spin parameters of our model are presented in Table \ref{results}. To construct it, initial scanning of 
parameter space needed an increased number of trial poles and iteration steps. Overall both spin 
solutions fit the visual lightcurves at a very good level of 0.012 mag. 


\subsubsection*{TPM analysis}

The two mirror pole solutions of Hekate are named AM 1 and AM 2. We used 52 infrared observations,
32 from IRAS (8 x 12 $\mu$m, 8 x 25 $\mu$m, 8 x 60 $\mu$m, 8 x 100 $\mu$m),
five from AKARI (2 x S9W, 3 x L18W),
and 15 from WISE (W4). We assumed $A=0.090$ for the Bond albedo. 

\begin{figure}
  \centering
  \includegraphics[width=0.80\linewidth]{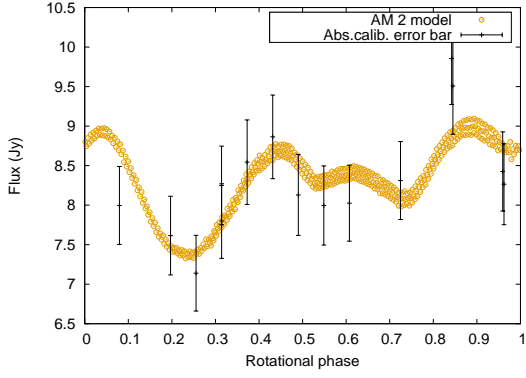}  
  \caption{
    (100) Hekate W4 data and model thermal lightcurves for shape model 2.
    Data error bars are 1-$\sigma$. Table~\ref{tab:tpm_100} summarises the TPM
    analysis.  
  } \label{fig:ThLC_100} 
\end{figure}

Both models provide formally acceptable fits, as do the corresponding spheres
(see Table~\ref{tab:tpm_100} and the $\chi^2$ plots in Fig. \ref{fig:100_OMR} in Appendix B). 
The fit to the WISE ``lightcurve'' seems
reasonable for both models, although not around the zero rotational phases shown
in Figs.~\ref{fig:ThLC_100}, \ref{fig:ThLC_100model1}, and \ref{fig:100_OMR} (third panel from the top). 
The IRAS data present a slope in the observation-to-model 
ratios (OMR) versus wavelength plot, which is shown in
the top panel of Fig.~\ref{fig:100_OMR}. 
Model 2 provides a better fit than 1 and the spheres so we select it as our
provisionally favoured solution. Nonetheless, these results could be biased by
our neglecting the dependence of thermal conductivity -- and therefore
thermal inertia -- with temperature (see Rozitis et al. 2018, and the review in Delbo et al. 2015, 
for instance), over the wide range of heliocentric distances sampled
(2.6 to 3.6 au). However, there is not enough data taken at long
heliocentric distances to make a conclusive statement. 

To summarise the TPM analysis: the surface roughness is not constrained, the diameter is 87$^{+5}_{-4}$\,km, and
geometric albedo, $p_V =$\,0.22$^{+0.03}_{-0.03}$. We would benefit from additional thermal light
curves at negative phase angles (i.e. after opposition) and higher heliocentric distances (to study
the conductivity dependence with temperatures). Low thermal inertias of  
approximately 5 SI units\footnote{[Jm$^{-2}$s$^{-0.5}$K$^{-1}$]}
and low to medium roughness fit the data better. 

\subsection{(109) Felicitas}
 Available photometric data for Felicitas came from the works of \citet{Zappala1983}, \citet{Harris1989}, 
 and from the SuperWASP archive (Grice et al. 2017). 
 We added to these two apparitions data from five more obtained within our campaign, and also data from the WISE 
 satellite obtained in W1 band, which is dominated by reflected light (thermal contribution in W1 for this target 
 was estimated at 8\% - 30\%). The latter dataset greatly helped to constrain the model, 
 in an approach first proposed by \citet{Durech2018b}.
 Felicitas lightcurves have been changing substantially between the apparitions, reaching peak-to-peak amplitudes 
 from $0.06$ to $0.22$ mag, with the synodic period around 13.194 hours 
 (Figs. \ref{109composit2015}, \ref{109composit2017}, and \ref{109composit2018} in the Appendix).


\subsection*{TPM analysis}

The two mirror solutions for Felicitas from Table \ref{results} are denoted AM 1 and AM 2. 
In the thermal approach we used 38 observations, 
15 from IRAS (5 x 12 $\mu$m, 7 x 25 $\mu$m, 3 x 60 $\mu$m), 
five from AKARI (3 x S9W, 2 x L18W), 
and 18 from WISE (9 x W3, 9 x W4). We assumed $A=0.025$ for the Bond albedo. 

\begin{figure}
  \centering
  \includegraphics[width=0.80\linewidth]{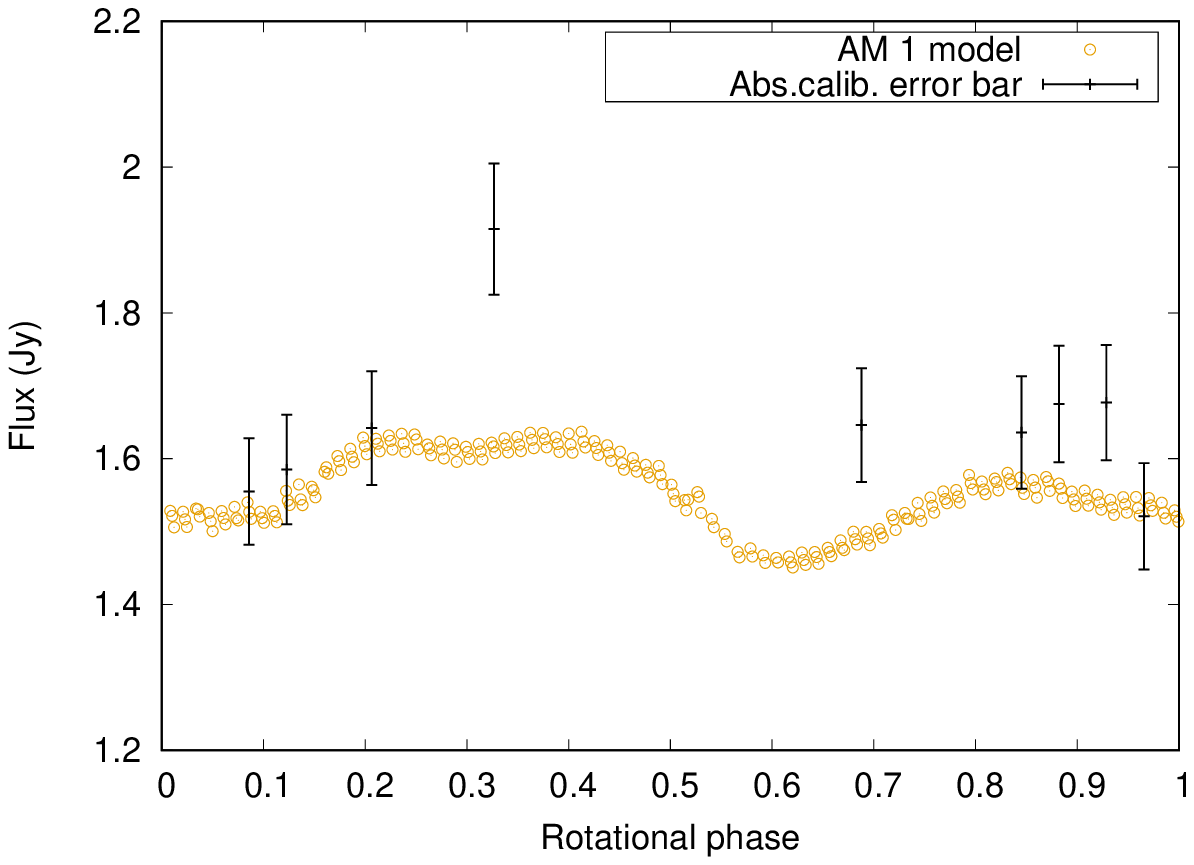}
  
  \includegraphics[width=0.80\linewidth]{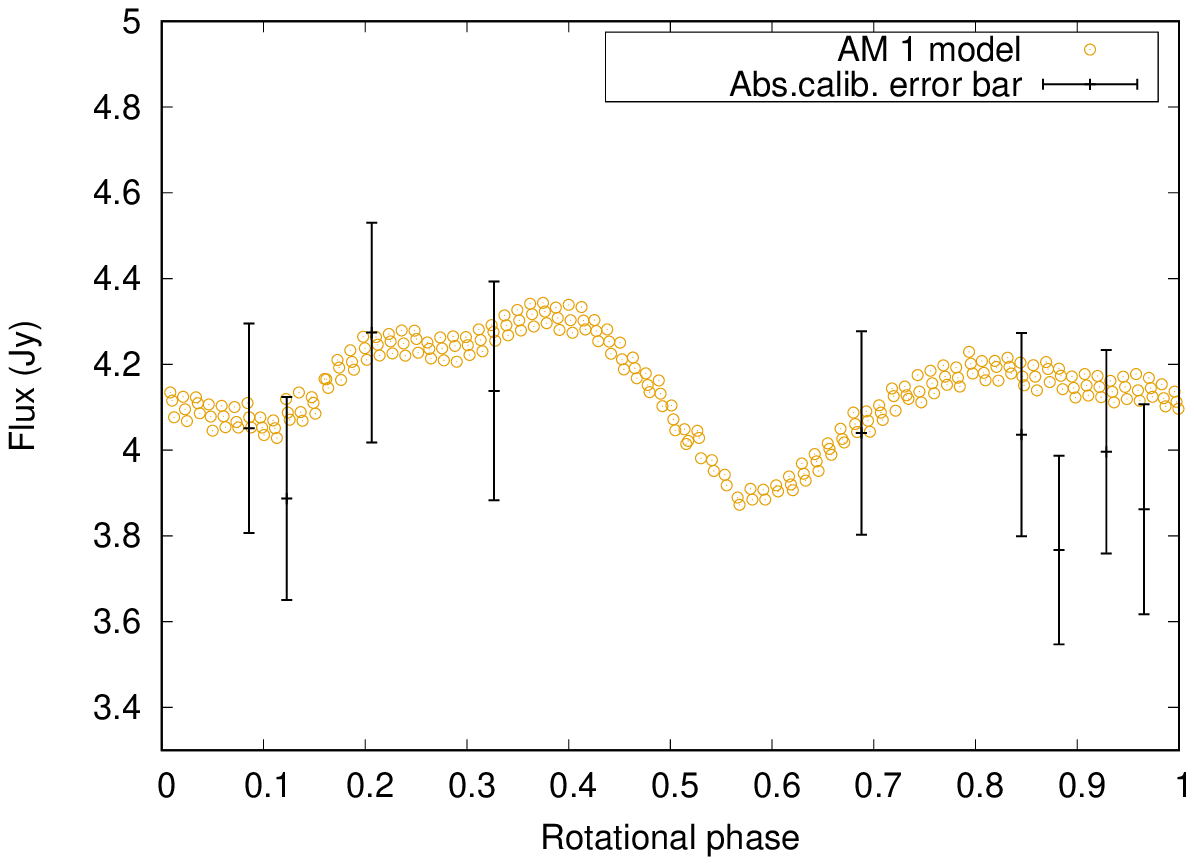}  

  \caption{
    Asteroid (109) Felicitas W3 (top) and W4 (bottom) data and model thermal light
    curves for shape model AM 1 (see Table~\ref{tab:tpm_109}). 
  } \label{fig:ThLC_109} 
\end{figure}

Unlike the AM 2 model, AM 1 provides a formally acceptable fit
(Table~\ref{tab:tpm_109}, and the $\chi^2$ plot in Fig. \ref{fig:109_OMR} in Appendix B), 
although the WISE bands are not fitted equally well:
the model overestimates the W4 data and underestimates the W3 (top panel in
Fig~\ref{fig:109_OMR}). Even though the model thermal lightcurves miss some W3
and W4 fluxes (Fig.~\ref{fig:ThLC_109}), we consider it as a preliminary
approximated solution given that it fits significantly better than the sphere.
Additional thermal curves could help confirm or reject this model and data at
negative phase angles could improve the constraints for the thermal inertia. 

As usual, the surface roughness is not constrained. From the diameter 
(85$^{+7}_{-5}$\,km), we obtain the albedo $p_V =$\,0.065$^{+0.008}_{-0.01}$.

\subsection{(195) Eurykleia}
 There was no prior publication presenting lightcurves of Eurykleia, but plots from three 
apparitions are available on the observers' web pages\footnote{http://obswww.unige.ch/$\sim$behrend/page\_cou.html, 
http://eoni.com/$\sim$garlitzj/Period.htm.}, indicating a long, 16.52 hours period. 
Here we publish all of these data, 
adding more recent observations from four more apparitions. During all the seven apparitions, Eurykleia displayed 
similarly shaped lightcurves with one minimum sharper than the other, which was accompanied by an additional bump or shelf
(see Figs. \ref{195composit2005} to \ref{195composit2017} in the Appendix). The amplitude was stable, being always around $0.24$ mag, 
which indicates a stable aspect angle due to the high inclination of the spin axis. 
As expected, the resulting model has a high value of $|\beta|$, and a formally large range of possible values of $\lambda$ 
(see Table \ref{results}). 

  
\subsection*{TPM analysis}  
The two mirror solutions from lightcurve inversion are labelled AM 1 and AM 2. We used a total of 57 infrared 
observations from IRAS (nine epochs x four filters), AKARI (4 S9W + 5 L18W) and WISE
W4 (12). The WISE W3 data were saturated for this object and the W4 data, with
reported magnitudes between 0.00 and 0.35, are near the -0.6 mag identified as
the onset of saturation for individual W4 images \citep{Cutri2012}. We assumed
$A=0.020$ for the Bond albedo.

Both shape models provide statistically similarly good fits to all the data, 
so the TPM cannot reject any of the mirror solutions in this case. Using all
the data to optimise the $\chi^2$ (i.e. considering $\nu=57-2=55$ degrees of
freedom), the minimum reduced chi-squared ($\bar{\chi}_m^2$) were 0.51 and
0.60 for the AM 1 and AM 2 models, respectively. These are significantly
better than the values obtained for spheres with the same respective spin
pole orientation (see Table~\ref{tab:tpm}). 

Although the thermal data seems to be fitted very well based on our small 
$\bar{\chi}^2_m$, the OMR of many IRAS data are
slightly but systematically above 1, the WISE OMRs slightly below 1, and the
AKARI ones are not fully horizontally aligned (Fig.~\ref{fig:195_OMR} in Appendix B,
upper panels). Figures~\ref{fig:W4ThLC} and \ref{fig:W4ThLCmodel2} in Appendix B  
show that both shape models fit the variation
of the W4 thermal lightcurve reasonably well. However, the fluxes predicted by the
TPM solution that best fits all the data (yellow open circles) systematically
overestimate most of the W4 observations. This is worse for the solution that
best fits the AKARI and the IRAS data (open triangles), because the fitted
diameter is larger in this case (third line in Table~\ref{tab:tpm}, first and third plot from 
the top in Fig. \ref{chi2_195}). However,
the OMR values for the IRAS and AKARI data do align horizontally
(Fig.~\ref{fig:195_OMR} in Appendix B, lower panels). Finally, we can fit the lightcurve
much better if we optimise the W4 data alone (filled squares), but the thermal
inertia is lower by a factor of three (fourth entry in Table~\ref{tab:tpm}, second and fourth 
plot in Fig. \ref{chi2_195}).

\begin{figure}
  \centering
  \includegraphics[width=0.80\linewidth]{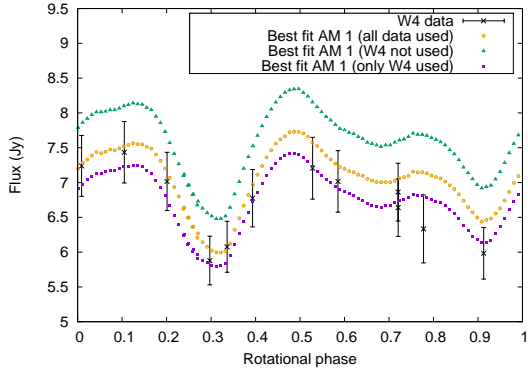}
  \caption{W4 data and model thermal lightcurves for (195) Eurykleia's shape
    model AM 1. The different models resulted from fitting different subsets of
    data. Table~\ref{tab:tpm} contains the corresponding thermo-physical
    parameters. 
  } \label{fig:W4ThLC} 
\end{figure}


Conclusions from TPM analysis are the following: 
both shapes seem to fit the data well, including the shape of the W4 thermal 
lightcurve (Fig.~\ref{fig:W4ThLC}), although the flux level of the W4 data is
systematically below our fitted model's fluxes. We do not have an
  explanation for such a systematic mismatch, but perhaps additional thermal IR
data could correct the TPM diameter of (195) Eurykleia to a higher value
closer to 90~km, or a lower value closer to 80 km. Thus, with
the current dataset, the possible diameters range from 78 to 98 km depending on
which subsets of data we fit (Table~\ref{tab:tpm}). As a compromise, we take
the mid-value in this range, $87^{+11}_{-9}$~km (3$\sigma$ level error bars,
$\sim$13\% relative error). 

On the other hand, the large $3\sigma$ range of possible thermal inertias does
not change when we fit different subsets of data, but the error bars are widely
asymmetric. We can only provide an upper limit of 70 SIu, with a 
best-fitting value of $\Gamma =$15~SIu. Surface roughness is not constrained
at the 3$\sigma$ level, but high roughness models fit the data better, as is
usually the case for main-belt asteroids. 

Finally, the geometric albedos derived from our radiometric diameters vary
slightly depending on the photometric phase correction applied and the source
of the absolute magnitudes and slope parameters. To compute our final value, we
use the $H$-$G_{12}$ values from the \citet{Oszkiewicz2011} table and calculate
the maximum and minimum possible $p_V$s using all the three sources listed above.
We find $p_V=0.06\pm0.02$.

\subsection{(301) Bavaria}

 The lightcurve from only one previous apparition of asteroid Bavaria has been reported in the literature 
\citep{Warner2004}. Our campaign built an extensive lightcurve dataset for this target, covering four viewing 
aspects and a wide range of phase angles. Our data covered sometimes four to five months within one apparition, and 
less than a year passed between consecutive observed apparitions. Such an observing strategy was defined 
as the most optimal for spin and shape reconstructions by \citet{Slivan2012}, and is confirmed by our experience. 
Bavaria during its 12.24-hours-long rotation displayed wide minima and amplitudes 
from 0.25 to 0.36 magnitudes (Figs. \ref{301composit2014} - \ref{301composit2018}  in the Appendix), 
exceeding our initial target selection criteria in the course of the campaign; 
nonetheless it was retained in the target list.
Although the solution for the sidereal period and two possible spin axis positions 
are very well defined, due to the high value of pole latitude (see Table \ref{results})  
the shape model vertical extent is poorly constrained.


\subsection*{TPM analysis}
The two mirror solutions from Table \ref{results} are AM 1 and AM 2. We used 36 thermal observations,
18 from IRAS (9 x 25 $\mu$m, 9 x 60 $\mu$m), 
six from AKARI (3 x S9W, 3 x L18W), 
and 12 from WISE (W4). We rejected IRAS 12- and 100-micron data because they
contain clear outliers (by a factor of several). 
We assumed $A=0.020$ for the Bond albedo. 

\begin{figure}
  \centering
  \includegraphics[width=0.80\linewidth]{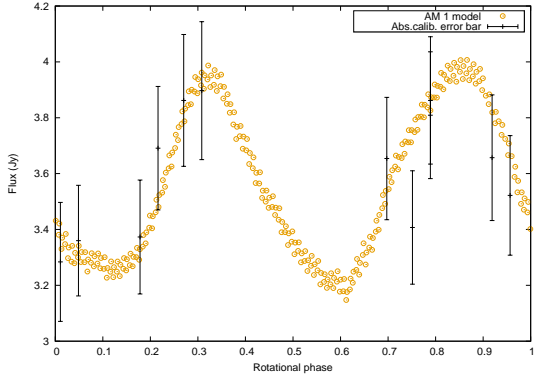}
  \caption{
    Asteroid (301) Bavaria AM 1 model thermal lightcurves and W4 data
    (see Table~\ref{tab:tpm_301}). 
  } \label{fig:ThLC_301} 
\end{figure}

Both models and the corresponding spheres provide a good fit (see Table
\ref{tab:tpm_301}, and the $\chi^2$ plots in Fig. \ref{fig:301_OMR} in Appendix B), 
although with slightly offset diameters and different
roughness (the spheres require low roughness rather than medium). The AM 1 OMR
plots are shown in Fig.~\ref{fig:301_OMR}, and the W4 model thermal lightcurves
with the data in Figs.~\ref{fig:ThLC_301} and \ref{fig:ThLC_301model2}.

As usual, the roughness is not constrained at the 3$\sigma$ level.
With a diameter 55$^{+2}_{-2}$\,km, we get $p_V =$\,0.047$^{+0.004}_{-0.003}$ and a
thermal inertia between 10 and 100 SI units.

\subsection{(335) Roberta}
We compiled a large dataset covering as many as nine apparitions of asteroid
Roberta, including data from \citet{Binzel1987}, \citet{Harris1992}, \citet{Lagerkvist1995}, \citet{Warner2007}, \citet{Pilcher2015}, 
SuperWASP archive (Grice et al. 2017), and our own data (presented in Figs. \ref{335composit2013}, 
\ref{335composit2015}, and \ref{335composit2017} in the Appendix).
This target showed rather rare, monomodal lightcurves of amplitudes from $0.13$ to $0.19$ mag, 
depending on the aspect. Its rotation period, 12.027 hours, commensurate with an Earth day, 
required observations from sites well spaced in longitude for full coverage.
Thanks to such an extensive dataset both spin and shape solutions are well constrained (Table \ref{results}), 
however different from those found by \citet{Blanco2000}, who reported four pole solutions, all 
much lower in $|\beta|$ than ours, and a sidereal period longer by 0.027 hours, which is a substantial difference.


\subsection*{TPM analysis}
The two mirror solutions are AM 1 and AM 2. Thermal data consisted of 44 observations,
22 from IRAS (6 x 12 $\mu$m, 6 x 25 $\mu$m, 6 x 60 $\mu$m, 4 x 100 $\mu$m),
nine from AKARI (4 x S9W and 5 x L18W),
and 13 from WISE (W4). We assumed $A=0.020$ for the Bond albedo. 

\begin{figure}
  \centering
  \includegraphics[width=0.80\linewidth]{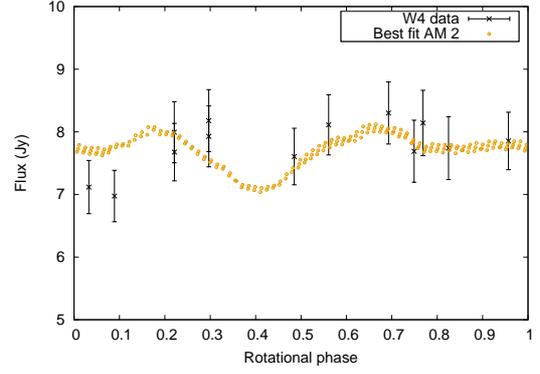} 
  \caption{
    Asteroid (335) Roberta's WISE data and model thermal lightcurves for shape model 
    2 and the corresponding best-fitting solutions (very low thermal
    inertia of 15 SIu). 
  } \label{fig:335_ThLC} 
\end{figure}

The best solutions for both models are comparably good, and even the spheres
provide a formally good fit with compatible results (see Table~\ref{tab:tpm_335}, 
and the $\chi^2$,  
and OMR plots in Fig.~\ref{fig:335_OMR} in Appendix B). The diameter is constrained at the 3$\sigma$-level
but with a relatively large relative error of $\sim$10\%. The fit to the WISE
lightcurve (Figs.~\ref{fig:335_ThLC} and \ref{fig:335_ThLCmodel1}) shows that the fit at phases between 0.0
and 0.10 might be improved with additional visible data for the inversion
(phase 0 corresponds to JD 2442489.835789).


The problem with this model is that the \chisq  \ versus $\Gamma$ curve is very
shallow at high $\Gamma$s, so the thermal inertia is basically unconstrained.
The aspect angles sampled by the thermal data are concentrated around equatorial
values, so additional data at other sub-observer latitudes and a dense thermal
lightcurve at pre-opposition (positive phase angle) would help. The diameter
constraint of 98$^{+10}_{-11}$\,km leads to an albedo of $p_V$ =
0.046$^{+0.014}_{-0.008}$.

\subsection{(380) Fiducia}
A previous lightcurve of Fiducia was published by \citet{Warner2004}. With this work we add  
six apparitions, confirming the synodic period around 13.72 hours. Fiducia's lightcurve shape varied substantially
from one apparition to another, reaching 0.32 mag amplitude in one, and then decreasing down to 0.04 in the other 
(see Figs. \ref{380composit2008} - \ref{380composit2018} in the Appendix). In the year 2014
Fiducia was observed exclusively on the 0.8m TJO (Telescopi Joan Or{\'o})
telescope in the Montsec Observatory in 
a mode similar to 'dense-sparse cadence' described by \citet{Warner2011}. 
In spite of the relative sparseness of datapoints, the lightcurve's general character is clearly outlined, 
and the synodic period from other apparitions is confirmed (Fig. \ref{380composit2014}).

In the apparition on the verge of the years 2015 and 2016, Fiducia was observed under 
an exceptionally large range of phase angles (Table \ref{obs}), which resulted in distinctive differences 
in the lightcurve character. Thus we present data from that apparition on two separate plots, one for small, 
and the other for large phase angles (Appendix, Figs. \ref{380composit2015small_phase_angle} and 
\ref{380composit2015large_phase_angle}, respectively).

The unique solution for the sidereal period could only be found in this case during denser scanning of parameter space. 
In spite of clear signatures of low inclination of the spin axis, the resulting pole solution 
is located at rather moderate latitudes (see Table \ref{results}). This might be a symptom of the lightcurve 
inversion method bias against low poles found by \citet{Cibulkova2016}, as small shape modifications can 
compensate for the shifted spin axis position of the model. 


\subsection*{TPM analysis}
The two obtained mirror solutions are AM 1 and AM 2. We had 54 thermal observations at our disposal,
38 from IRAS (10 x 12 $\mu$m, 10 x 25 $\mu$m, 10 x 60 $\mu$m, 8 x 100 $\mu$m),
seven from AKARI (5 x S9W and 2 x L18W),
and nine from WISE (W4). We assumed $A=0.020$ for the Bond albedo. 

\begin{figure}
  \centering
  \includegraphics[width=0.80\linewidth]{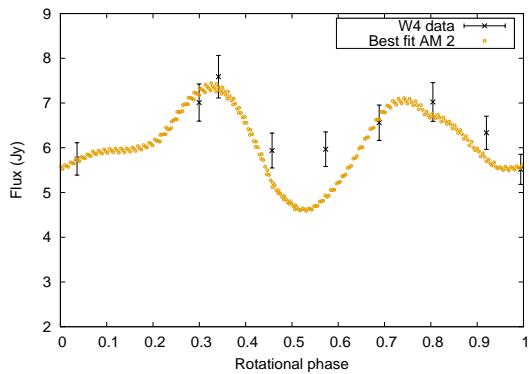} 
  \caption{
    Asteroid (380) Fiducia's WISE data and model thermal lightcurves for shape model 
    2 and the corresponding best-fitting solutions (very low thermal inertia of
    10 SIu; see Table \ref{tab:tpm_380})). 
  } \label{fig:ThLC_380} 
\end{figure}

The best solutions for both models have comparable \chisq and very shallow
minima in the \chisq \ versus $\Gamma$ plots. Thermal inertias between 0 and 150
SIu fit the data (3$\sigma$ limits), but the best fit is
$\Gamma$=10 SI units (see Table~\ref{tab:tpm_380}, the $\chi^2$ plot, 
and OMR plot in Fig.~\ref{fig:380_OMR} in Appendix B).
The corresponding spheres give similar diameters but fit the data better with
higher thermal inertia ($\Gamma\approx$200 SIu). Only the most extreme roughness
values produce fits outside the 3$\sigma$ range, and low roughness solutions
(rms$<$0.30) fit the data better. The shapes reasonably reproduce the W4 lightcurve 
(Fig.~\ref{fig:ThLC_380}, and \ref{fig:ThLC_380model1}), but the fits could probably be improved with
``less boxy'' shape models. 


From TPM analysis one can conclude that the roughness is unconstrained, but thermal inertias are lower than 150 at the
3$\sigma$ level (the best fit is for $\Gamma = 10$\,SIu). The diameter range
of 72$^{+9}_{-5}$\,km leads to an albedo $p_V=0.057^{+0.009}_{-0.012}$. 

\subsection{(468) Lina}
Lightcurves from two previous apparitions of Lina have been published by \citet{Tedesco1979} and 
\citet{Buchheim2007}. Our visual observations of this target spanned six apparitions, and we found it 
displaying complex lightcurves, with wide, wavy maxima and narrow minima 
(Figs. \ref{468composit2006} to \ref{468composit2016} in the Appendix). 
Peak-to-peak amplitudes ranged between $0.13$ and $0.18$ mag within its 16.48 hours period.
Model spin parameters are given in Table \ref{results}. Shape model stretch along the spin axis is somewhat 
uncertain. 


\subsection*{TPM analysis}
The two mirror solutions are named AM 1 and AM 2. We used 33 observations,
seven from IRAS (2 x 12 $\mu$m, 2 x 25 $\mu$m, 2 x 60 $\mu$m, 1 x 100 $\mu$m),
eight from AKARI (2 x S9W and 6 x L18W),
and 20 from WISE (W4), and assumed $A=0.020$ for the Bond albedo.

\begin{figure}
  \centering
    \includegraphics[width=0.80\linewidth]{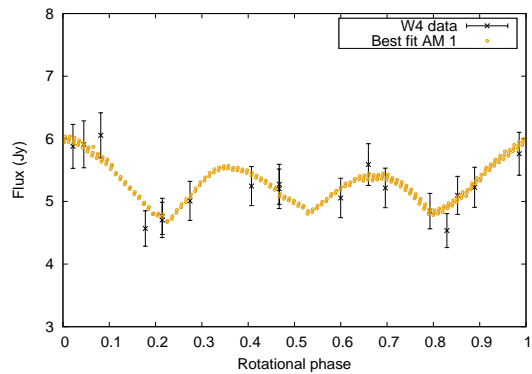}  
  \caption{
    Asteroid (468) Lina's WISE data and model thermal lightcurves for shape model 
    1's corresponding best-fitting solutions (see Table  \ref{tab:tpm_468}).
  } \label{fig:ThLC_468} 
\end{figure}

The comparatively fewer data points are available for this object. The best-fitting
roughness and thermal inertias are low (rms $\sim$ 0.3 and $\Gamma$=20 SIu;
see Table~\ref{tab:tpm_468}, and the $\chi^2$ plots in Fig. \ref{fig:468_OMR} in Appendix B), 
even for the spherical models. The fit to the
WISE lightcurve is reasonably good (Fig.~\ref{fig:ThLC_468}, and \ref{fig:ThLC_468model2}), but the \chisq is
1.20, which is not optimal. The data sample widely different heliocentric
distances, from 2.5 to 3.5 au (Fig.~\ref{fig:468_OMR}), but the residuals do not
present a strong trend.

The analysis points to thermal inertias lower than 300 SIu and low roughness,
but this object requires more thermal data to constrain these properties better.
Even the diameter has a relatively large error bar by TPM standards:
69$^{+11}_{-4}$\,km. The corresponding albedo is 0.052$^{+0.006}_{0.014}$.

\subsection{(538) Friederike}

Data previously published came from one apparition and clearly displayed a very long, 46.728 hours period for this target 
\citep{Pilcher2013}. 
In addition to it we used previously unpublished data from the years 2003 and 2006, with only partial coverage though 
(Table \ref{obs}), and own data from five more apparitions. Those with good phase coverage are shown 
in Figs. \ref{538composit2014} - \ref{538composit2018} in the Appendix. Lightcurves of Friederike had minima of different widths and amplitudes 
from $0.20$ to $0.25$ mag.

In spite of a large and varied dataset, only with the addition of WISE W1 data could the unique solution for period, spin axis, and shape be found. These data proved to be a necessary, continuous basis lasting a few tens of hours, while other 
lightcurves covered only at most 20\% of the full rotation. Such extremely long period targets are especially 
challenging for ground-based studies, being better targets for wide-field, space-borne observatories. 
Due to this long period and relatively short time span of the observational dataset, the precision of the sidereal period determination 
is lower than in the case of other targets (Table \ref{results}).

\subsection*{TPM analysis}

 The two mirror spin-shape solutions are AM 1 (168$\deg$,-58$\deg$, 46.73928 h)
  and AM 2 (328$\deg$,-59$\deg$, 46.73985 h). Friederike is well observed at
  thermal wavelengths: two four-band IRAS measurements, ten measurements by
  AKARI (4 x S9W, 6 x L18W), taken at different epochs before and after
  opposition, and 23 WISE data points (at W3 and W4), also at two
  epochs before and after opposition.

  Solution AM 1 provides a borderline acceptable fit to all thermal data simultaneously,
  as do the corresponding spheres (see Table \ref{tab:tpm_538}). However, the AM 2 solution seems
  to match the WISE W3 and W4 lightcurves a bit better. The residuals in
  the observation-divided-by-model plots (see Fig. \ref{fig:538_OMR}) indicate that the
  spin-shape solutions are not perfect or -- alternatively -- that there are
  surface variegations that influence the thermal fluxes. The WISE W3 and
  W4 lightcurves favour AM 2, but do not help to settle the spin ambiguity
  completely. 
  Overall, in this case high and extremely high surface roughness worked
  very well and all the radiometric solutions point to low values for
  the thermal inertia well below $\Gamma$ = 20 SIunits (see $\chi^2$ plots in Fig. \ref{fig:538_OMR}), 
  with a trend to higher inertias at shorter heliocentric distance (2.6 au) and lower values
  at the largest heliocentric distance ($>$ 3.5 au). The heliocentric
  influence on thermal inertia is also visible in the radiometric
  solutions for the individual datasets (see Table \ref{tab:tpm_538}): all WISE
  data are taken at r$_{helio}$ $>$ 3.4 au, while the AKARI and IRAS
  data are taken well below 3.0 au.
The overall best radiometric solution for AM 2 and extremely high surface
  roughness produces $\Gamma$ = 20 SIunits, a diameter of 76$^{+4}_{-2}$
  , and a geometric albedo of 0.06$\pm$0.01.
\begin{figure}
  \centering
  
  \includegraphics[width=0.80\linewidth]{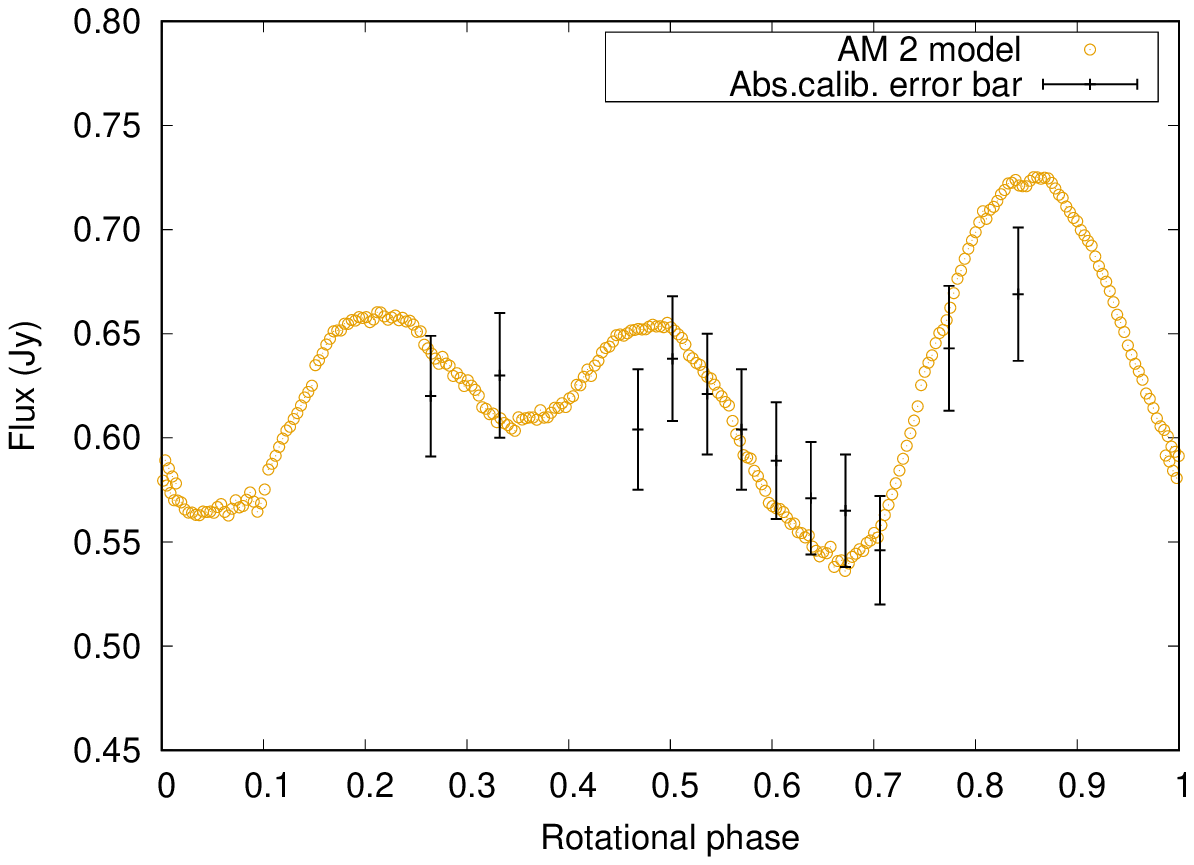}
  
  \includegraphics[width=0.80\linewidth]{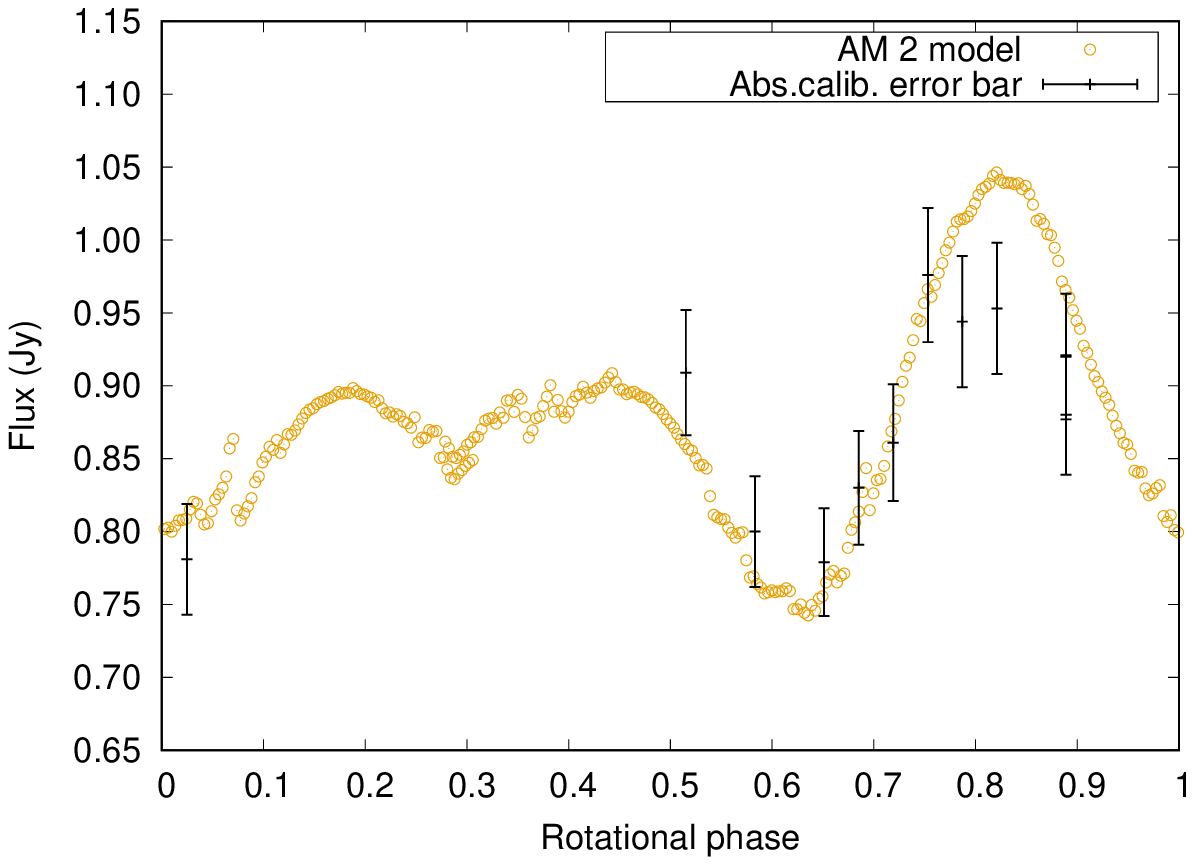}

  \includegraphics[width=0.80\linewidth]{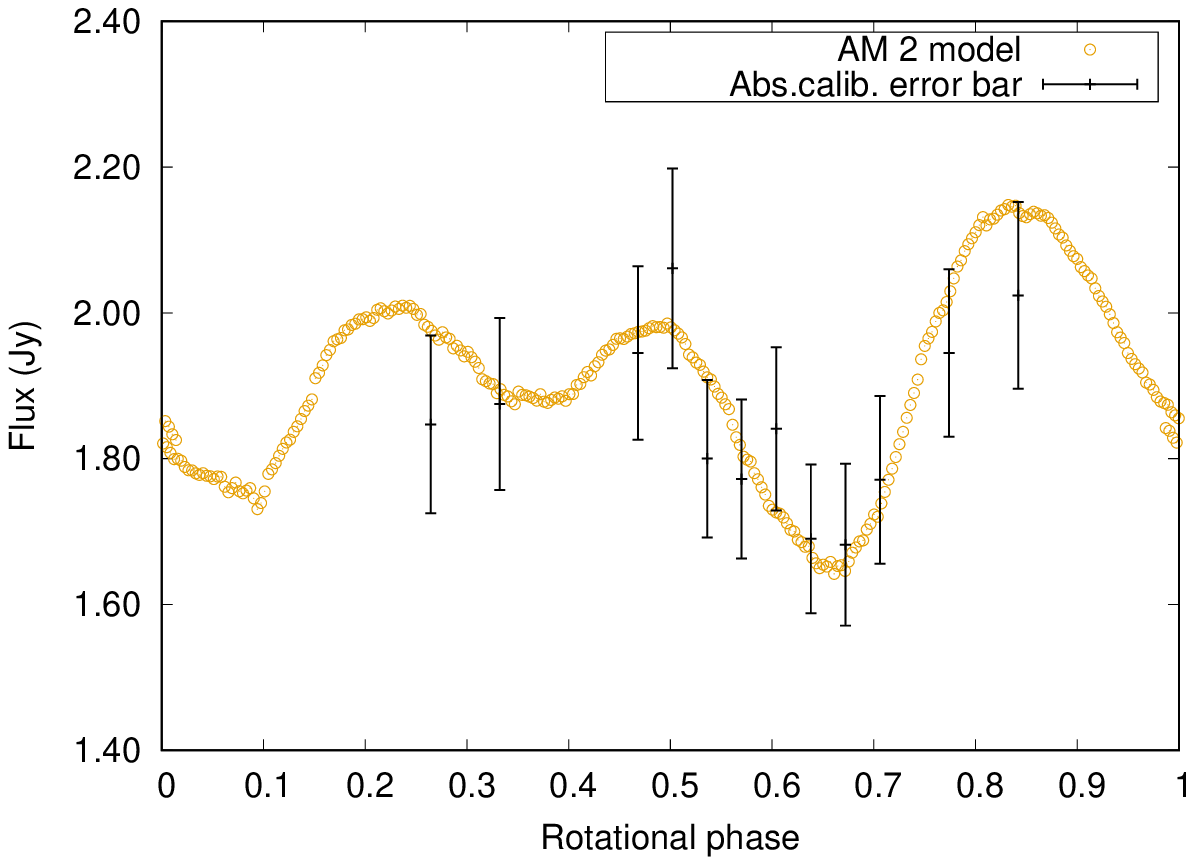}  

  \includegraphics[width=0.80\linewidth]{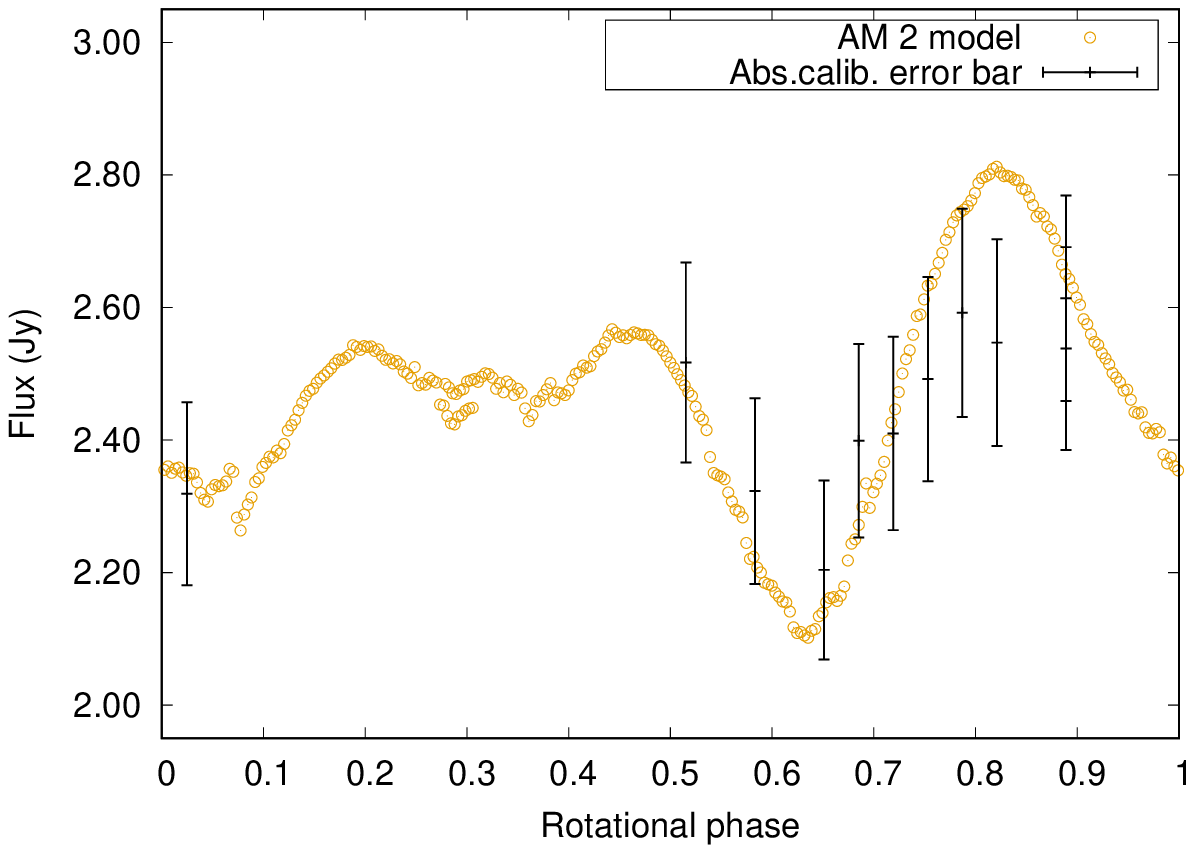} 

  \caption{
    Asteroid (538) Friederike's WISE W3 (first two plots from the top) and W4 data (last two plots) taken at two epochs, 
    and thermal lightcurves for shape model
     AM 2 and the corresponding best-fitting solutions obtained based on all thermal data. 
  } \label{fig:ThLC_538} 
\end{figure}

\subsection{(653) Berenike}

Berenike displayed small amplitudes in the range of 0.05 - 0.16 (Figs. \ref{653composit2005} - \ref{653composit2016} in the Appendix), 
also in the archival observations by \citet{Binzel1987} and \citet{Galad2008}.
However, in the last apparition observed in this work the amplitude unexpectedly rose to 0.38 mag, 
being over two times larger than ever, which is confirmed by two independent observing runs (Fig. \ref{653composit2018}). 
That apparition must have provided the only viewing geometry when signatures of full elongation of this target could be visible. 
This case shows the importance of probing a wide range of geometries for correct reproduction of the spin axis position 
and shape elongation. The synodic period was around 12.485 hours, and the full dataset spanned seven apparitions. 

The shape model is indeed elongated in the equatorial dimensions, and this time the pole position is low, 
as expected (see Table \ref{results}). Still, the shape model extent along the spin axis is poorly constrained, 
probably due to the lack of data from geometries where intermediate amplitudes could be observed.


\subsection*{TPM analysis}
We denote the two mirror solutions as AM 1 and AM 2. We used 49 thermal observations,
24 from IRAS (8 x 12 $\mu$m, 8 x 25 $\mu$m, 8 x 60 $\mu$m), 
eight from AKARI (4 x S9W, 4 x L18W), 
and 17 from WISE (8 x W3, 9 x W4). 
The Bond albedo was assumed to be $A=0.070$. 

\begin{figure}
  \centering
  \includegraphics[width=0.80\linewidth]{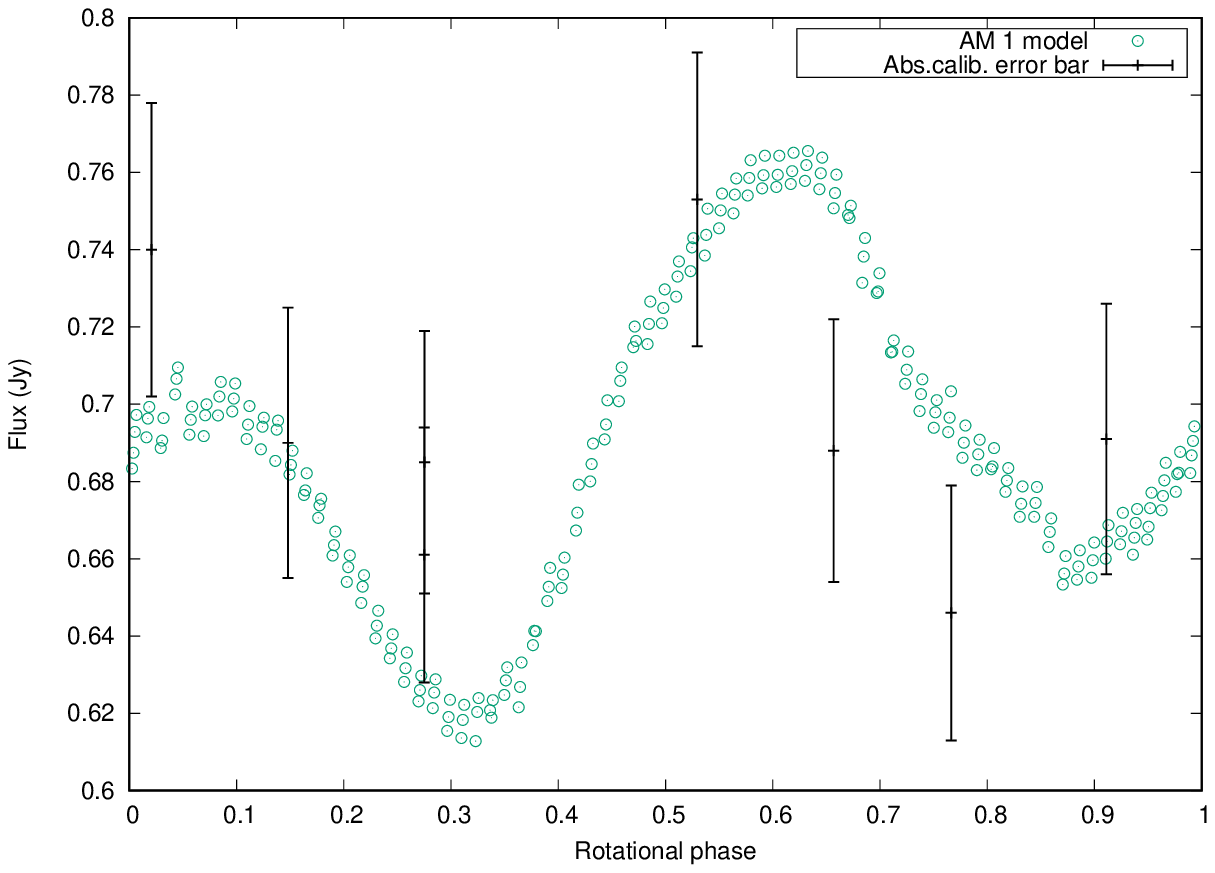}
  
  \includegraphics[width=0.80\linewidth]{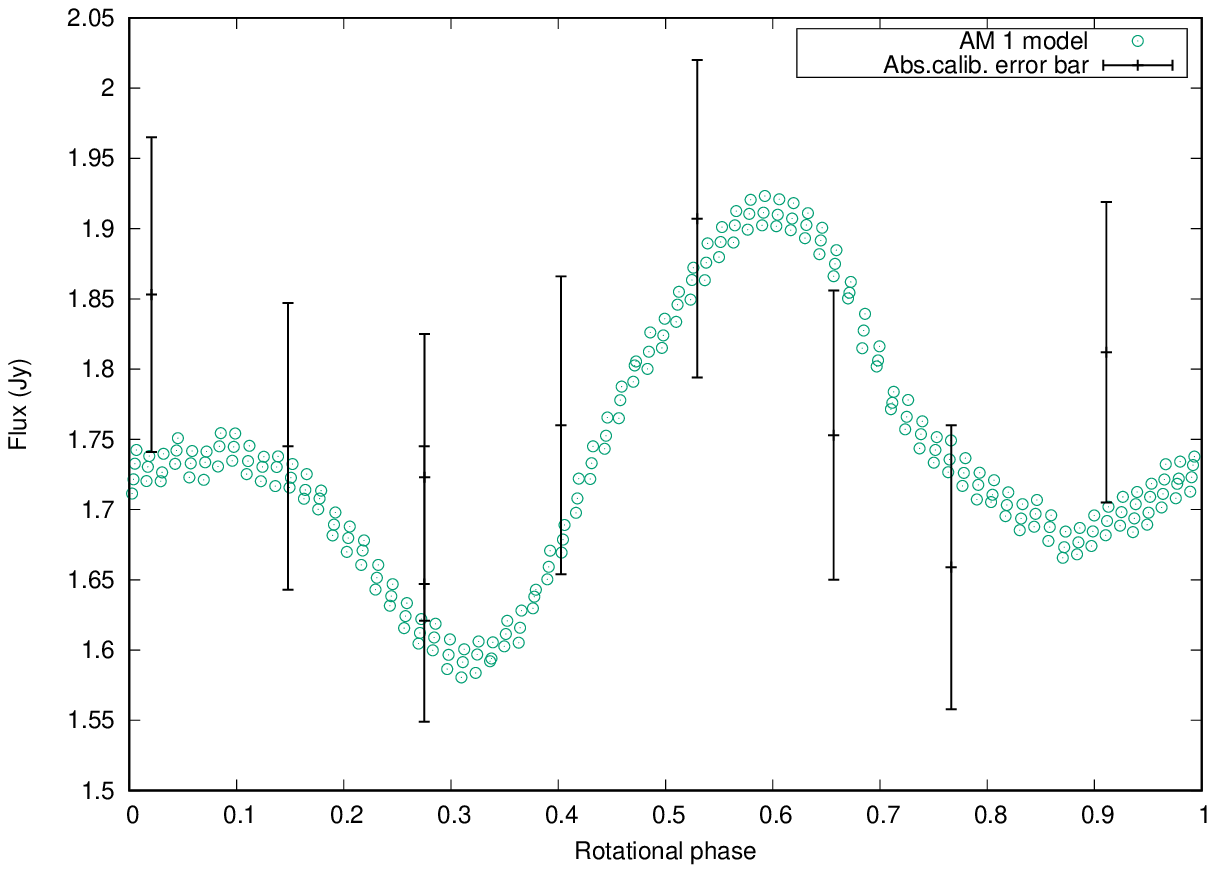}  

  \caption{
    Asteroid (653) Berenike WISE data (W3 top, W4 bottom) and the AM 1 model's thermal
    lightcurves (see Table~\ref{tab:tpm_653}). 
  } \label{fig:ThLC_653} 
\end{figure}

Model AM 1 is significantly better than AM 2 and the spheres in this case,
although the reduced $\chi^2$ is 1.1, slightly over unity
(Table \ref{tab:tpm_653}, and the $\chi^2$ plot in Fig. \ref{fig:653_OMR} in  Appendix B). 
The IRAS data residuals show some scatter and 
the WISE data are reasonably well fitted but still present some waviness in
the OMR versus rotational phase plot (Figs.~\ref{fig:ThLC_653} and 
\ref{fig:653_OMR}).
The latter could indicate that there is room for improvement of the shape model.
Very high roughness solutions fit the data better, and an rms lower than 0.3 can
be rejected at the 3$\sigma$ level.


We conclude that additional densely sampled thermal lightcurves and an improved shape model
could improve the constraints on the thermal inertia and surface roughness. 
With a diameter 46$^{+4}_{-2}$\,km, we get $p_V =$\,0.18$^{+0.02}_{-0.03}$,
and the best fit is obtained for medium values of thermal inertia. 

\subsection{(673) Edda}
The first published lightcurve of Edda was obtained within our project \citep{Marciniak2016}, 
and was highly asymmetric, with maxima unequally spaced in time. 
Here we present data from five more apparitions, confirming the period of 22.34 hours, and 
non-typical, asymmetric lightcurve behaviour (see Figs. \ref{673composit2005} to \ref{673composit2017} in the Appendix).
This dataset was complemented by one more apparition with partial coverage by data from the SuperWASP archive 
(Grice et al. 2017). lightcurve amplitudes ranged from 0.13 to 0.23 mag.
The shape model from lightcurve inversion is somewhat angular, and the fit to some of the lightcurves 
of smallest amplitude is imperfect. Both spin solutions are well constrained and are given in Table \ref{results}.


\subsection*{TPM analysis}
The two mirror solutions are named AM 1 and AM 2. At our disposal there were 54 infrared observations,
20 from IRAS (7 x 12 $\mu$m, 7 x 25 $\mu$m, 6 x 60 $\mu$m), six from AKARI (3 x S9W and
3 x L18W), and 28 WISE (14 x W3 and 14 x W4). We assumed $A=0.047$ for the Bond albedo.

\begin{figure}
  \centering
  \includegraphics[width=0.80\linewidth]{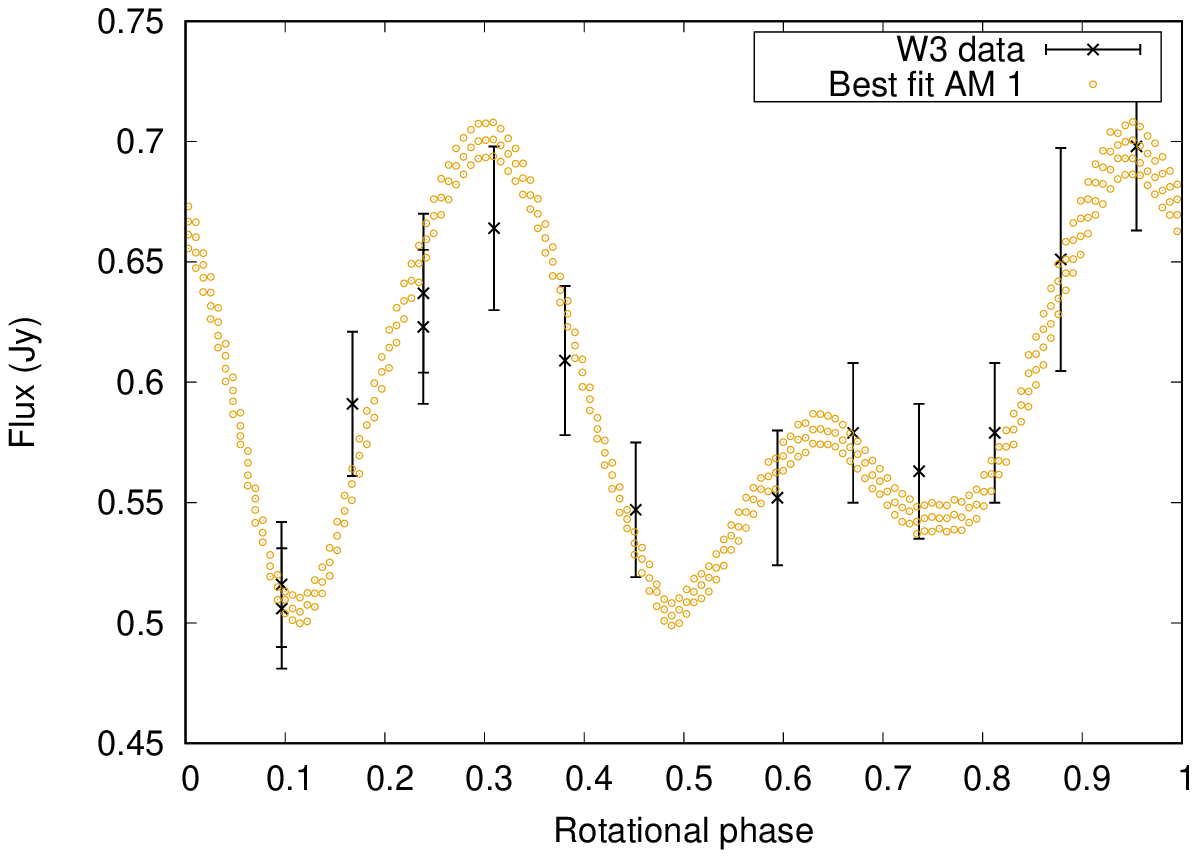}

  \includegraphics[width=0.80\linewidth]{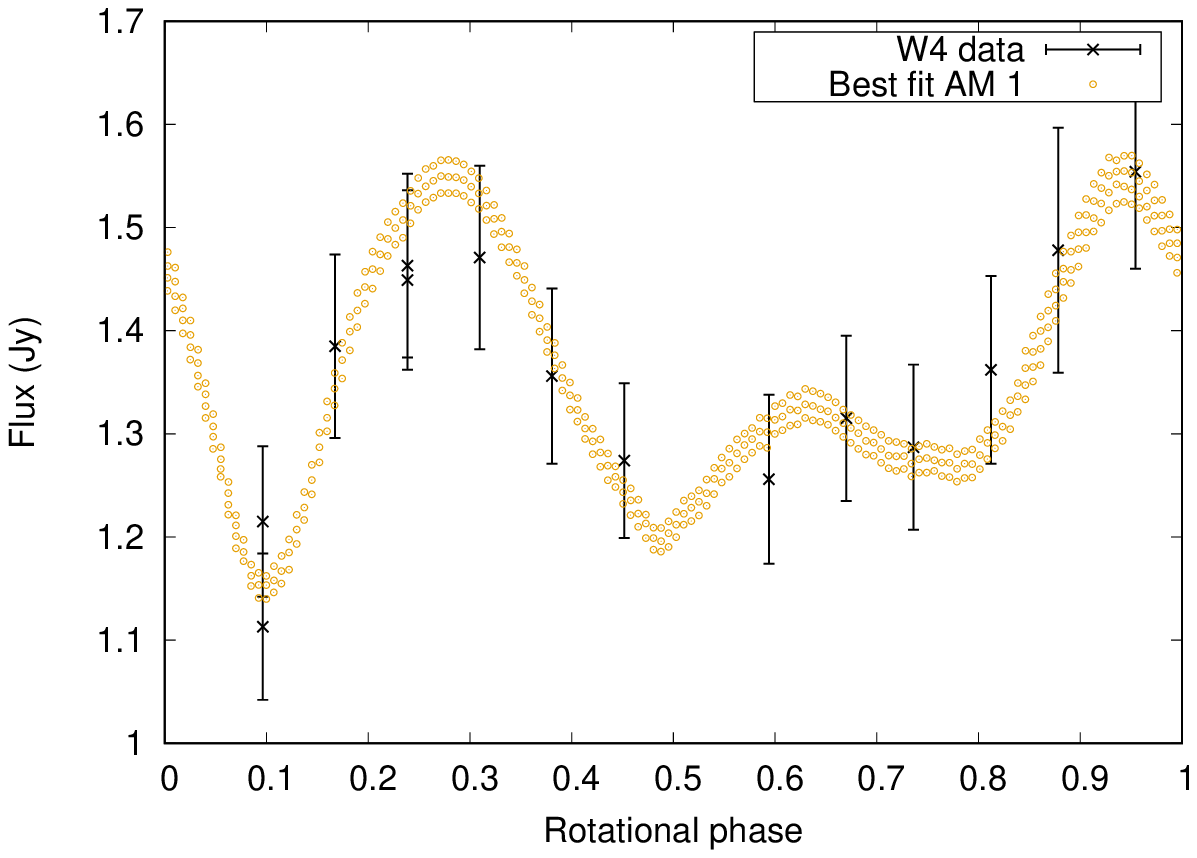}  

  \caption{
    Asteroid (673) Edda's WISE data and model thermal lightcurves for shape model 1's
    best-fitting solution (very low thermal inertia of 3 SI units).
  } \label{fig:ThLC_673} 
\end{figure}

The best solutions for both models have comparable and very low $\bar\chi^2_m$
with very low thermal inertia of about 3 SI units (see Table~\ref{tab:tpm_673}
and the $\chi^2$ plots in Fig. \ref{fig:673_OMR} in the Appendix B). 
The low $\bar{\chi}^2_m$s suggest the error bars might be overestimated, so we
normalise the $\bar\chi^2$ curves to have the minimum at 1 in order to compute
the uncertainties of the parameters (see discussion in \citealt{Hanus2015}). 
The shapes fit the WISE data well (Fig.~\ref{fig:ThLC_673} and \ref{fig:ThLC_673model2}), and unlike (195)
Eurykleia's case, the best-fitting solution also fits the AKARI data with a
reasonably flat OMR versus wavelength plot (Fig.~\ref{fig:673_OMR}). 


To conclude, although high thermal inertias of about 70 SI units are still
allowed, the best solutions seem to point to very low thermal inertias of
around 3 SI units. The diameters are constrained to be 38$^{+6}_{-2}$~km
(3$\sigma$ error bars), which lead to a visible geometric albedo of
$p_V=0.13^{+0.03}_{-0.05}$. This value is somewhat low for a typical S-type
(classified based on a visible SMASS spectrum, 
Small Main-Belt Asteroid Spectroscopic Survey \citealt{Bus2002a,Bus2002b}).   

\subsection{(834) Burnhamia}

\begin{figure}
  \centering
  \includegraphics[width=0.80\linewidth]{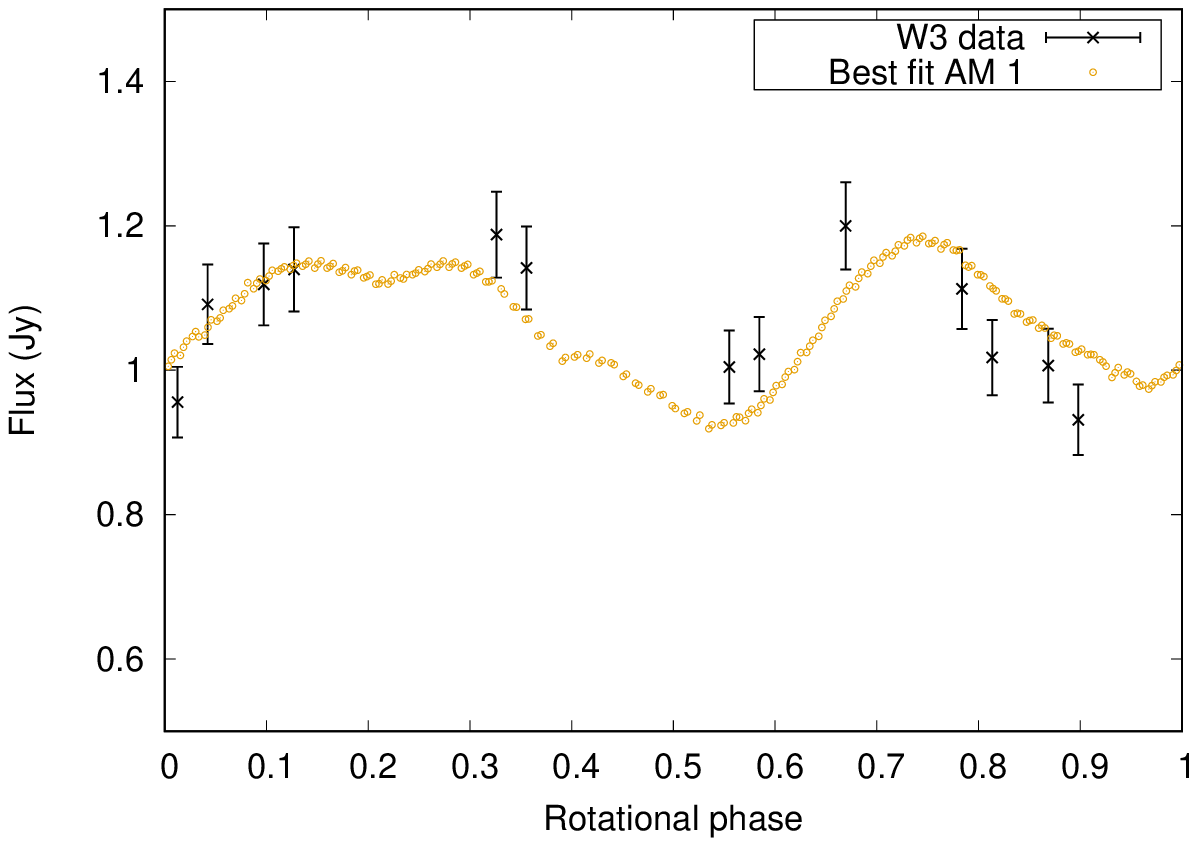}  

  \includegraphics[width=0.80\linewidth]{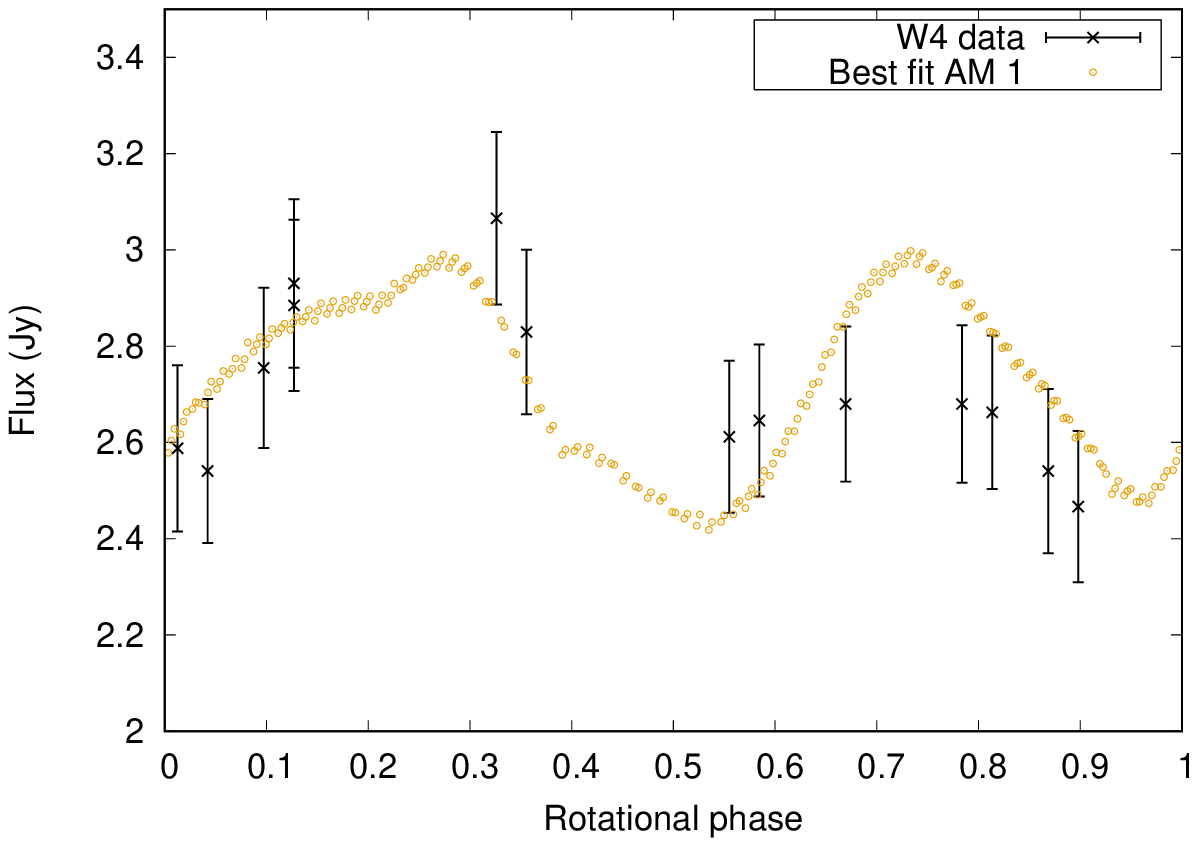}  

  \caption{
    Asteroid (834) Burnhamia's WISE data and model thermal lightcurves for shape model
    1's best-fitting solution (see Table \ref{tab:tpm_834}). 
  } \label{fig:ThLC_834} 
\end{figure}

Previously observed by \citet{Buchheim2007} and in our campaign, 
Burnhamia displayed lightcurves of extrema at unequal levels, 
13.87 hour period, and amplitudes from $0.15$ to $0.25$ mag. 
We accumulated data from six apparitions well spread in longitude, and present them in 
Figs. \ref{834composit2005} to \ref{834composit2017} in the Appendix. 
In the lightcurve inversion we obtained two well-defined pole solutions (Table \ref{results}) with somewhat unconstrained 
shape models both in the vertical dimension and in the level of smoothness of the shape. 


\subsection*{TPM analysis}
The two mirror solutions are denoted AM 1 and AM 2. We used 70 thermal observations,
42 from IRAS (11 x 12 $\mu$m, 11 x 25 $\mu$m, 11 x 60 $\mu$m, 9 x 100 $\mu$m),
one from AKARI (L18W),
and 26 from WISE (13 x W3, 13 x W4). We assumed $A=0.035$ for the Bond albedo. 

It turned out to be yet another slow rotator with seemingly low thermal inertia. The three-sigma upper
limit on 50 SI is quite clear. The \chisq is $\sim$ 0.8 for both models
(Table \ref{tab:tpm_834}, and the $\chi^2$ plots in Fig. \ref{fig:834_OMR} in  Appendix B), 
so the fit is good but there is room for
improvement. The WISE data are reasonably well fitted (Fig.\ref{fig:ThLC_834}, and \ref{fig:ThLC_834model2}),
but the 100-micron  IRAS data are not (these data carry less weigh for the fit,
however). 


As usual, the roughness is not constrained, the diameter is 67$^{+8}_{-6}$\,km and
$p_V =$\,0.074$^{+0.014}_{-0.016}$. We could benefit from additional thermal light
curves at positive phase angles (pre-opposition) and lower heliocentric
distance  although there is no apparent trend in the lowest panel of
Fig. \ref{fig:834_OMR}.

\section{Discussion and conclusions}

We present here 11 new spin and shape models of slowly rotating asteroids from
lightcurve inversion on dense lightcurves, with determined sizes, albedos, and
thermal parameters. Together with five models from our previous study \citep{Marciniak2018}, 
we have 16 shape models within our sample of slow-rotators with applied
  TPM. This number substantially enlarges the available pool of well-studied
slow rotators. Models obtained in this work provide good resolution, ``smooth''
shape representations for a multitude of future applications, including
  calibration in the infrared. 

\begin{figure}
\vspace{1cm}
  \centering
    \includegraphics[width=\linewidth]{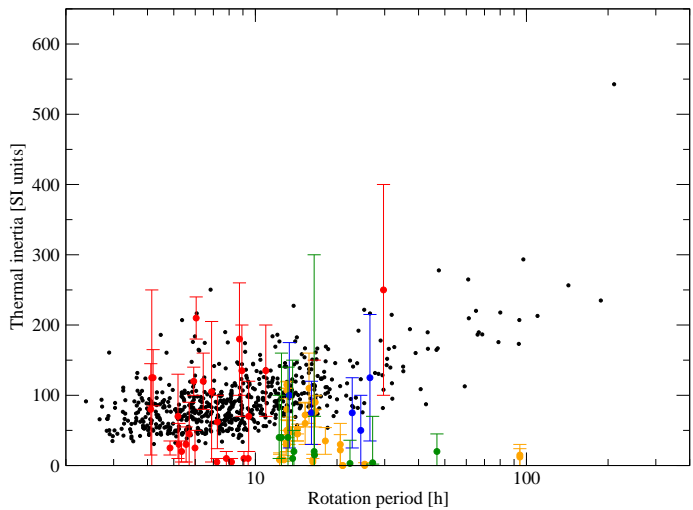}
  \caption{
Updated plot from \citet{Harris2016} of thermal inertia dependence on rotation period.
Black dots are thermal inertias estimated from $\eta$ values by \citet{Harris2016}, 
red dots from detailed TPM compiled by \citet{Delbo2015}, 
orange are slow rotators from ``Varied-Shape TPM'' by \citet{Hanus2018b},  
blue are from TPM results of \citet{Marciniak2018}, and green dots from this work. 
Our new results do not confirm the growing trend.
  } \label{HD}
\end{figure}

One of our aims was to verify whether the thermal inertia indeed increases with
the rotation period \citep{Harris2016}. Our initial sample 
\citep{Marciniak2018} seemed to be in line with this finding, while our current
sample is dominated by targets with lower thermal inertias. To ensure
this different result is not related to the fact that we used a different code
for this work\footnote{We used the TPM code of \citet{Mueller2002} in
  \citet{Marciniak2018}, whereas here we used the TPM code of \citet{Delbo2002}
  modified in \citet{Ali-Lagoa2014} mentioned in Sect.~\ref{sec:model}.}, we
modelled one of the targets from \citet{Marciniak2018} and reproduced the
results. This is expected given that we use the same model approach and
approximations. 
We also cross-checked our results with those estimated by
  \citet{Harris2016} and found one overlapping target, 487 Venetia, with
  consistent thermal inertia values: 96 SIu in \citet{Harris2016}, and 100
  $\pm$ 75 SIu in \citet{Marciniak2018}. 
In Fig. \ref{HD} we superimposed our results on the plot from 
\citet{Harris2016}. 

  Taking both of our samples together \citep[from this work and from][]{Marciniak2018},
  we find diverse values of thermal inertia, ranging from 3 up to
  125 SI units. To make a quantitative comparison, we collected the thermal
  inertias of all targets with sizes 30 < D < 200 km from \citet{Hanus2018b},
  \citet{Delbo2015}, and our values and split them into two groups: slow
  ($P>12$ hrs) and fast ($P<12$ hrs) rotators. Neither the two-sample
  Kolmogorov-Smirnov test nor the k-sample Anderson-Darling test rule out the
  null hypothesis that both samples (with 33 and 36 values, respectively) are
  drawn from the same distribution. The p-values are higher than 0.3, far even
  from a lax threshold of statistical significance of 5\%. Perhaps a larger
  sample could lead to a firmer conclusion in the future. 

This does not necessarily deny the hypothesis of $\Gamma$ growth with period;
instead it might indicate various levels of regolith development on the surface,
connected to age and/or collisional history, and space weathering.
  This could be further tested if we had sufficiently large samples of objects
  belonging to young and old collisional families, since younger surfaces are
  expected to have less developed regoliths \citep[e.g.][]{Delbo2014}. The
  interpretation of our results might also be complicated by the temperature
  dependence of the thermal conductivity, 
 which we account for by assuming the scaling relation between $\Gamma$ 
 and heliocentric distance given for example in Eq. 13 of \citet{Delbo2015} 
 to normalise the values of $\Gamma$ to 1 au (these values are given in the last column 
 of Table \ref{TPMresults}). 
 This scaling is related to the $T^3$ dependence of the thermal conductivity 
 \citep[e.g.][]{Vasavada1999}, and translates into a thermal inertia dependence 
 of $\Gamma \propto T^{3/2}$.  However, we note that Centaurs and TNOs (trans-Neptunian objects)
 follow a different 
 behaviour \citep[$\Gamma \propto T^2$;][]{Lellouch2013}, and \citet{Rozitis2018} 
 have found more extreme dependencies in two near-Earth asteroids ($\Gamma \propto T^{4.4}$ 
 and $\Gamma \propto T^{2.92}$), although the fitted exponents have large error bars. 
 More work is certainly needed in this direction.

Our large observing campaign, targeted at about 100 asteroids with slow rotation and small amplitudes, 
already resulted in the gathering of around 10,000 hours of photometric data, where hundreds of hours are needed 
for a unique lightcurve inversion solution for each of such targets. 
There are a few factors further limiting the number of targets with all the parameters uniquely determined, 
like shape model imperfections, or thermal data limited geometries. 
In the future we are planning to add more photometric data to our datasets accumulated already including 
photometry from the TESS (Transiting Exoplanet Survey Satellite) 
mission, which should improve the uniqueness of many spin and shape solutions for targets not presented here. 
The TESS mission is observing in a similar cadence to the Kepler Space Telescope. Moreover, unlike in the case of Kepler, 
 full frames are going to be downlinked, so TESS data are going to be perfect for extensive studies of slow rotators 
 \citep{Pal2018}.
It is also possible that the shape models that best fit visual lightcurves were not the best possible ones 
from the thermal point of view, 
so the varied-shape TPM method \citep{Hanus2015}, or simultaneous fitting of visual and thermal data 
using a convex inversion TPM \citep{Durech2017} could help to resolve issues with unconstrained solutions.
To overcome problems with limited thermal data we are planning 
proposals to VLT/VISIR (Very Large Telescope with the VLT spectrometer and imager for the mid-infrared) 
and SOFIA (Stratospheric Observatory For Infrared Astronomy) 
infrared facilities, carefully choosing targets and 
observing geometries to best complement the existing datasets in terms of aspect, pre- and post-opposition 
geometry, and heliocentric distance.

\begin{acknowledgements}
       This work was supported by the National Science Centre, Poland, through grant no. 2014/13/D/ST9/01818. 
       The research leading to these results has received funding from the European Union's
       Horizon 2020 Research and Innovation Programme, under Grant Agreement no 687378 (SBNAF).
       The research of VK was supported by a grant from the Slovak Research and Development Agency, number APVV-15-0458.
       R. Duffard acknowledges financial support from the State Agency for Research of the Spanish MCIU through the 
       ''Center of Excellence Severo Ochoa'' award for the Instituto de Astrof{\'i}sica de Andaluc{\'i}a(SEV-2017-0709). 
       The Joan Or{\'o} Telescope (TJO) of the Montsec Astronomical Observatory (OAdM) 
       is owned by the Catalan Government and operated by the Institute for Space Studies of Catalonia (IEEC).
       This article is based on observations made in the Observatorios de Canarias del IAC with the 
       0.82m IAC80 telescope operated on the island of Tenerife by the Instituto de Astrof{\'i}sica de Canarias (IAC) 
       in the Observatorio del Teide. 
This article is based on observations made with the SARA telescopes (Southeastern Association for Research in Astronomy),
whose nodes are located at the Observatorios de Canarias del IAC on the island of La Palma in the Observatorio 
del Roque de los Muchachos; Kitt Peak, AZ under the auspices of the National Optical Astronomy Observatory (NOAO);  
and Cerro Tololo Inter-American Observatory (CTIO) in La Serena, Chile.
This project uses data from the SuperWASP archive. The WASP project is currently funded and operated by Warwick University 
and Keele University, and was originally set up by Queen's University Belfast, the Universities of Keele, St. Andrews, 
and Leicester, the Open University, the Isaac Newton Group, the Instituto de Astrofisica de Canarias, 
the South African Astronomical Observatory, and by STFC.

Funding for the Kepler and K2 missions is
provided by the NASA Science Mission Directorate.
The data presented in this paper were
obtained from the Mikulski Archive for Space
Telescopes (MAST). STScI is operated by the
Association of Universities for Research in Astronomy,
Inc., under NASA contract NAS5-
26555. Support for MAST for non-HST data is
provided by the NASA Office of Space Science
via grant NNX09AF08G and by other grants
and contracts.
\end{acknowledgements}

\bibliographystyle{aa}
\bibliography{bibliography}

\begin{appendix}
\clearpage
\section{Visible photometry}
Details of the observing runs (Table \ref{obs}) 
and composite lightcurves of asteroids with spin and shape models presented here 
(Figs. \ref{100composit2006} - \ref{834composit2017}).
\begin{table}[h!]
\begin{scriptsize}
\noindent 
\begin{tabularx}{\textwidth}{llp{15mm}lp{10mm}p{40mm}l}
\hline \hline
 Date & $\lambda$ & Phase angle & Duration  & $\sigma$ & Observer & Site \\
      &   [deg]   &   [deg]     & [hours]   &  [mag]   &          &      \\
\hline
&&&&&&\\                      
\multicolumn{7}{l}{{\bf (100) Hekate}}\\
&&&&&&\\                      
\hline
&&&&&&\\                      
 2006 Nov 27.2 & 118.8 & 13.1 & 5.3 & 0.030 & - & SuperWASP \\
 2006 Nov 29.2 & 118.7 & 12.7 & 5.5 & 0.026 & - & SuperWASP \\
 2006 Nov 30.2 & 118.6 & 12.5 & 5.3 & 0.023 & - & SuperWASP \\
 2006 Dec 14.2 & 117.2 &  9.3 & 5.7 & 0.021 & - & SuperWASP \\
 2006 Dec 15.2 & 117.1 &  9.0 & 4.9 & 0.019 & - & SuperWASP \\
 2006 Dec 16.2 & 117.0 &  8.7 & 5.8 & 0.018 & - & SuperWASP \\
 2006 Dec 23.1 & 115.9 &  6.7 & 4.8 & 0.016 & - & SuperWASP \\
 2006 Dec 30.0 & 114.7 &  4.6 & 4.1 & 0.032 & - & SuperWASP \\
&&&&&&\\
 2015 Jul 11.0  & 314.5 & 10.0 & 0.5 & 0.004 & S. Fauvaud                & Bardon, France\\
 2015 Jul 12.1  & 314.3 &  9.6 & 3.3 & 0.004 & S. Fauvaud                & Bardon, France\\
 2015 Jul 13.1  & 314.2 &  9.2 & 2.9 & 0.005 & S. Fauvaud                & Bardon, France\\
 2015 Jul 14.1  & 314.0 &  8.8 & 3.9 & 0.003 & S. Fauvaud                & Bardon, France\\
 2015 Jul 15.1  & 313.9 &  8.4 & 3.4 & 0.003 & S. Fauvaud                & Bardon, France\\
 2015 Jul 16.0  & 313.7 &  8.0 & 2.9 & 0.010 & W. Og{\l}oza              & Suhora, Poland\\
 2015 Jul 16.0  & 313.7 &  8.0 & 0.4 & 0.006 & J. J. Sanabria            & Torreaguila, Spain\\
 2015 Jul 16.1  & 313.7 &  8.0 & 3.5 & 0.004 & S. Fauvaud                & Bardon, France\\
 2015 Jul 21.0  & 312.9 &  5.9 & 3.3 & 0.006 & W. Og{\l}oza              & Suhora, Poland\\
 2015 Aug  7.1  & 309.6 &  1.9 & 3.7 & 0.004 & S. Fauvaud                & Bardon, France\\
 2015 Sep  1.9  & 305.6 & 12.6 & 4.3 & 0.005 & W. Og{\l}oza              & Suhora, Poland\\
 2015 Sep 14.9  & 304.8 & 16.6 & 1.9 & 0.011 & -                         & Montsec, CAT, Spain\\
 2015 Sep 19.9  & 304.8 & 17.8 & 3.5 & 0.008 & -                         & Montsec, CAT, Spain\\
 2015 Sep 20.9  & 304.8 & 18.1 & 3.6 & 0.012 & -                         & Montsec, CAT, Spain\\
&&&&&&\\
 2016 Oct  3.1  &  63.1 & 15.2 & 4.5 & 0.003 & S. Fauvaud                & Bardon, France\\
 2016 Oct  7.0  &  62.9 & 14.3 & 4.7 & 0.004 & S. Fauvaud                & Bardon, France\\
 2016 Oct  8.0  &  62.8 & 14.1 & 2.2 & 0.004 & S. Fauvaud                & Bardon, France\\
 2016 Oct 28.1  &  60.3 &  8.3 & 7.9 & 0.004 & S. Fauvaud                & Bardon, France\\
 2016 Oct 30.2  &  60.0 &  7.6 & 2.5 & 0.004 & S. Fauvaud                & Bardon, France\\
 2016 Oct 31.0  &  59.8 &  7.4 & 3.5 & 0.005 & S. Fauvaud                & Bardon, France\\
 2016 Nov  2.3  &  59.4 &  6.5 & 9.3 & 0.006 & T. Polakis, B. Skiff      & Tempe, AZ, USA\\
 2016 Nov  5.4  &  58.8 &  5.5 & 8.6 & 0.005 & T. Polakis, B. Skiff      & Tempe, AZ, USA\\
 2016 Nov  6.3  &  58.6 &  5.2 & 9.0 & 0.005 & T. Polakis, B. Skiff      & Tempe, AZ, USA\\
 2016 Nov  7.3  &  58.4 &  4.9 & 9.0 & 0.006 & T. Polakis, B. Skiff      & Tempe, AZ, USA\\
 2016 Nov  8.3  &  58.2 &  4.6 & 7.5 & 0.006 & T. Polakis, B. Skiff      & Tempe, AZ, USA\\
 2016 Nov  9.3  &  58.0 &  4.3 & 8.7 & 0.005 & T. Polakis, B. Skiff      & Tempe, AZ, USA\\
 2016 Dec  4.2  &  52.9 &  6.5 & 8.0 & 0.005 & T. Polakis, B. Skiff      & Tempe, AZ, USA\\
 2016 Dec  5.2  &  52.7 &  6.8 & 7.9 & 0.007 & T. Polakis, B. Skiff      & Tempe, AZ, USA\\
 2017 Jan 10.8  &  49.7 & 15.5 & 4.5 & 0.013 & A. Marciniak              & Borowiec, Poland\\
&&&&&&\\
 2018 Feb  1.1  & 135.2 & 14.5 &18.1 & 0.0004 & -        & Kepler Space Telescope\\
 2018 Feb  2.1  & 135.2 & 14.4 &27.5 & 0.0004 & -        & Kepler Space Telescope\\
 2018 Feb  3.3  & 135.2 & 14.2 & 8.8 & 0.0004 & -        & Kepler Space Telescope\\
 2018 Feb  4.0  & 135.2 & 14.1 &23.5 & 0.0004 & -        & Kepler Space Telescope\\
 2018 Feb  5.0  & 135.2 & 14.0 &23.5 & 0.0004 & -        & Kepler Space Telescope\\
 2018 Feb  5.9  & 135.2 & 13.9 &17.6 & 0.0004 & -        & Kepler Space Telescope\\
 2018 May 19.1  & 121.2 & 13.4 &24.0 & 0.0010 & -        & Kepler Space Telescope\\
 2018 May 19.8  & 121.2 & 13.6 &23.5 & 0.0011 & -        & Kepler Space Telescope\\
 2018 May 20.8  & 121.2 & 13.7 &23.5 & 0.0012 & -        & Kepler Space Telescope\\
 2018 May 21.8  & 121.1 & 13.9 &23.5 & 0.0011 & -        & Kepler Space Telescope\\
 2018 May 22.5  & 121.1 & 14.0 &10.8 & 0.0013 & -        & Kepler Space Telescope\\
                &       &      &{\bf 404.1} hours total &&&\\
\hline \hline
\end{tabularx}
\caption{Observation details: Mid-time observing date, ecliptic longitude of the target,
sun-target-observer phase angle, duration of the observing run, photometric scatter,
observer and site name.  See Marciniak et al. (2018) for telescope and site details. 
The remaining part of the table for the rest of the targets, together with the 
photometric data from all individual nights are available through the CDS archive, {\tt http://cdsarc.u-strasbg.fr.}}
\label{obs}
\end{scriptsize}
\end{table}

\clearpage
\vspace{0.5cm}

    \begin{table*}[ht]
    \centering
\begin{tabularx}{\linewidth}{XX}
\includegraphics[width=0.35\textwidth,angle=270]{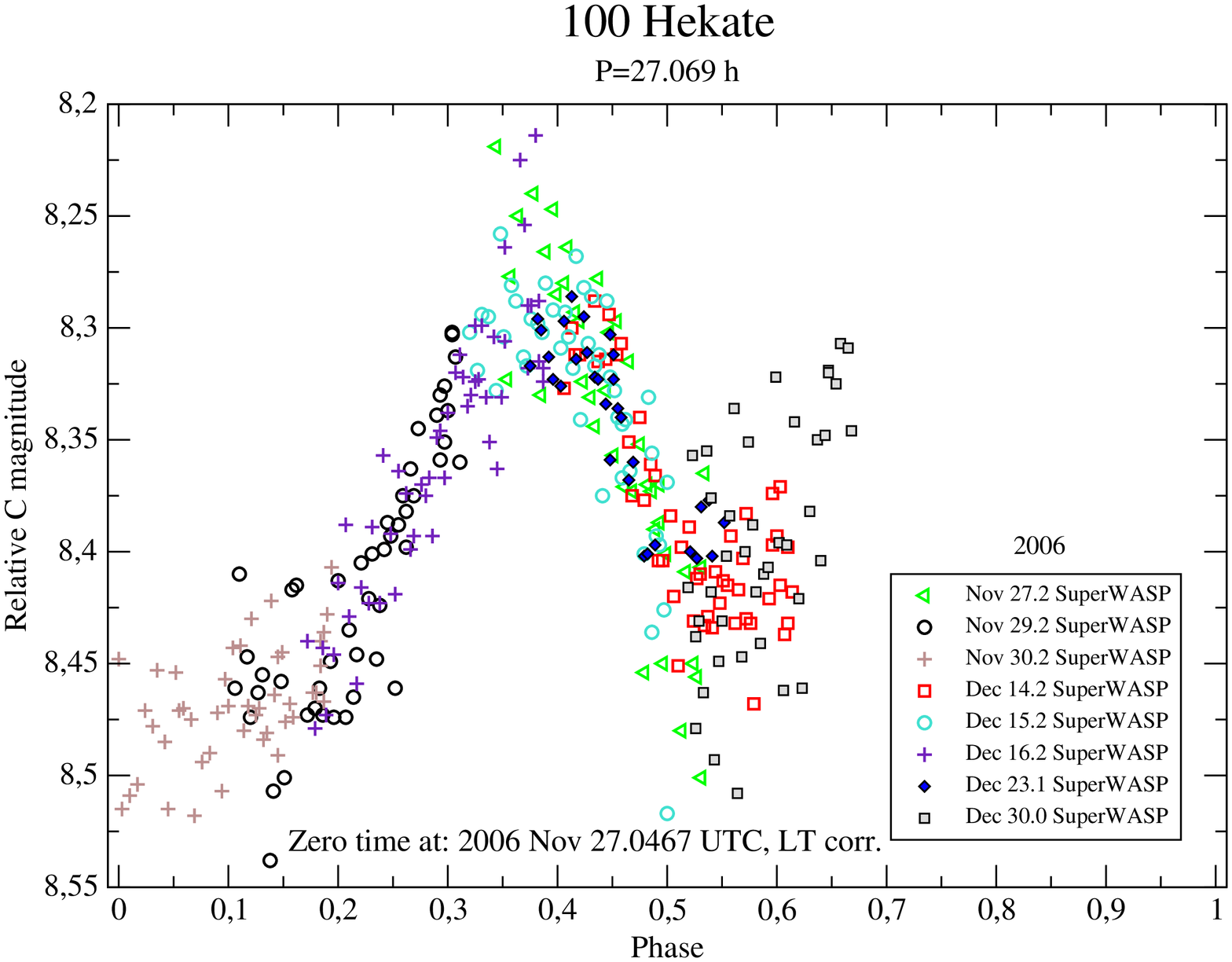} 
\captionof{figure}{Composite lightcurve of (100) Hekate from the year 2006.}
\label{100composit2006}
&
\includegraphics[width=0.35\textwidth,angle=270]{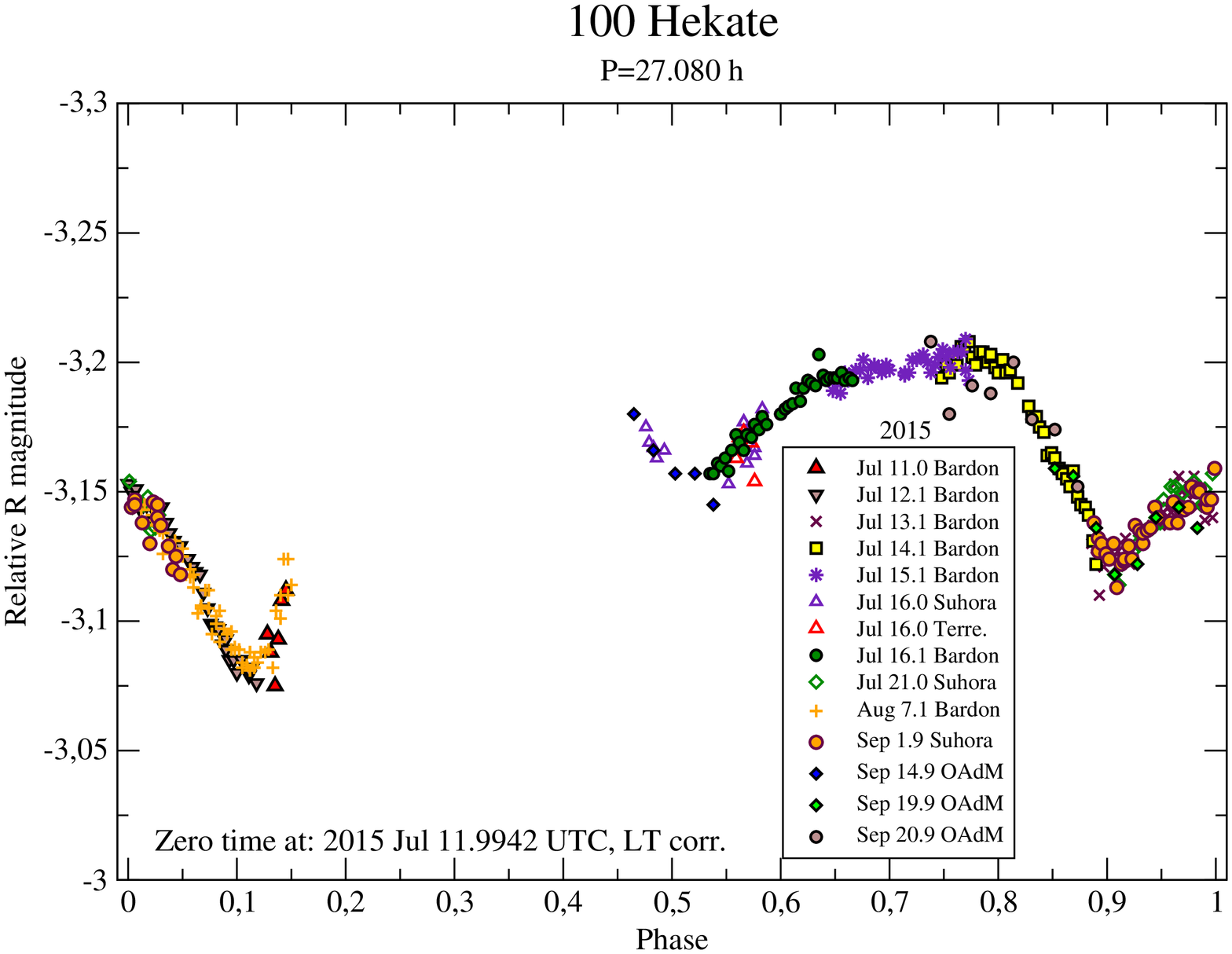} 
\captionof{figure}{Composite lightcurve of (100) Hekate from the year 2015.}
\label{100composit2015}
\\
\includegraphics[width=0.35\textwidth,angle=270]{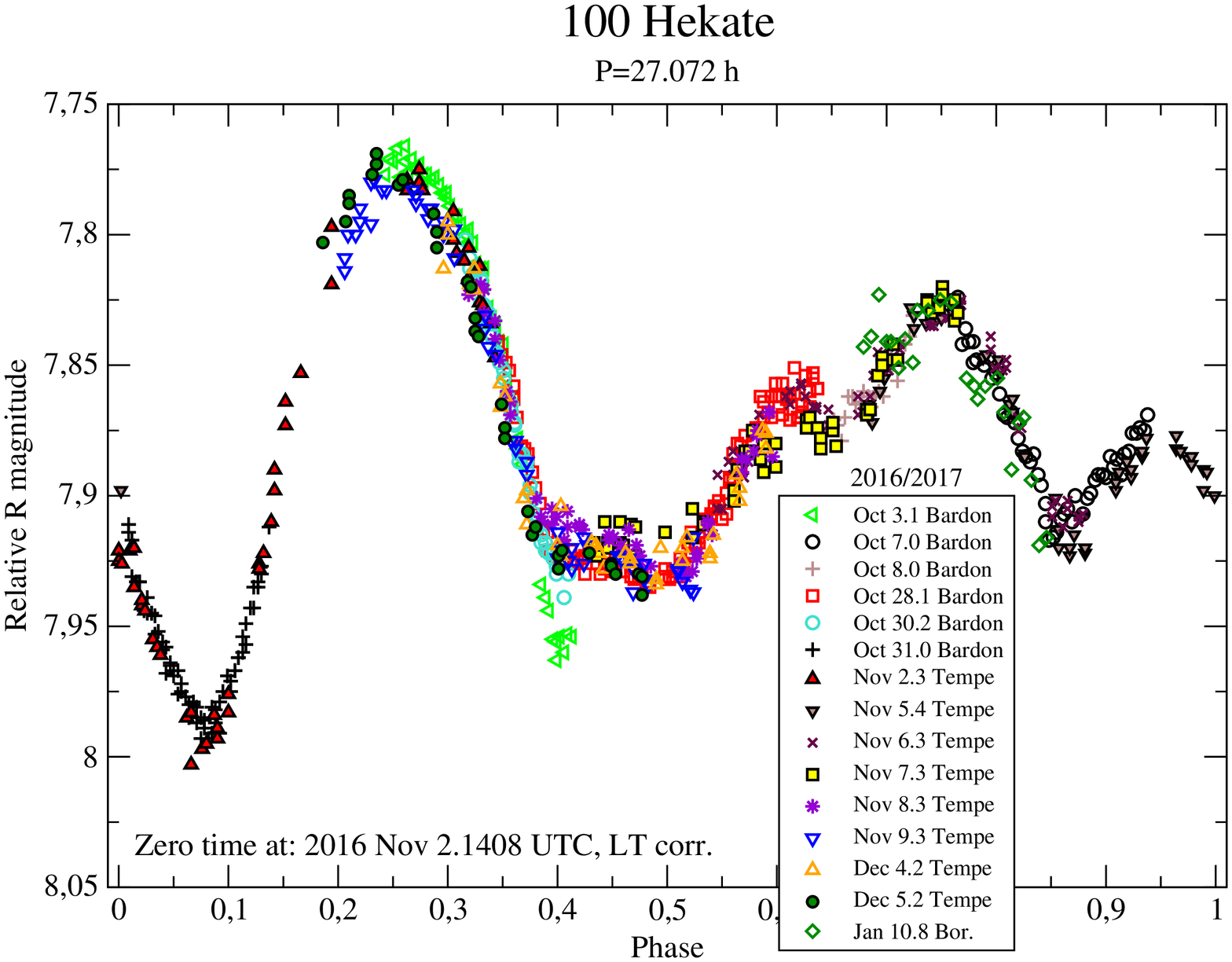} 
\captionof{figure}{Composite lightcurve of (100) Hekate from the years 2016-2017.}
\label{100composit2016}
&
\includegraphics[width=0.35\textwidth,angle=270]{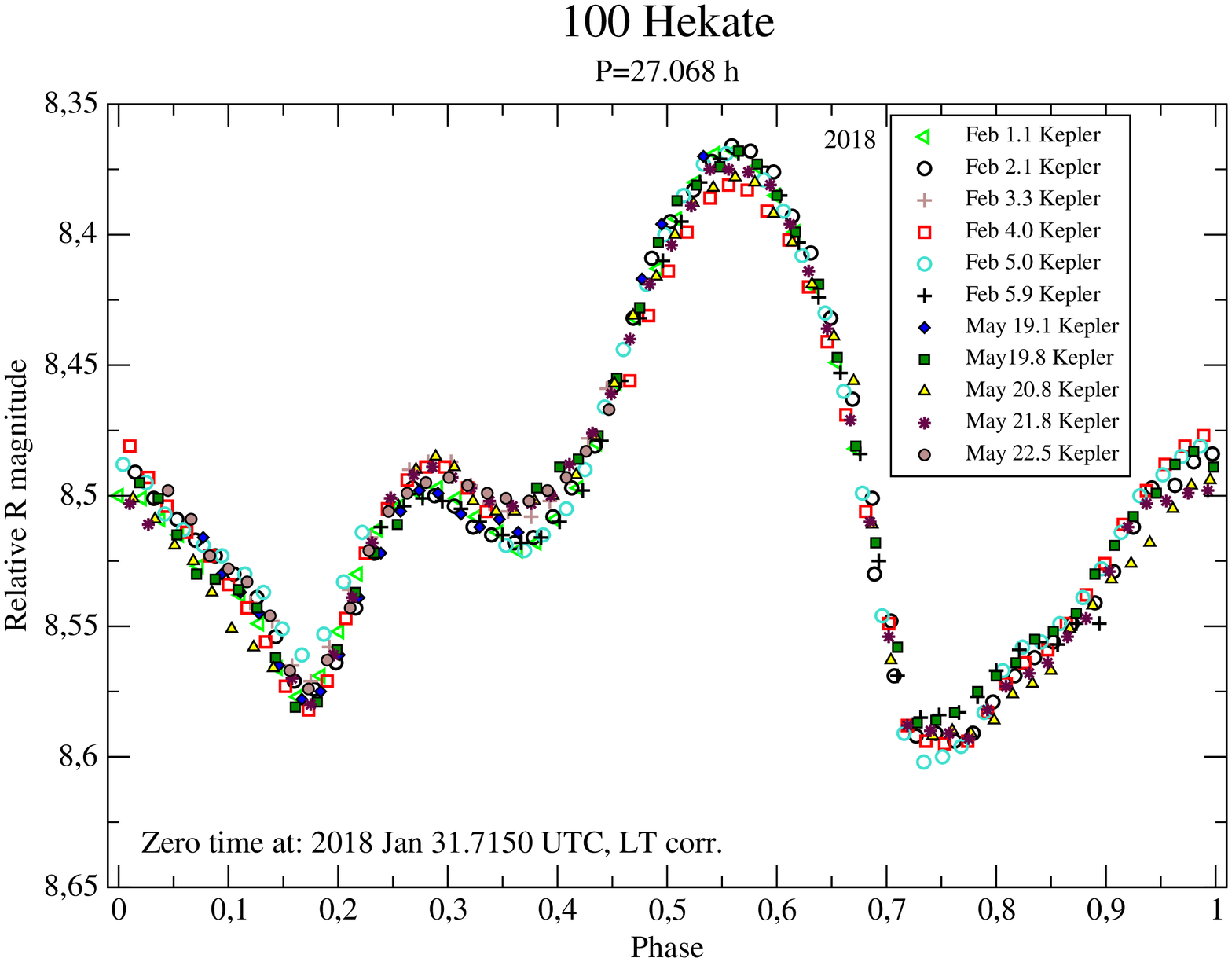} 
\captionof{figure}{Composite lightcurve of (100) Hekate based on Kepler K2 data from the year 2018.}
\label{100composit2018}
\\
\includegraphics[width=0.35\textwidth,angle=270]{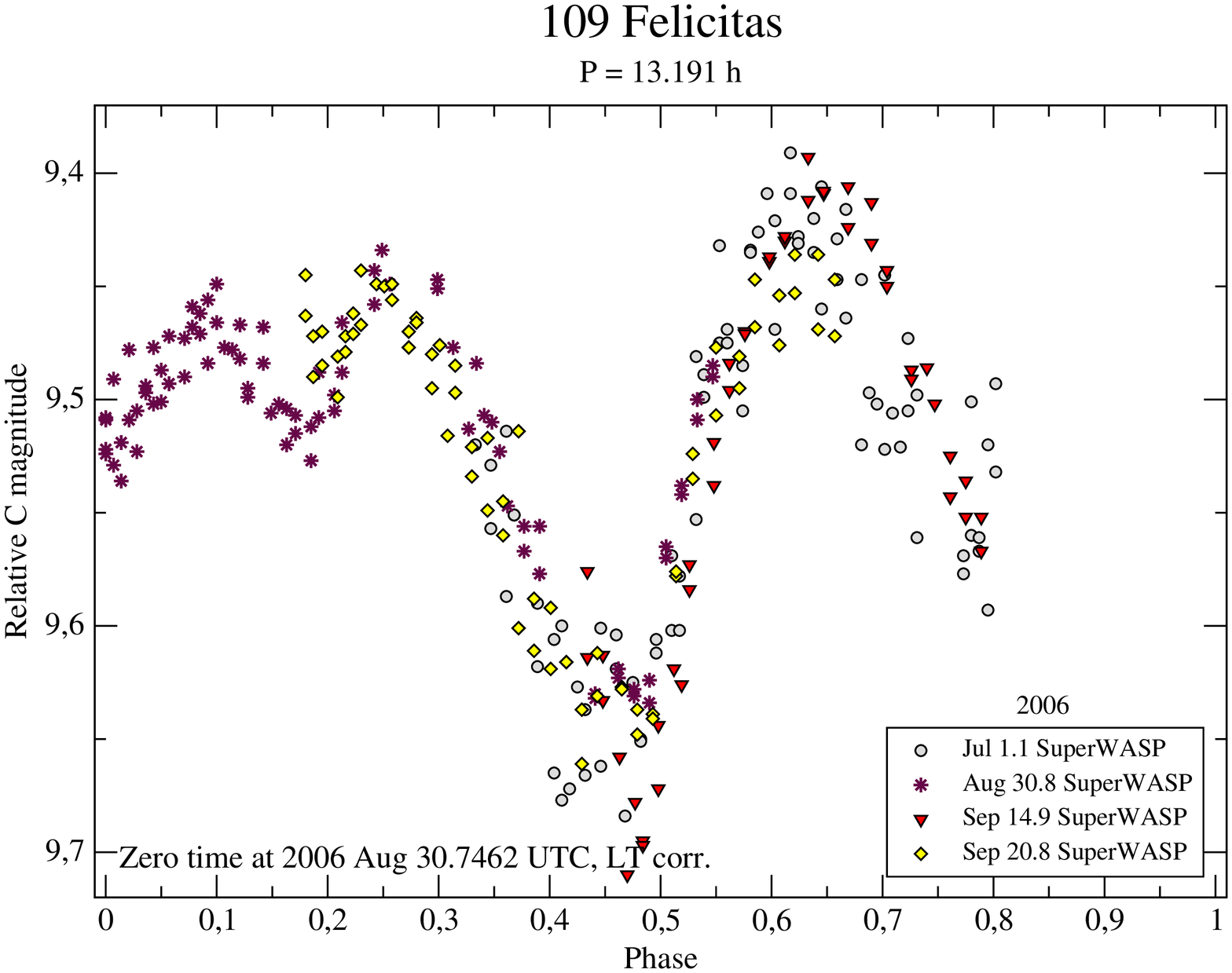} 
\captionof{figure}{Composite lightcurve of (109) Felicitas from the year 2006.}
\label{109composit2006}
&
\includegraphics[width=0.35\textwidth,angle=270]{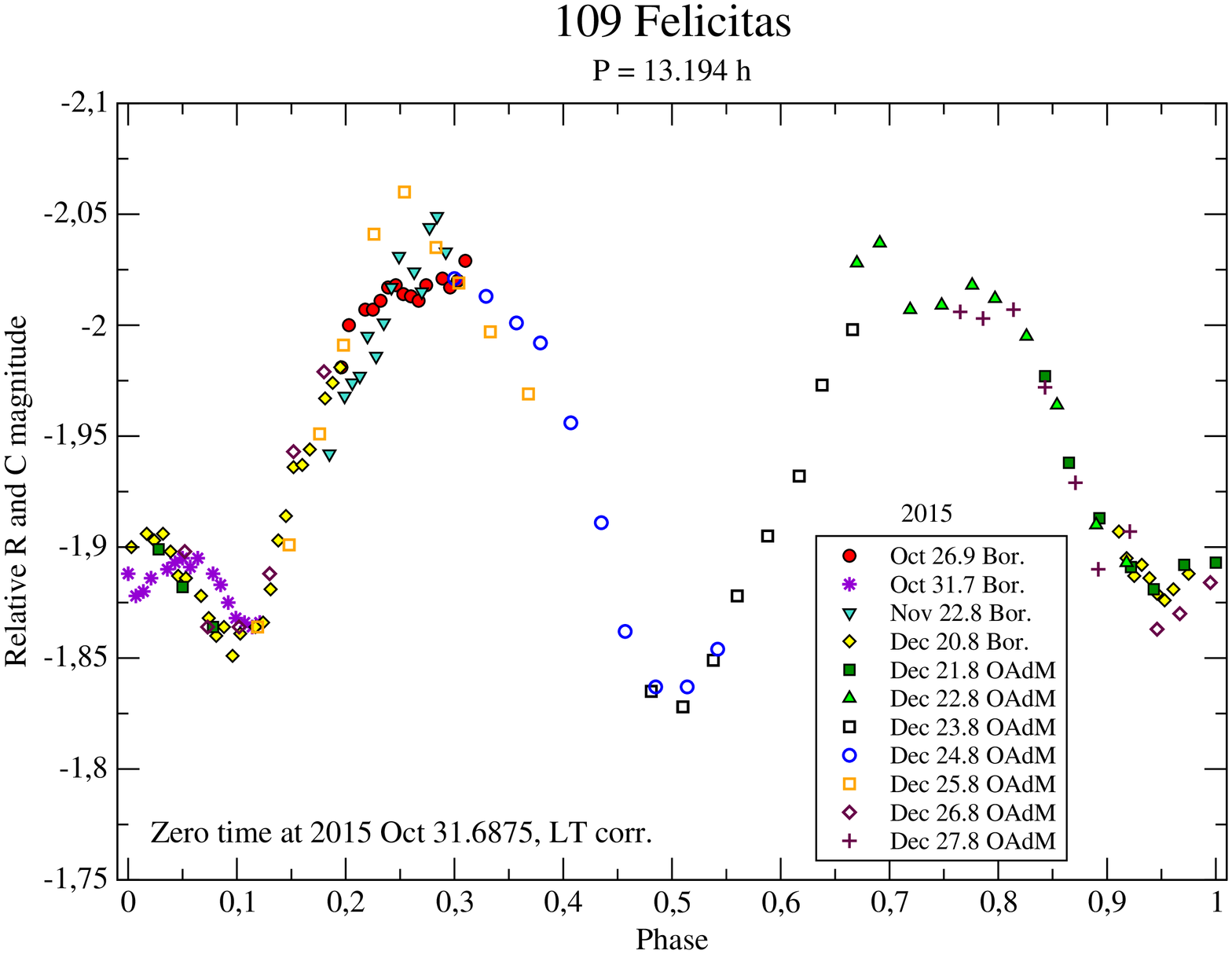} 
\captionof{figure}{Composite lightcurve of (109) Felicitas from the year 2015.}
\label{109composit2015}
\\
\end{tabularx}
    \end{table*}%

\clearpage
\vspace{0.5cm}

    \begin{table*}[ht]
    \centering
\begin{tabularx}{\linewidth}{XX}
\includegraphics[width=0.35\textwidth,angle=270]{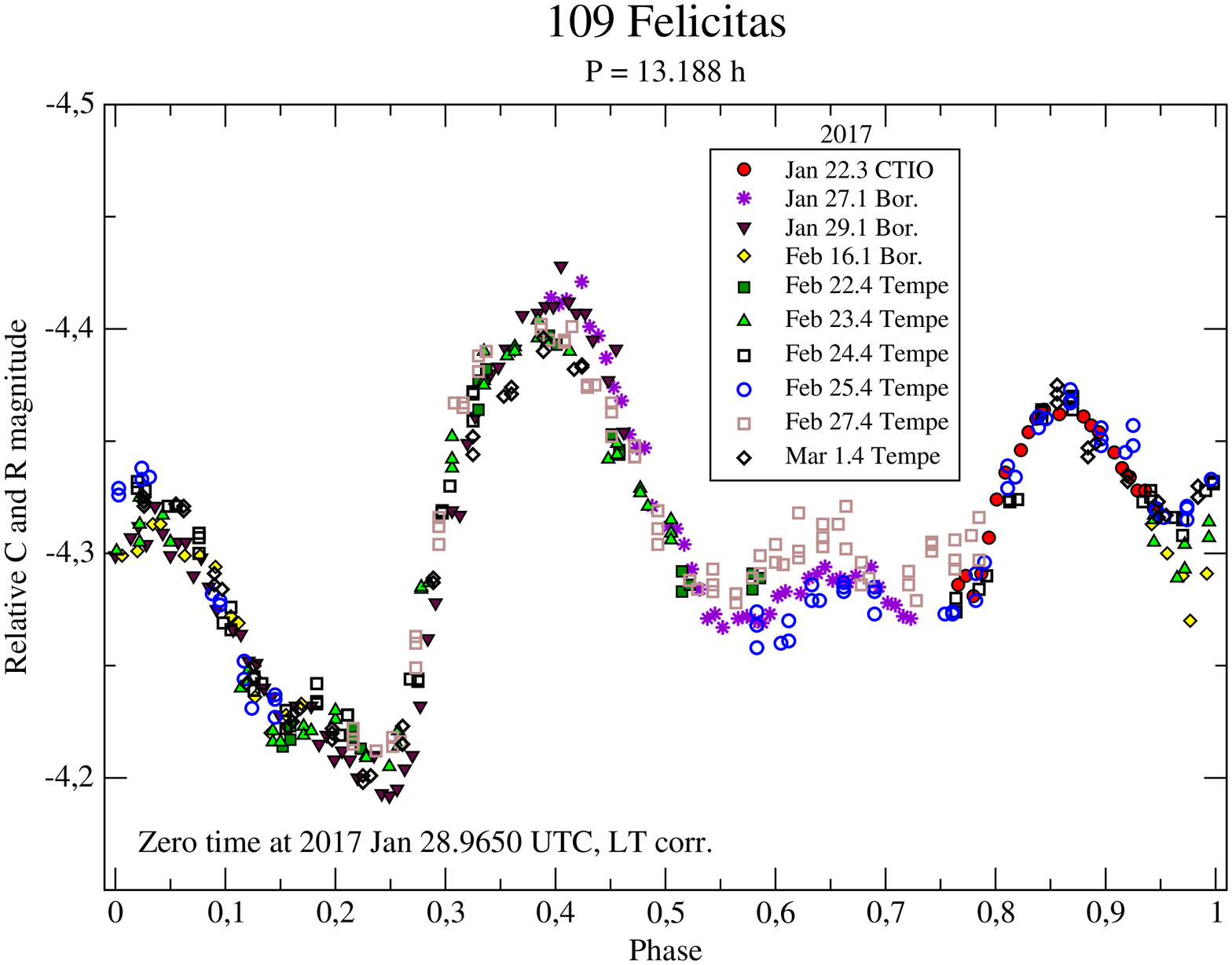} 
\captionof{figure}{Composite lightcurve of (109) Felicitas from the year 2017.}
\label{109composit2017}
&
\includegraphics[width=0.35\textwidth,angle=270]{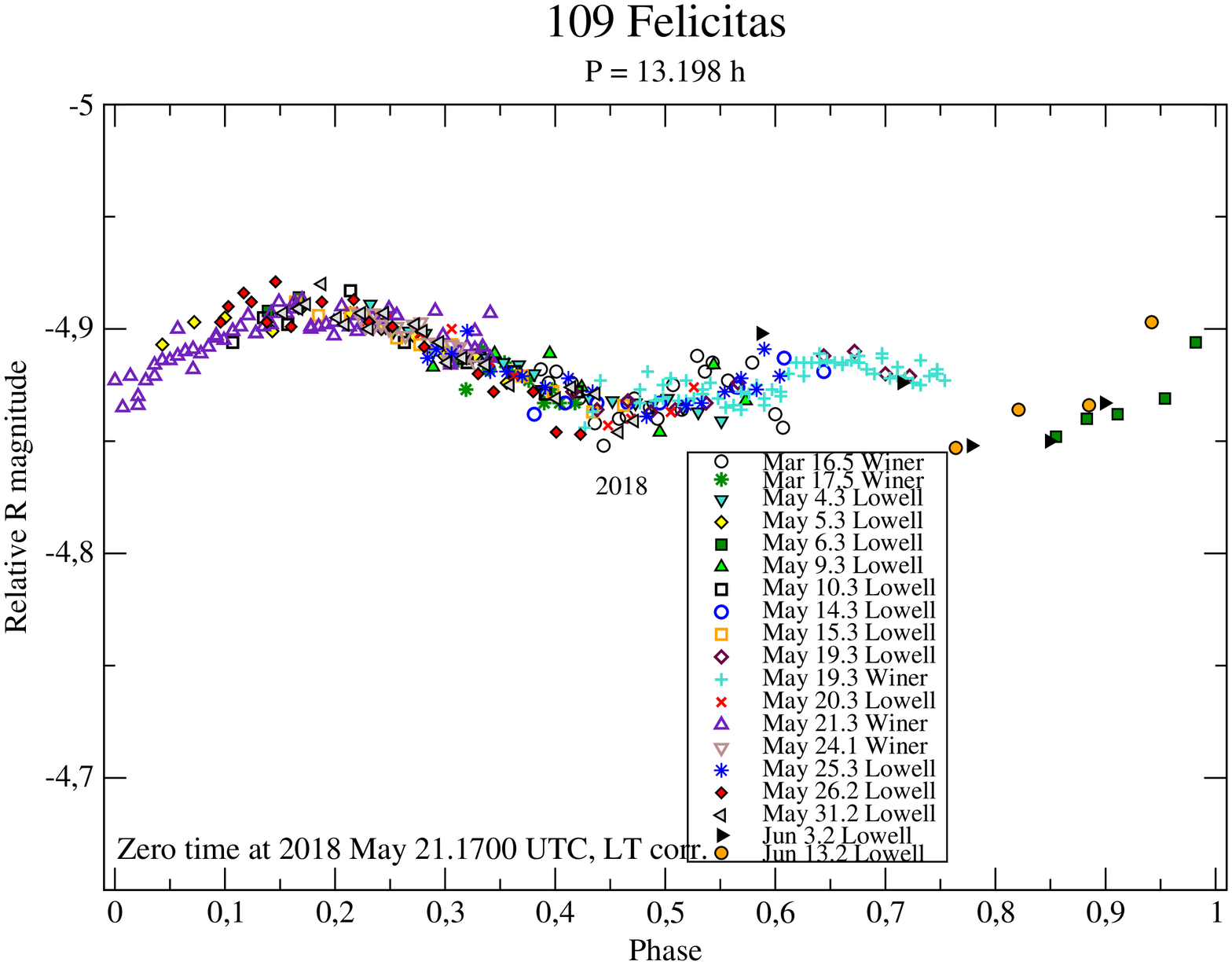} 
\captionof{figure}{Composite lightcurve of (109) Felicitas from the year 2018.}
\label{109composit2018}
\\
\includegraphics[width=0.35\textwidth,angle=270]{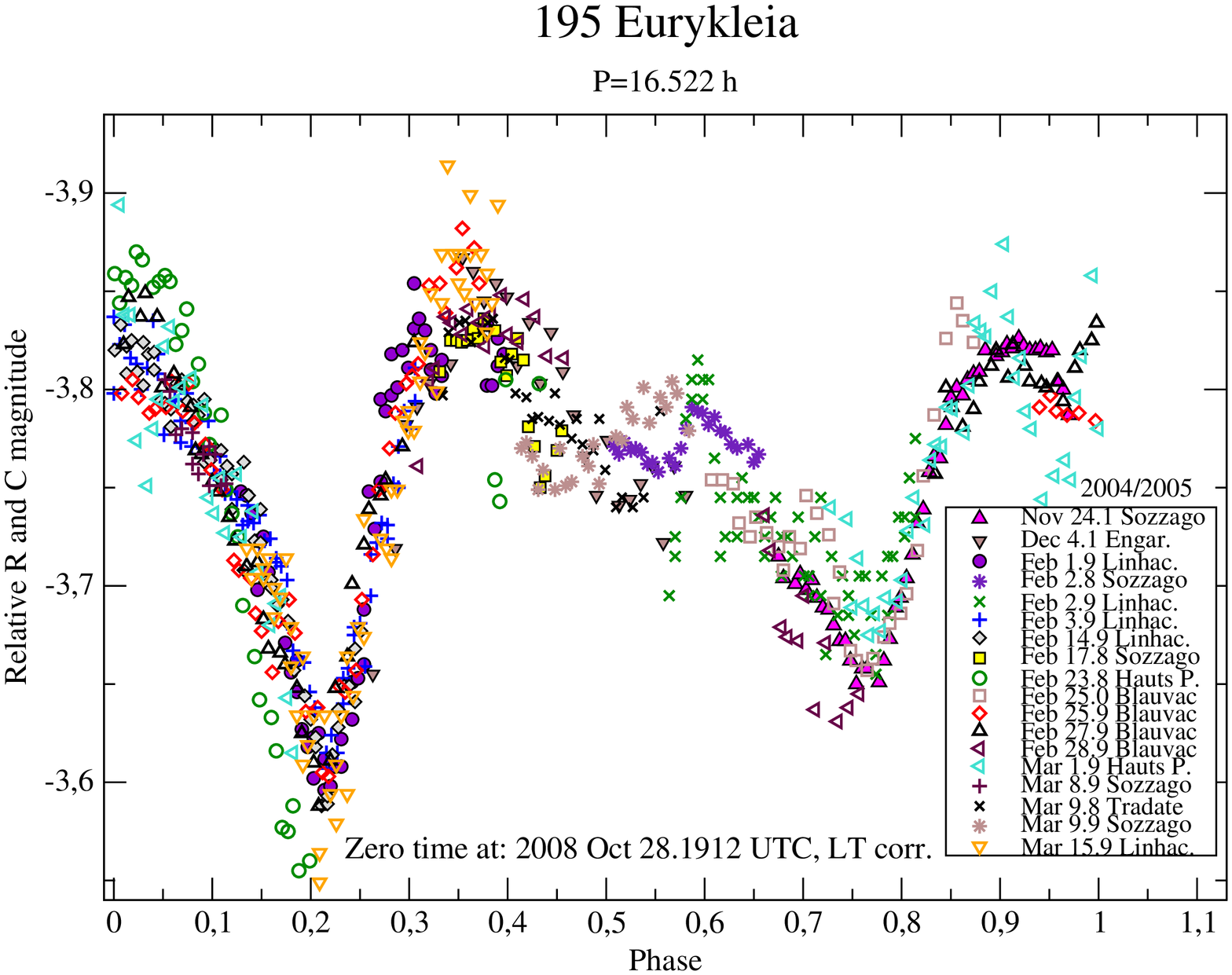} 
\captionof{figure}{Composite lightcurve of (195) Eurykleia from the years 2004-2005.}
\label{195composit2005}
&
\includegraphics[width=0.35\textwidth,angle=270]{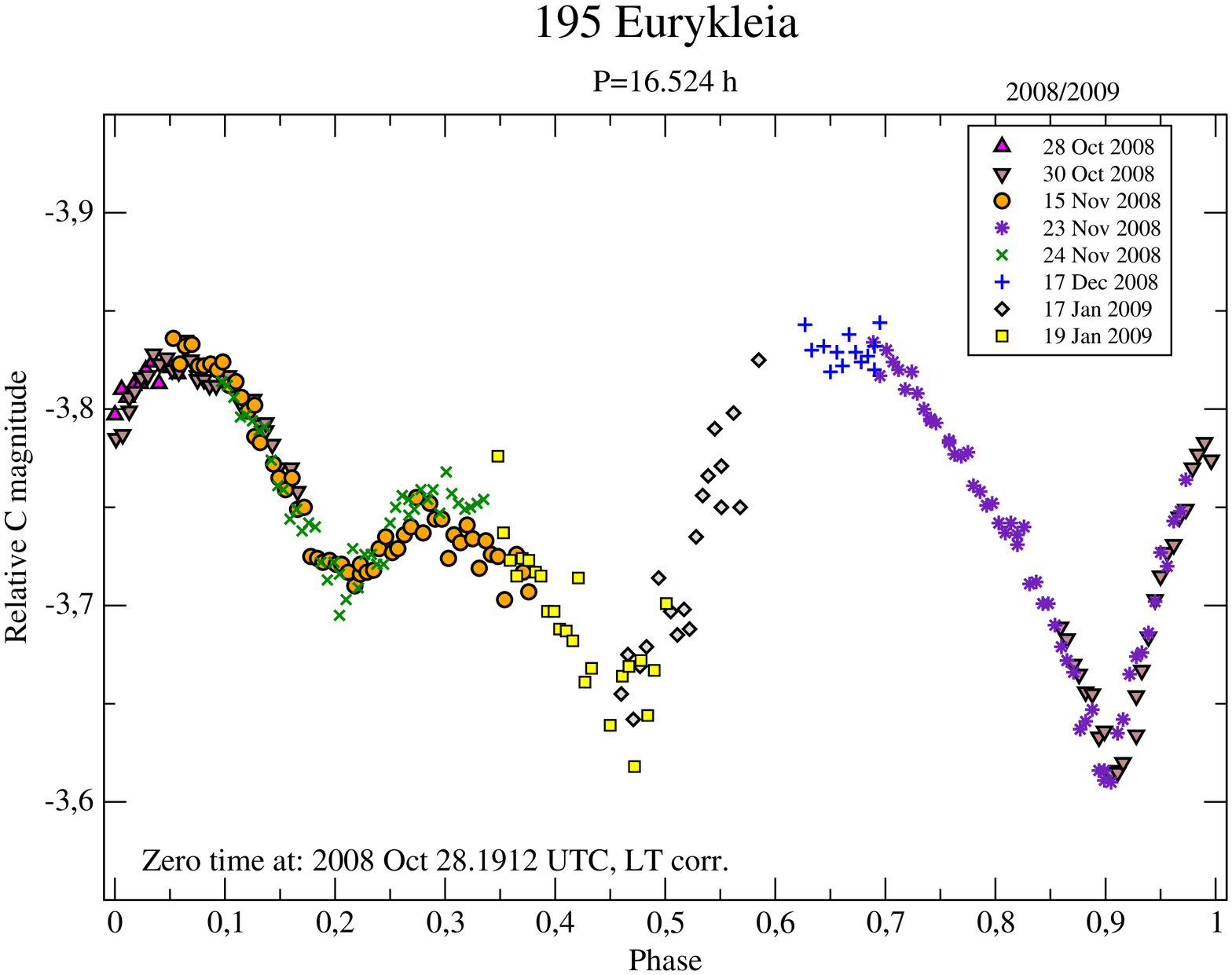} 
\captionof{figure}{Composite lightcurve of (195) Eurykleia from the year 2008.}
\label{195composit2008}
\\
\includegraphics[width=0.35\textwidth,angle=270]{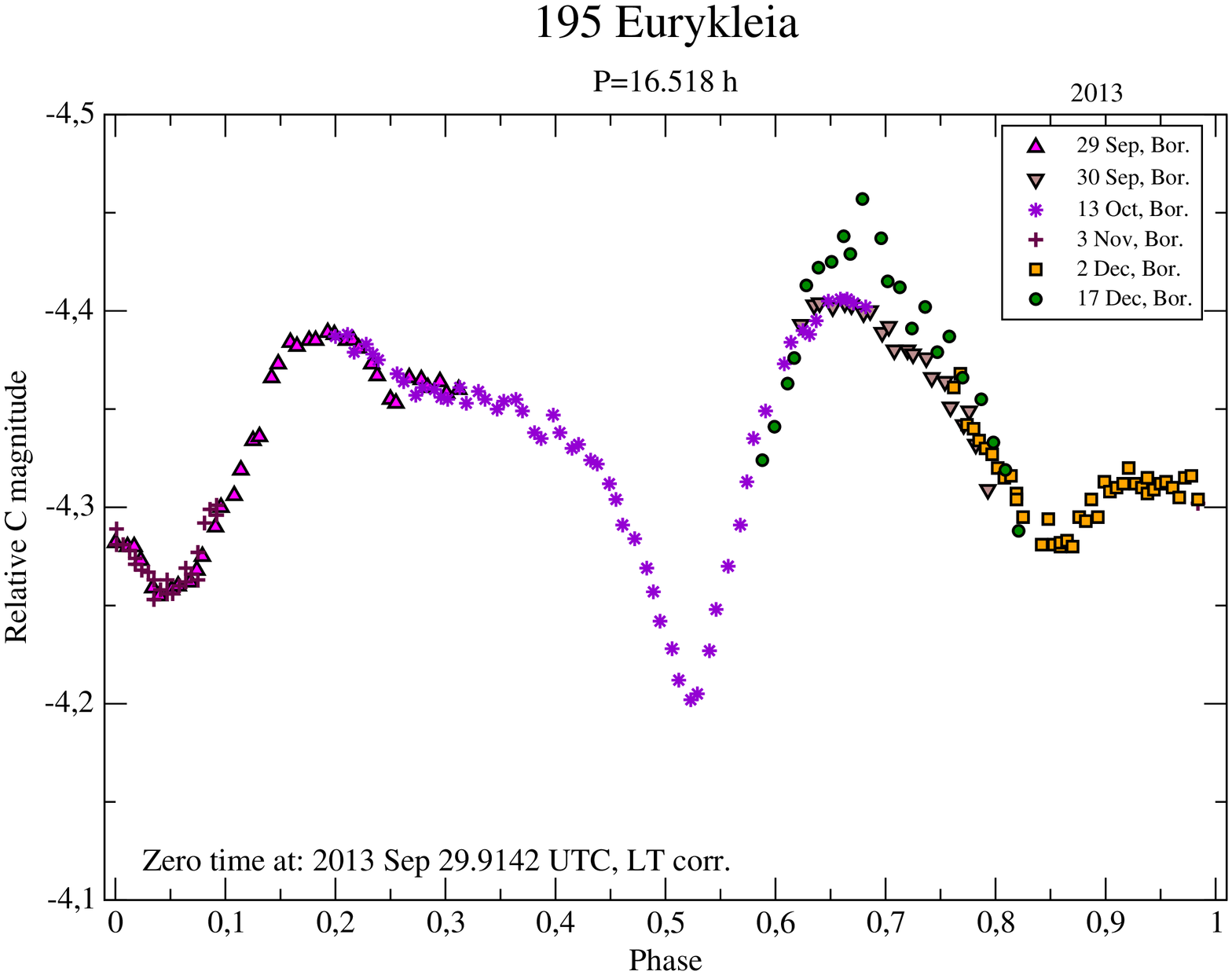} 
\captionof{figure}{Composite lightcurve of (195) Eurykleia from the year 2013.}
\label{195composit2013}
&
\includegraphics[width=0.35\textwidth,angle=270]{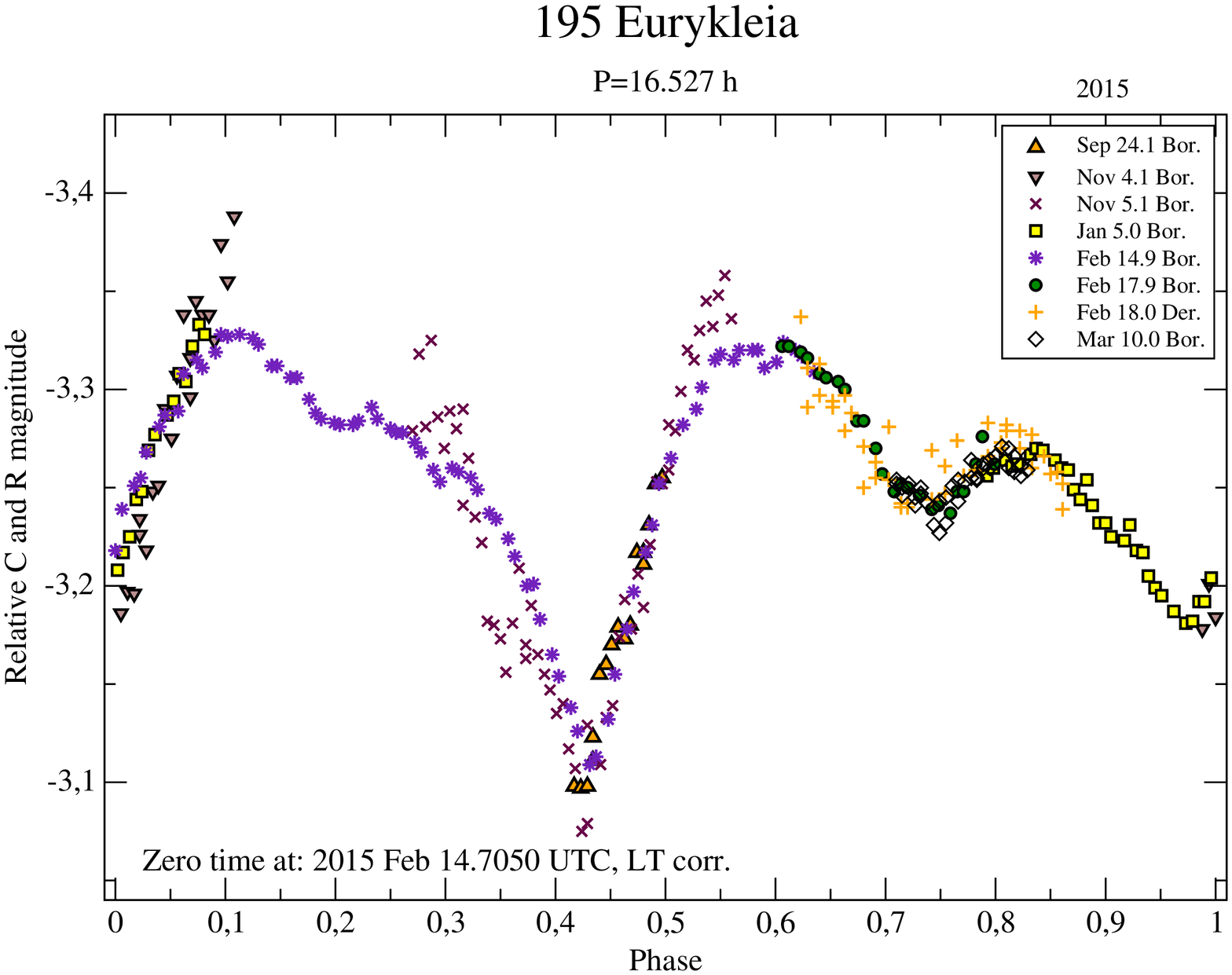} 
\captionof{figure}{Composite lightcurve of (195) Eurykleia from the year 2015.}
\label{195composit2015}
\\
\end{tabularx}
    \end{table*}%

\clearpage
\vspace{0.5cm}

    \begin{table*}[ht]
    \centering
\begin{tabularx}{\linewidth}{XX}
\includegraphics[width=0.35\textwidth,angle=270]{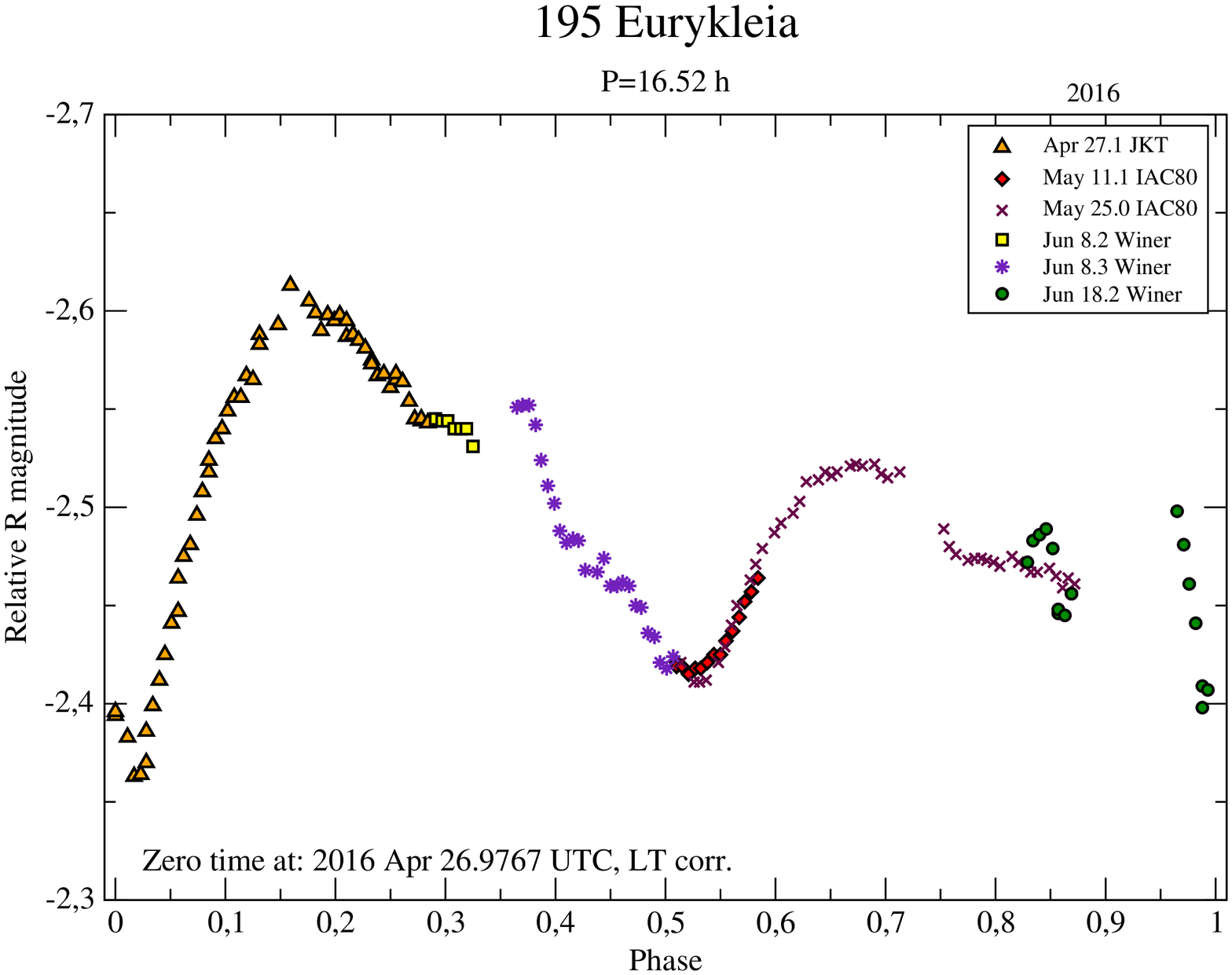} 
\captionof{figure}{Composite lightcurve of (195) Eurykleia from the year 2016.}
\label{195composit2016}
&
\includegraphics[width=0.35\textwidth,angle=270]{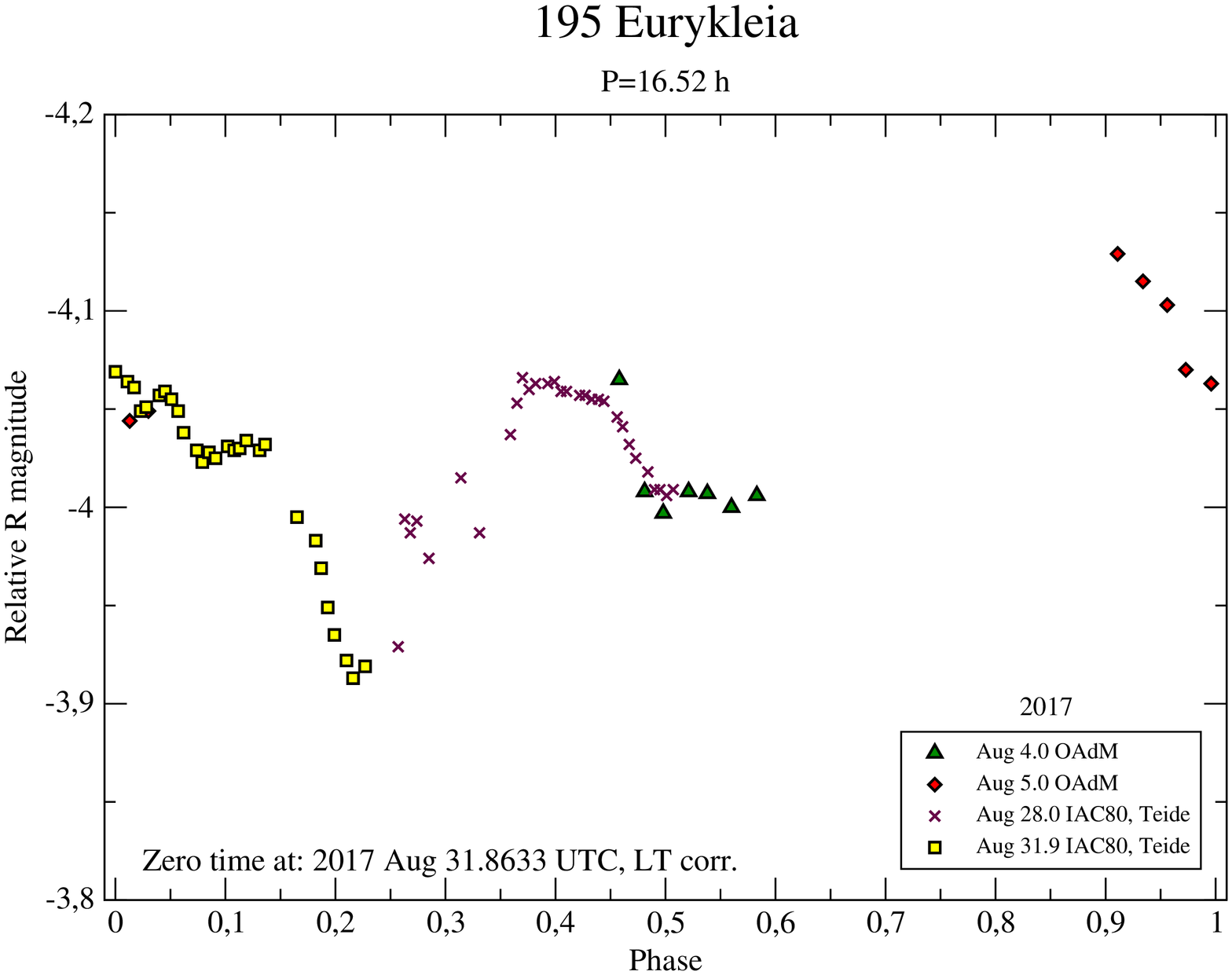} 
\captionof{figure}{Composite lightcurve of (195) Eurykleia from the year 2017.}
\label{195composit2017}
\\
\includegraphics[width=0.35\textwidth,angle=270]{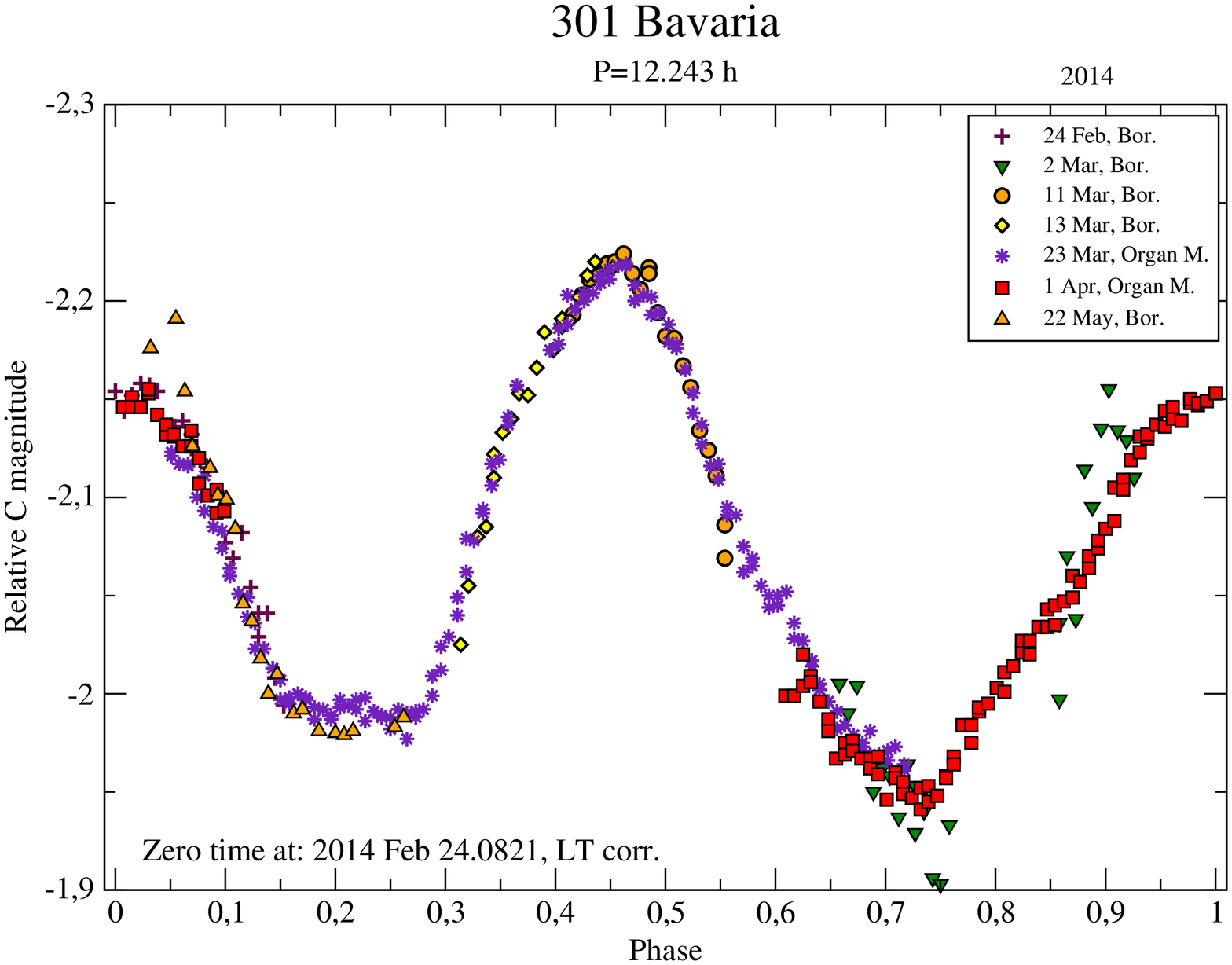} 
\captionof{figure}{Composite lightcurve of (301) Bavaria from the year 2014.}
\label{301composit2014}
&
\includegraphics[width=0.35\textwidth,angle=270]{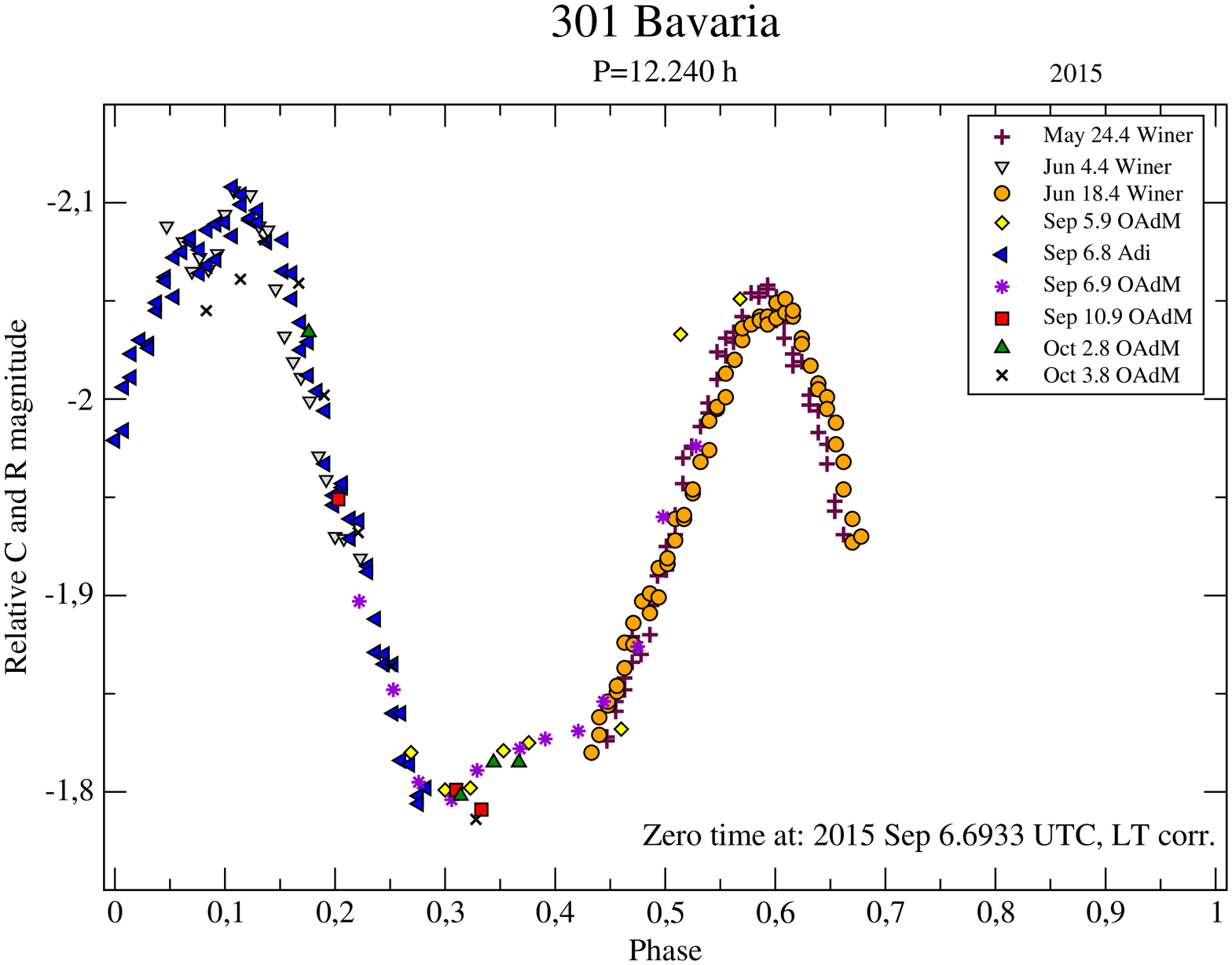} 
\captionof{figure}{Composite lightcurve of (301) Bavaria from the year 2015.}
\label{301composit2015}
\\
\includegraphics[width=0.35\textwidth,angle=270]{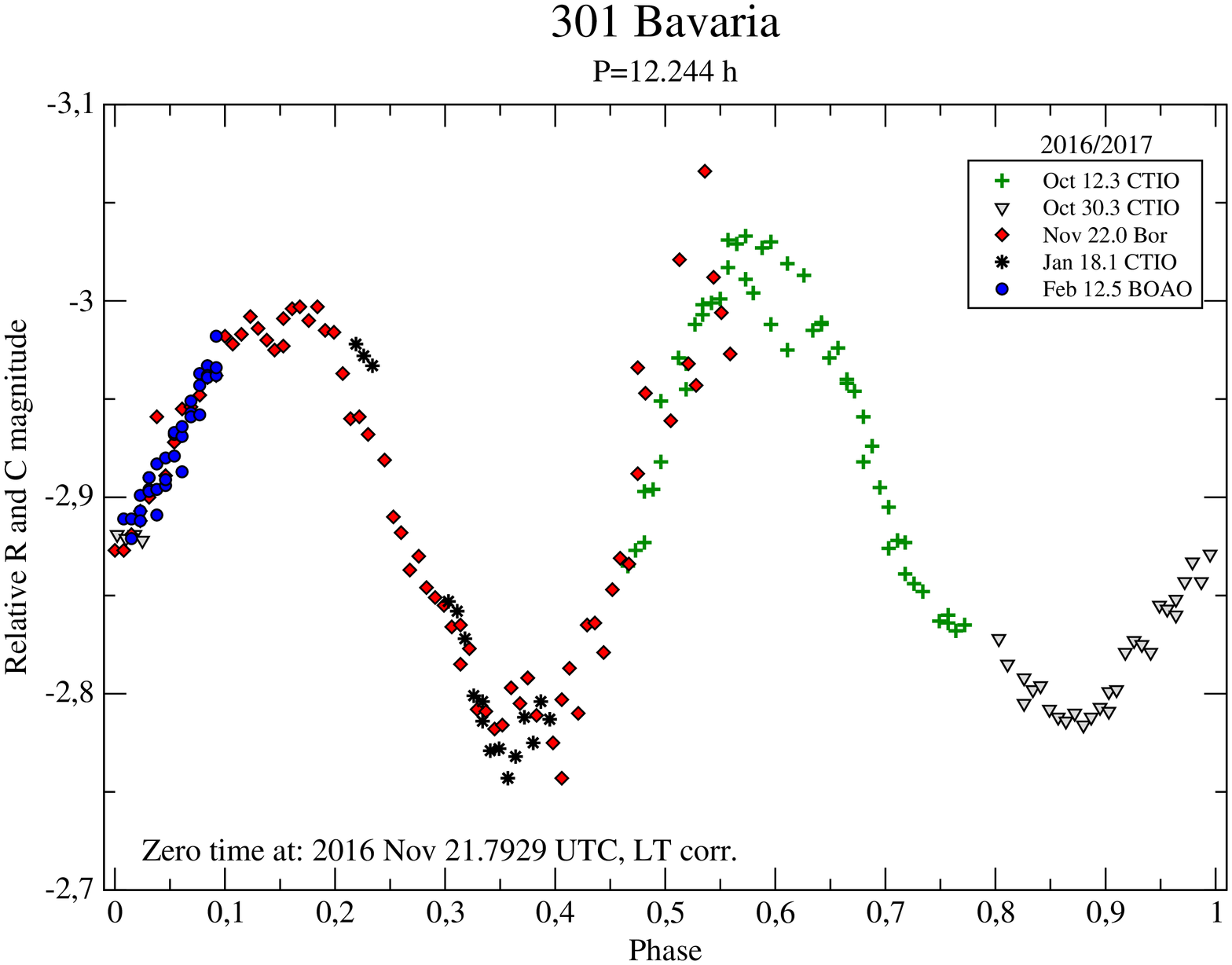} 
\captionof{figure}{Composite lightcurve of (301) Bavaria from the years 2016-2017.}
\label{301composit2016}
&
\includegraphics[width=0.35\textwidth,angle=270]{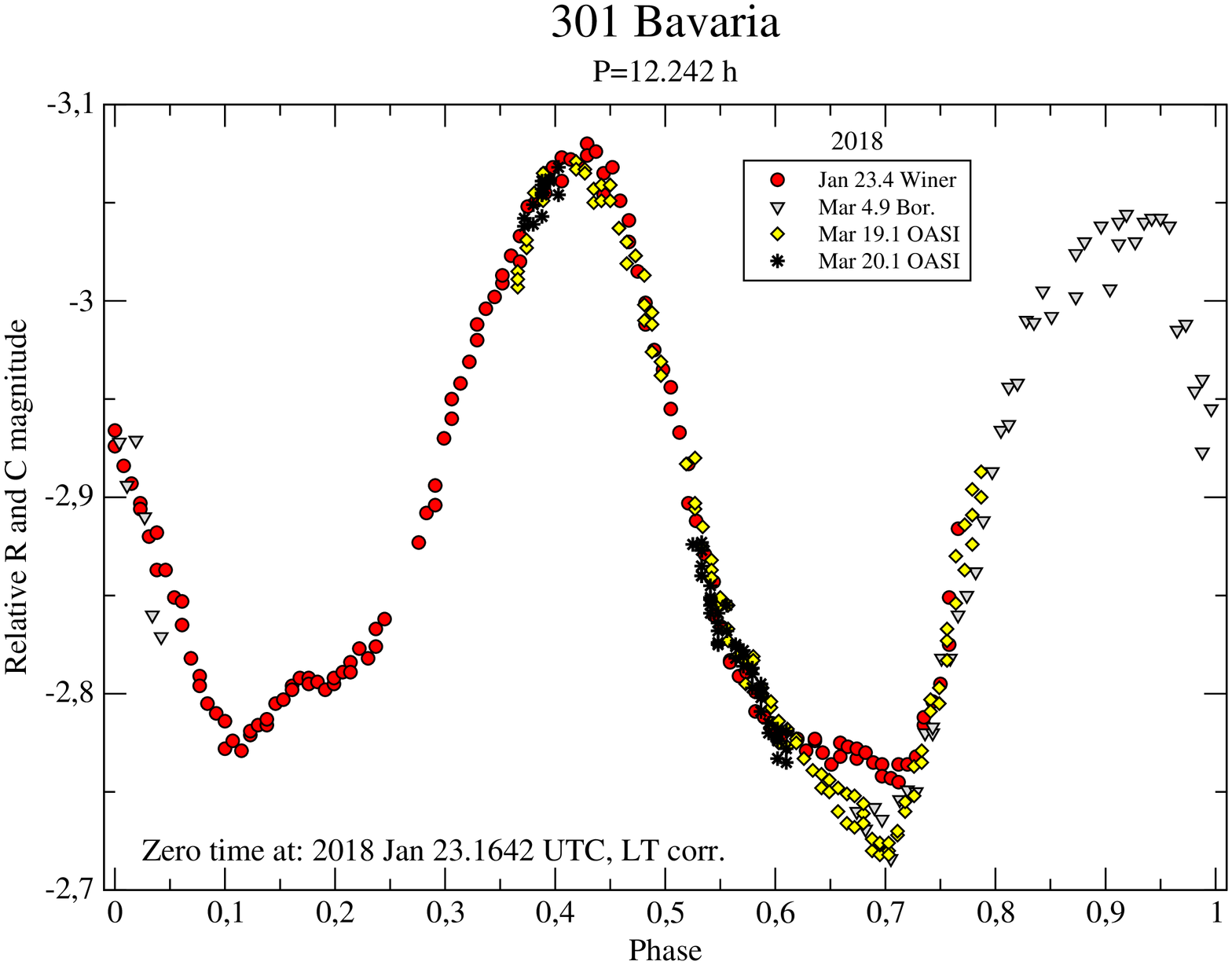} 
\captionof{figure}{Composite lightcurve of (301) Bavaria from the year 2018.}
\label{301composit2018}
\\
\end{tabularx}
    \end{table*}%

\clearpage
\vspace{0.5cm}

    \begin{table*}[ht]
    \centering
\begin{tabularx}{\linewidth}{XX}
\includegraphics[width=0.35\textwidth,angle=270]{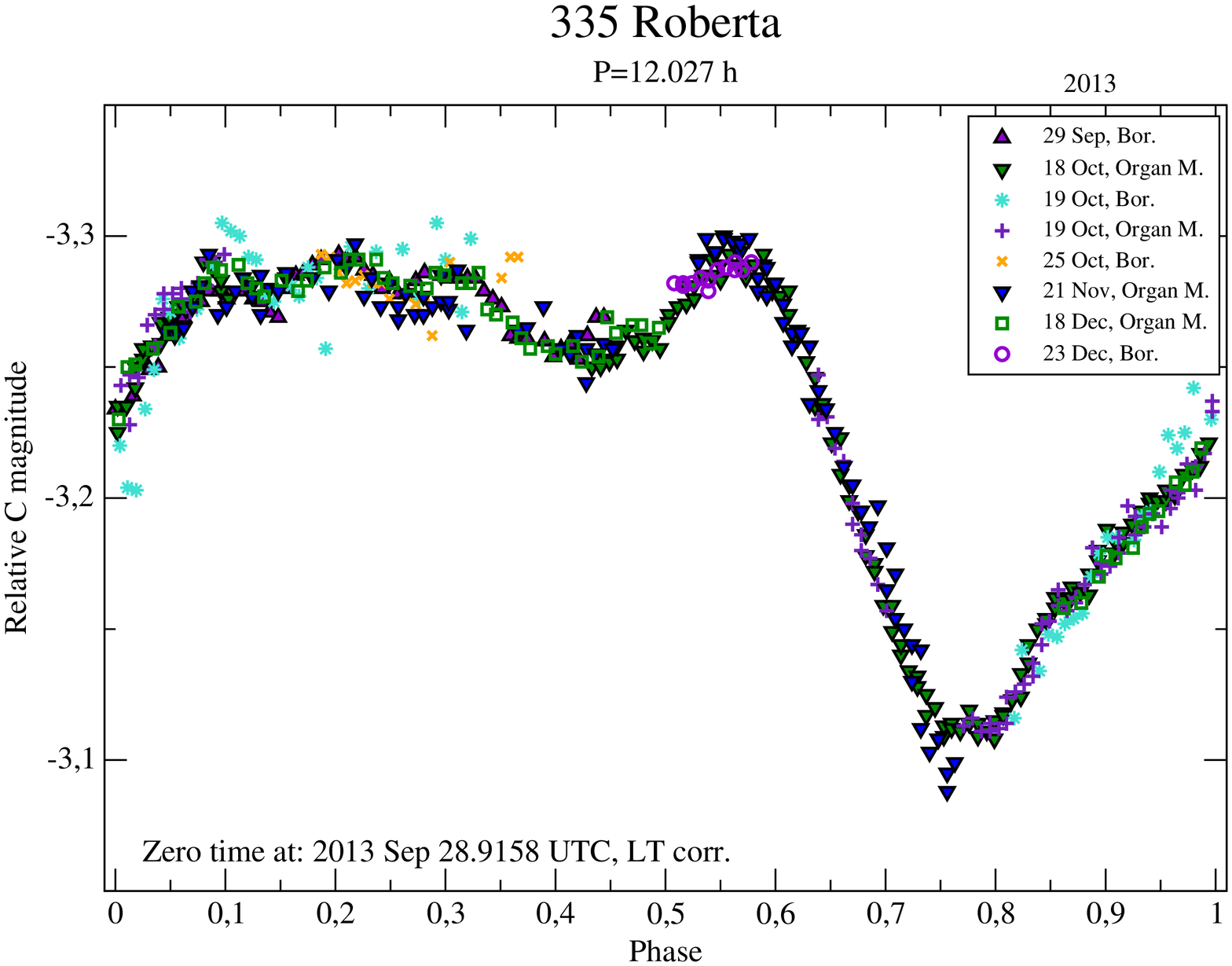} 
\captionof{figure}{Composite lightcurve of (335) Roberta from the year 2013.}
\label{335composit2013}
&
\includegraphics[width=0.35\textwidth,angle=270]{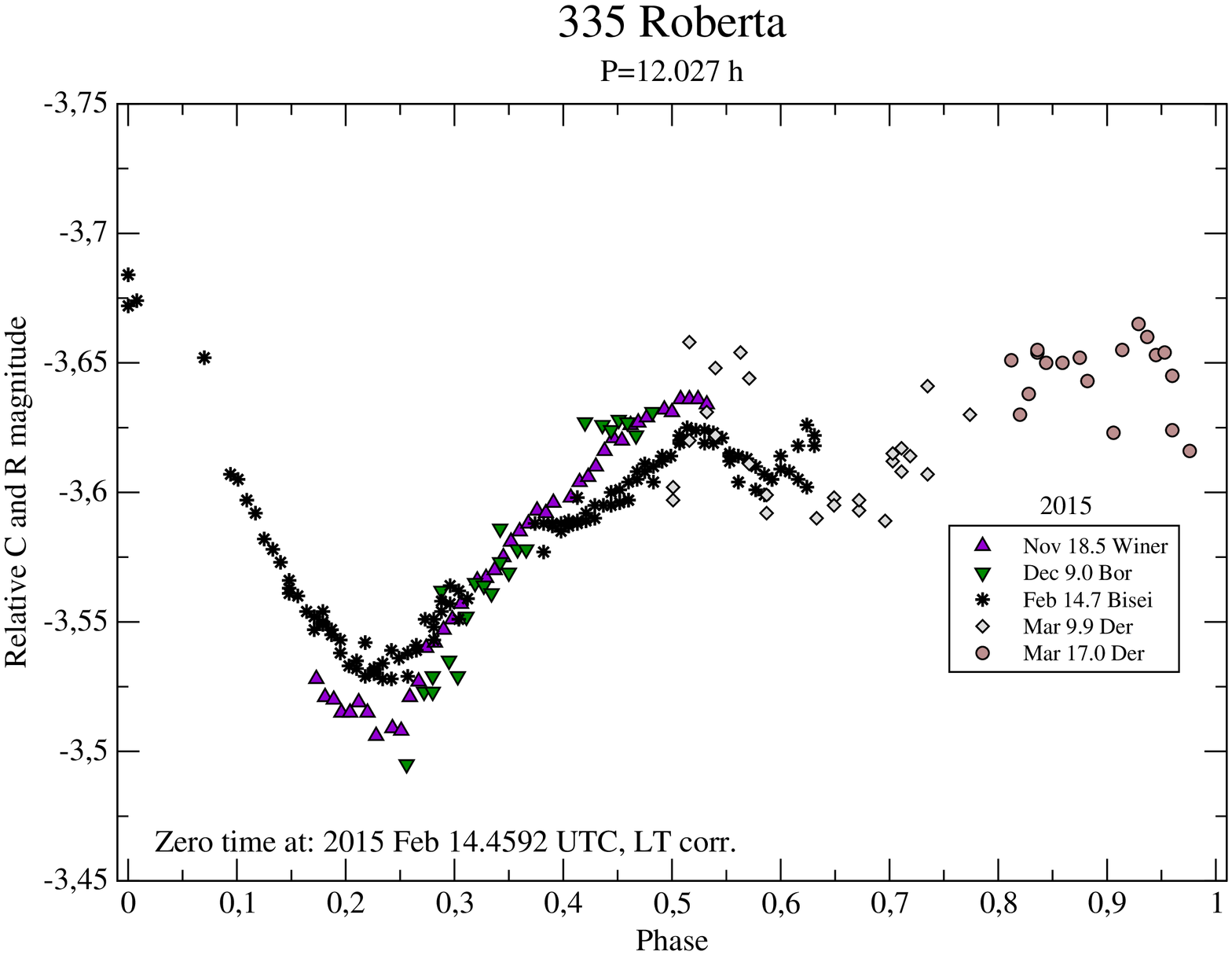} 
\captionof{figure}{Composite lightcurve of (335) Roberta from the year 2015.}
\label{335composit2015}
\\
\includegraphics[width=0.35\textwidth,angle=270]{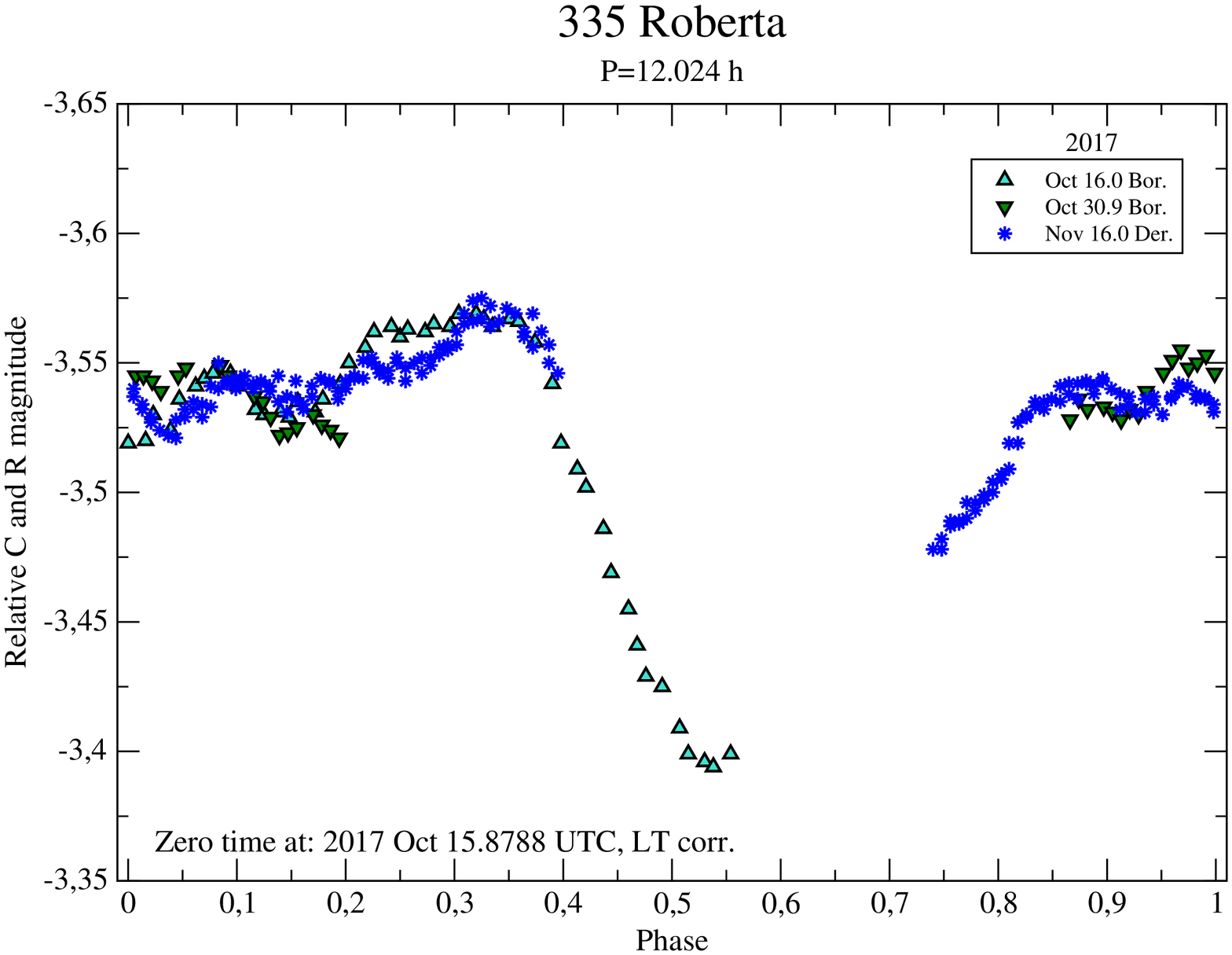} 
\captionof{figure}{Composite lightcurve of (335) Roberta from the year 2017.}
\label{335composit2017}
&
\includegraphics[width=0.35\textwidth,angle=270]{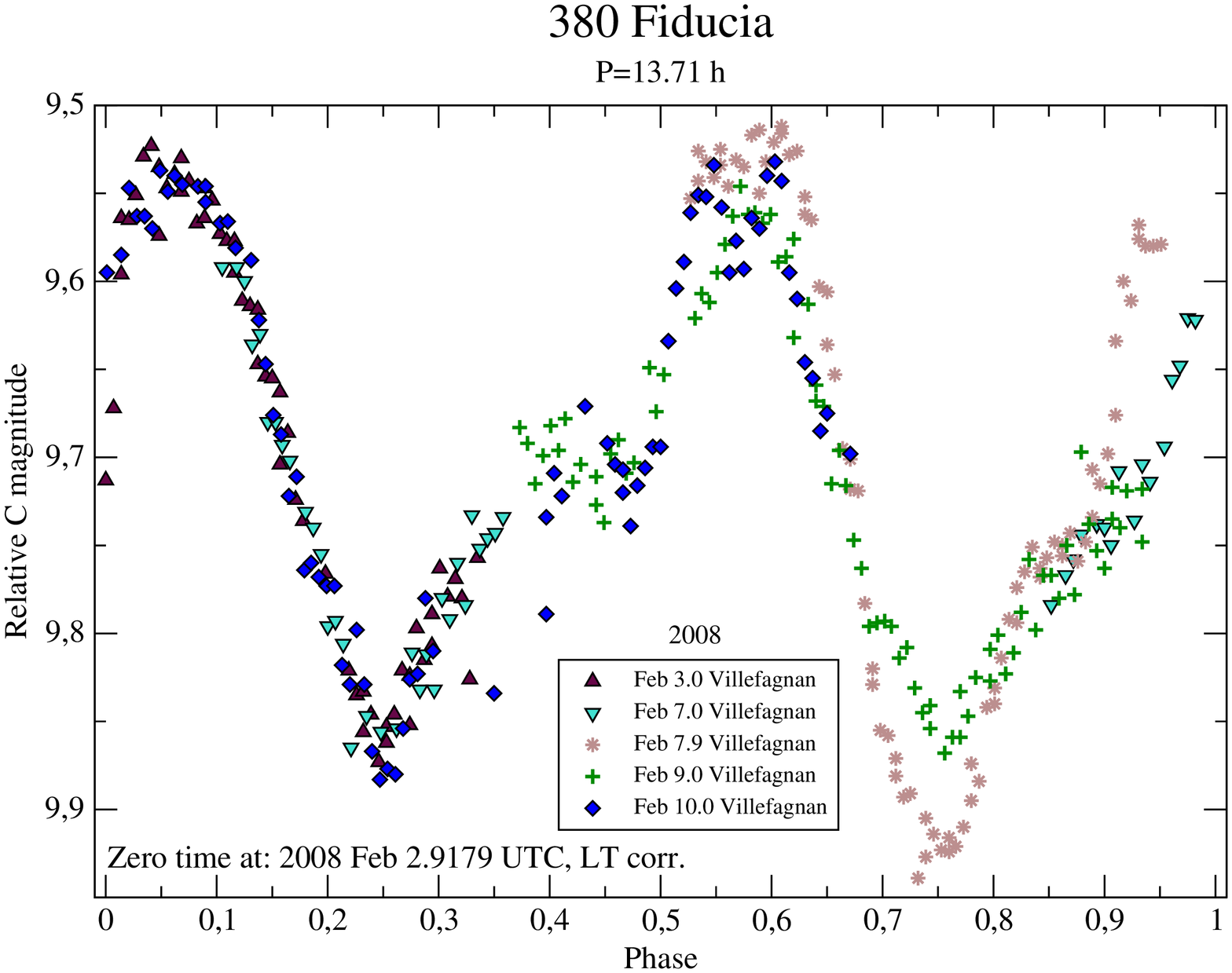} 
\captionof{figure}{Composite lightcurve of (380) Fiducia from the year 2008.}
\label{380composit2008}
\\
\includegraphics[width=0.35\textwidth,angle=270]{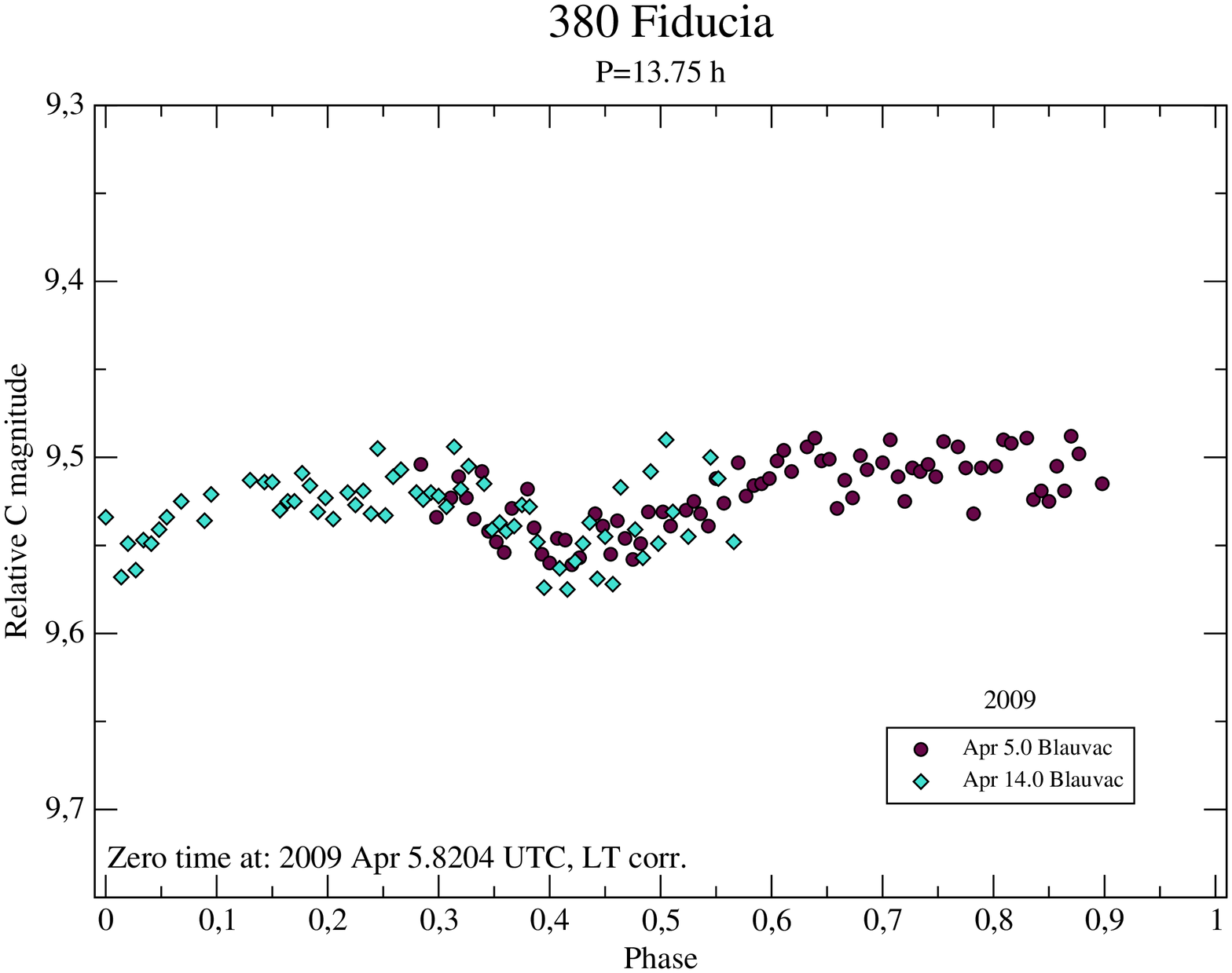} 
\captionof{figure}{Composite lightcurve of (380) Fiducia from the year 2009.}
\label{380composit2009}
&
\includegraphics[width=0.35\textwidth,angle=270]{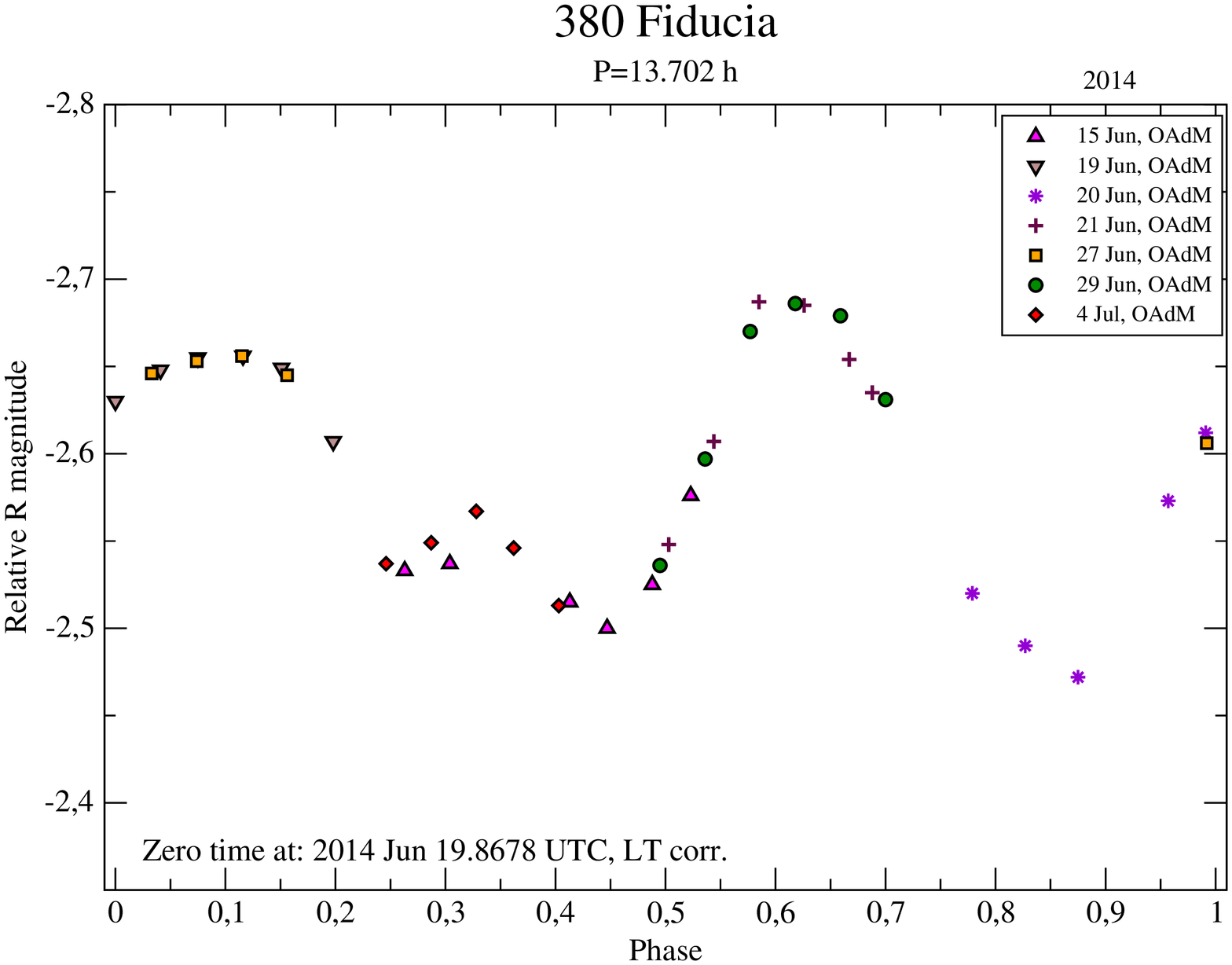} 
\captionof{figure}{Composite lightcurve of (380) Fiducia from the year 2014.}
\label{380composit2014}
\\
\end{tabularx}
    \end{table*}%

\clearpage
\vspace{0.5cm}

    \begin{table*}[ht]
    \centering
\begin{tabularx}{\linewidth}{XX}
\includegraphics[width=0.35\textwidth,angle=270]{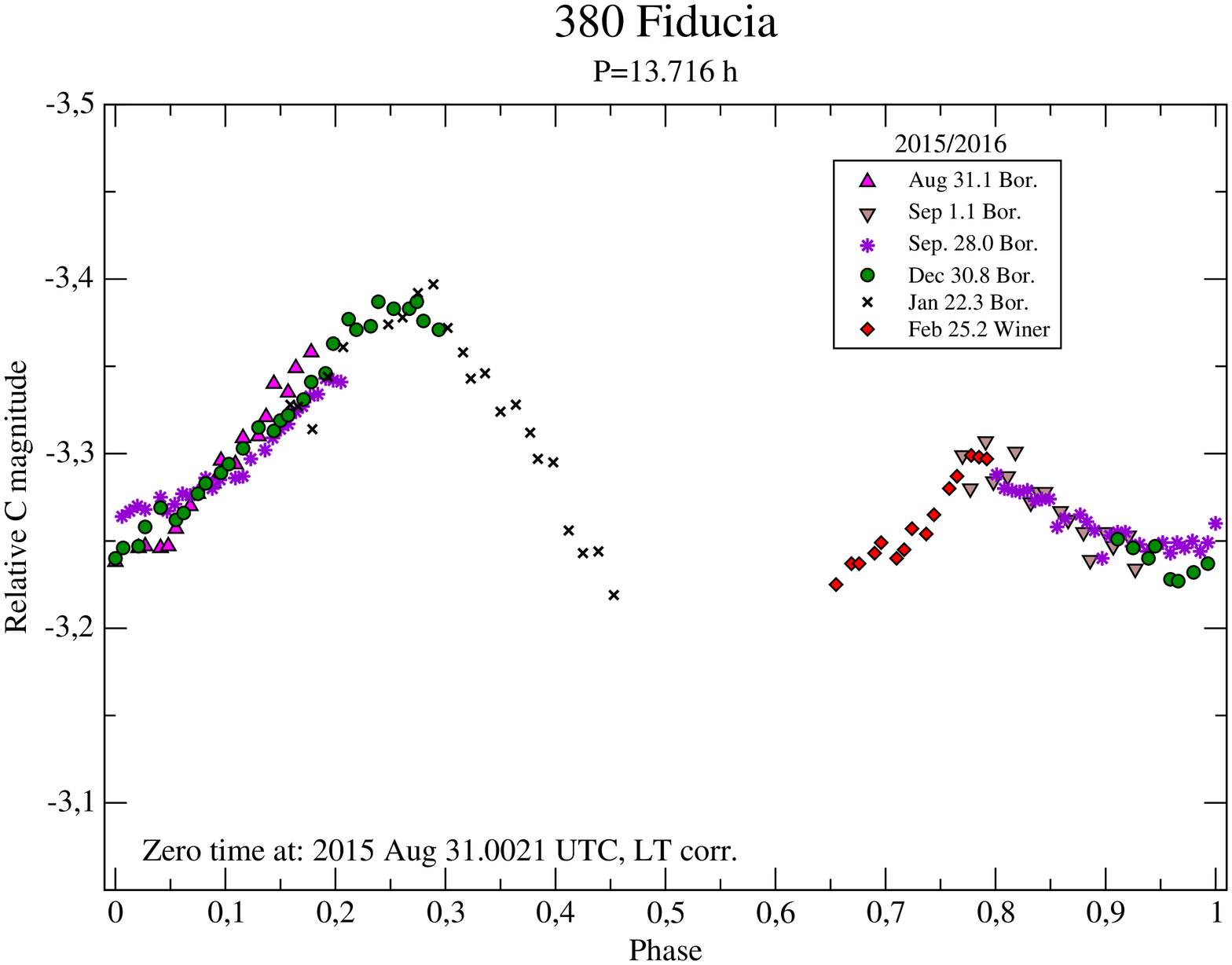} 
\captionof{figure}{Composite lightcurve of (380) Fiducia from the years 2015-2016 under large phase angle.}
\label{380composit2015large_phase_angle}
&
\includegraphics[width=0.35\textwidth,angle=270]{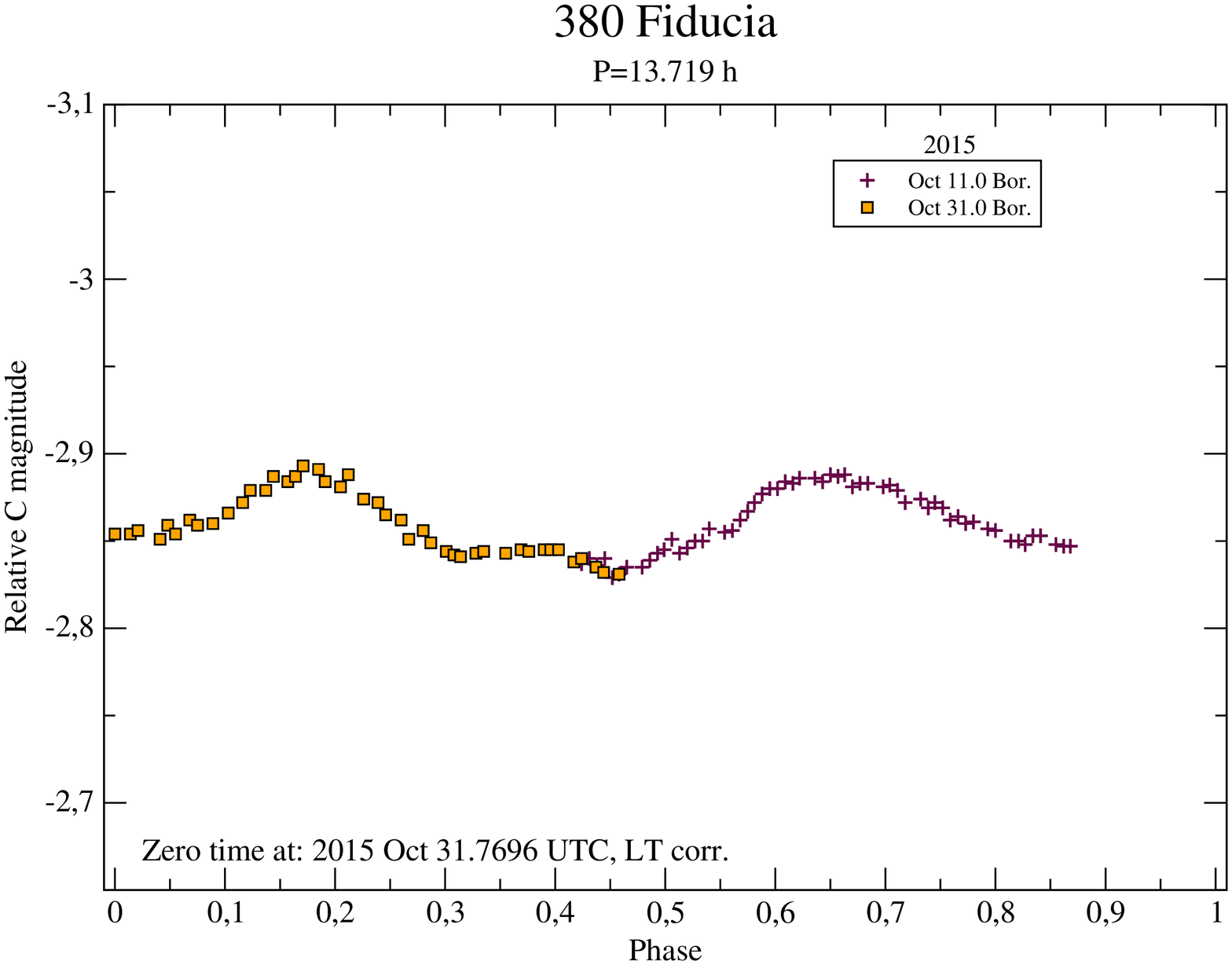} 
\captionof{figure}{Composite lightcurve of (380) Fiducia from the year 2015 under small phase angle.}
\label{380composit2015small_phase_angle}
\\
\includegraphics[width=0.35\textwidth,angle=270]{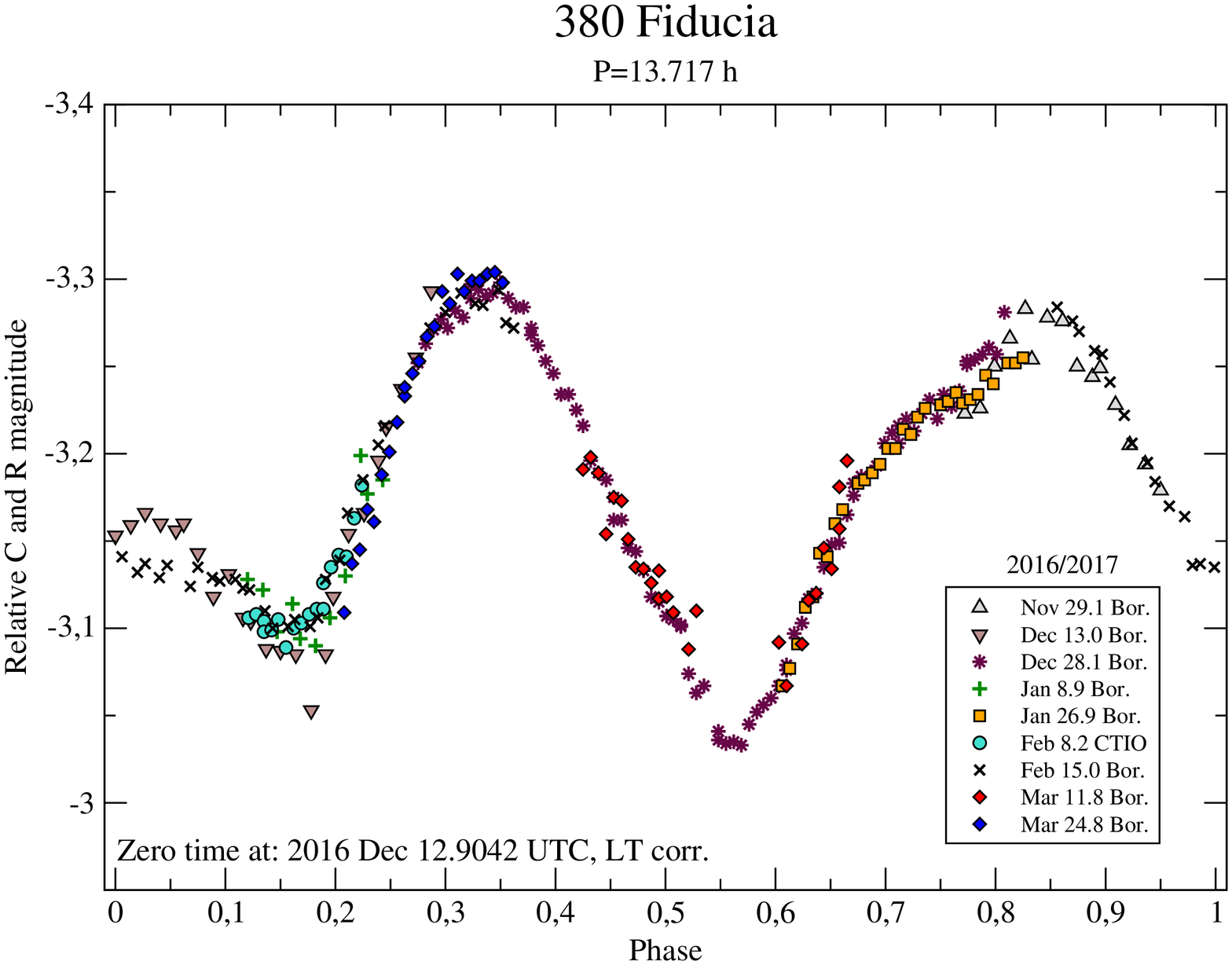} 
\captionof{figure}{Composite lightcurve of (380) Fiducia from the years 2016-2017.}
\label{380composit2017}
&
\includegraphics[width=0.35\textwidth,angle=270]{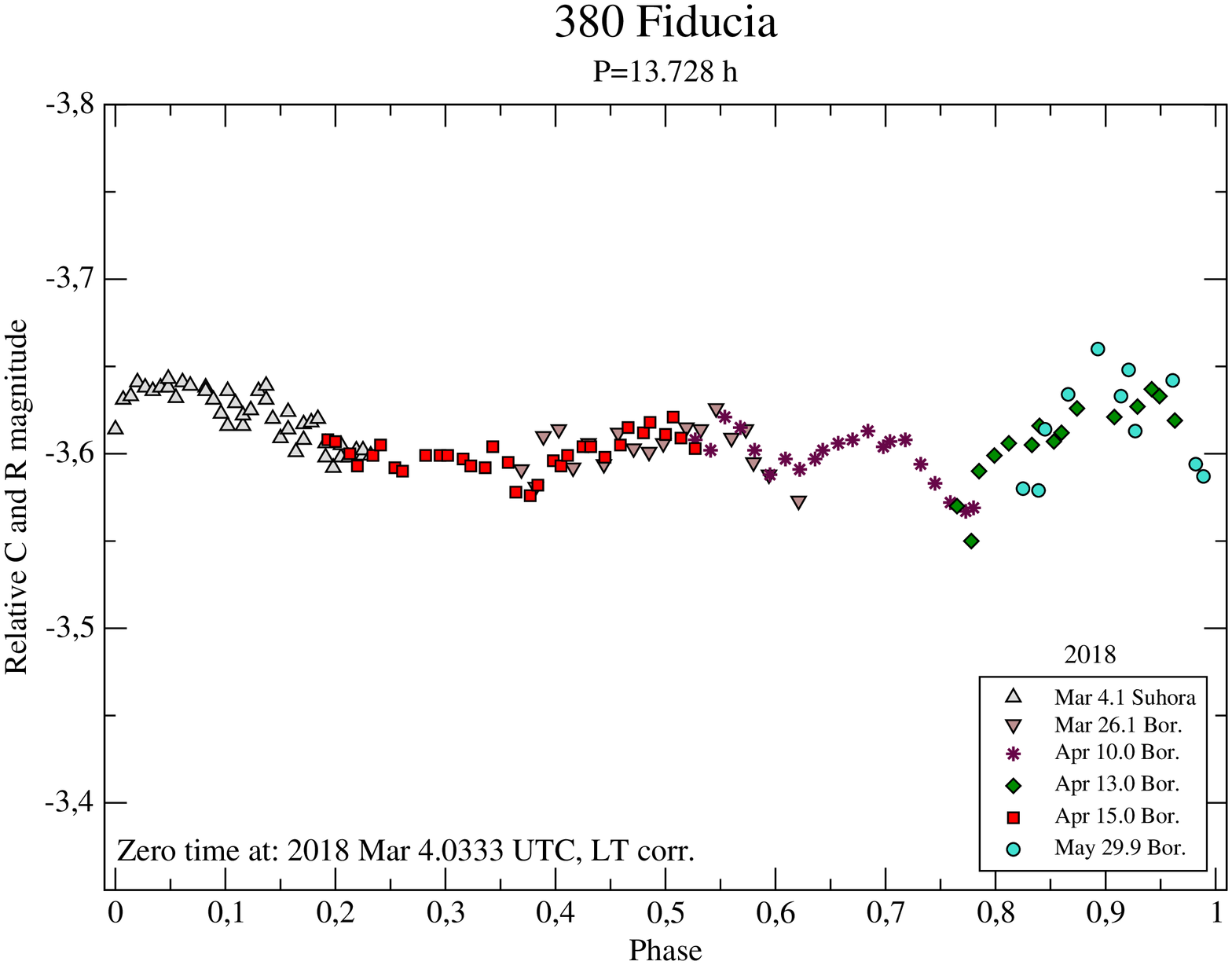} 
\captionof{figure}{Composite lightcurve of (380) Fiducia from the year 2018.}
\label{380composit2018}
\\
\includegraphics[width=0.35\textwidth,angle=270]{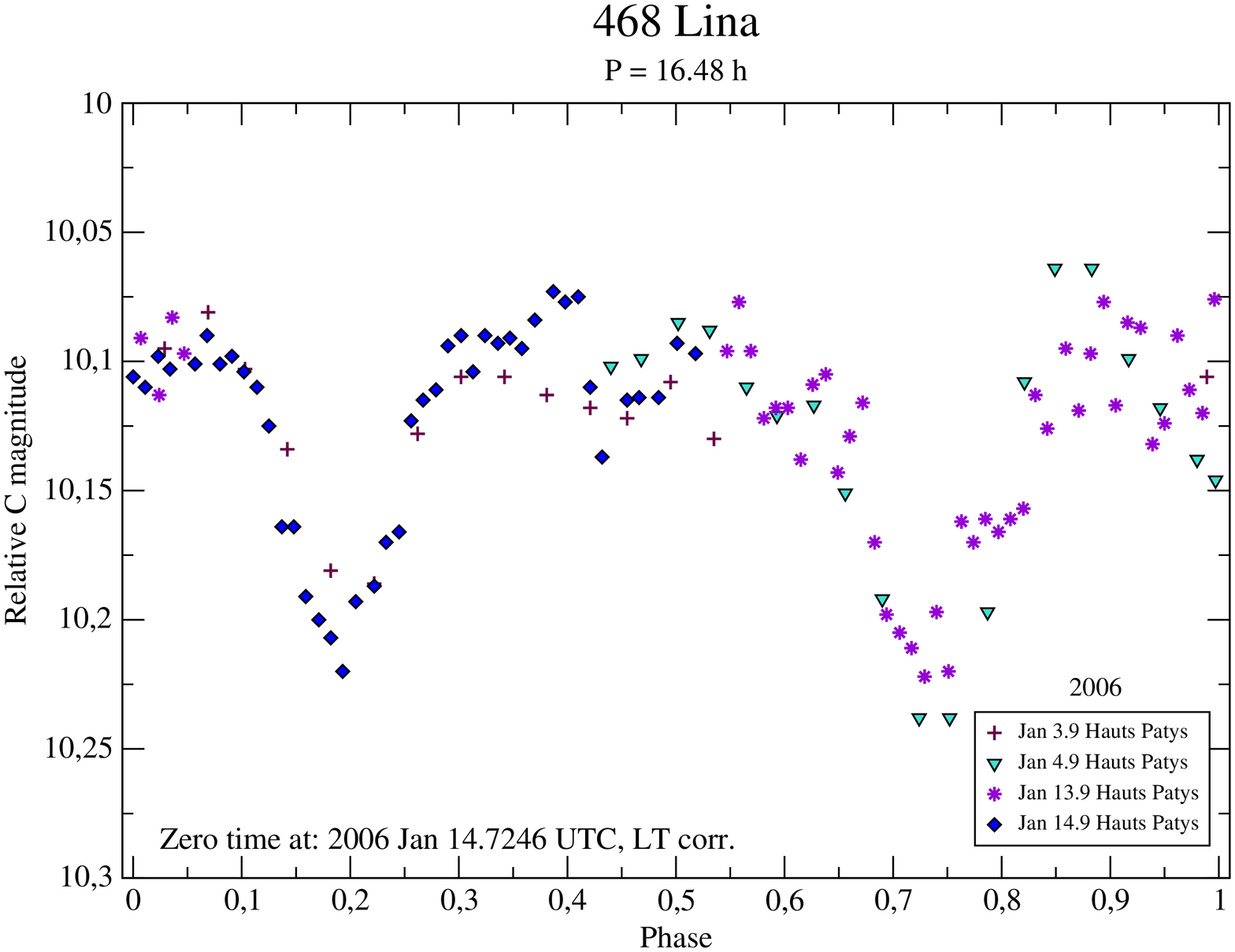} 
\captionof{figure}{Composite lightcurve of (468) Lina from the year 2006.}
\label{468composit2006}
&
\includegraphics[width=0.35\textwidth,angle=270]{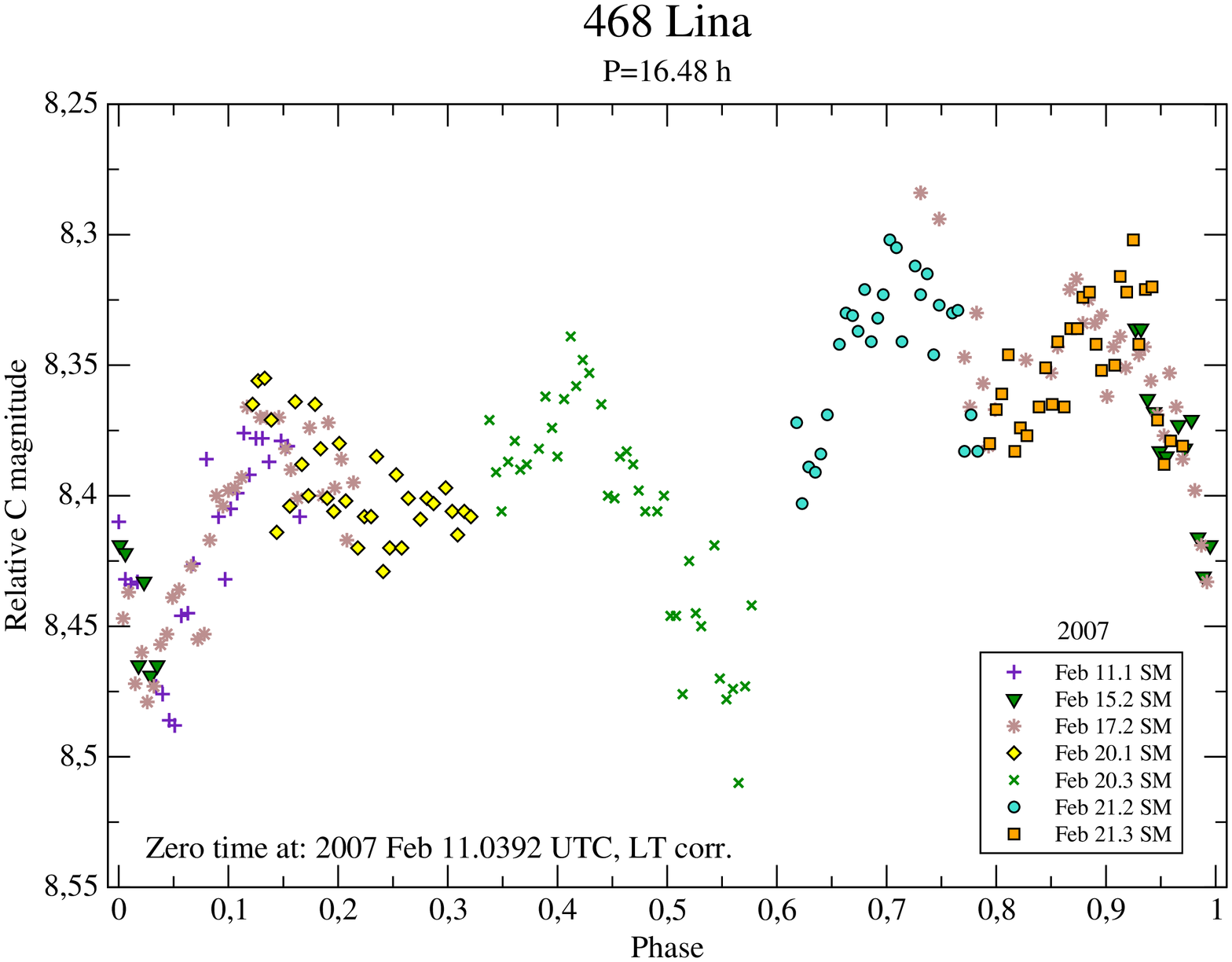} 
\captionof{figure}{Composite lightcurve of (468) Lina from the year 2007.}
\label{468composit2007}
\\
\end{tabularx}
    \end{table*}%

\clearpage
\vspace{0.5cm}

    \begin{table*}[ht]
    \centering
\begin{tabularx}{\linewidth}{XX}
\includegraphics[width=0.35\textwidth,angle=270]{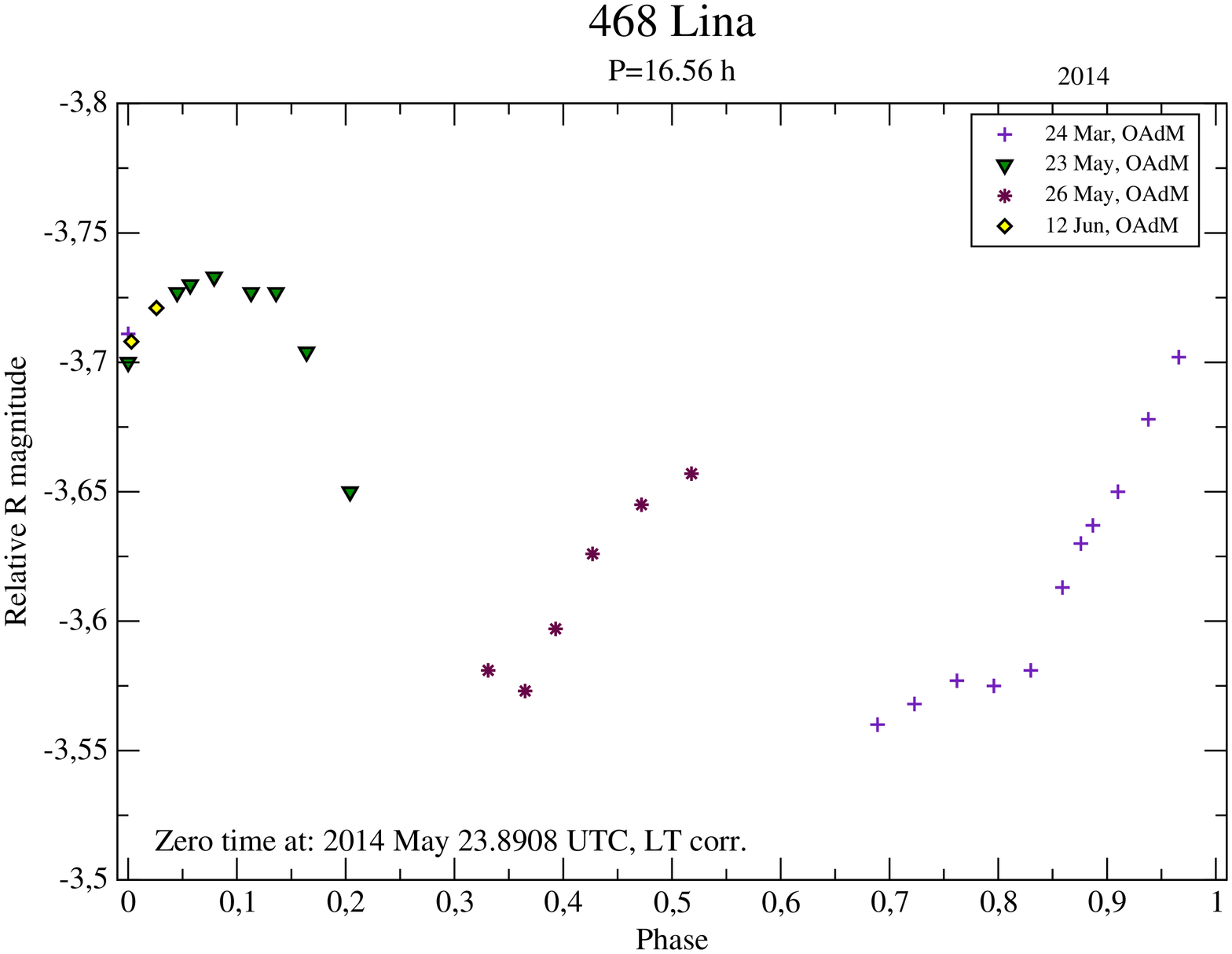} 
\captionof{figure}{Composite lightcurve of (468) Lina from the year 2014.}
\label{468composit2014}
&
\includegraphics[width=0.35\textwidth,angle=270]{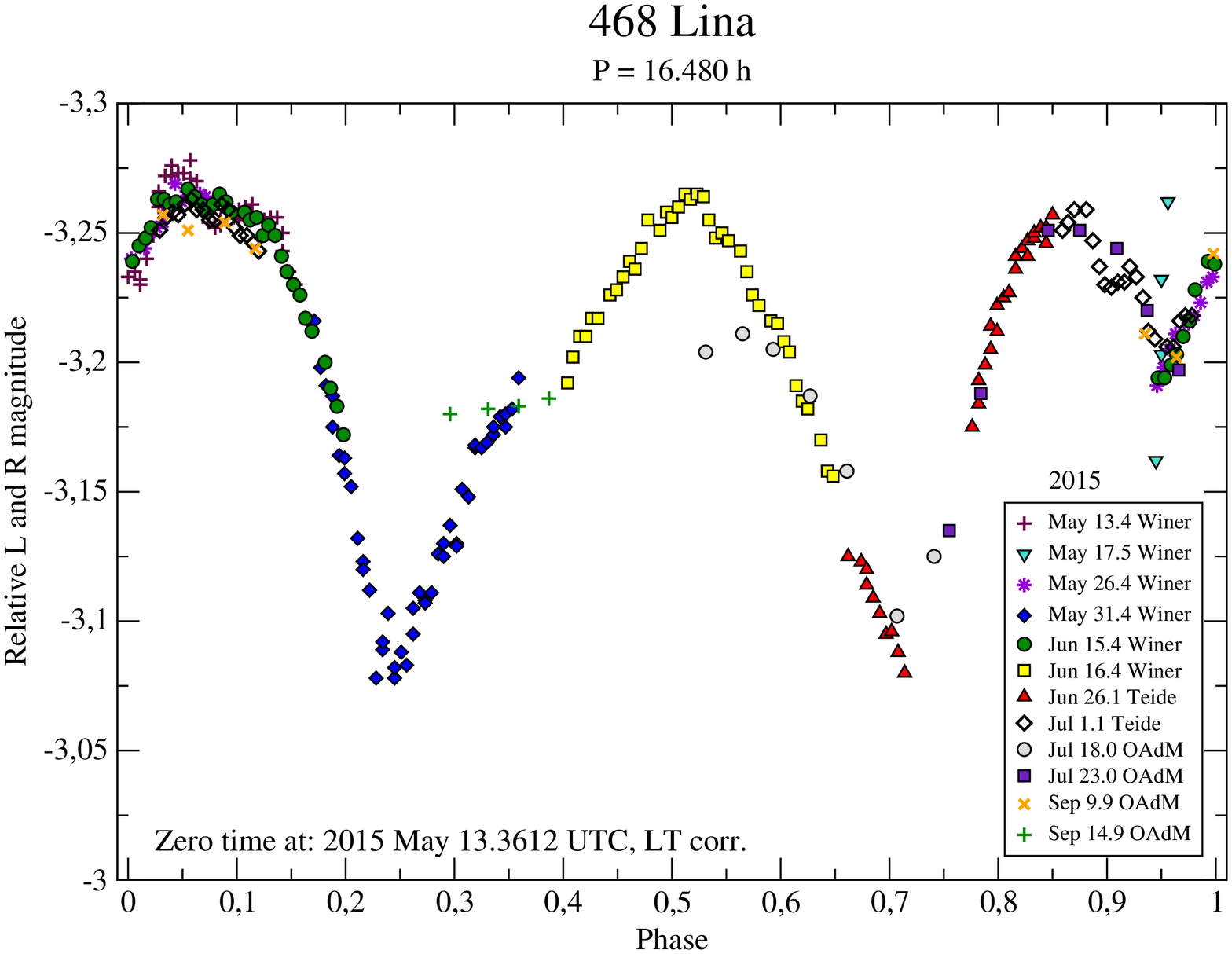} 
\captionof{figure}{Composite lightcurve of (468) Lina from the year 2015.}
\label{468composit2015}
\\
\includegraphics[width=0.35\textwidth,angle=270]{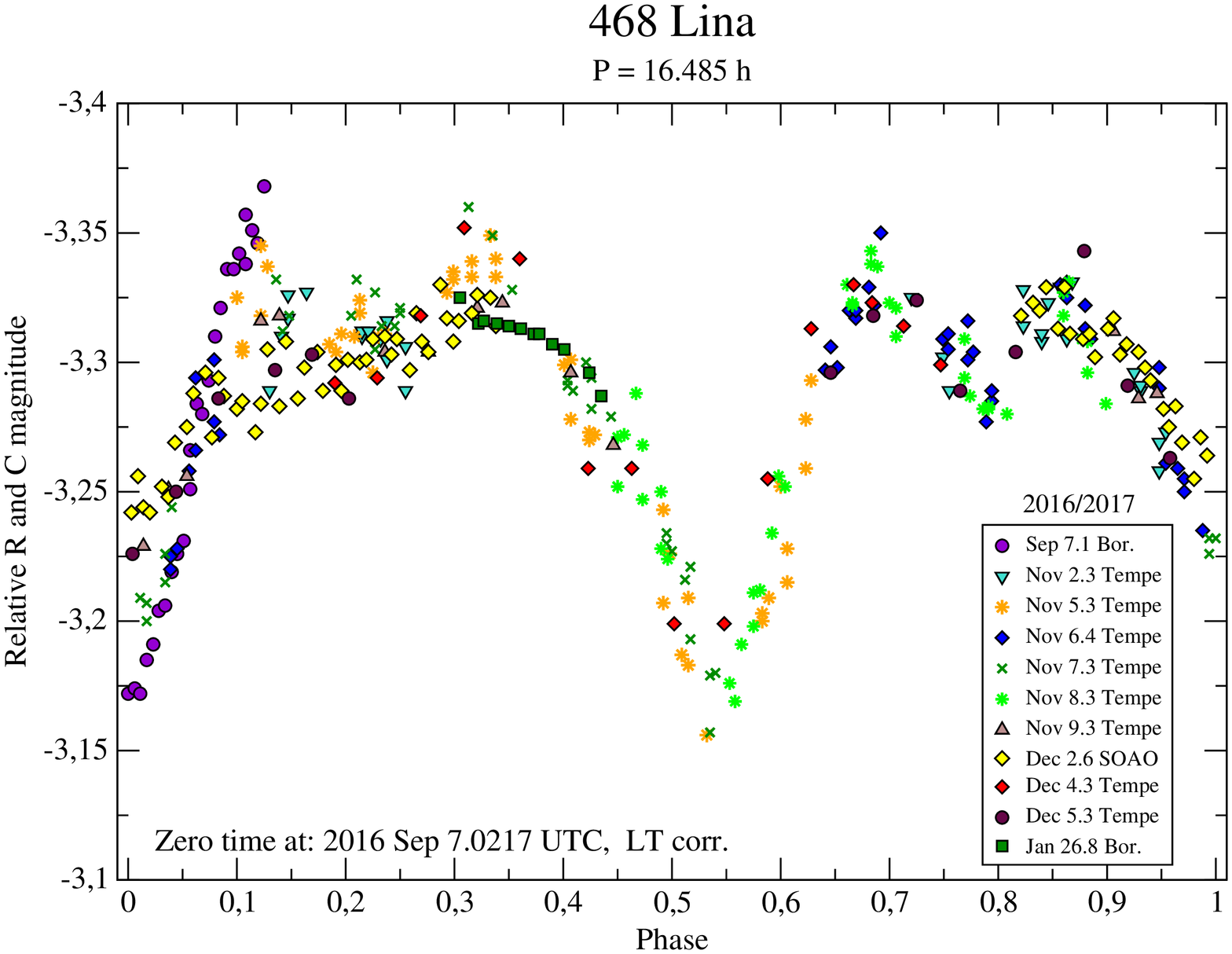} 
\captionof{figure}{Composite lightcurve of (468) Lina from the years 2016-2017.}
\label{468composit2016}
&
\includegraphics[width=0.35\textwidth,angle=270]{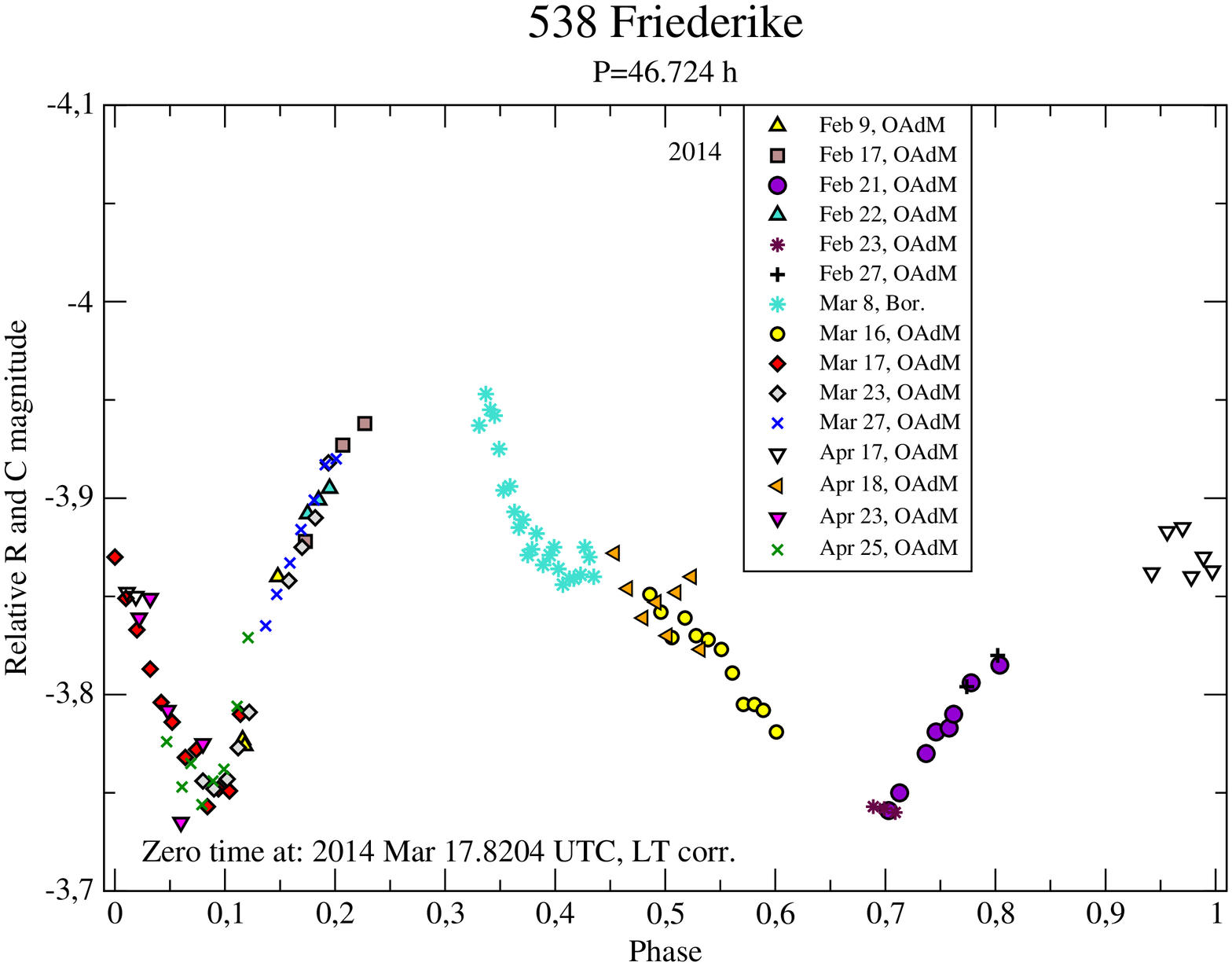} 
\captionof{figure}{Composite lightcurve of (538) Friederike from the year 2014.}
\label{538composit2014}
\\
\includegraphics[width=0.35\textwidth,angle=270]{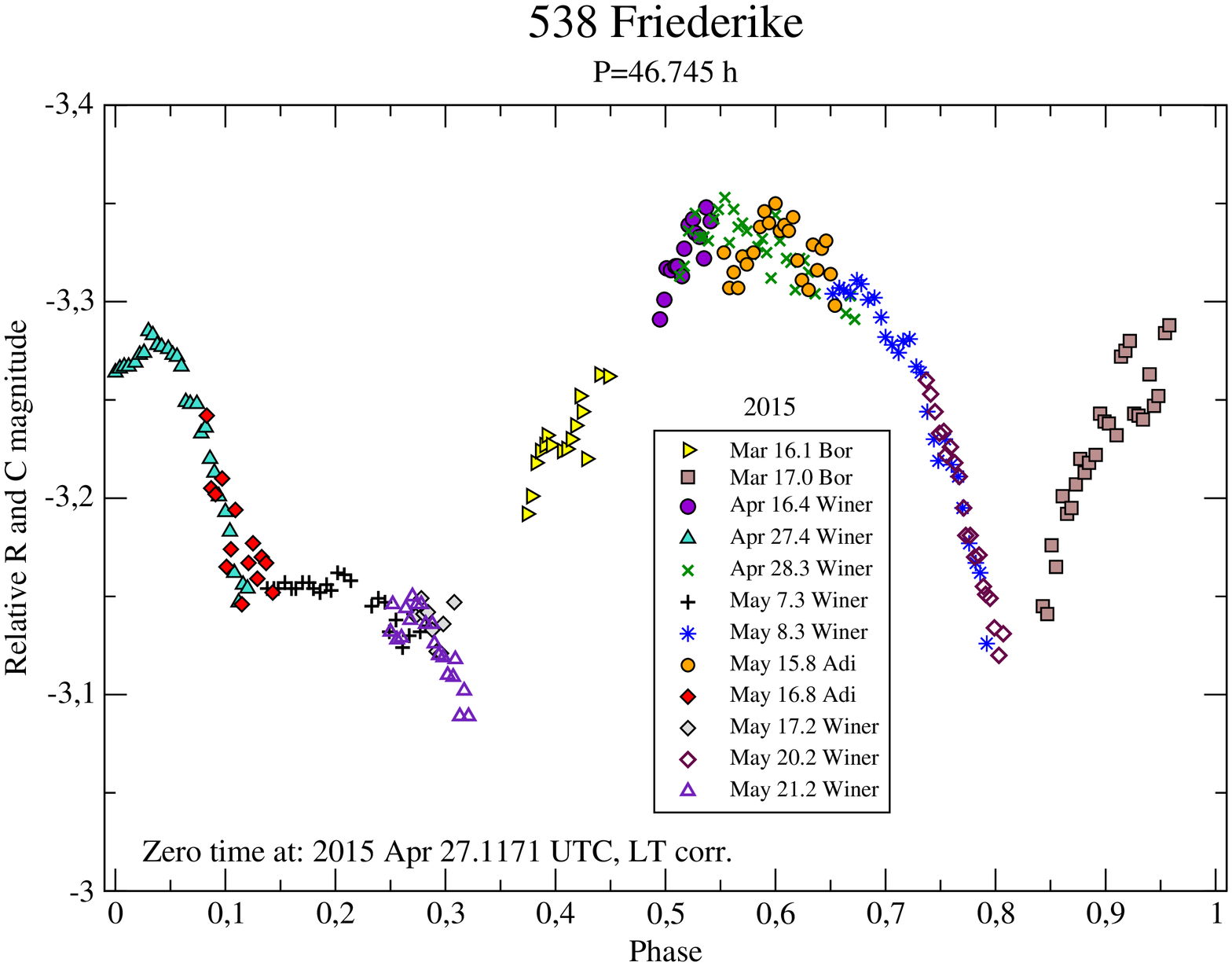} 
\captionof{figure}{Composite lightcurve of (538) Friederike from the year 2015.}
\label{538composit2015}
&
\includegraphics[width=0.35\textwidth,angle=270]{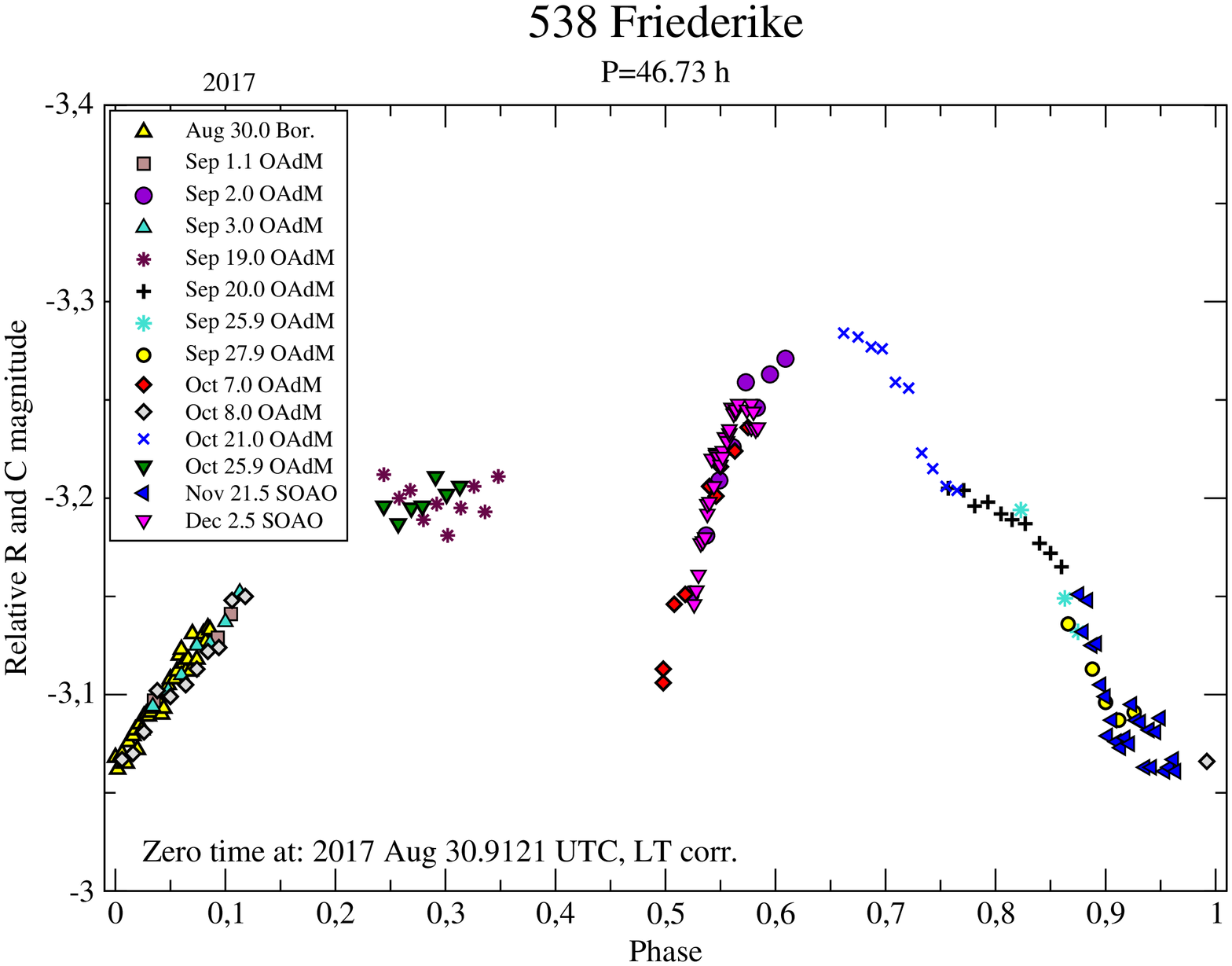} 
\captionof{figure}{Composite lightcurve of (538) Friederike from the year 2017.}
\label{538composit2017}
\\
\end{tabularx}
    \end{table*}%

\clearpage
\vspace{0.5cm}

    \begin{table*}[ht]
    \centering
\begin{tabularx}{\linewidth}{XX}
\includegraphics[width=0.35\textwidth,angle=270]{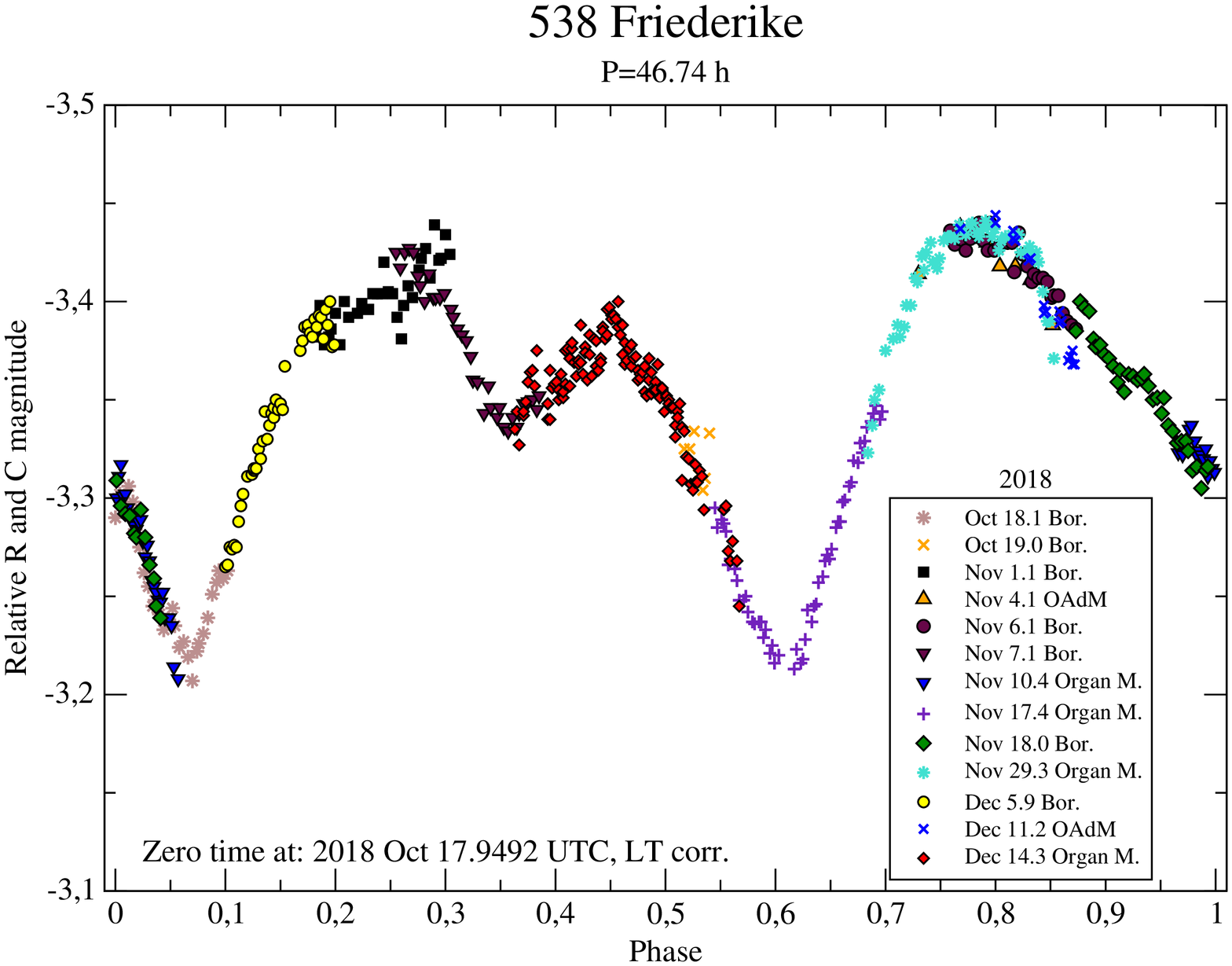} 
\captionof{figure}{Composite lightcurve of (538) Friederike from the year 2018.}
\label{538composit2018}
&
\includegraphics[width=0.35\textwidth,angle=270]{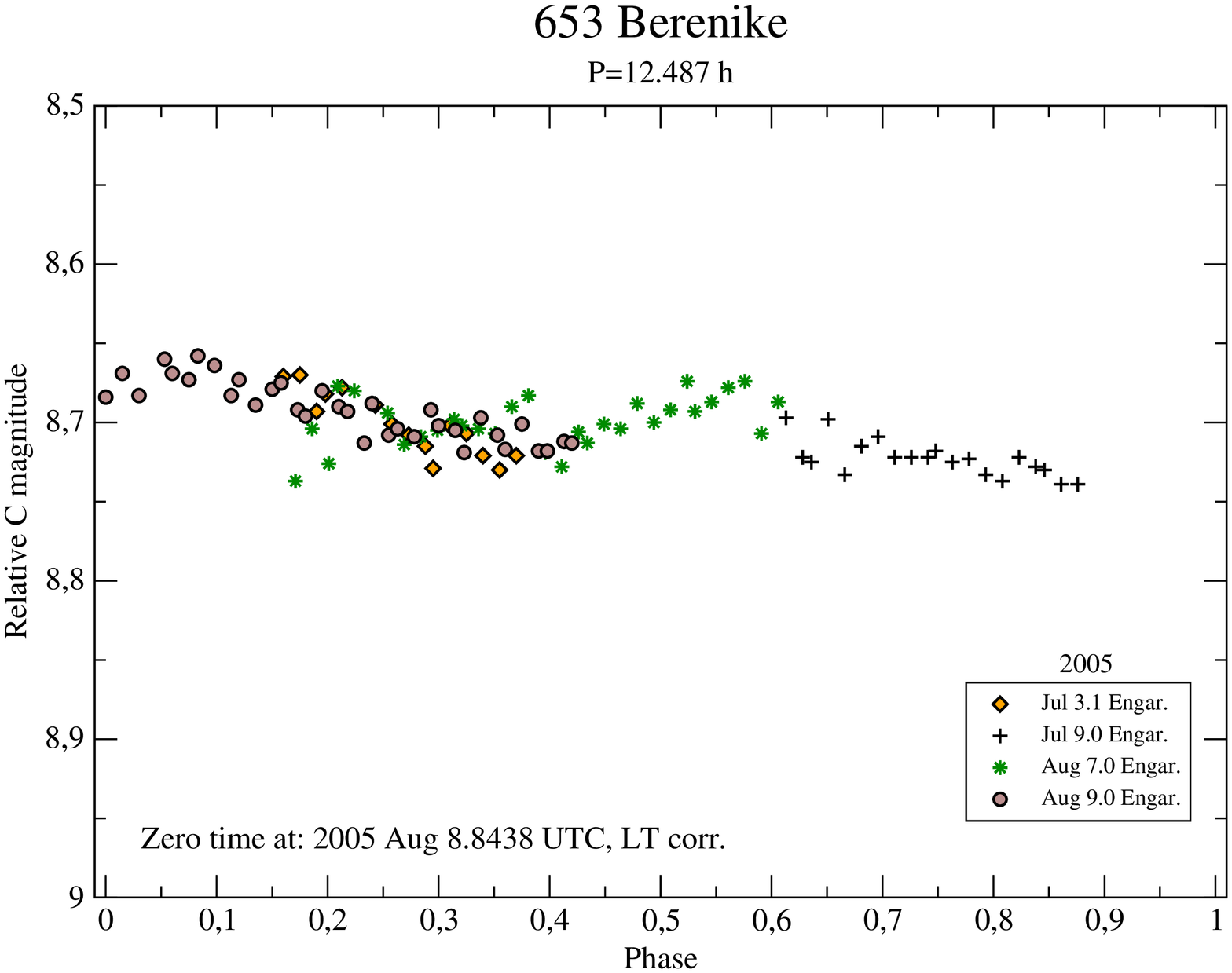} 
\captionof{figure}{Composite lightcurve of (653) Berenike from the year 2005.}
\label{653composit2005}
\\
\includegraphics[width=0.35\textwidth,angle=270]{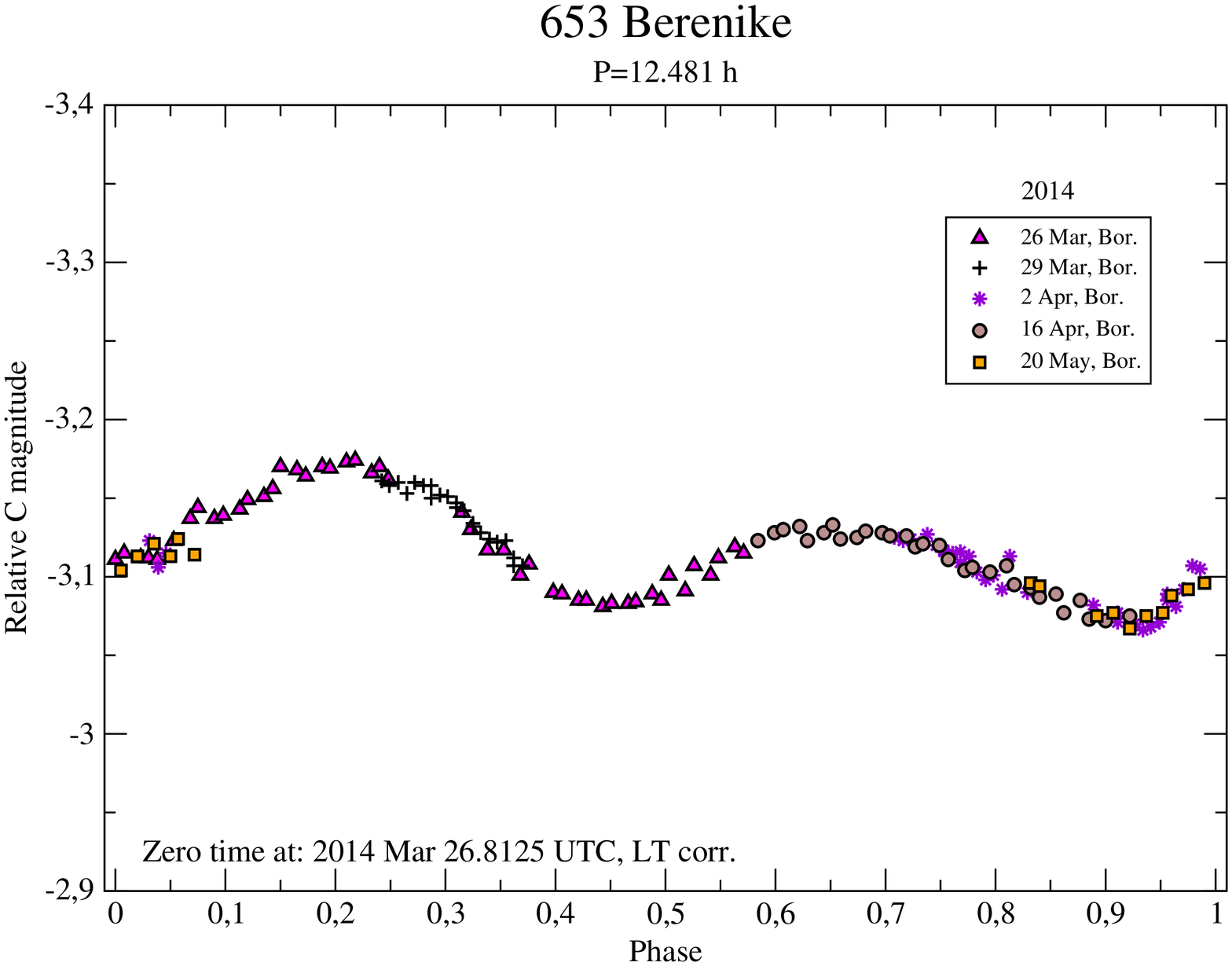} 
\captionof{figure}{Composite lightcurve of (653) Berenike from the year 2014.}
\label{653composit2014}
&
\includegraphics[width=0.35\textwidth,angle=270]{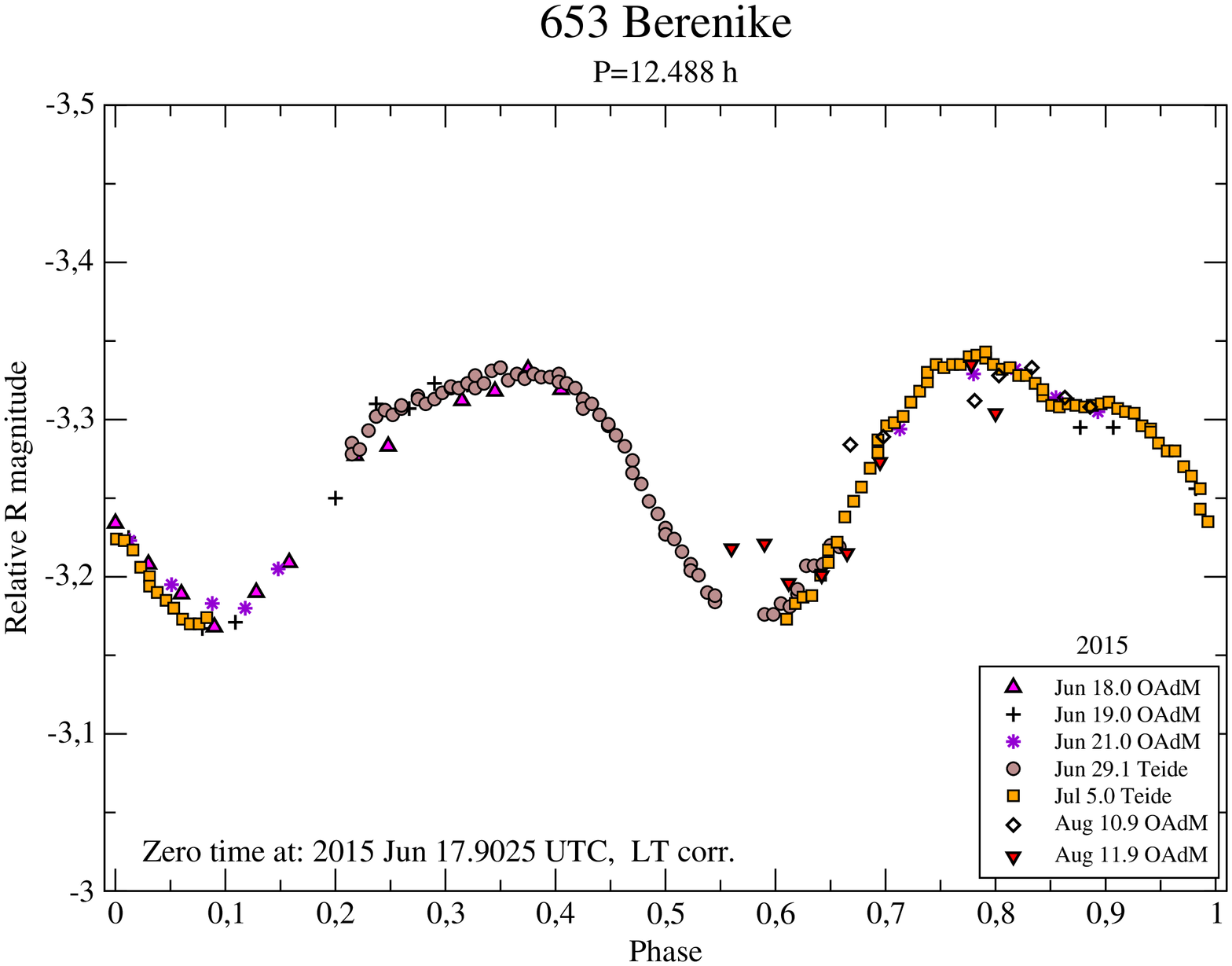} 
\captionof{figure}{Composite lightcurve of (653) Berenike from the year 2015.}
\label{653composit2015}
\\
\includegraphics[width=0.35\textwidth,angle=270]{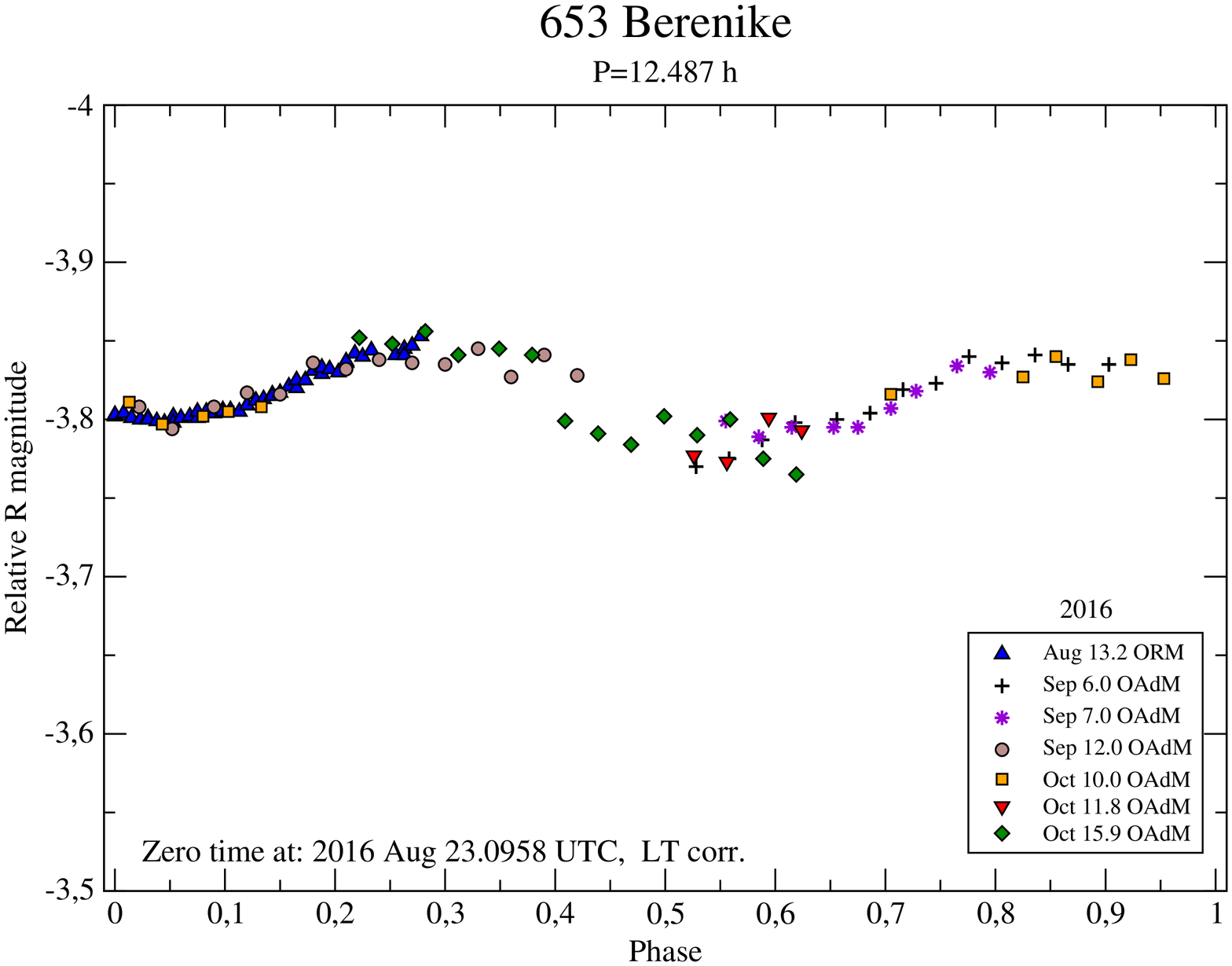} 
\captionof{figure}{Composite lightcurve of (653) Berenike from the year 2016.}
\label{653composit2016}
&
\includegraphics[width=0.35\textwidth,angle=270]{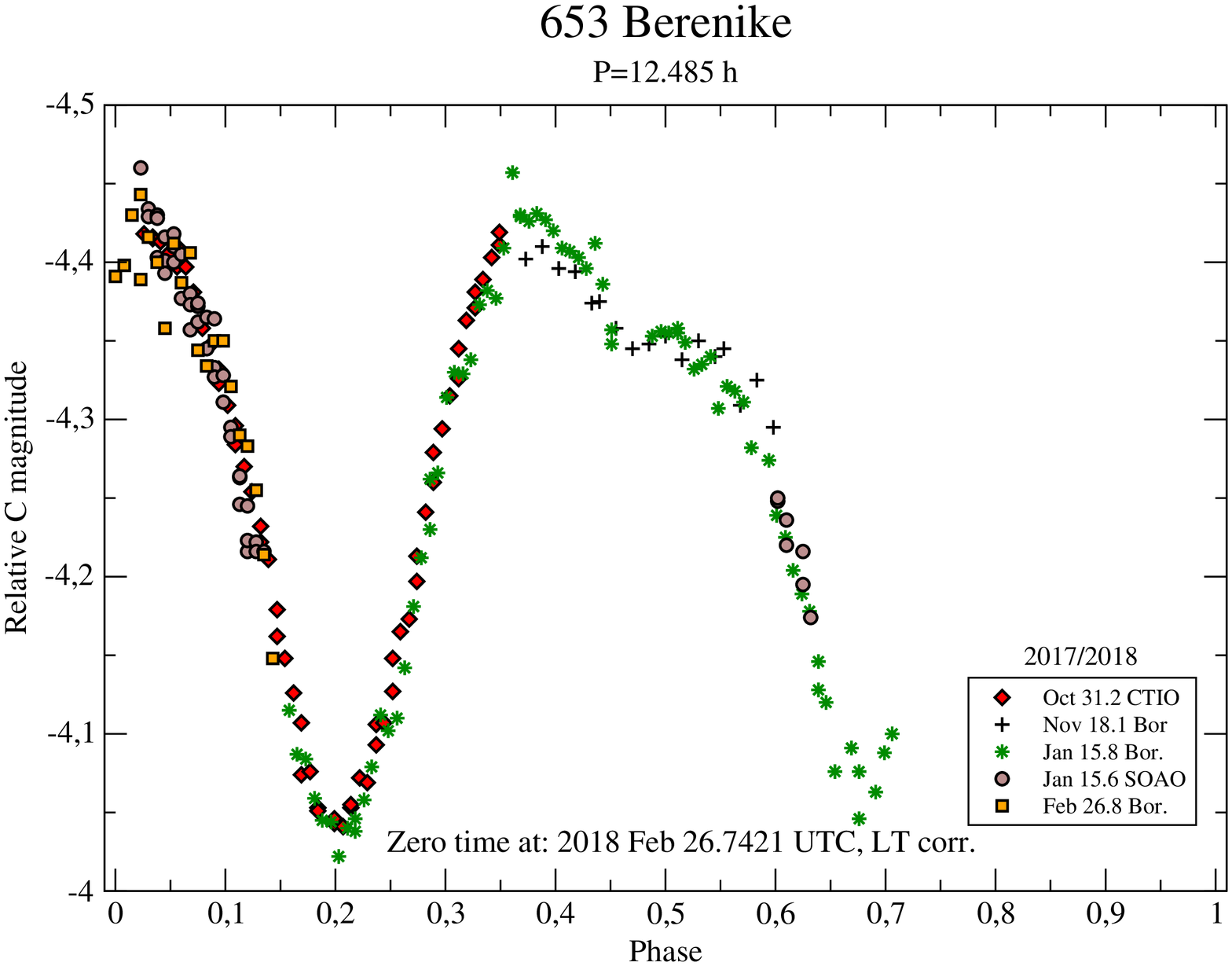} 
\captionof{figure}{Composite lightcurve of (653) Berenike from the years 2017-2018.}
\label{653composit2018}
\\
\end{tabularx}
    \end{table*}%

\clearpage
\vspace{0.5cm}

    \begin{table*}[ht]
    \centering
\begin{tabularx}{\linewidth}{XX}
\includegraphics[width=0.35\textwidth,angle=270]{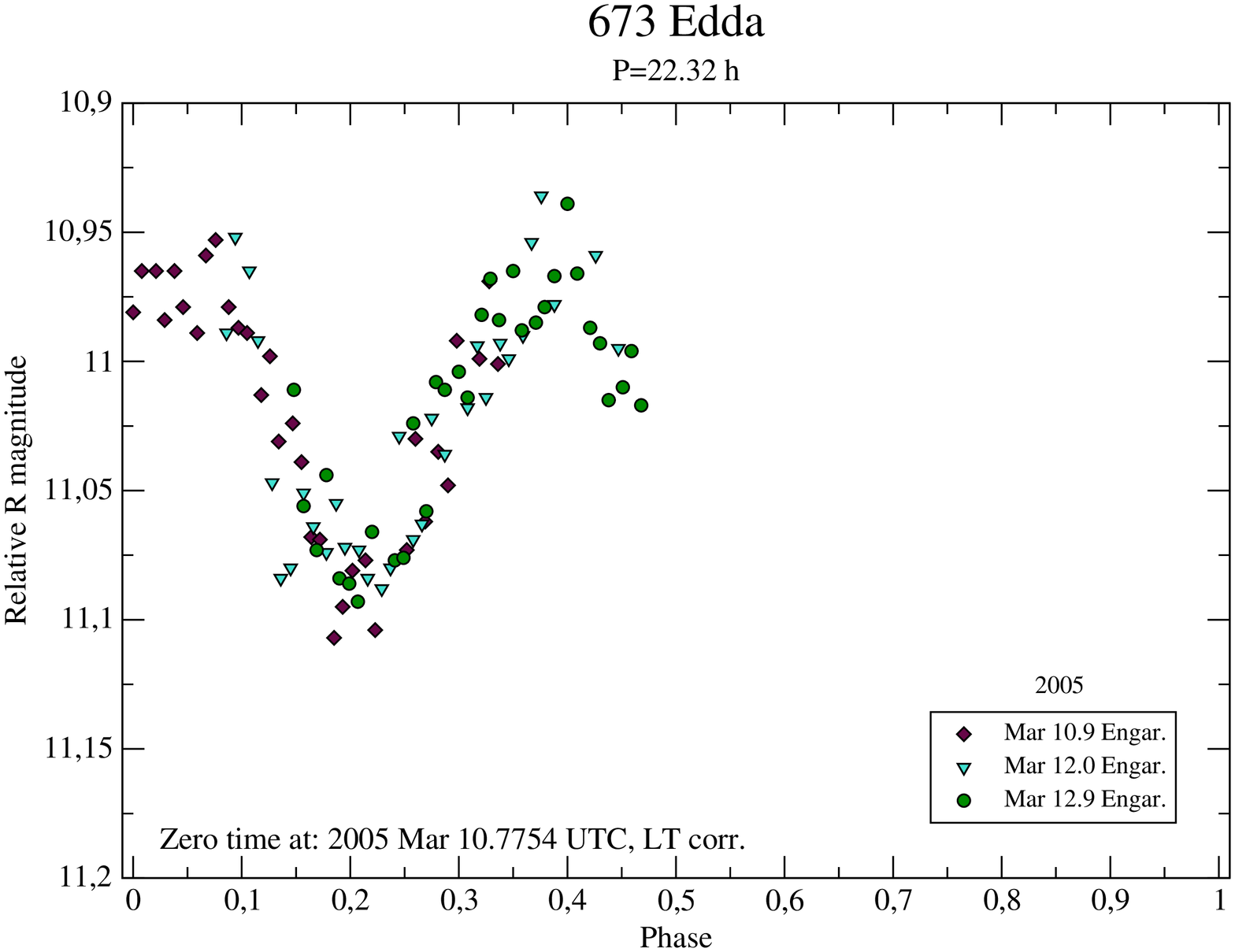} 
\captionof{figure}{Composite lightcurve of (673) Edda from the year 2005.}
\label{673composit2005}
&
\includegraphics[width=0.35\textwidth,angle=270]{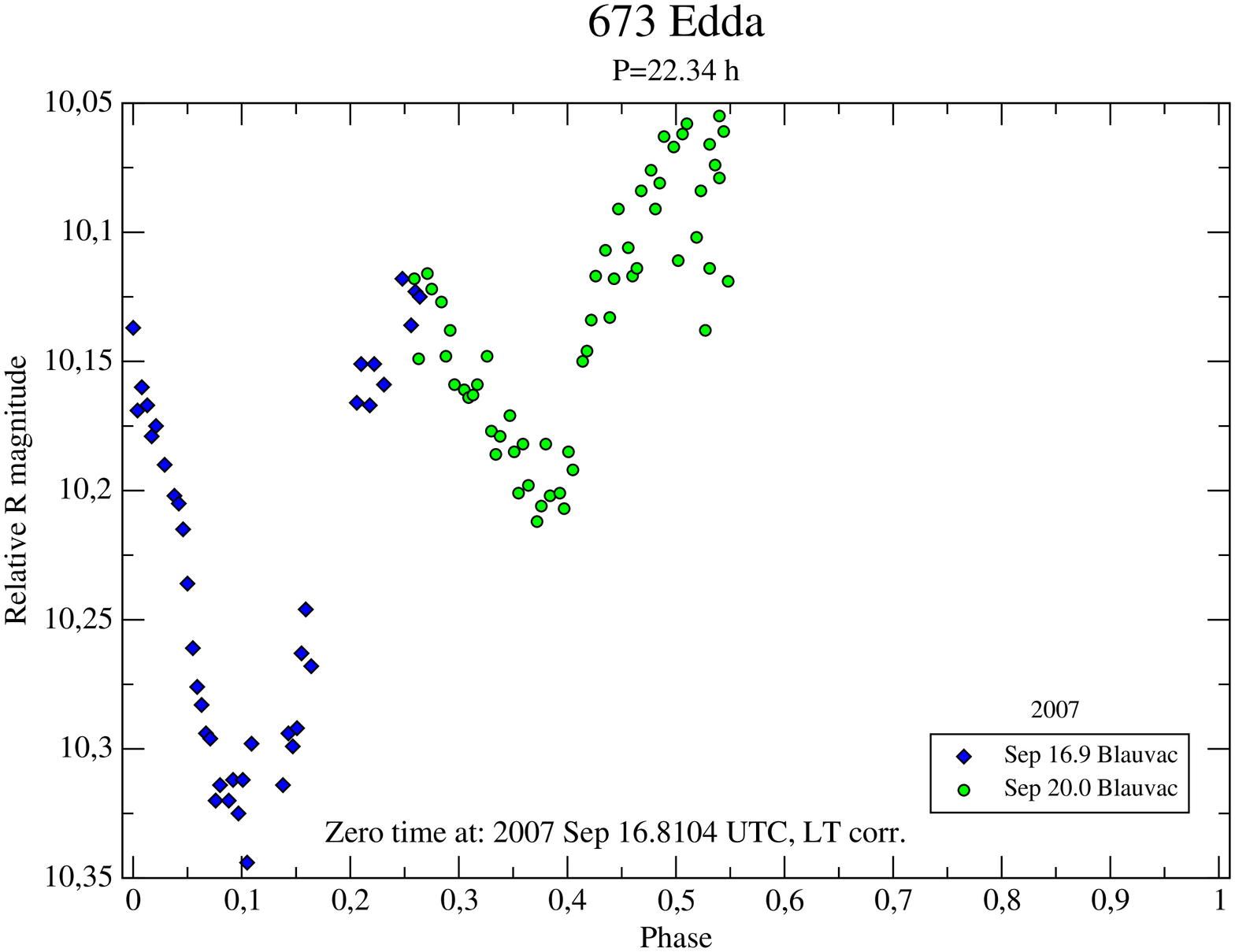} 
\captionof{figure}{Composite lightcurve of (673) Edda from the year 2007.}
\label{673composit2007}
\\
\includegraphics[width=0.35\textwidth,angle=270]{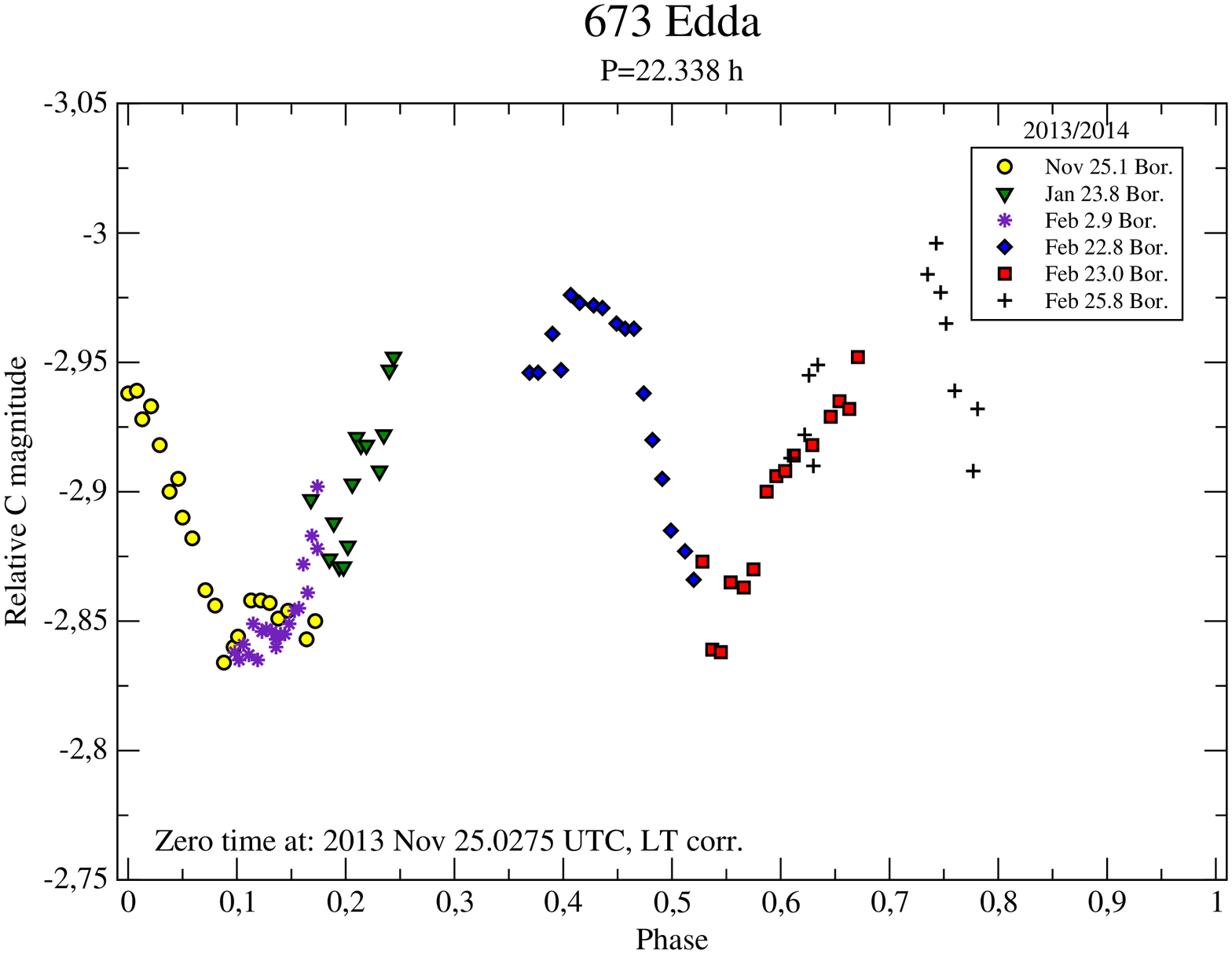} 
\captionof{figure}{Composite lightcurve of (673) Edda from the years 2013-2014.}
\label{673composit2014}
&
\includegraphics[width=0.35\textwidth,angle=270]{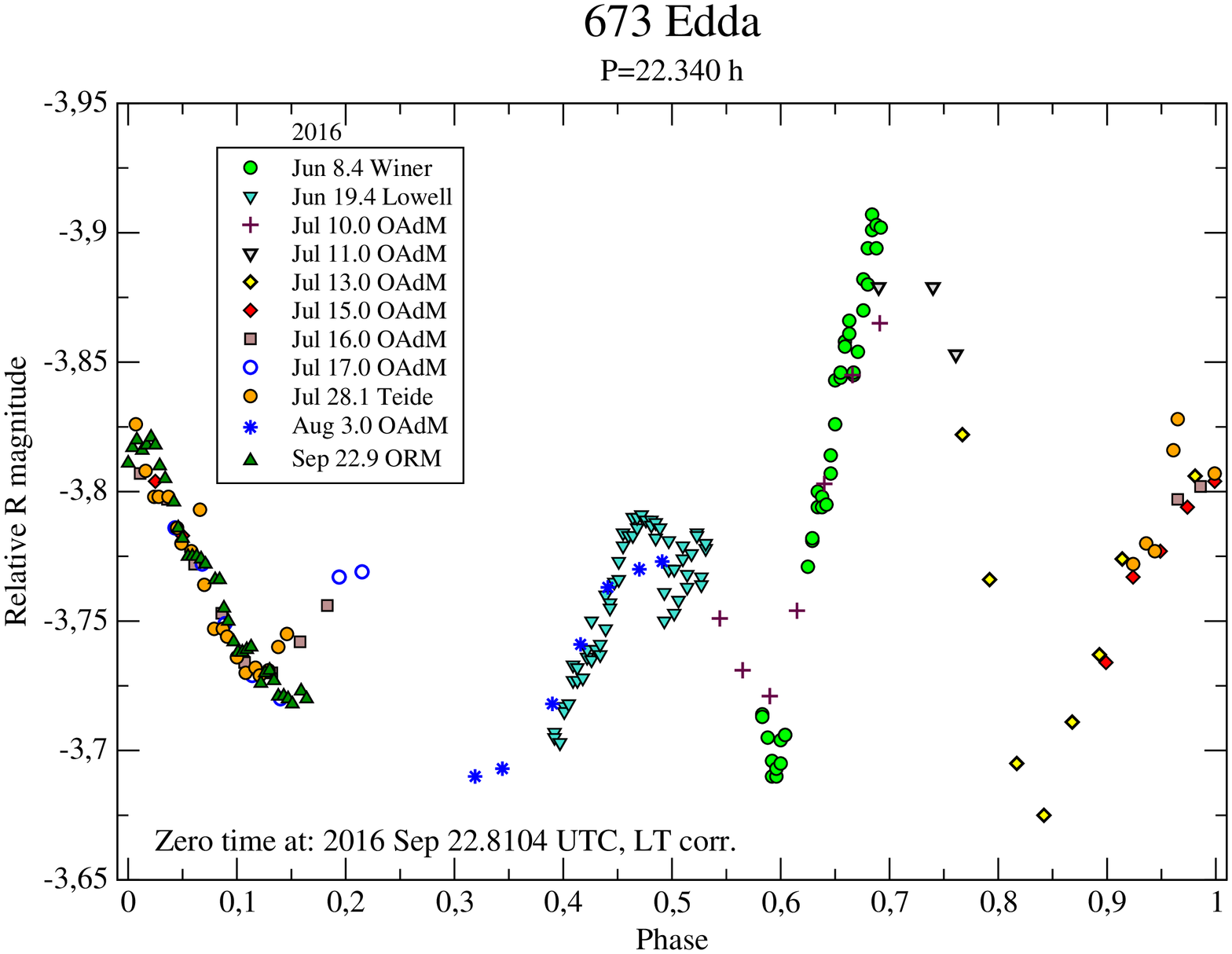} 
\captionof{figure}{Composite lightcurve of (673) Edda from the year 2016.}
\label{673composit2016}
\\
\includegraphics[width=0.35\textwidth,angle=270]{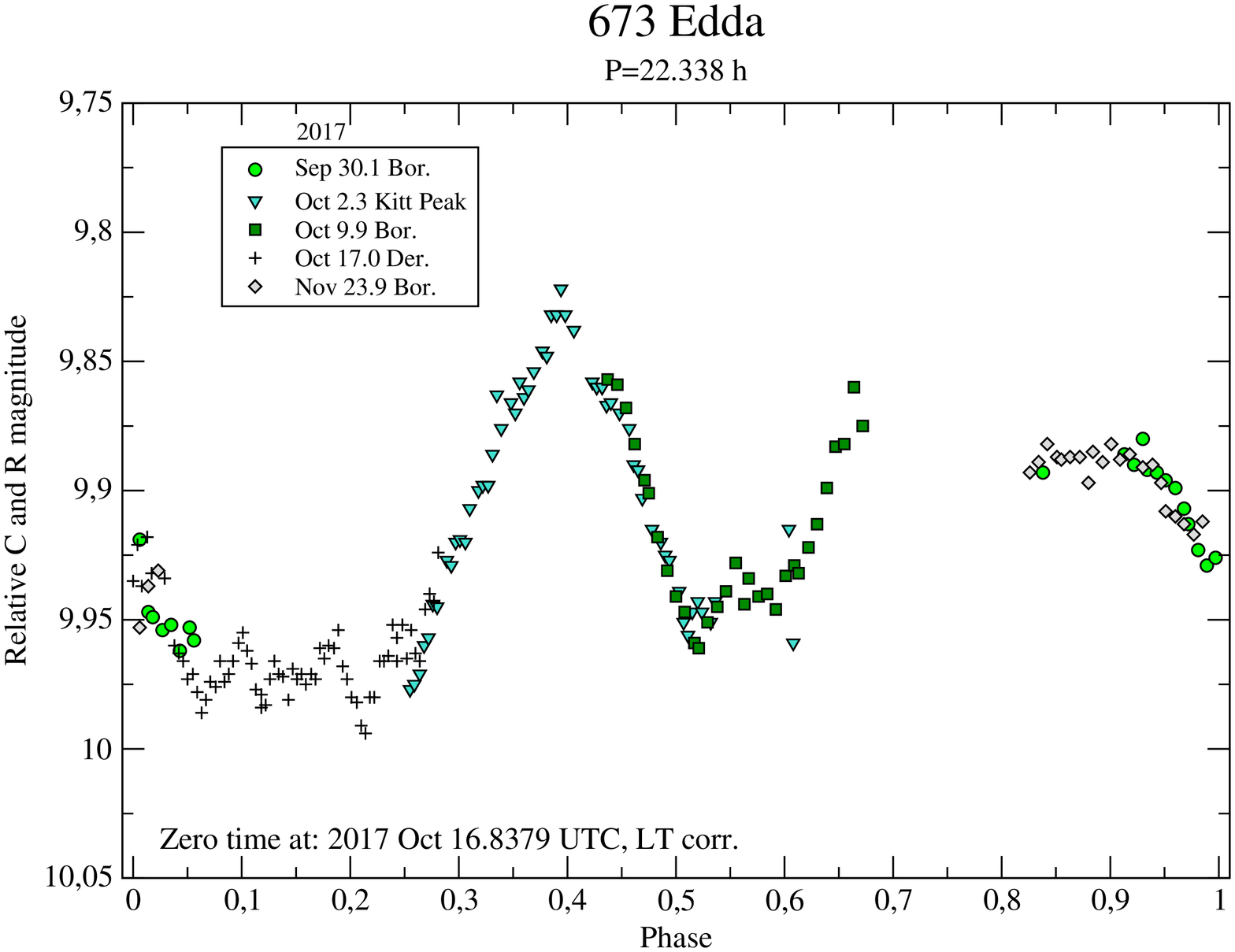} 
\captionof{figure}{Composite lightcurve of (673) Edda from the year 2017.}
\label{673composit2017}
&
\includegraphics[width=0.35\textwidth,angle=270]{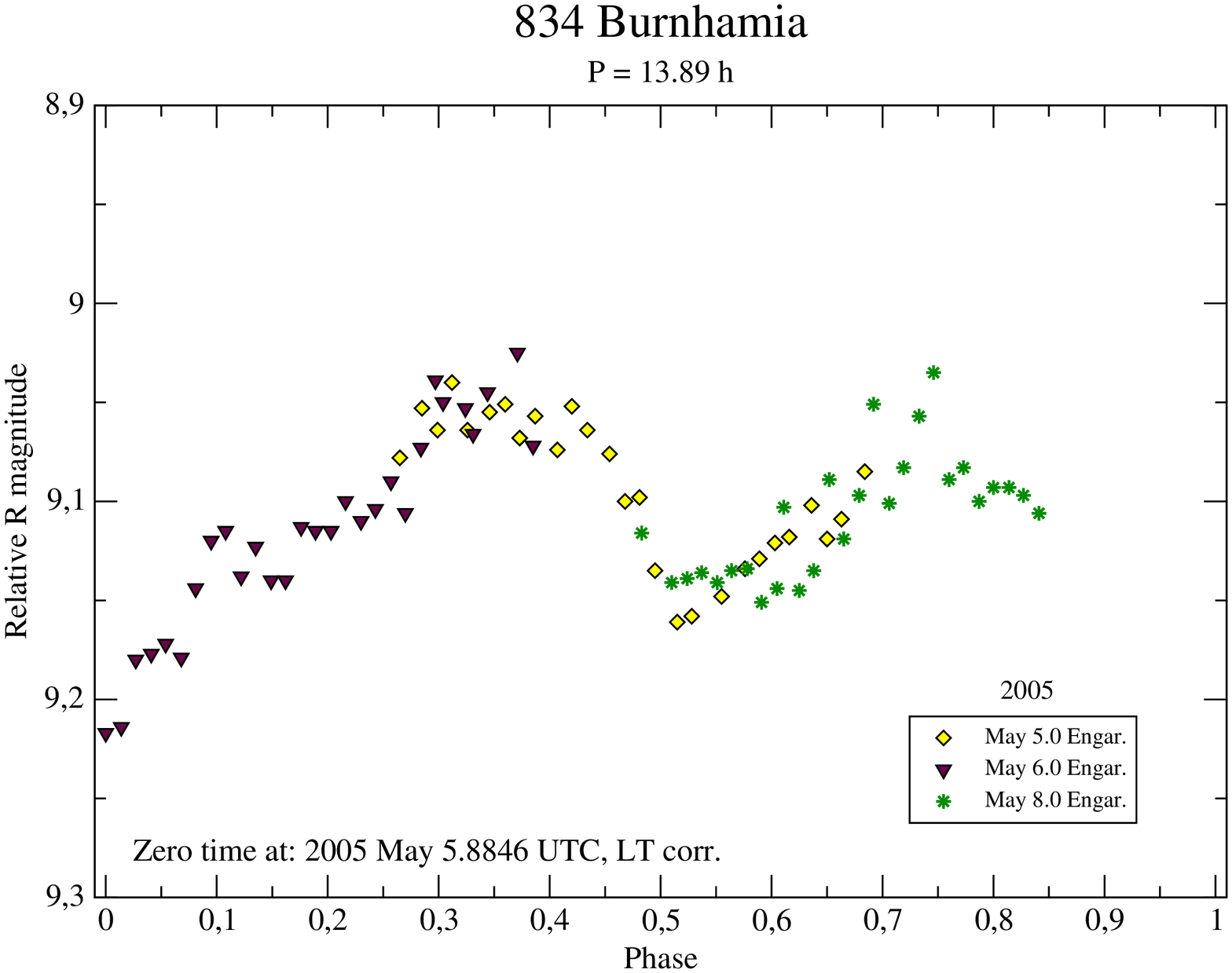} 
\captionof{figure}{Composite lightcurve of (834) Burnhamia from the year 2005.}
\label{834composit2005}
\\
\end{tabularx}
    \end{table*}%

\clearpage
\vspace{0.5cm}

    \begin{table*}[ht]
    \centering
\begin{tabularx}{\linewidth}{XX}
\includegraphics[width=0.35\textwidth,angle=270]{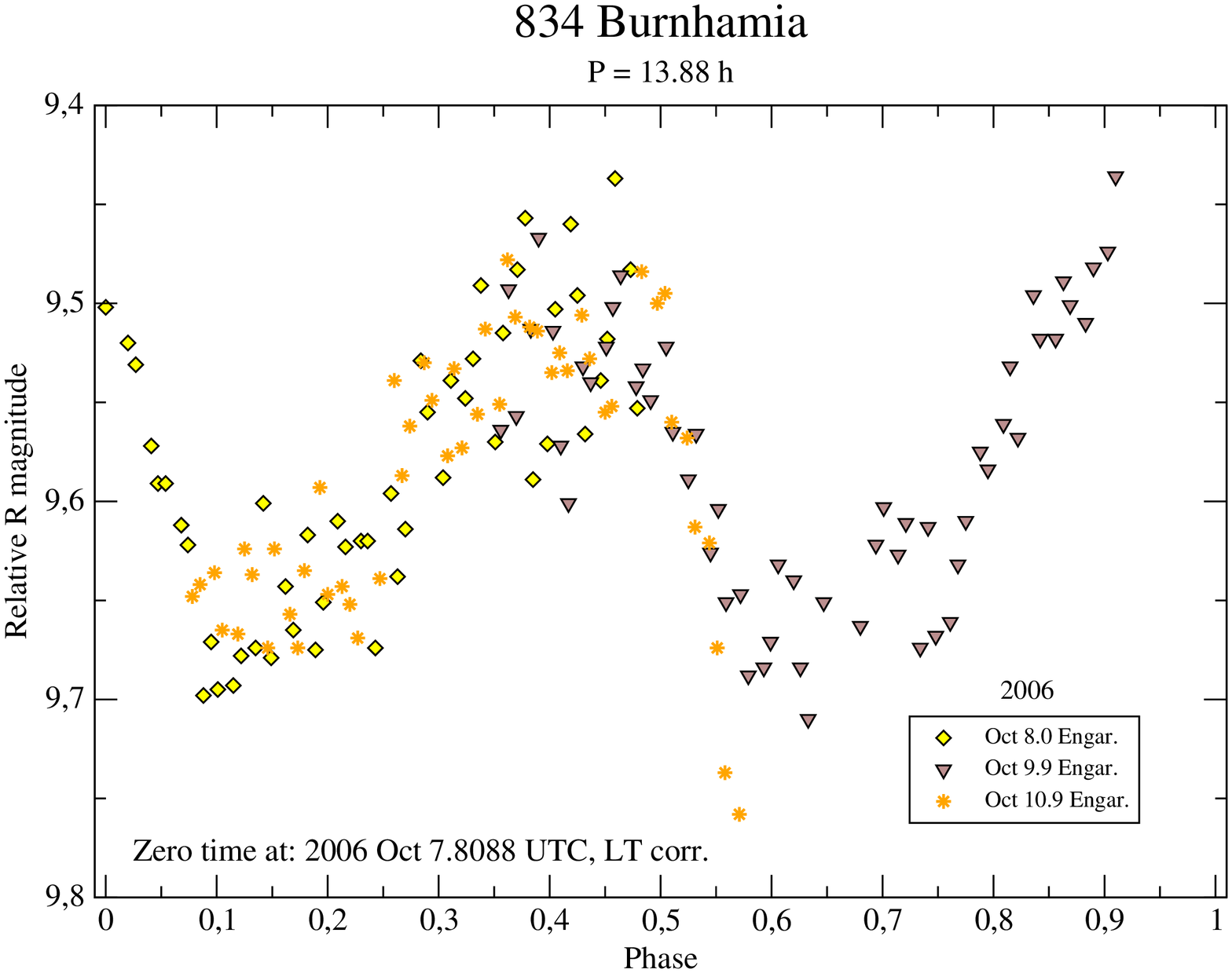} 
\captionof{figure}{Composite lightcurve of (834) Burnhamia from the year 2006.}
\label{834composit2006}
&
\includegraphics[width=0.35\textwidth,angle=270]{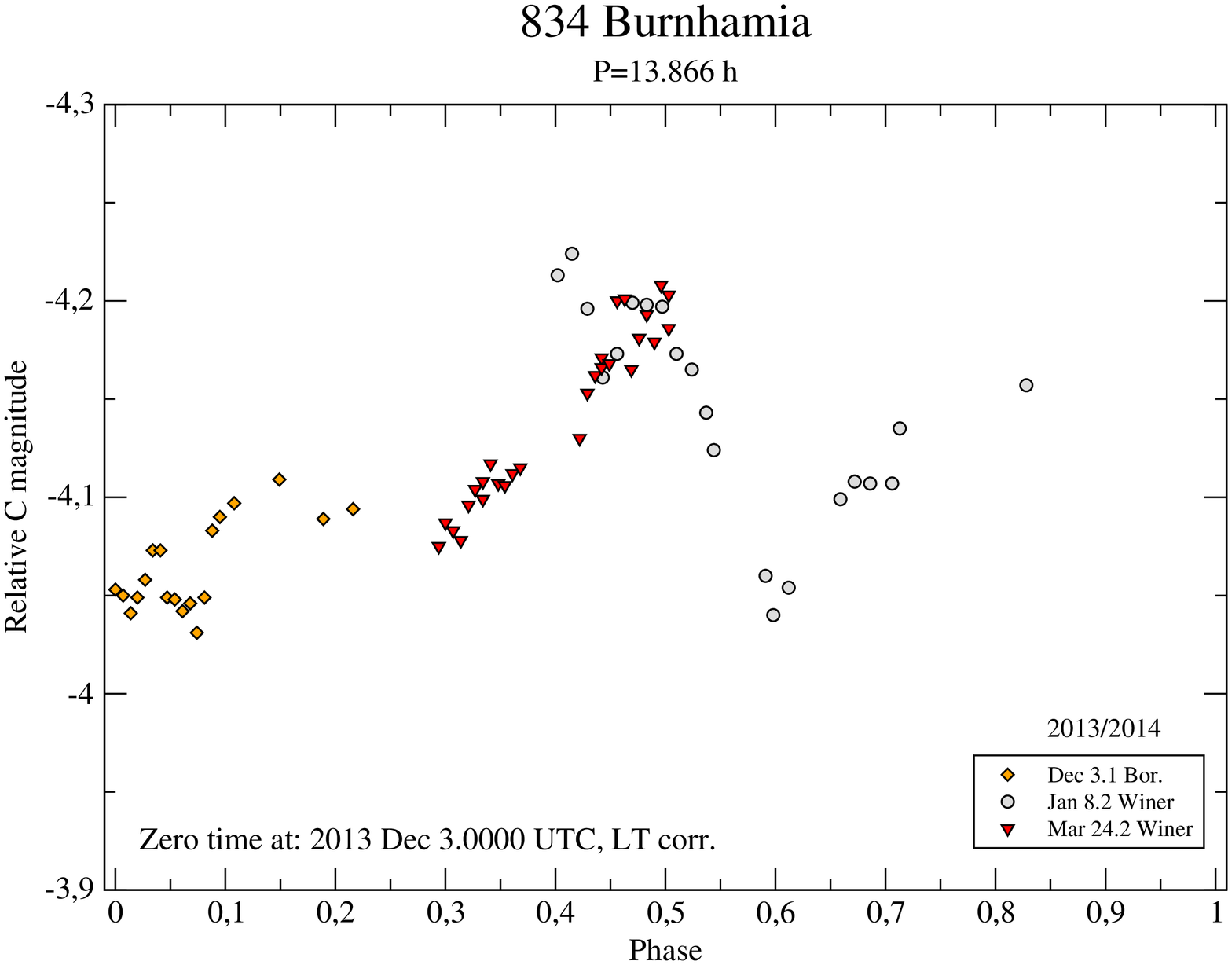} 
\captionof{figure}{Composite lightcurve of (834) Burnhamia from the years 2013-2014.}
\label{834composit2014}
\\
\includegraphics[width=0.35\textwidth,angle=270]{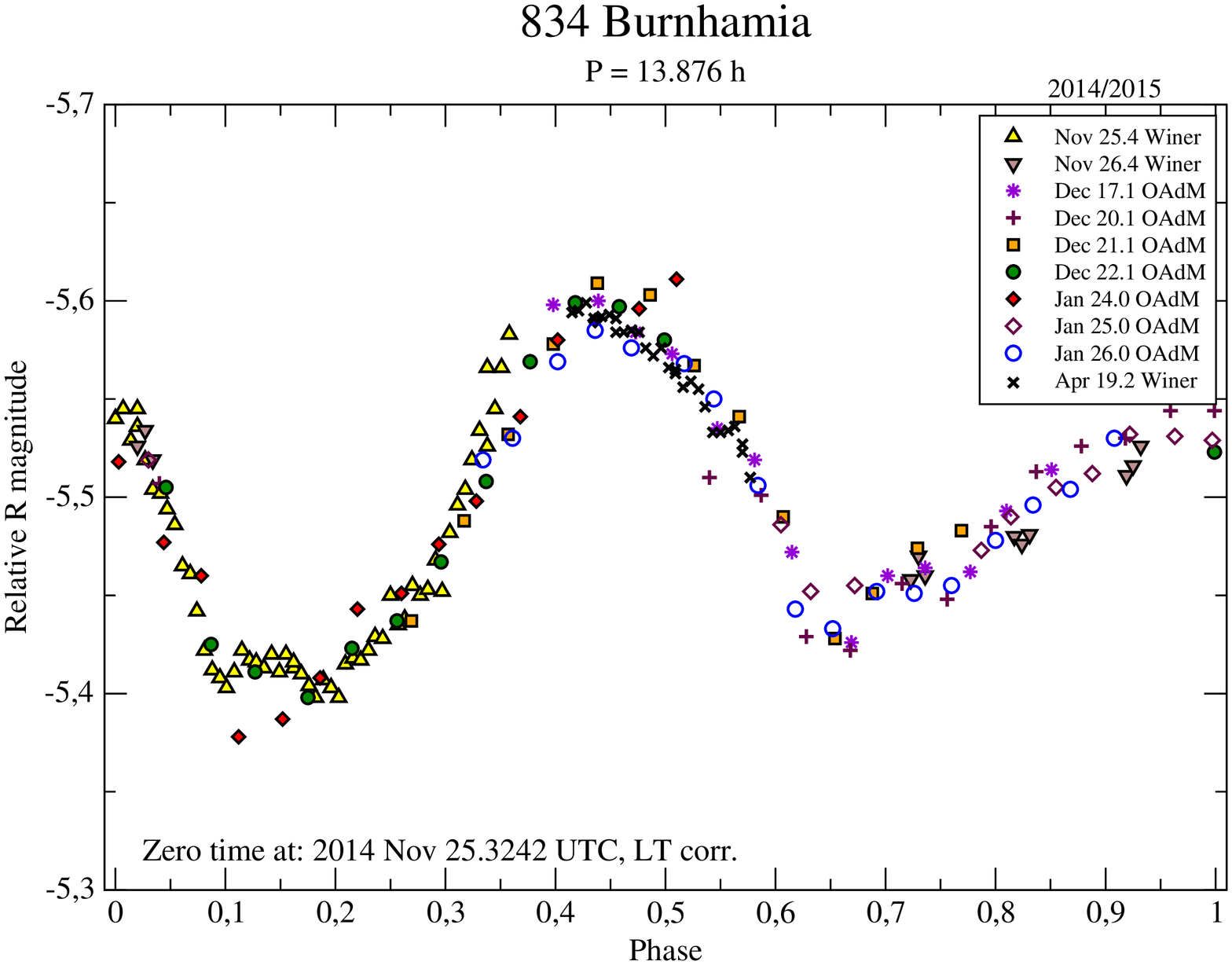} 
\captionof{figure}{Composite lightcurve of (834) Burnhamia from the years 2014-2015.}
\label{834composit2015}
&
\includegraphics[width=0.35\textwidth,angle=270]{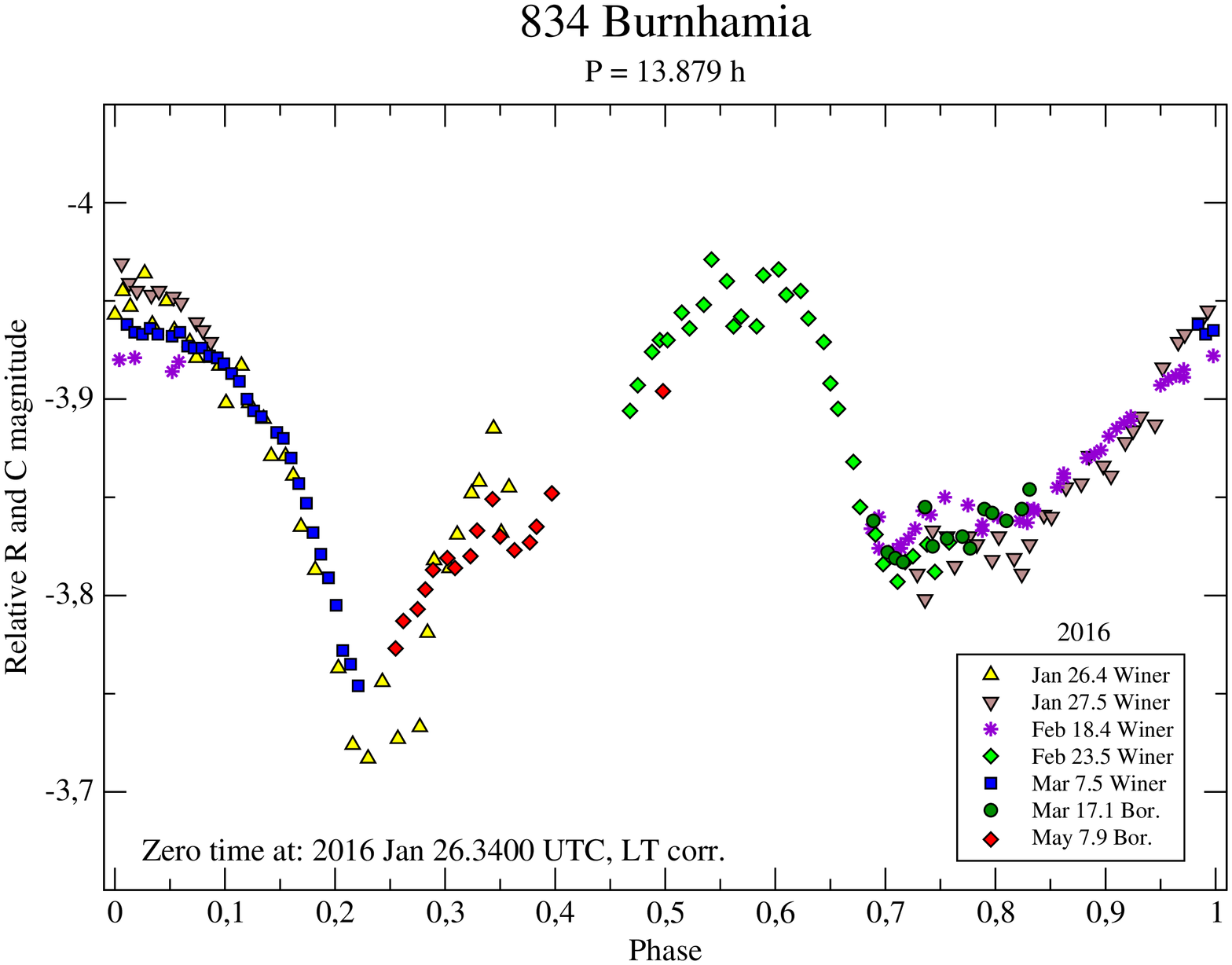} 
\captionof{figure}{Composite lightcurve of (834) Burnhamia from the year 2016.}
\label{834composit2016}
\\
\includegraphics[width=0.35\textwidth,angle=270]{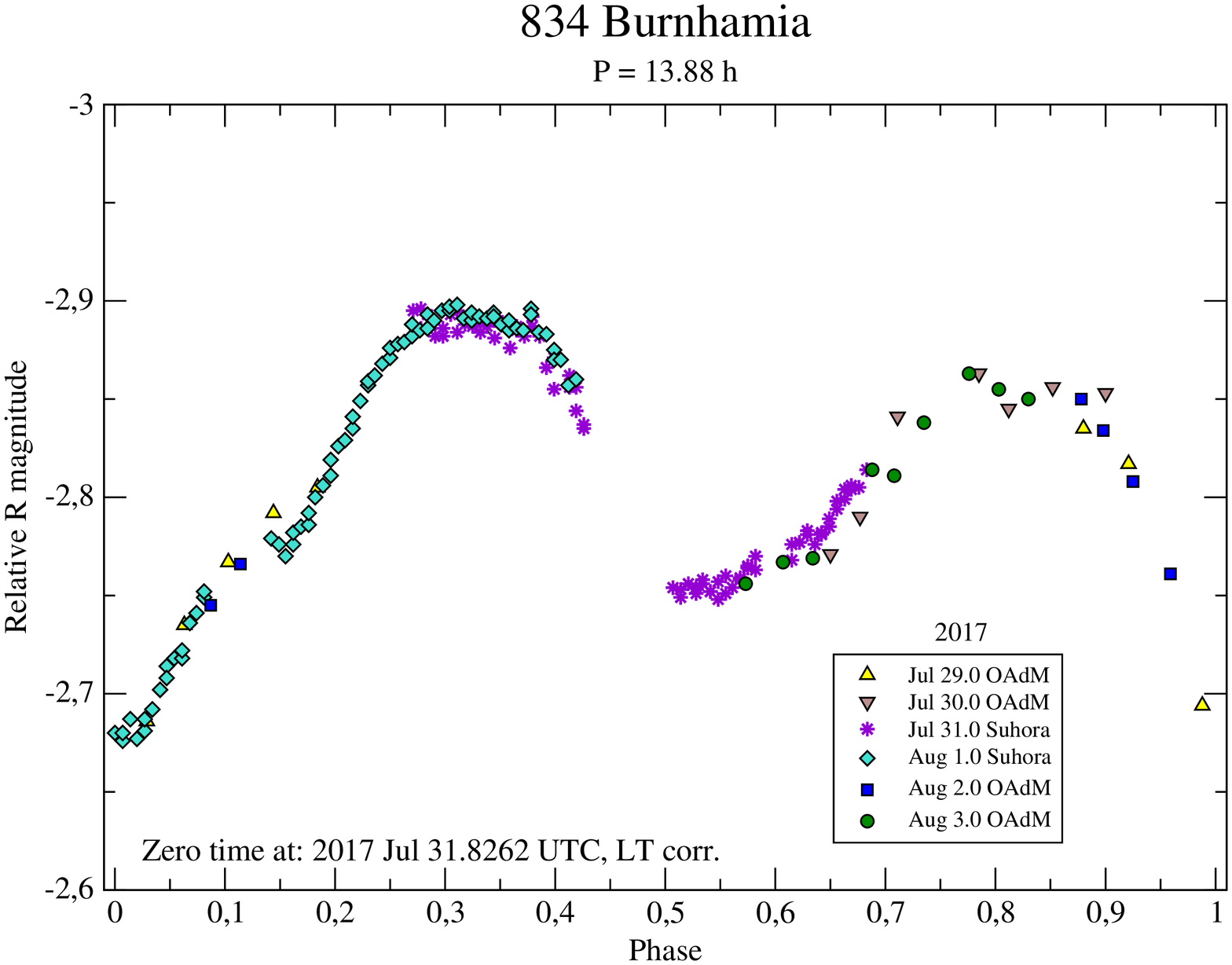} 
\captionof{figure}{Composite lightcurve of (834) Burnhamia from the year 2017.}
\label{834composit2017}
\\
\end{tabularx}
    \end{table*}%


\clearpage
\section{Thermophysical model details}
  In Sect.~\ref{sec:model} in the main text, we provided an
  overview of our modelling approach.  Below, we give a few
  more details about how we determine the Bond albedo ($A$)
  for the TPM and how we estimate our error bars. We also 
  include all the tables (Tables \ref{tab:tpm_100} to
  \ref{tab:tpm_834}), observations-to-model-ratio plots
  (Figs. \ref{fig:100_OMR} to \ref{fig:834_OMR}), and
  $\chi^2$ versus $\Gamma$ plots that we used to give full
  information on the TPM analysis.

  We fix the value of the Bond albedo to run the model (let us call it $A_i$).
  Our TPM diameters ($D$) do not always lead to the same value of Bond
  albedo (say, $A_o$) when we use the $H$-$G_{12}$ values. However, the 
  differences are not large enough to be meaningful in practice as long as
  $A\sim 0.1$ or lower. This is because flux is proportional to $T^4$, and
  that in turn is proportional to $(1-A)$, so for example wrongly assuming $A=0.05$
  instead of $A=0.10$ would lead to a $\sim$5\% offset in the flux, which is
  still comparable to the absolute calibration uncertainties. On the other
  hand, the effect is indeed strong for objects with higher $A$ (e.g. Vesta,
  with $A\sim0.2$). 

  To give numbers in our case, all our targets -- except one -- have $A$s
  in the range 0.02--0.07, and the differences between $A_i$ and $A_o$ never
  lead to systematic model flux differences $>$0.5\%, so the effect is
  negligible. For our highest albedo target, (100) Hekate, we used $A_i$ =
  0.10 but got $A_o$=0.12. This 20\% offset leads nonetheless to a $\sim$2\%
  systematic difference in the model fluxes, which is not ideal but it is
  well within the absolute calibration flux uncertainties. More
  quantitatively, we re-ran the TPM for model 2 with $A_i$=0.12 and got
  virtually the same size and thermal inertia but a higher surface roughness
  (rms = 0.45 instead of rms = 0.4). In this case, the slightly lower
  temperature due to the higher albedo seems to be compensated by a small
  increase in the roughness.

  To estimate $D$ and $\Gamma$ error bars we have followed standard
  procedures \citep[more details and discussion can be found in][]{Ali-Lagoa2014,Hanus2015}.
  Namely, if we have \chisq $\approx$1, the $3\sigma$ error bars are given by
  the range of $\Gamma$s and $D$s of those models with \chisq$<3\sqrt{2/\nu}$,
  where $\nu$ is the effective number of degrees of freedom
  \citep[e.g.][]{Press1986}. When \chisq is much lower than 1 or up to 1.5,
  we assume we can scale the $\chi^2$ curves to have the minimum at 1 and apply
  the same formula. This is not rigourous mathematically but it is a working
  assumption to have some estimate of the error bars \citep{Hanus2015}. After
  all, a very low \chisq could be due to some error bars being overestimated
  (the IRAS data have the largest ones). If \chisq is about 1.5, we assume it
  is still a reasonable fit given the model simplification (constant parameters
  over the surface, wavelength-independent emissivity) or other sources of 
  errors. For instance, we do not consider the uncertainties in shape (no shape 
  modelling scheme so far provides shape uncertainties), rotation period, or
  spin orientation explicitly. These are coupled parameters themselves in the
  shape modelling so we cannot simply vary them separately within their
  uncertainties to explore the effects in TPM. Hanus et al. (2015) proposed
  bootstrapping the visible photometry to obtain $\sim$30 shape and spin models
  that were subsequently input to the TPM (the so-called varied-shape TPM).
  The values of acceptable $\Gamma$s spanned were significantly larger than the
  classical approach only in some cases, not systematically.

\begin{table*}
  \centering
  \caption{Summary of TPM results for (100) Hekate. 
  }\label{tab:tpm_100}    
  \begin{tabular}{|l | l c c c  c l|}
    
    \hline
    Shape model  & IR data subset & $\bar{\chi}^2_{m}$ &   $D \pm 3\sigma$ (km) & $\Gamma \pm 3\sigma$ (SIu)  & Roughness (rms)  &  Comments                 \\
    \hline\hline
    & & & & & & \\
    AM 1 & All data & 0.8 & 90$^{+6}_{-4}$ & 6$^{+114}_{-4}$ & Low/Medium (0.30) & rms unconstrained at the 3$\sigma$ level \\
    & & & & & & \\
    AM 1 sphere & All data & 0.8 & 86$^{+6}_{-5}$ & 20$^{+160}_{-20}$ & Medium/high (0.6) &  rms unconstrained at the 3$\sigma$ level\\
    & & & & & & \\
    \hline
    & & & & & & \\
    AM 2 & All data & 0.65 & 87$^{+5}_{-4}$ & 4$^{+66}_{-2}$ & Medium (0.40) & rms unconstrained at the 3$\sigma$ level \\
    & & & & & & \\
    AM 2 sphere & All data & 0.8 & 86$^{+8}_{-8}$ & 16$^{+144}_{-16}$ & Medium/high ($\sim$0.6) & rms unconstrained at the 3$\sigma$ level\\
    & & & & & & \\
    \hline
  \end{tabular}
\end{table*}

\begin{table*}
  \centering
  \caption{Summary of TPM results for (109) Felicitas. 
  }\label{tab:tpm_109}    
  \begin{tabular}{|l | l c c c  c l|}
    
    \hline
    Shape model  & IR data subset & $\bar{\chi}^2_{m}$ &   $D \pm 3\sigma$ (km) & $\Gamma \pm 3\sigma$ (SIu)  & Roughness (rms)  &  Comments                 \\
    \hline\hline
    & & & & & & \\
    AM 1 & All data & 1.1 & 85$^{+7}_{-5}$ & 40$^{+100}_{-40}$ & Ext. high ($\sim$1.0) & rms unconstrained at the 3$\sigma$ level \\
    & & & & & & \\
    AM 1 sphere & All data & 2.2 & 97 & 200 & Ext. high ($\sim$1.0) & Bad fit \\
    & & & & & & \\
    \hline
    & & & & & & \\
    AM 2 & All data & 2.0 & 81 & 4 & Ext. high ($\sim$1.0) & Bad fit \\
    & & & & & & \\
    AM 2 sphere & All data & 3.4 & 101 & 250 & Ext. high ($\sim$1.0) & Bad fit \\
    & & & & & &\\
    \hline
  \end{tabular}
\end{table*}

\begin{table*}[h!tb]
  \centering
  \caption{Summary of TPM results for (195) Eurykleia fitting different
    subsets of data. For example, the ``Excluding W4'' label refers to
    the fact that the W4 data were not used to optimise the $\chi^2$ in
    that case. }\label{tab:tpm}    
  \begin{tabular}{|l | l c c c  c l|}
    
    \hline
    Shape model  & IR data subset & $\bar{\chi}^2_{m}$ &   $D \pm 3\sigma$ (km) & $\Gamma \pm 3\sigma$ (SIu)  & Roughness (rms)  &  Comments                 \\
    \hline\hline
    & & & & & & \\
    AM 1 & All data & 0.51 & 82$^{+5}_{-4}$ & $15^{+37}_{-15}$ & High (0.65) & rms$>$0.34 at the 1$\sigma$ level, \\
    & & & & & & but otherwise unconstrained  \\
    AM 1 sphere & All data & 1.23 & 83$^{+7}_{-4}$ & 25$^{+65}_{-20}$ & Extr. high ($>$1.0) & Roughness unconstrained \\
    & & & & & & \\
    AM 1 & Excluding W4 & 0.50 & 87$^{+5}_{-6}$ & 15$^{+35}_{-15}$  & Medium (0.45) & Roughness unconstrained \\
    & & & & & & \\
    AM 1 & Only W4 & 0.23 & 81 & 5 & Low (0.29) & Artificially low $\bar{\chi}^2_m$. Single-epoch data  \\
    & & & & & & are insufficient to constrain params.\\
    \hline
    & & & & & & \\
    AM 2 & All data & 0.60 & 86$^{+5}_{-4}$ & 20$^{+40}_{-20}$ & Extr. high ($\sim$1.0) & Roughness rms$>$0.34 \\
    & & & & & & at the 3$\sigma$ level\\
    AM 2 sphere & All data & 1.17 & 84$^{+8}_{-3}$ & 30$^{+60}_{-25}$  & Extr. high ($>$1.0) &  Roughness unconstrained \\
    & & & & & & \\
    AM 2 & Excluding W4 & 0.58 & 91$^{+7}_{-6}$ & 15$^{+50}_{-15}$ & Med.-high (0.5)  & Fitted with rel. calibration error bars \\
    & & & & & & \\
    AM 2 & Only W4 & 0.47 & 80 & 5 & Extr. high ($>$1.0) & Artificially low $\bar{\chi}^2_m$. Single-epoch data \\
    & & & & & & are insufficient to constrain params.\\
    \hline
  \end{tabular}
\end{table*}

\begin{table*}
  \centering
  \caption{Summary of TPM results for (301) Bavaria. 
  }\label{tab:tpm_301}    
  \begin{tabular}{|l | l c c c  c l|}
    
    \hline
    Shape model  & IR data subset & $\bar{\chi}^2_{m}$ &   $D \pm 3\sigma$ (km) & $\Gamma \pm 3\sigma$ (SIu)  & Roughness (rms)  &  Comments                 \\
    \hline\hline
    & & & & & & \\
    AM 1 & All data & 0.34 & 55$^{+2}_{-2}$ & 40$^{+60}_{-30}$ & Med.-high (0.5) & rms unconstrained at the 3$\sigma$ level \\
    & & & & & & \\
    AM 1 sphere & All data & 0.7 & 57$^{+3}_{-3}$ & 30$^{+90}_{-25}$ & Low (0.20) & rms unconstrained at the 3$\sigma$ level \\
    & & & & & & \\
    \hline
    & & & & & & \\
    AM 2 & All data & 0.36 & 55$^{+2}_{-2}$ & 50$^{+50}_{-35}$ & Medium/high (0.60) & rms unconstrained at the 3$\sigma$ level \\
    & & & & & & \\
    AM 2 sphere & All data & 0.7 & 57$^{+3}_{-3}$ & 30$^{+70}_{-25}$ & Low (0.20) & rms unconstrained at the 3$\sigma$ level \\
    & & & & & &\\
    \hline
  \end{tabular}
\end{table*}

\begin{table*}
  \centering
  \caption{Summary of TPM results for (335) Roberta. 
  }\label{tab:tpm_335}    
  \begin{tabular}{|l | l c c c  c l|}
    
    \hline
    Shape model  & IR data subset & $\bar{\chi}^2_{m}$ &   $D \pm 3\sigma$ (km) & $\Gamma \pm 3\sigma$ (SIu)  & Roughness (rms)  &  Comments                 \\
    \hline\hline
    & & & & & & \\
    AM 1 & All data & 0.70 & 100$^{+10}_{-11}$ & $70^{+1500}_{-55}$ & Very low (0.21) & Roughness and $\Gamma$ unconstrained \\
    & & & & & & at the 3$\sigma$ level \\
    AM 1 sphere & All data & 0.71 & 97$^{+8}_{-11}$ & 150$^{+1350}_{-125}$ & Extr. high ($\sim$1.0) &  Compatible solution but requires \\
    & & & & & & extremely high roughness\\
    \hline
    & & & & & & \\
    AM 2 & All data & 0.60 & 98$^{+10}_{-8}$ & 90$^{+1410}_{-85}$ & Low (0.34) & Roughness and $\Gamma$ unconstrained  \\
    & & & & & & at the 3$\sigma$ level\\
    AM 2 sphere & All data & 0.68 & 97$^{+8}_{-11}$ & 150$^{+1350}_{-125}$ & Extr. high ($\sim$1.0) & Very similar to ``sphere AM 1'' fit\\
    & & & & & & \\
    \hline
  \end{tabular}
\end{table*}

\begin{table*}
  \centering
  \caption{Summary of TPM results for (380) Fiducia.
  }\label{tab:tpm_380}    
  \begin{tabular}{|l | l c c c  c l|}
    
    \hline
    Shape model  & IR data subset & $\bar{\chi}^2_{m}$ &   $D \pm 3\sigma$ (km) & $\Gamma \pm 3\sigma$ (SIu)  & Roughness (rms)  &  Comments                 \\
    \hline\hline
    & & & & & & \\
    AM 1 & All data & 0.91 & 73$^{+8}_{-5}$ & 15$^{+135}_{-15}$ & Low (0.25) & rms$<$0.90 at the 3$\sigma$ level \\
    & & & & & & \\
    AM 1 sphere & All data & 1.33 & 78 & 200 & Extr. high ($\sim$1.0) & Borderline acceptable $\bar\chi^2_m$ \\
    & & & & & & but unconstrained thermal props.\\
    \hline
    & & & & & & \\
    AM 2 & All data & 0.59 & 72$^{+9}_{-5}$ & 10$^{+140}_{-10}$ & Low (0.25) & rms$<$0.90 at the 3$\sigma$ level \\
    & & & & & & \\
    AM 2 sphere & All data & 1.41 & 72 & 200 & Extr. high ($\sim$1.0) & Borderline acceptable $\bar\chi^2_m$ \\
    & & & & & & but unconstrained thermal props.\\
    \hline
  \end{tabular}
\end{table*}

\begin{table*}
  \centering
  \caption{Summary of TPM results for (468) Lina. 
  }\label{tab:tpm_468}    
  \begin{tabular}{|l | l c c c  c l|}
    
    \hline
    Shape model  & IR data subset & $\bar{\chi}^2_{m}$ &   $D \pm 3\sigma$ (km) & $\Gamma \pm 3\sigma$ (SIu)  & Roughness (rms)  &  Comments                 \\
    \hline\hline
    & & & & & & \\
    AM 1 & All data & 1.20 & 69$^{+11}_{-4}$ & 20$^{+280}_{-20}$ & Very low (0.20) & rms unconstrained at the 3$\sigma$ level \\
    & & & & & & \\
    AM 1 sphere & All data & 2.0 & 65. & 35 & Low ($\sim$0.30) & Bad $\bar\chi^2_m$ but \\
    & & & & & & similar thermal properties \\
    & & & & & & \\
    AM 2 & All data & 1.22 & 70$^{+11}_{-5}$ & 20$^{+280}_{-20}$ & Very low (0.2) & rms not constrained at the 3$\sigma$ level \\
    & & & & & & \\
    AM 2 sphere & All data & 2.0 & 65 & 30 & Low ($\sim$0.30) & Bad $\bar\chi^2_m$ but\\
    & & & & & & similar thermal properties\\
    \hline
  \end{tabular}
\end{table*}

\begin{table*}
  \centering
  \caption{
    Summary of TPM results for (538) Friederike.
  }\label{tab:tpm_538}   
  \begin{tabular}{|l | l c c c  c l|}
    
    \hline
    Shape model  & IR data subset & $\bar{\chi}^2_{m}$ &   $D \pm 3\sigma$ (km) & $\Gamma \pm 3\sigma$ (SIu)  & Roughness (rms)  &  Comments                 \\
    \hline\hline
    & & & & & & \\
AM 1        &  all data & 1.65  &   74$^{+2}_{-1}$ &    $<$ 15      & Extr. high (0.9)  &   Bad fit \\
    & & & & & & \\
AM 1 sphere &  all data & 1.52  &   71$^{+3}_{-1}$ &    $<$ 28      & Extr. high $\sim$ 1 &  Bad fit  \\
    & & & & & & \\
AM 1        & WISE only & 0.99  &   73$^{+3}_{-2}$ & 10$^{+34}_{-6}$    &   Extr. high $\sim$ 1 & rms > 0.40 \\
    & & & & & & \\
    \hline
    & & & & & & \\
AM 2        &  all data & 2.04  &   75$^{+2}_{-1}$ &    $<$ 14      &     High (0.75)  & Bad fit  \\
    & & & & & & \\
AM 2 sphere &  all data & 1.52  &   71$^{+3}_{-1}$ &    $<$ 28      &  Extr. high $\sim$ 1   & Bad fit  \\
    & & & & & & \\
AM 2        & WISE only &  0.72  &  76$^{+5}_{-2}$ &  20$^{+25}_{-20}$   &   Extr. high $\sim$ 1  &  rms > 0.45  \\
    & & & & & & \\
    \hline
  \end{tabular}
\end{table*}

\begin{table*}
  \centering
  \caption{Summary of TPM results for (653) Berenike. 
  }\label{tab:tpm_653}    
  \begin{tabular}{|l | l c c c  c l|}
    
    \hline
    Shape model  & IR data subset & $\bar{\chi}^2_{m}$ &   $D \pm 3\sigma$ (km) & $\Gamma \pm 3\sigma$ (SIu)  & Roughness (rms)  &  Comments                 \\
    \hline\hline
    & & & & & & \\
    AM 1 & All data & 1.1 & 46$^{+4}_{-2}$ & 40$^{+120}_{-40}$ & Med.-high (0.5) & rms$<$0.3 rejected at the 3$\sigma$ level \\
    & & & & & & \\
    AM 1 sphere & All data & 2.0 & 52 & 100 & Extr. high (1.00) & Bad fit \\
    & & & & & & \\
    \hline
    & & & & & & \\
    AM 2 & All data & 2.0 & 52 & 120 & Extr. high (1.00) & Bad fit \\
    & & & & & & \\
    AM 2 sphere & All data & 2.6 & 52 & 100 & Extr. high(1.00) & Bad fit\\
    & & & & & &\\
    \hline
  \end{tabular}
\end{table*}
\begin{table*}
  \centering
  \caption{Summary of TPM results for (673) Edda. 
  }\label{tab:tpm_673}    
  \begin{tabular}{|l | l c c c  c l|}
    
    \hline
    Shape model  & IR data subset & $\bar{\chi}^2_{m}$ &   $D \pm 3\sigma$ (km) & $\Gamma \pm 3\sigma$ (SIu)  & Roughness (rms)  &  Comments                 \\
    \hline\hline
    & & & & & & \\
    AM 1 & All data & 0.47 & 38$^{+6}_{-2}$ & 3$^{+33}_{-3}$ & Med.-high (0.5) & rms$>$0.34 at the 3$\sigma$ level \\
    & & & & & & \\
    AM 1 sphere & All data & 1.83 & 38 & 5 & Medium (0.4) & Bad fit \\
    & & & & & & \\
    \hline
    & & & & & & \\
    AM 2 & All data & 0.59 & 38$^{+2}_{-2}$ & 3$^{+37}_{-3}$ & Extr. high ($\sim$1.0) & rms$>$0.34 at the 3$\sigma$ level \\
    & & & & & & \\
    AM 2 sphere & All data & 1.76 & 38 & 10. & Medium (0.45) & Bad fit \\
    & & & & & & \\
    \hline
  \end{tabular}
\end{table*}

\begin{table*}
  \centering
  \caption{Summary of TPM results for (834) Burnhamia. 
  }\label{tab:tpm_834}    
  \begin{tabular}{|l | l c c c  c l|}
    
    \hline
    Shape model  & IR data subset & $\bar{\chi}^2_{m}$ &   $D \pm 3\sigma$ (km) & $\Gamma \pm 3\sigma$ (SIu)  & Roughness (rms)  &  Comments                 \\
    \hline\hline
    & & & & & & \\
    AM 1 & All data & 0.78 & 67$^{+8}_{-6}$ & 22$^{+23}_{-22}$ & Extr. high (0.9) & rms unconstrained at the 3$\sigma$ level \\
    & & & & & & \\
    AM 1 sphere & All data & 1.12 & 65$^{+6}_{-5}$ & 40$^{+45}_{-40}$ & Extr. high ($\sim$1.0) & Acceptable $\bar\chi^2_m$ and \\
    & & & & & & similar thermal properties \\
    \hline
    & & & & & & \\
    AM 2 & All data & 0.80 & 66$^{+5}_{-4}$ & 20$^{+30}_{-20}$ & High (0.6) & rms not constrained at the 3$\sigma$ level \\
    & & & & & & \\
    AM 2 sphere & All data & 1.12 & 65$^{+6}_{-5}$ & 40$^{+45}_{-40}$ & Extr. high ($\sim$1.0) & Acceptable $\bar\chi^2_m$ but\\
    & & & & & & similar thermal properties\\
    \hline
  \end{tabular}
\end{table*}

\clearpage

\begin{figure}
  \centering
  \includegraphics[width=0.80\linewidth]{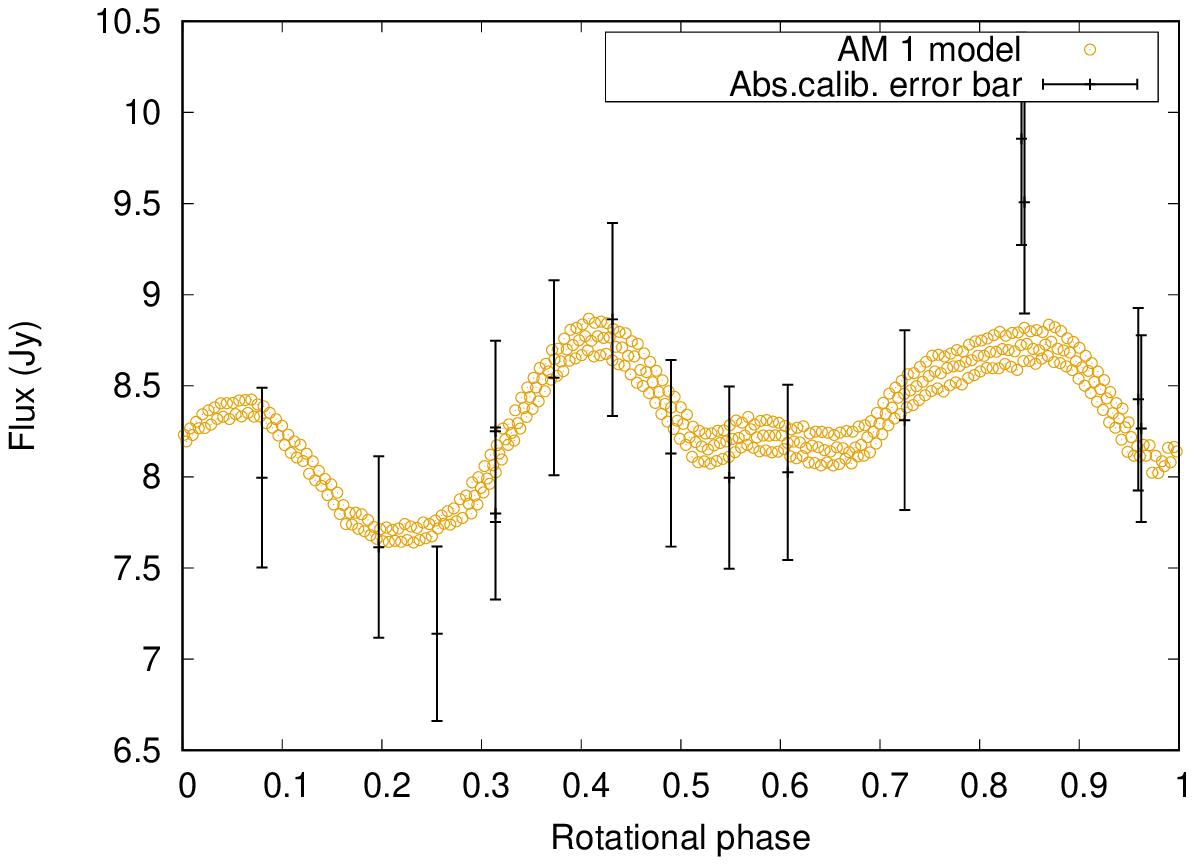}
  \caption{
    Asteroid (100) Hekate W4 data and model thermal lightcurves for shape model 1.
    Data error bars are 1-$\sigma$. Table~\ref{tab:tpm_100} summarises the TPM
    analysis.  
  } \label{fig:ThLC_100model1} 
\end{figure}

\begin{figure}
  \centering
  \includegraphics[width=0.80\linewidth]{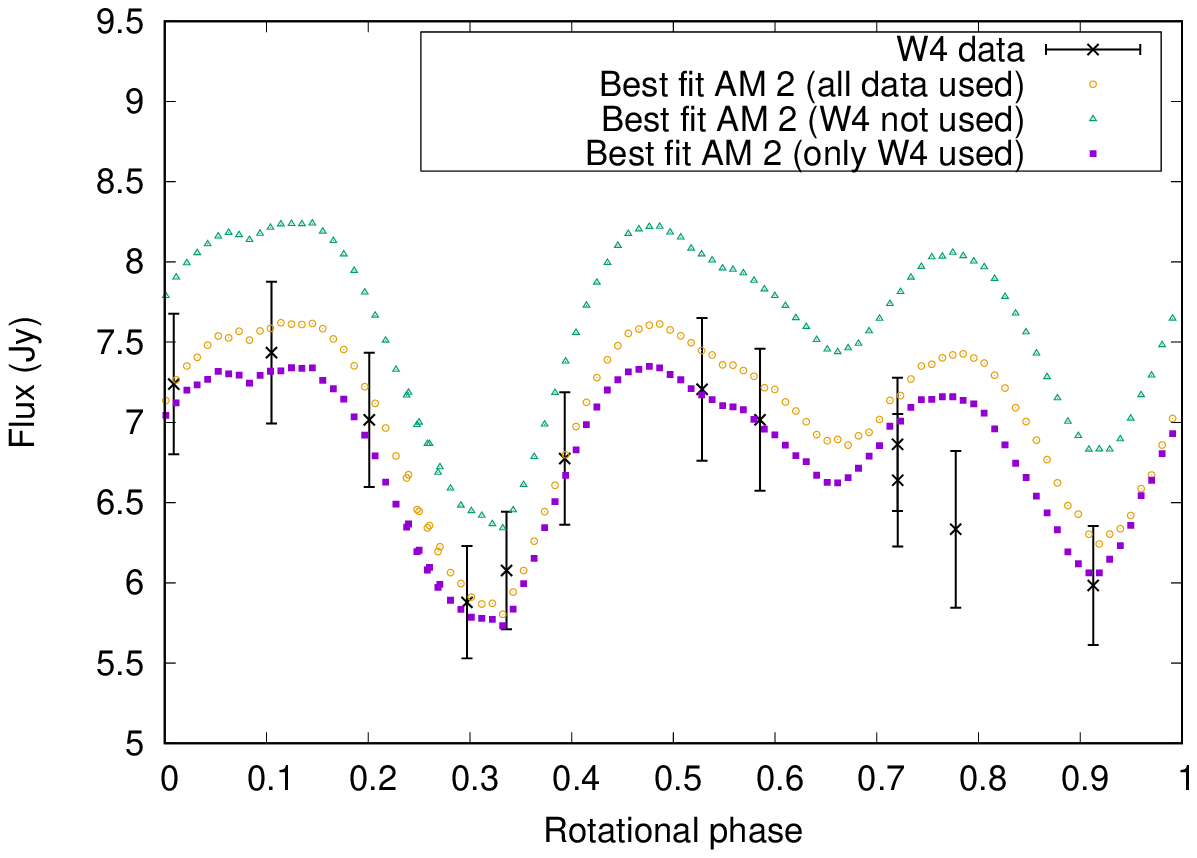}
  \caption{W4 data and model thermal lightcurves for (195) Eurykleia's shape
    model AM 2. The different models resulted from fitting different subsets of
    data. Table~\ref{tab:tpm} contains the corresponding thermo-physical
    parameters. 
  } \label{fig:W4ThLCmodel2} 
\end{figure}

\begin{figure}
  \centering
  \includegraphics[width=0.80\linewidth]{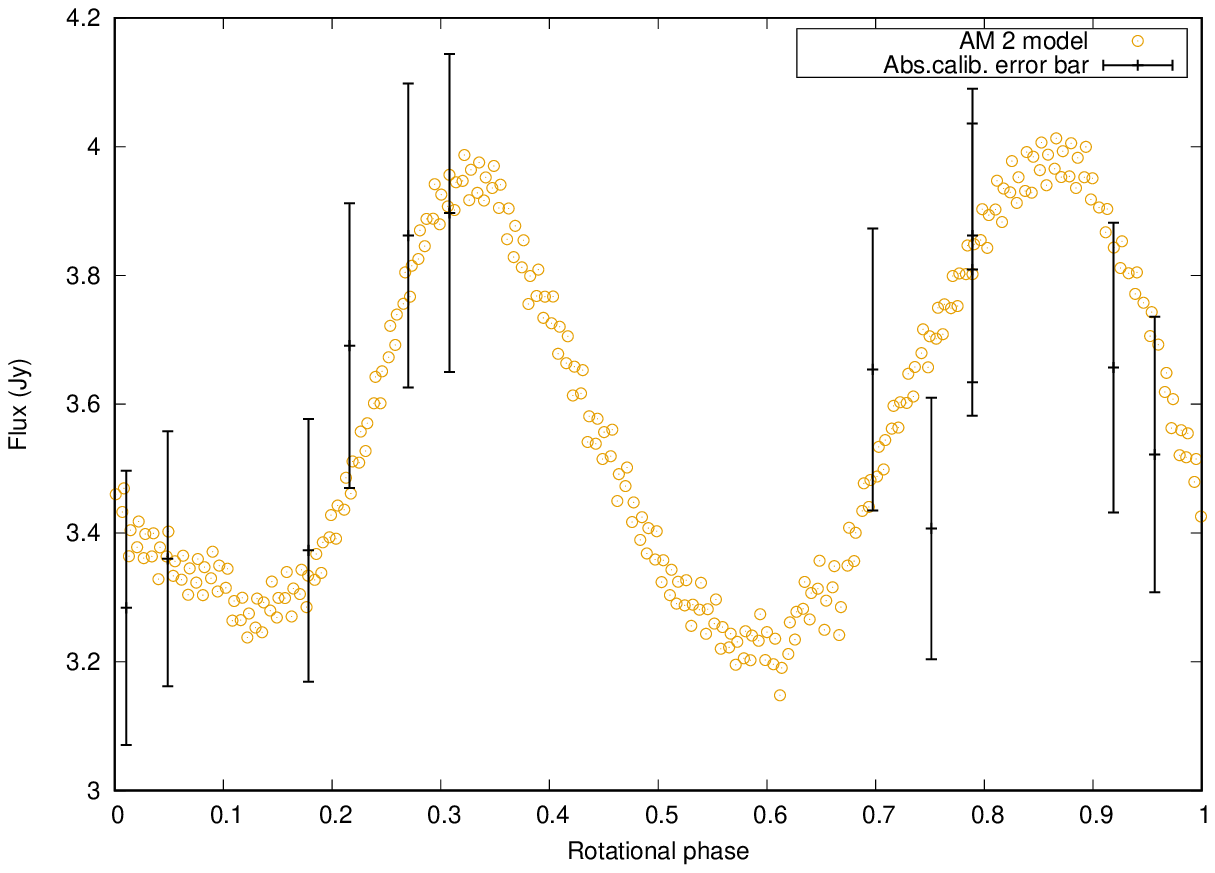}  
  \caption{
    Asteroid (301) Bavaria AM 2 model thermal lightcurves and W4 data
    (see Table~\ref{tab:tpm_301}). 
  } \label{fig:ThLC_301model2} 
\end{figure}

\begin{figure}
  \centering
  \includegraphics[width=0.80\linewidth]{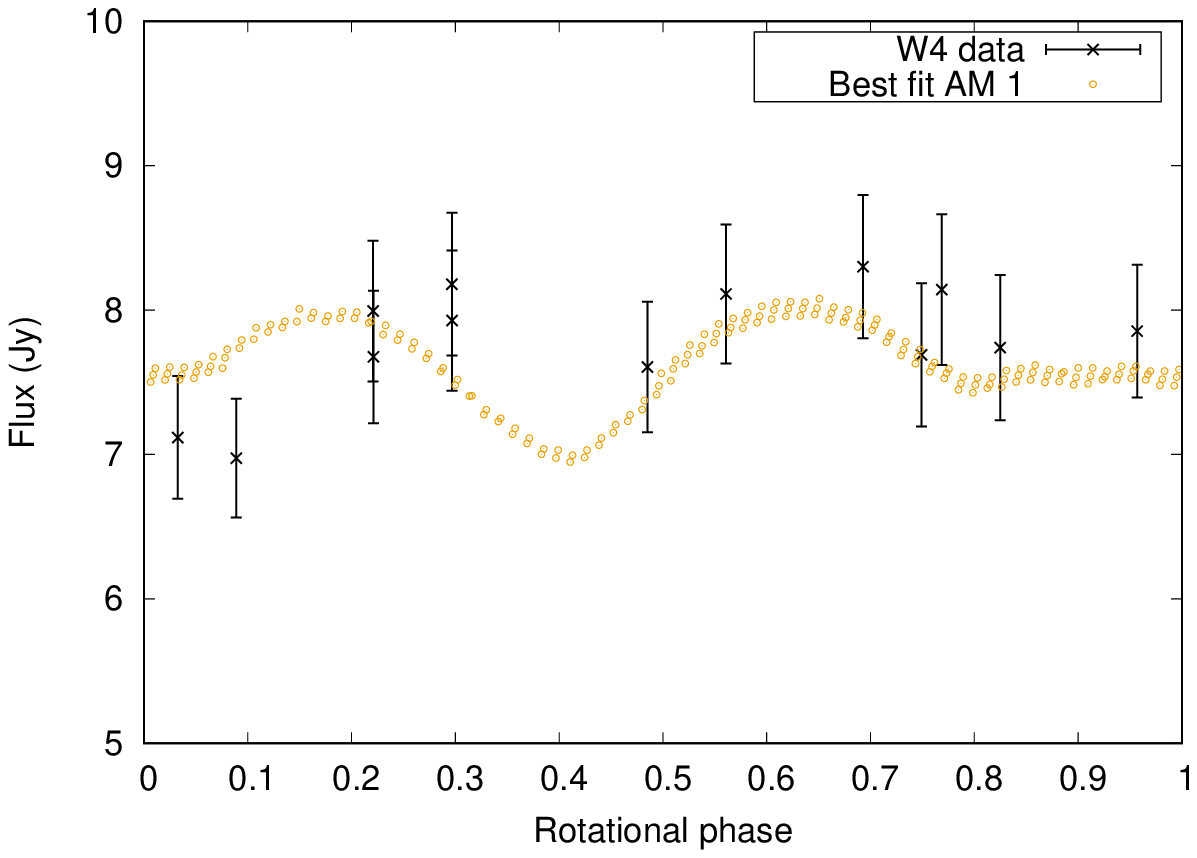}
  \caption{
    Asteroid (335) Roberta's WISE data and model thermal lightcurves for shape model 
    1 and the corresponding best-fitting solutions (very low thermal
    inertia of 15 SIu). 
  } \label{fig:335_ThLCmodel1} 
\end{figure}

\begin{figure}
  \centering
  \includegraphics[width=0.80\linewidth]{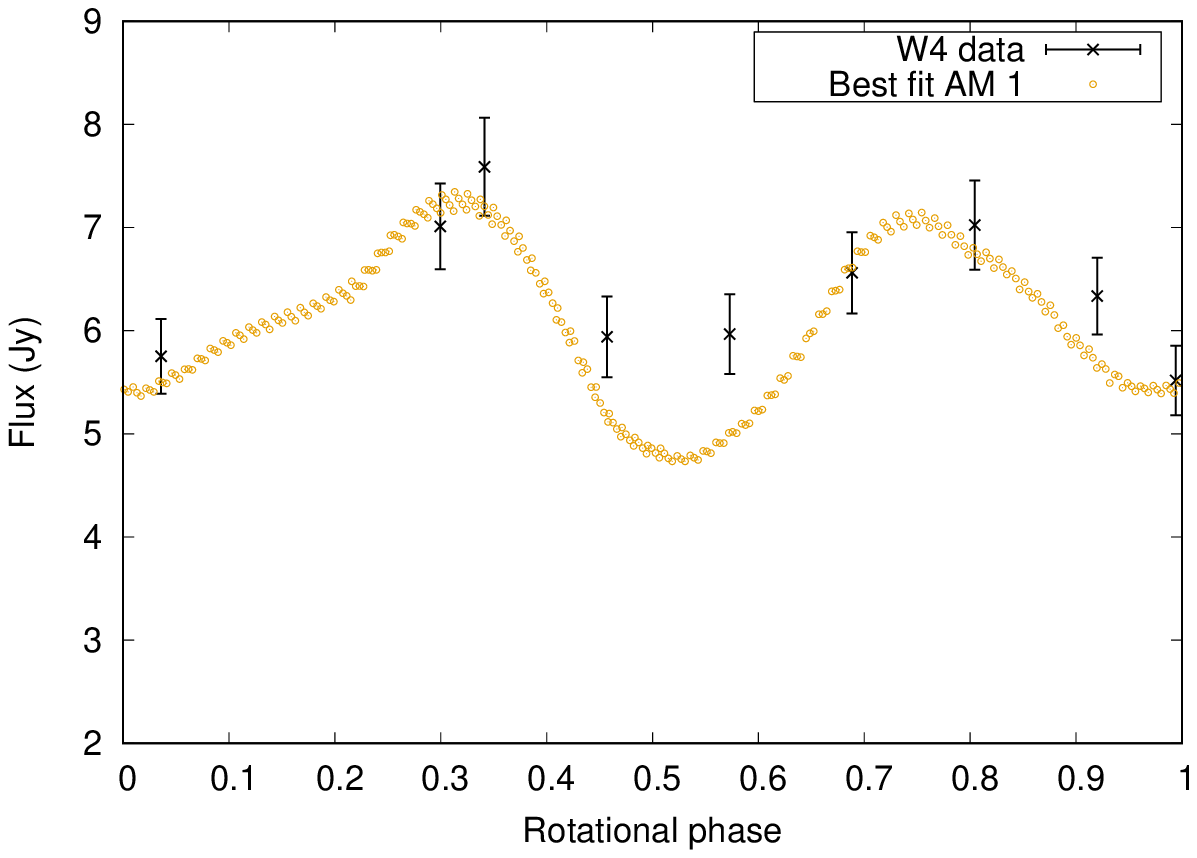}  
  \caption{
    Asteroid (380) Fiducia's WISE data and model thermal lightcurves for shape model 
    1 and the corresponding best-fitting solutions (very low thermal inertia of
    15 SIu; see Table \ref{tab:tpm_380}). 
  } \label{fig:ThLC_380model1} 
\end{figure}

\begin{figure}
  \centering
  \includegraphics[width=0.80\linewidth]{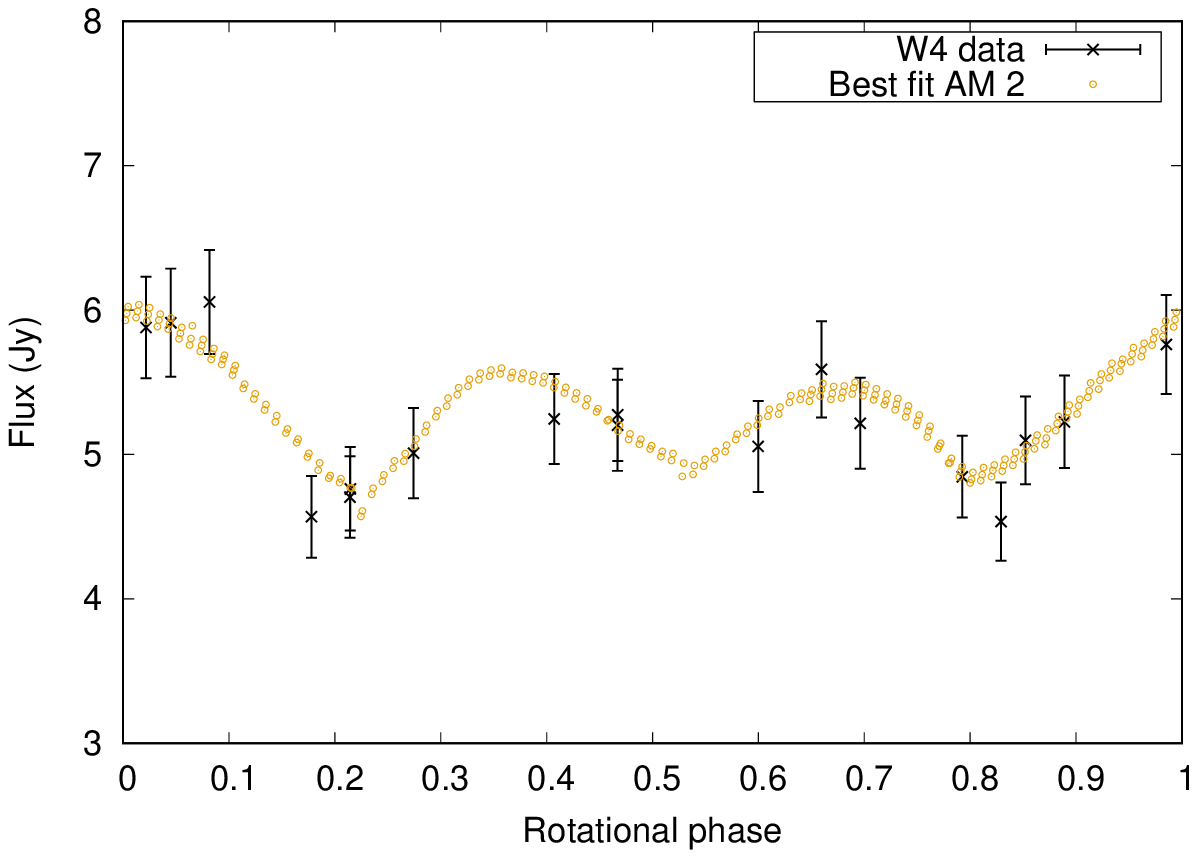} 
  \caption{
    Asteroid (468) Lina's WISE data and model thermal lightcurves for shape model 
    2's corresponding best-fitting solutions (see Table  \ref{tab:tpm_468}).
  } \label{fig:ThLC_468model2} 
\end{figure}

\begin{figure}
  \centering
    \includegraphics[width=0.80\linewidth]{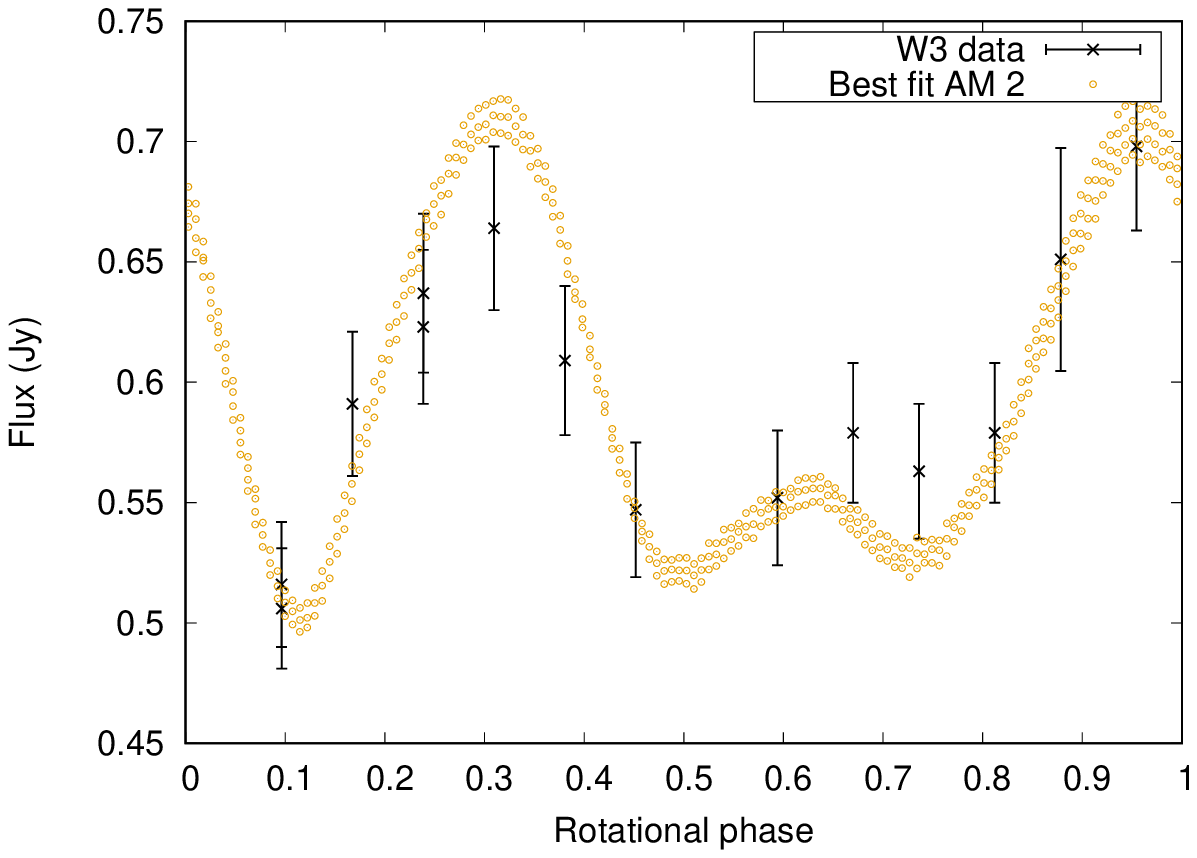}

    \includegraphics[width=0.80\linewidth]{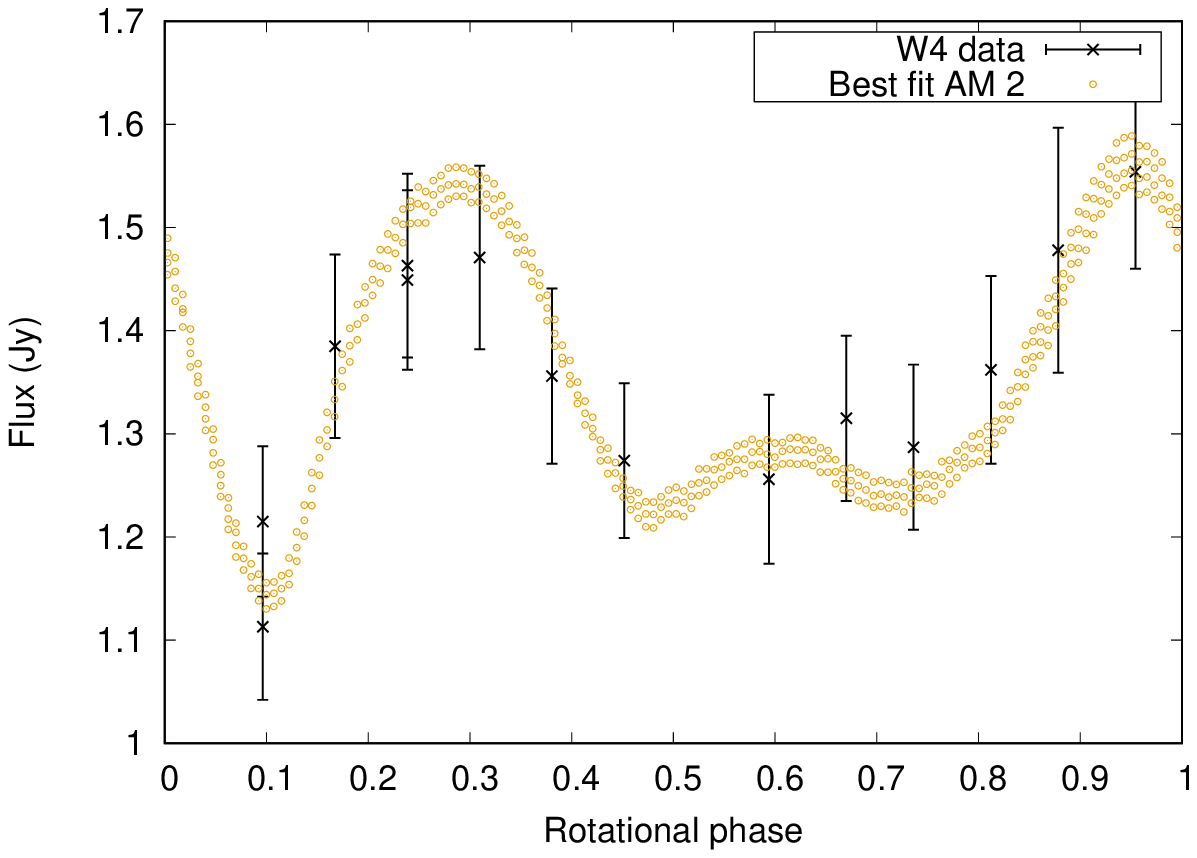} 

  \caption{
    Asteroid (673) Edda's WISE data and model thermal lightcurves for shape model 2's
    best-fitting solution (very low thermal inertia of 3 SI units).
  } \label{fig:ThLC_673model2} 
\end{figure}

\begin{figure}
  \centering
  \includegraphics[width=0.80\linewidth]{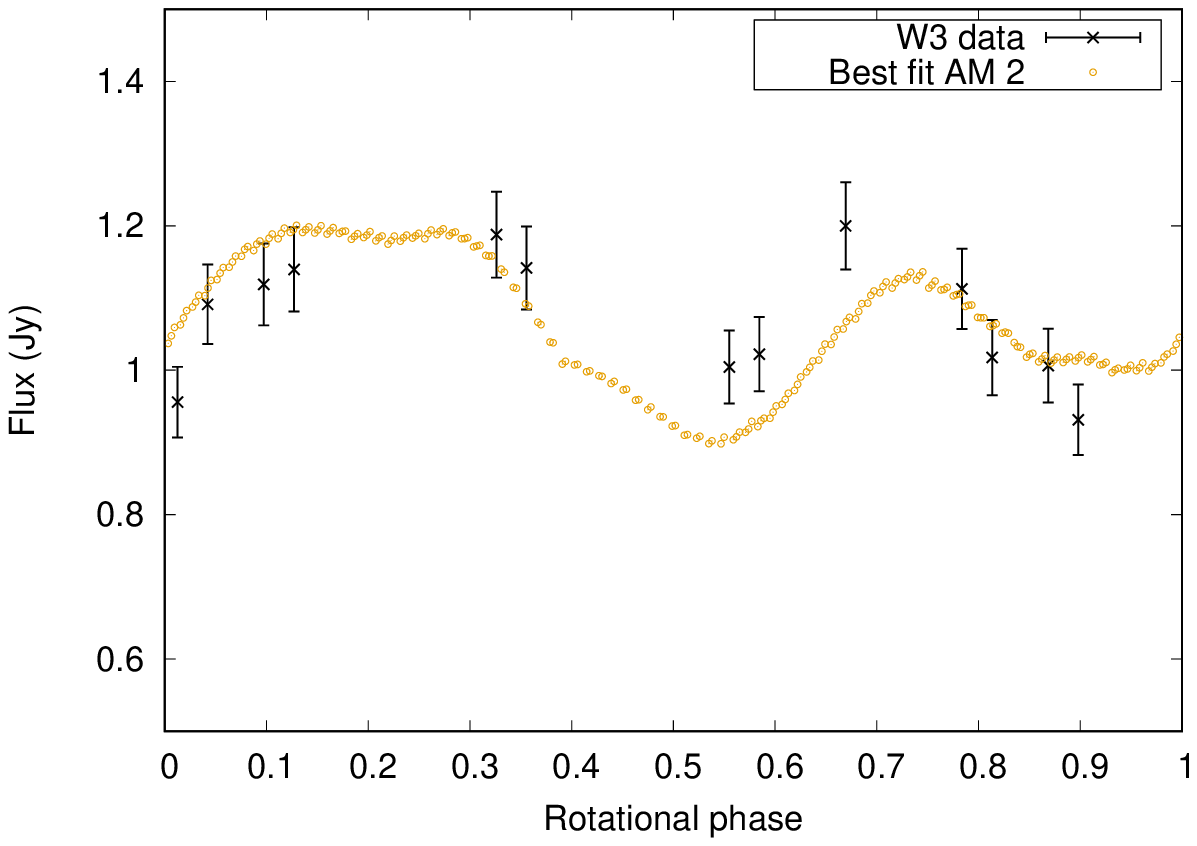} 

  \includegraphics[width=0.80\linewidth]{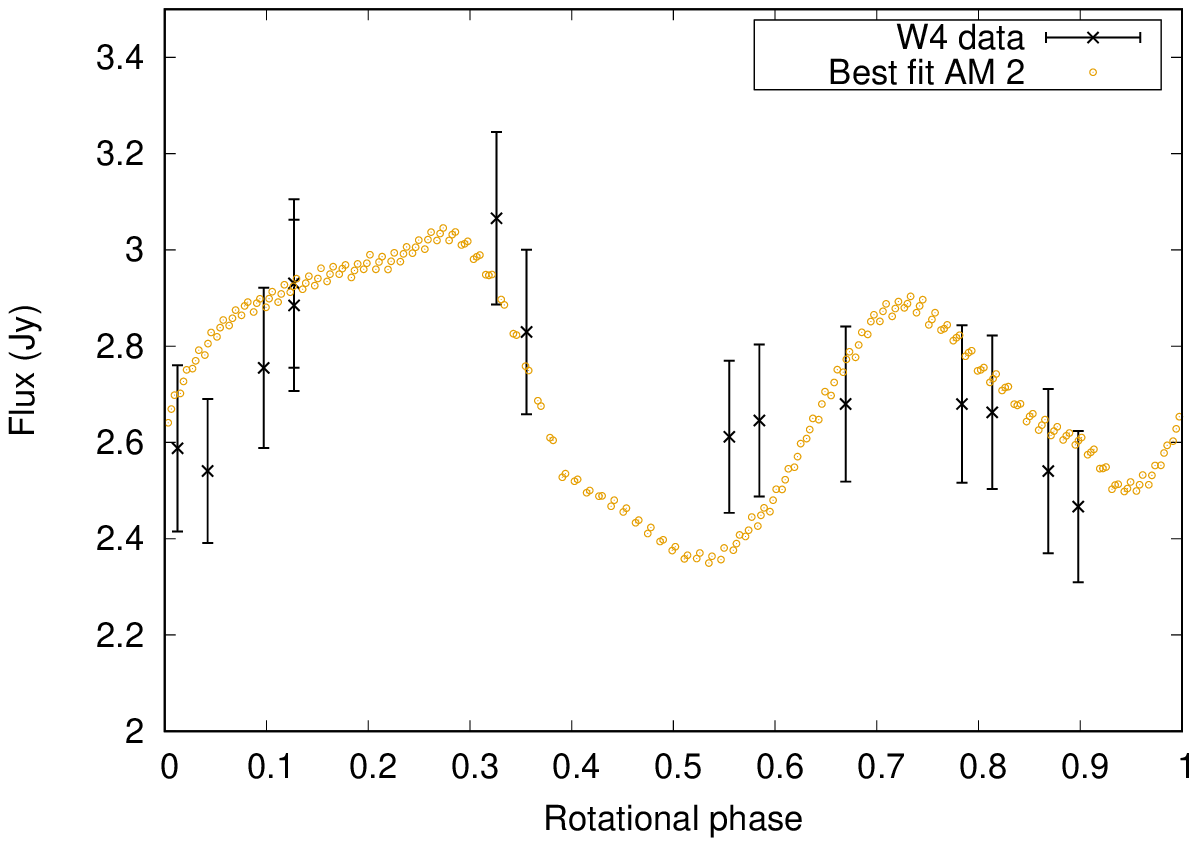} 

  \caption{
    Asteroid (834) Burnhamia's WISE data and model thermal lightcurves for shape model
    2's best-fitting solution (see Table \ref{tab:tpm_834}). 
  } \label{fig:ThLC_834model2} 
\end{figure}

\clearpage

    \begin{table*}[ht]
    \centering
\begin{tabularx}{\linewidth}{XX}
  \includegraphics[width=0.90\linewidth]{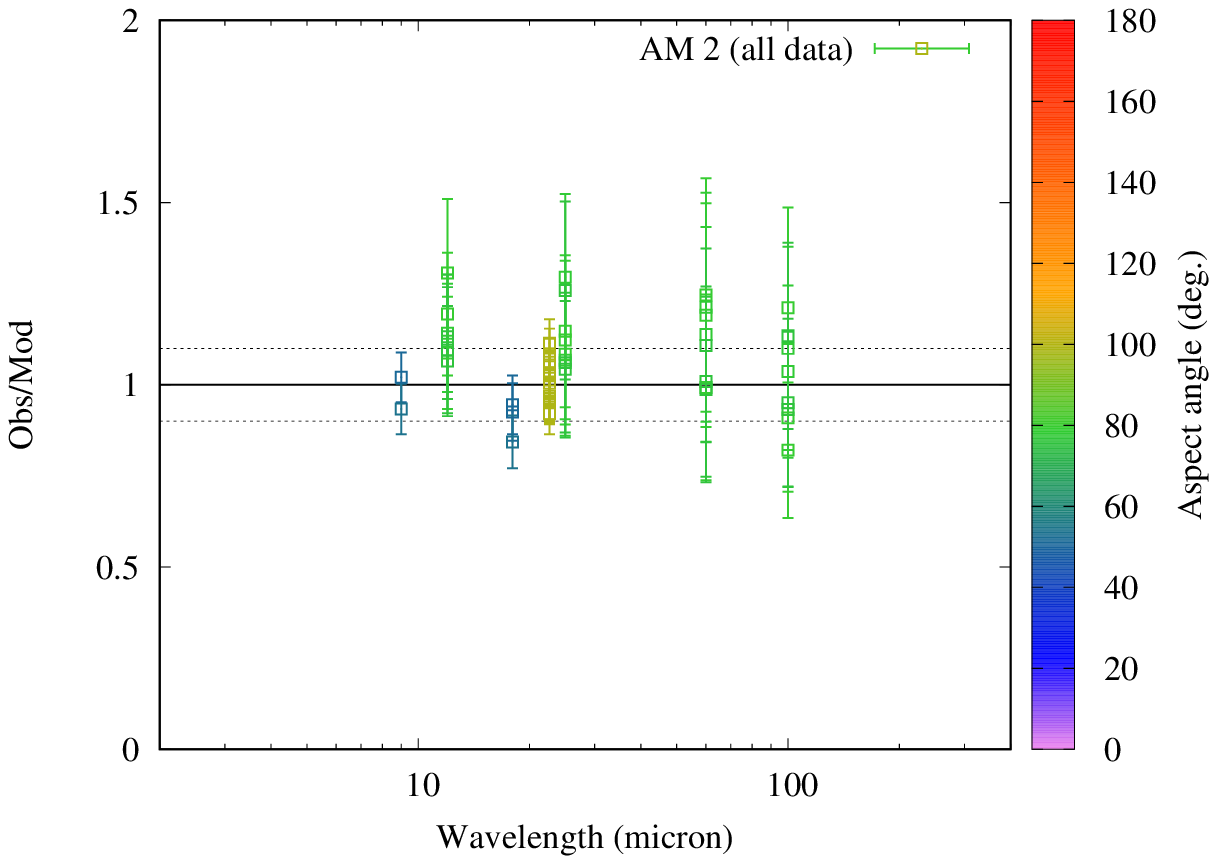}
&
  \includegraphics[width=0.90\linewidth]{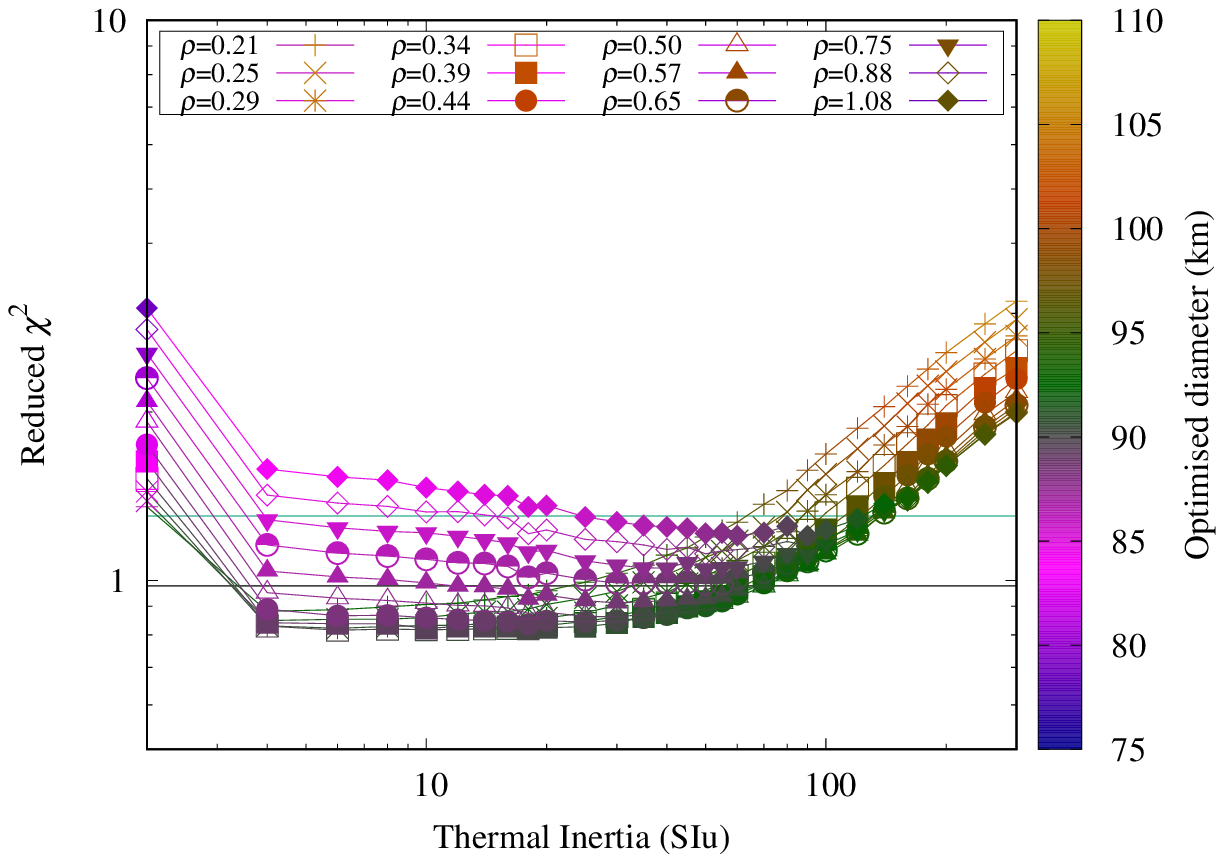}
\\
  \includegraphics[width=0.90\linewidth]{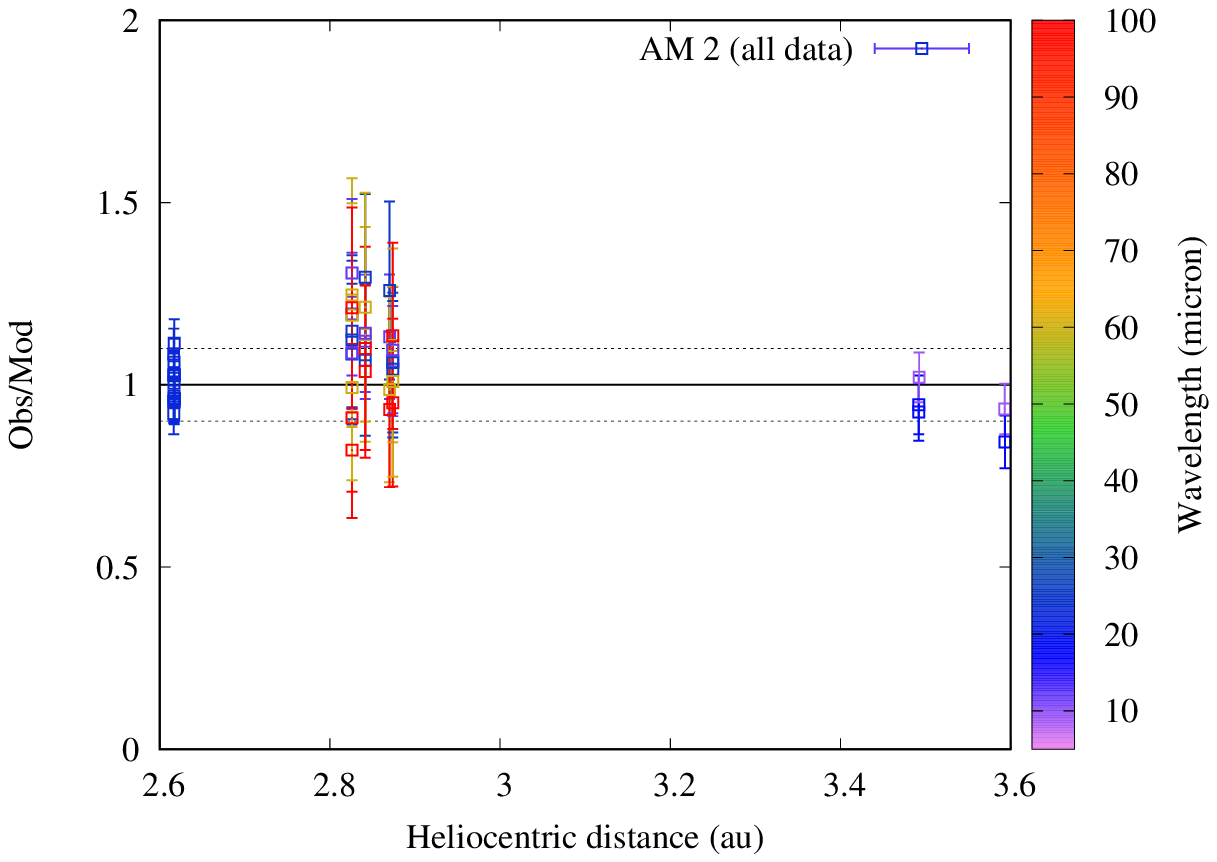}
&
  \includegraphics[width=0.90\linewidth]{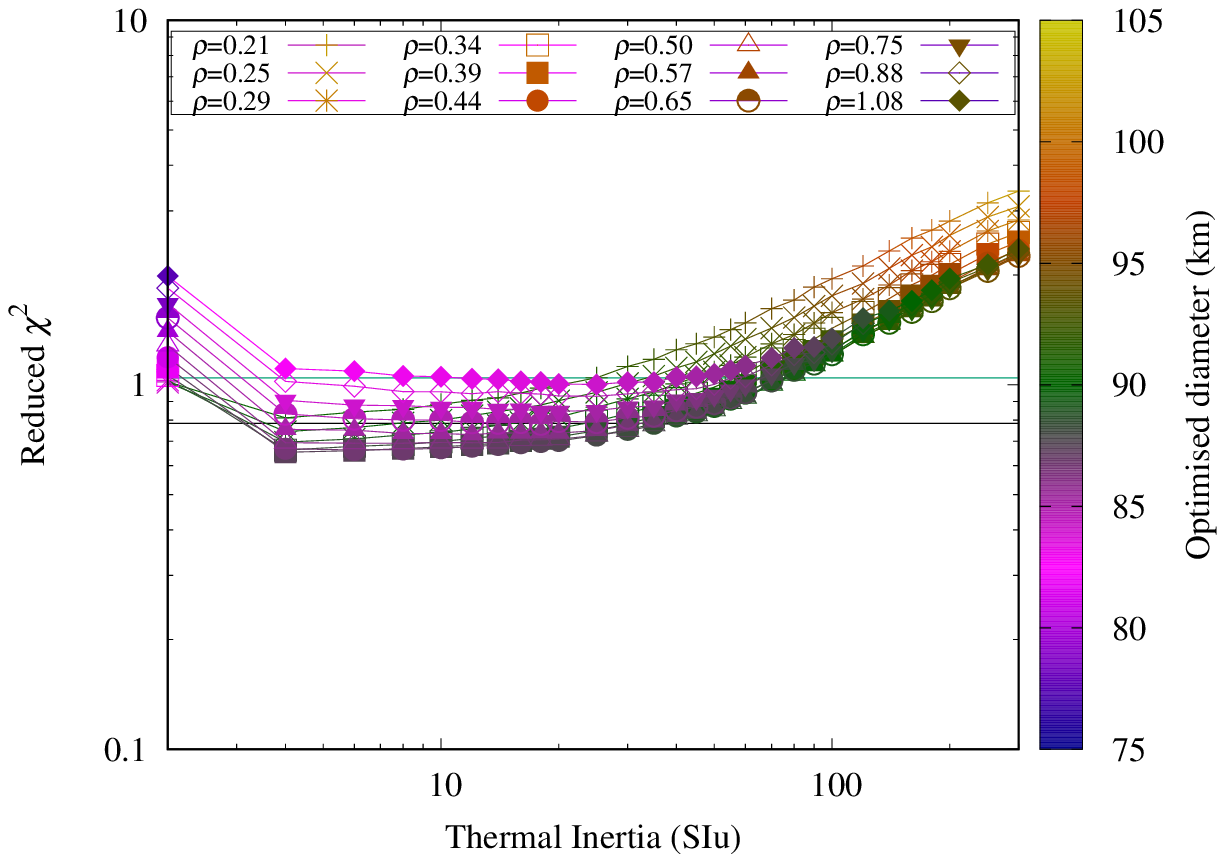}  
\\
  \includegraphics[width=0.90\linewidth]{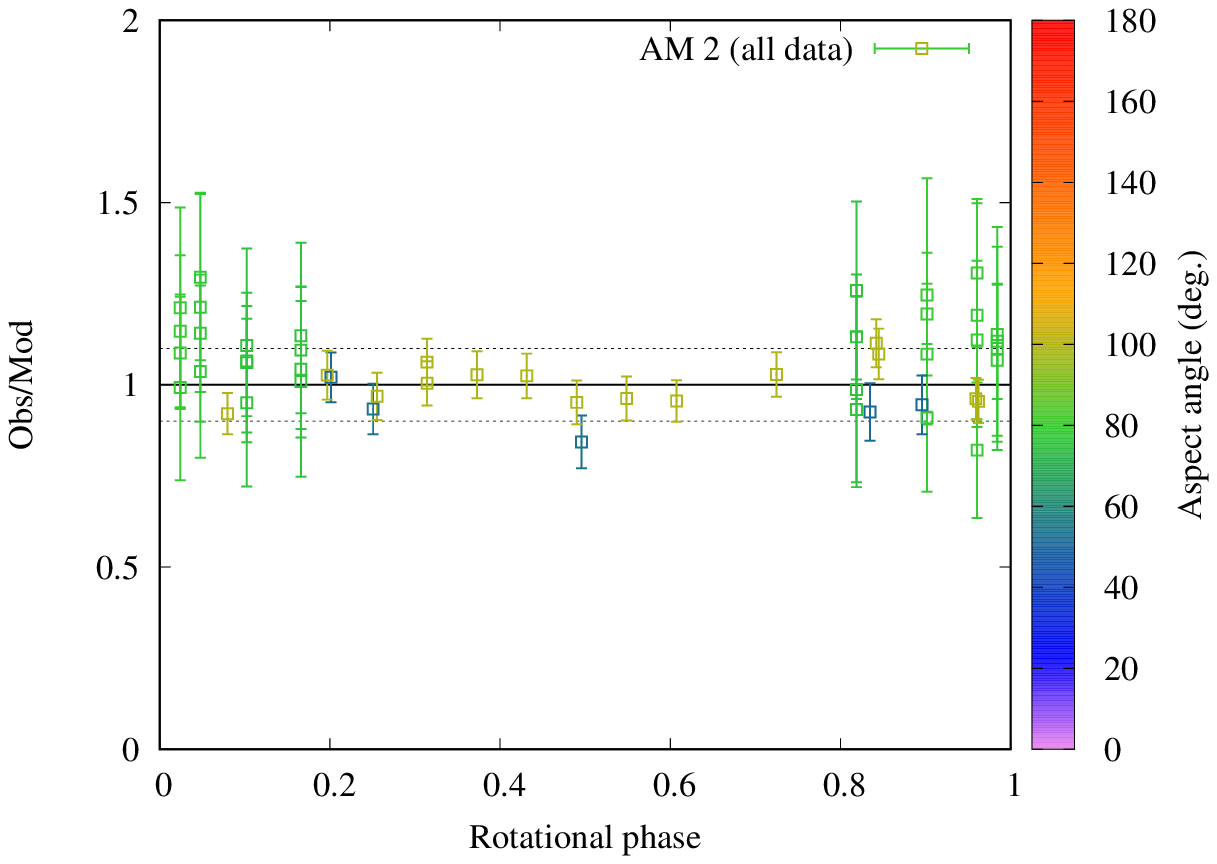}
\\
  \includegraphics[width=0.90\linewidth]{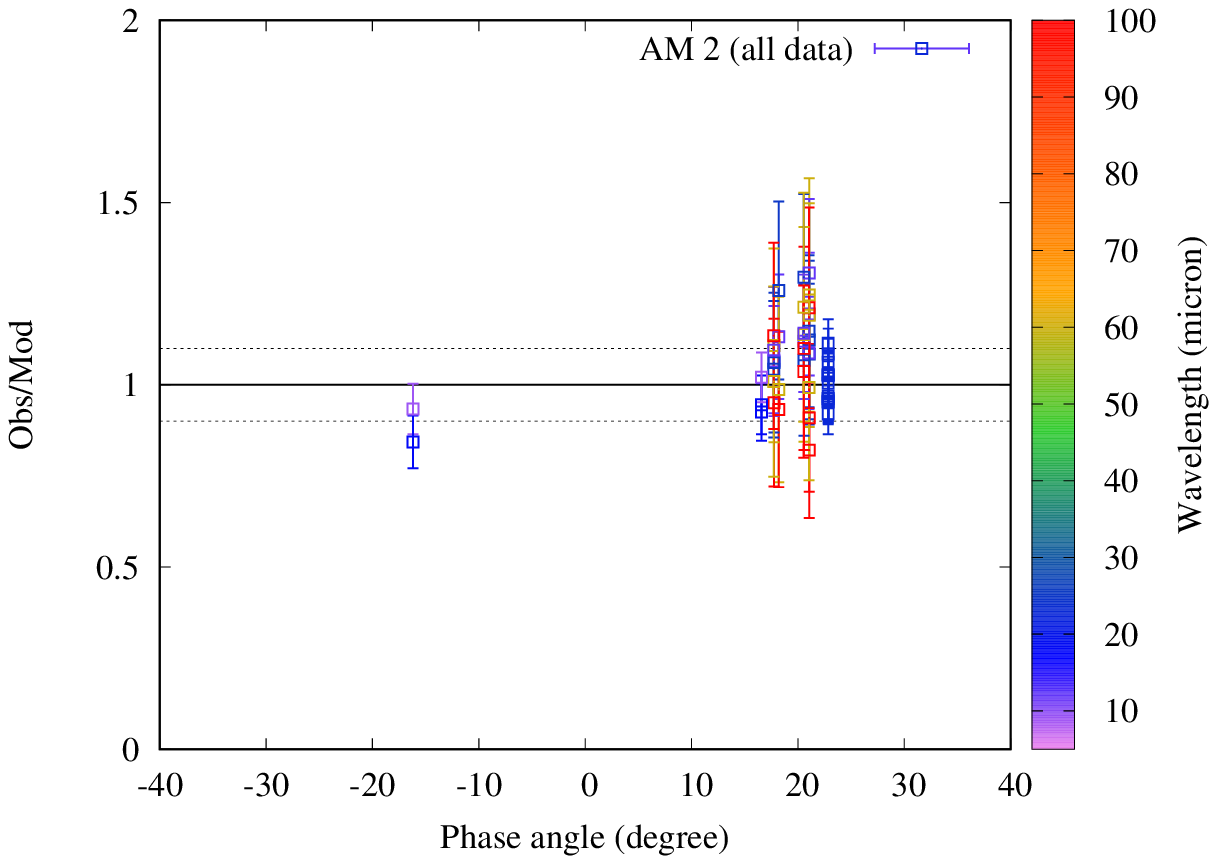}
\captionof{figure}{Asteroid (100) Hekate, left from top to bottom: observation to model ratios versus
    wavelength, heliocentric distance, rotational phase, and phase angle. 
Depending on the plot, the colour indicates the aspect angle or wavelength at which the observations were taken 
(see the legend on the right). The plots for AM 1 looked similar. 
   Right: $\chi^2$ versus thermal inertia curves for (100) Hekate model 1 (top) and model 2 (bottom). 
    In all the subsequent figures of this kind, the colours denote optimised diameter, 
    while various symbols show various levels of surface roughness. }
\label{fig:100_OMR}
\\
\end{tabularx}
    \end{table*}

\clearpage

    \begin{table*}[ht]
    \centering
\begin{tabularx}{\linewidth}{XX}
  \includegraphics[width=0.90\linewidth]{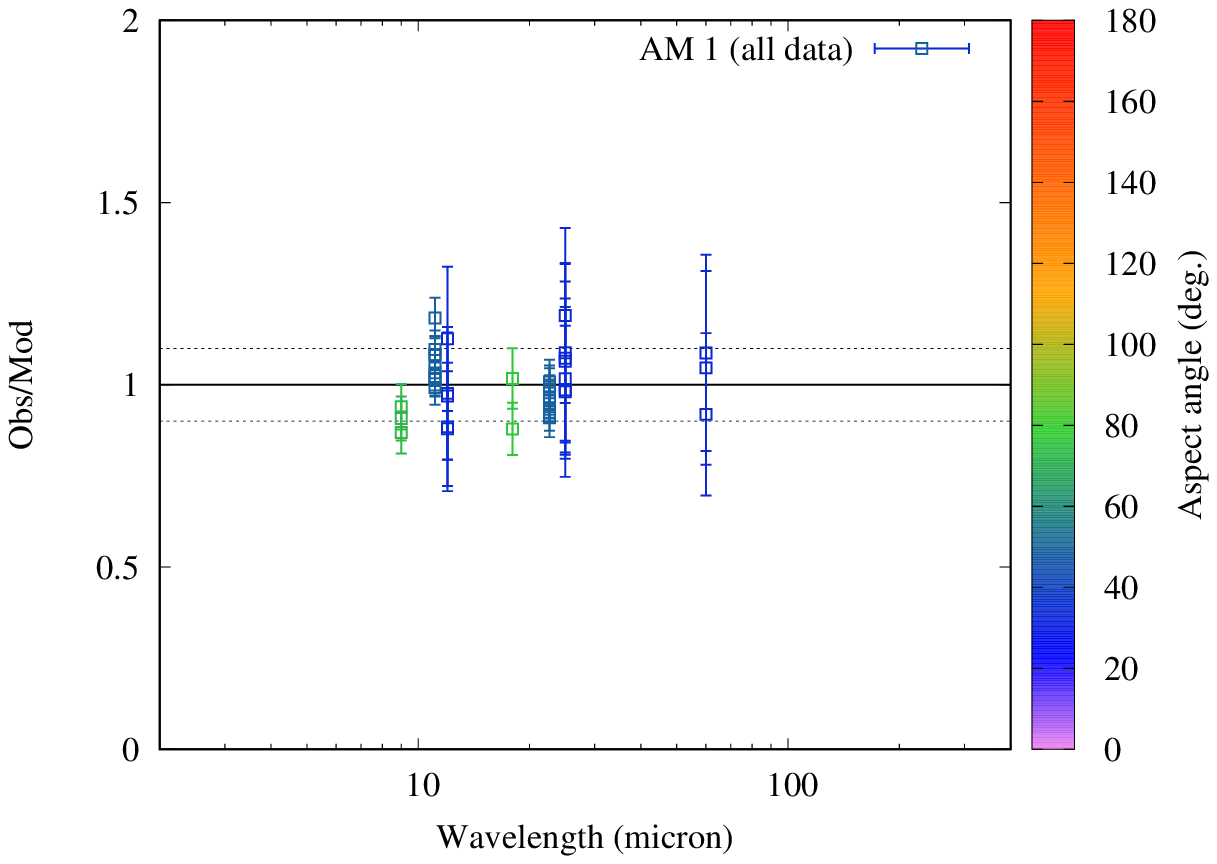}
&
  \includegraphics[width=0.90\linewidth]{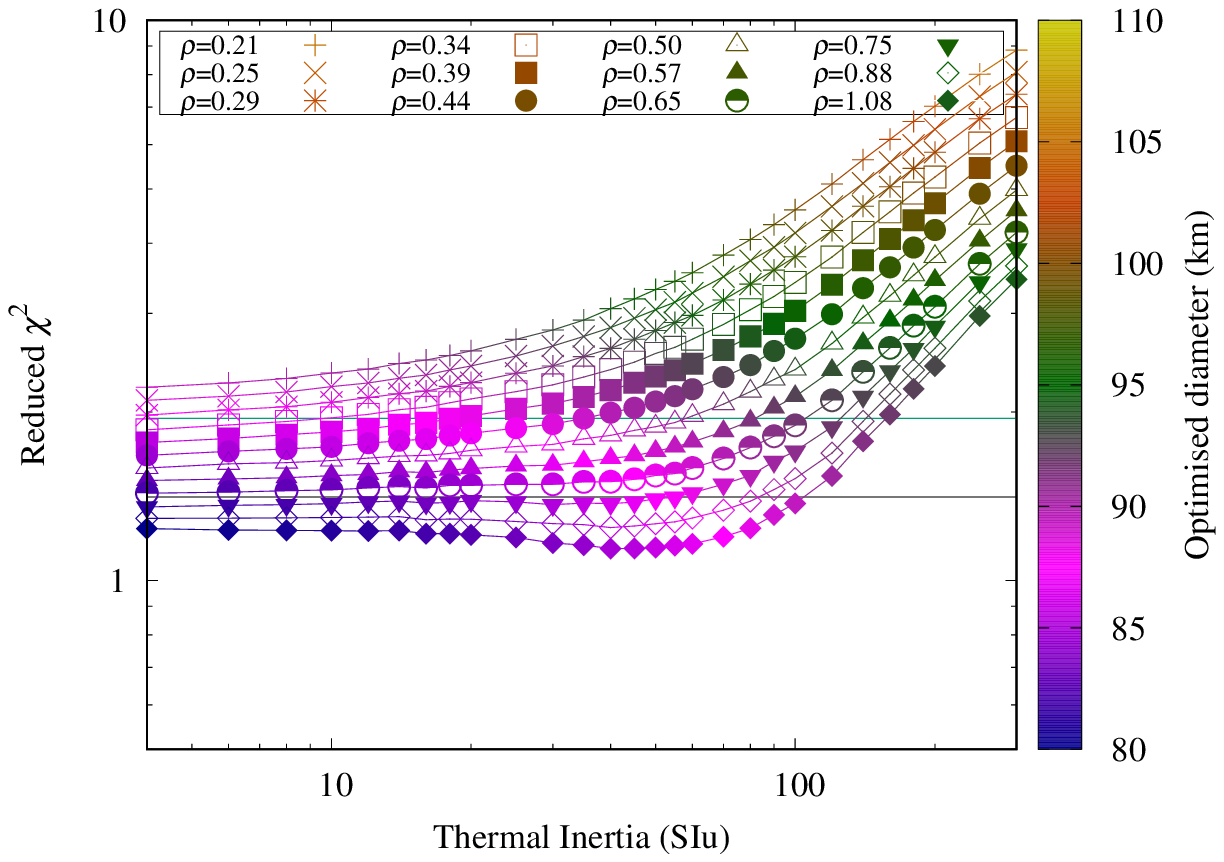}
\\
  \includegraphics[width=0.90\linewidth]{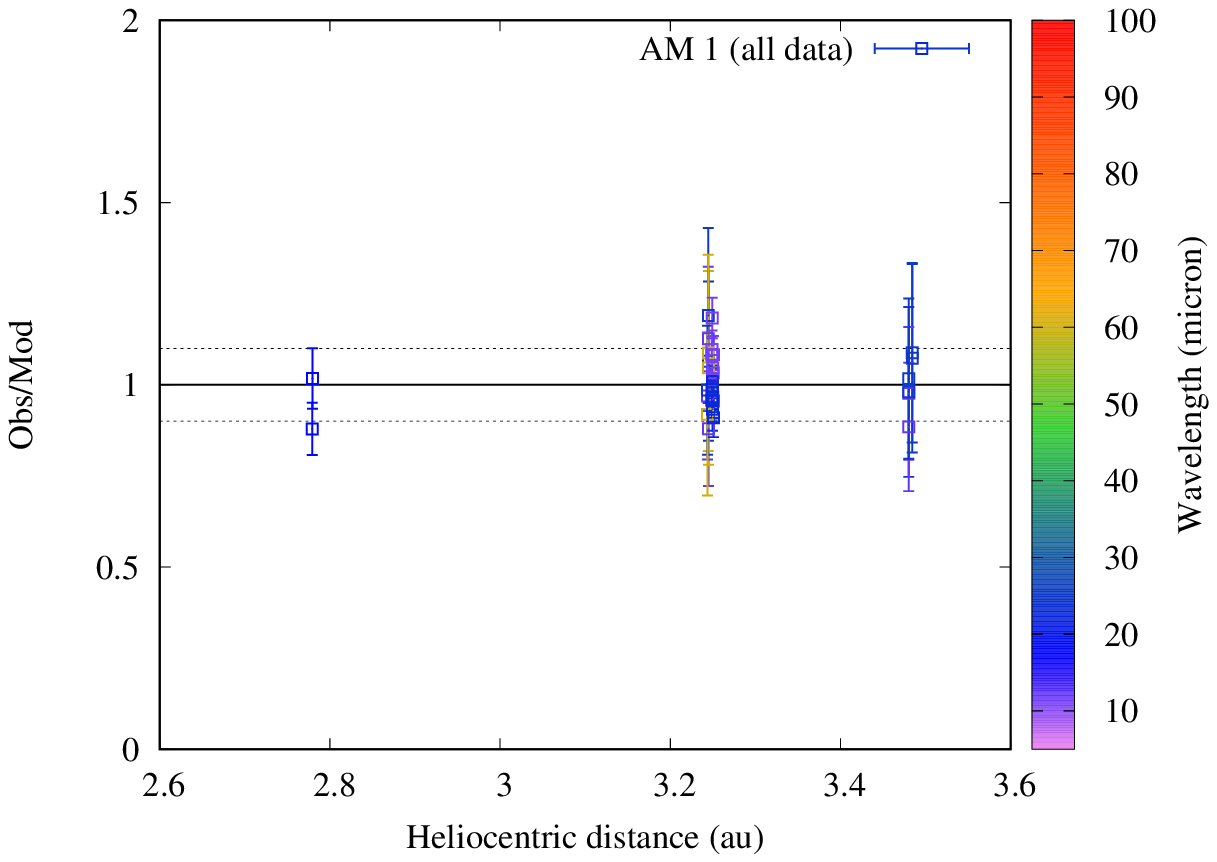}
\\
  \includegraphics[width=0.90\linewidth]{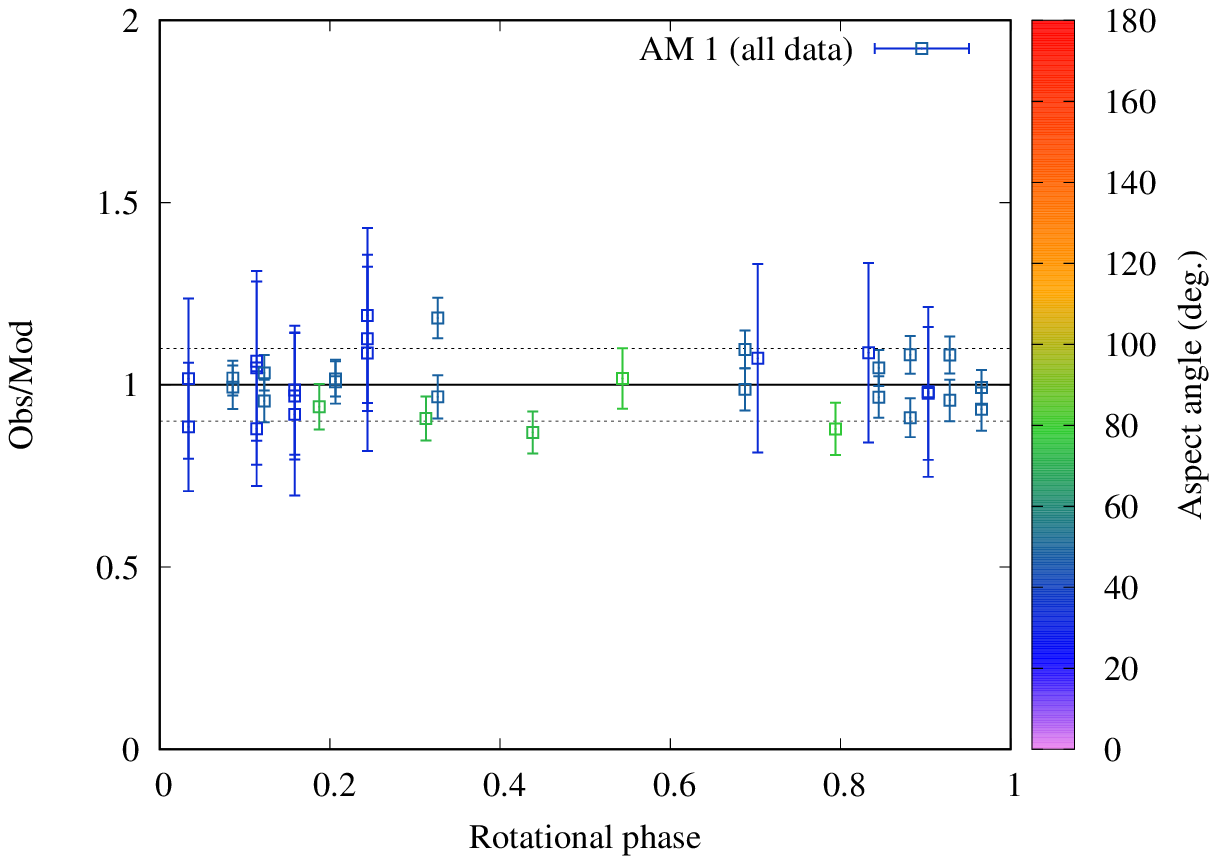}
\\
  \includegraphics[width=0.90\linewidth]{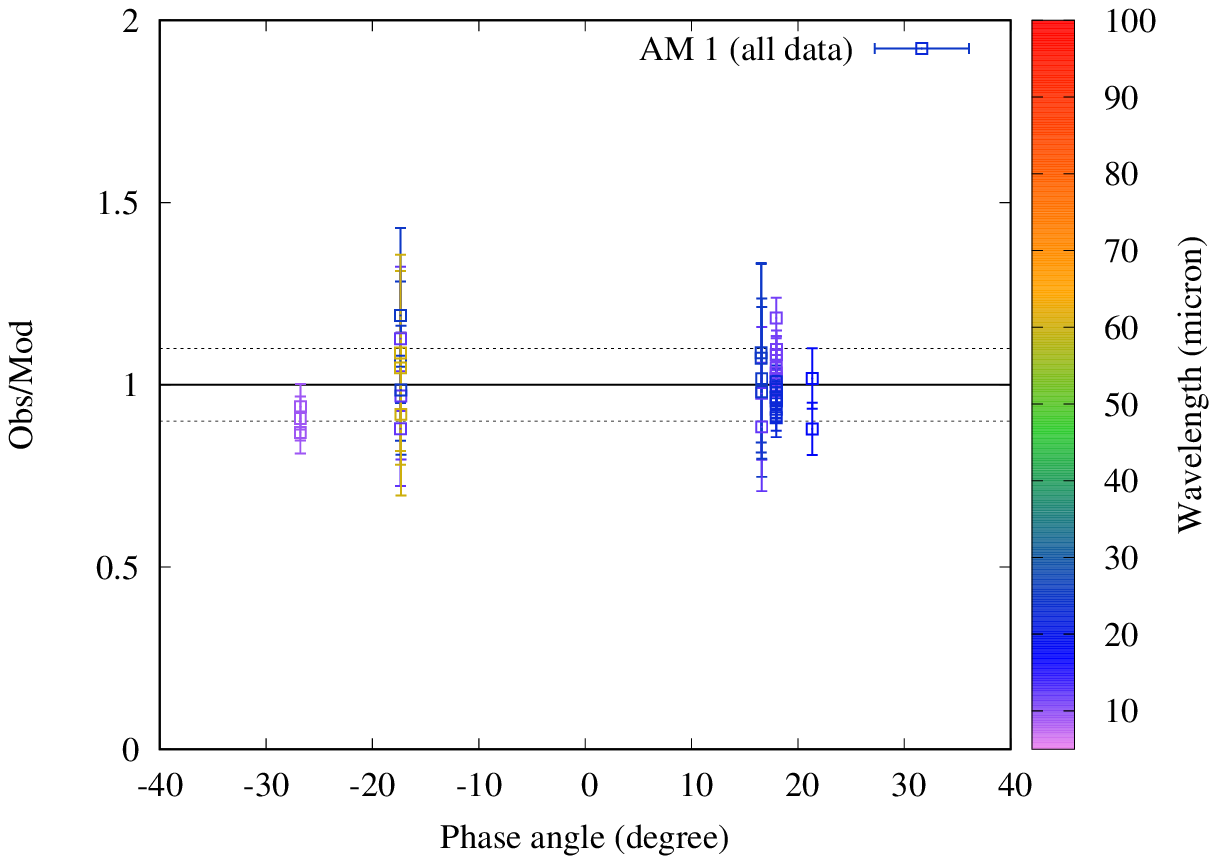}
\captionof{figure}{Asteroid (109) Felicitas, left, from top to bottom: observation to model ratios
    versus wavelength, heliocentric distance, rotational phase, and phase
    angle for model AM 1 (AM 2 was rejected). 
    Right: $\chi^2$ versus thermal inertia curves for model AM1.} 
  \label{fig:109_OMR}
\\
\end{tabularx}  
    \end{table*}

\clearpage

    \begin{table*}[ht]
    \centering
\begin{tabularx}{\linewidth}{XX}
  \includegraphics[width=0.90\linewidth]{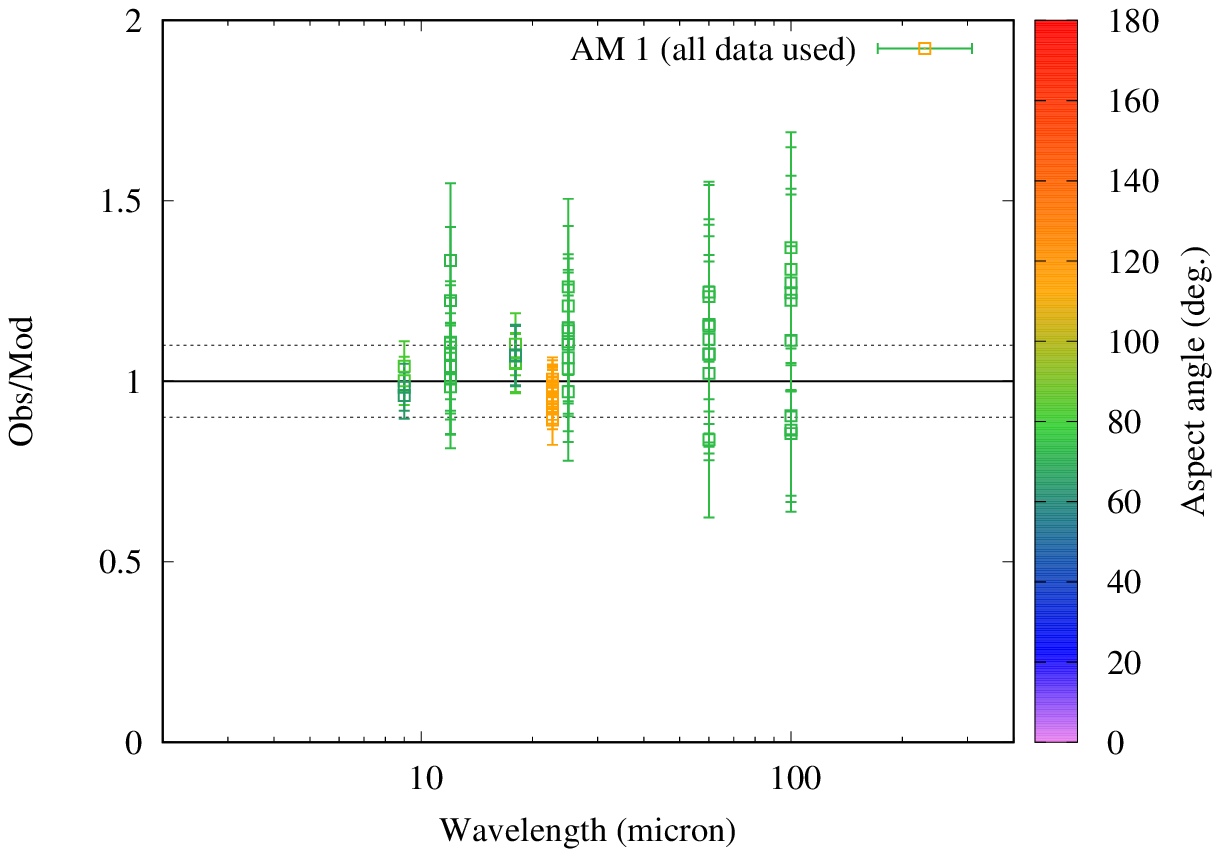}  
&
  \includegraphics[width=0.90\linewidth]{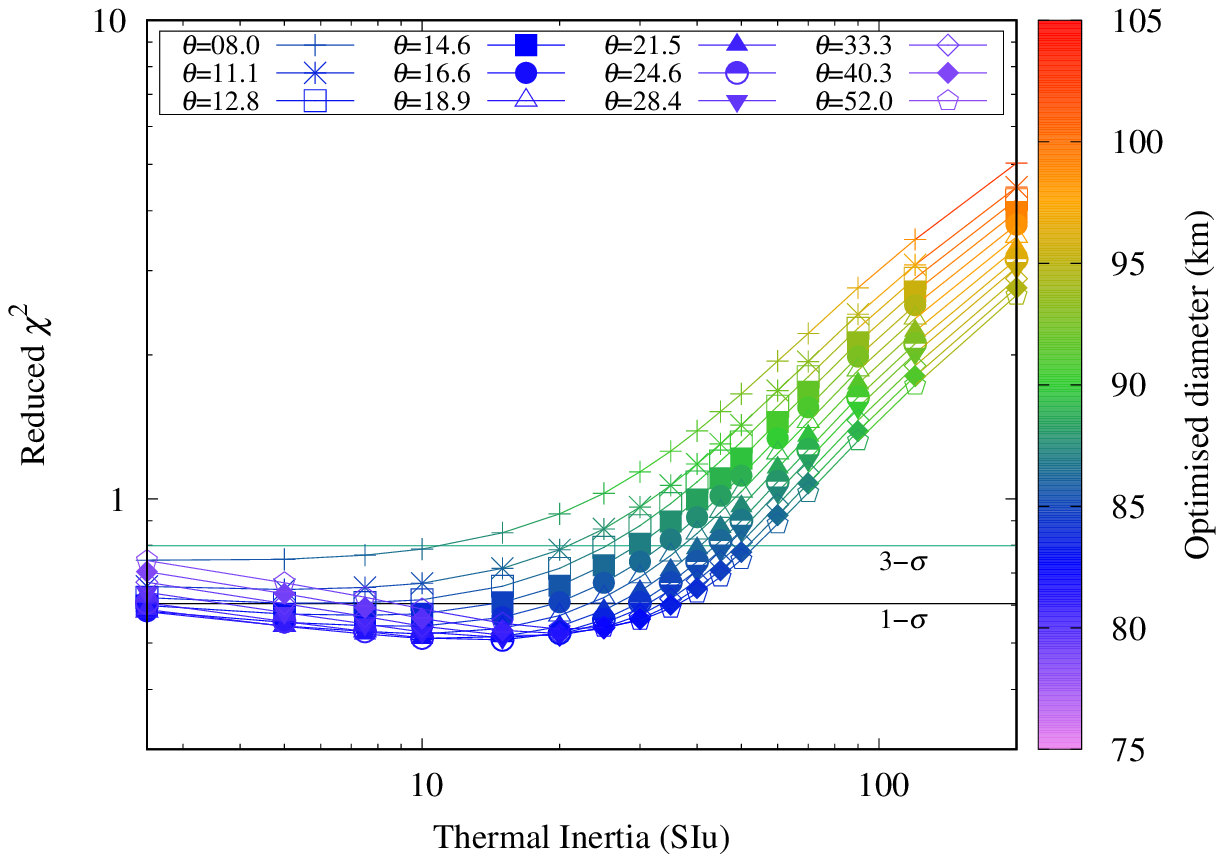}
\\
  \includegraphics[width=0.90\linewidth]{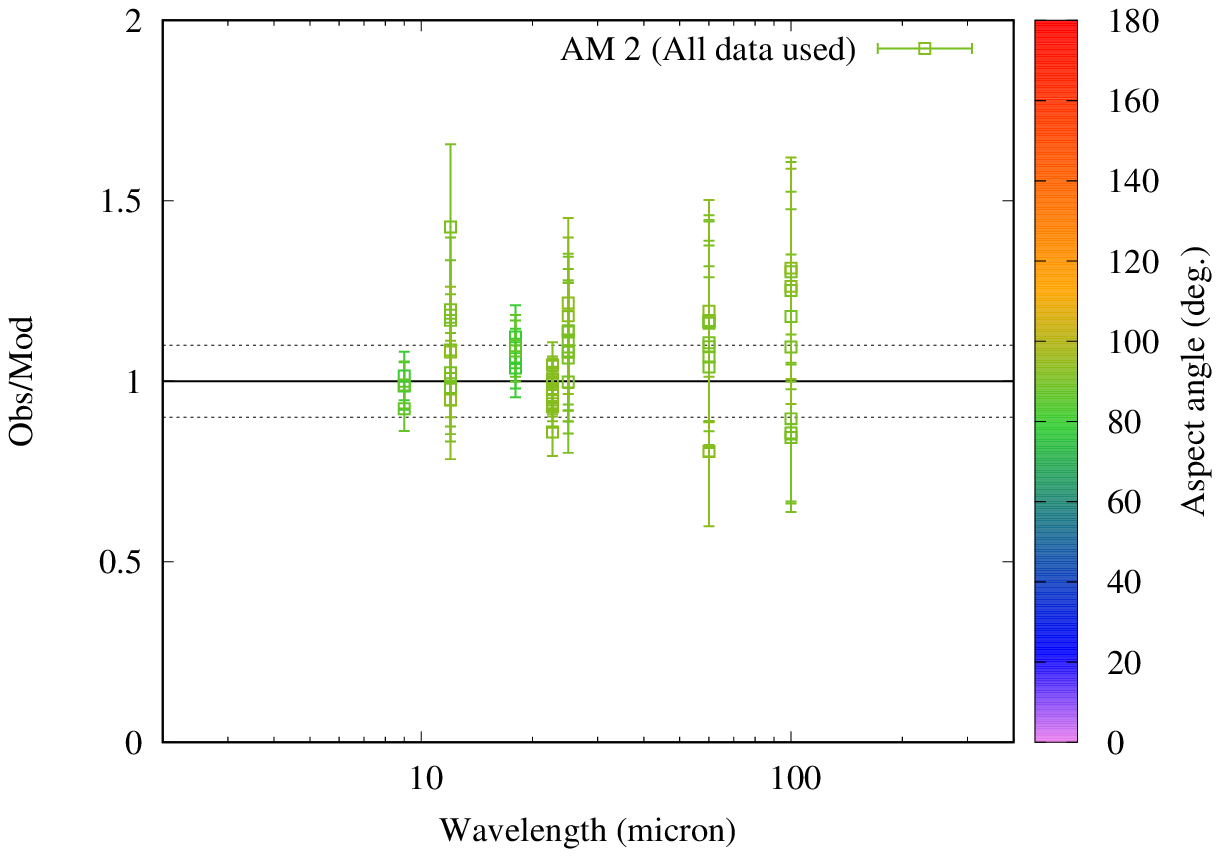}
&
  \includegraphics[width=0.90\linewidth]{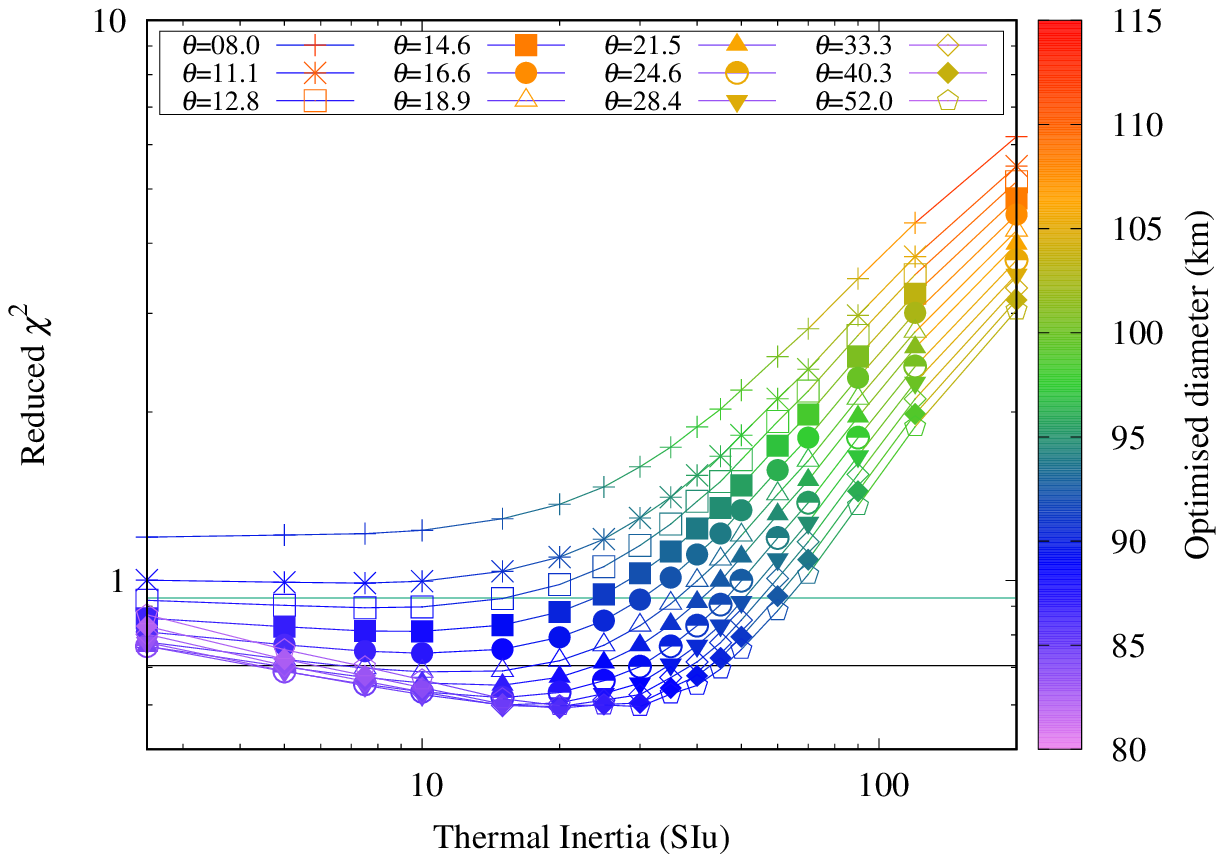}
\\
  \includegraphics[width=0.90\linewidth]{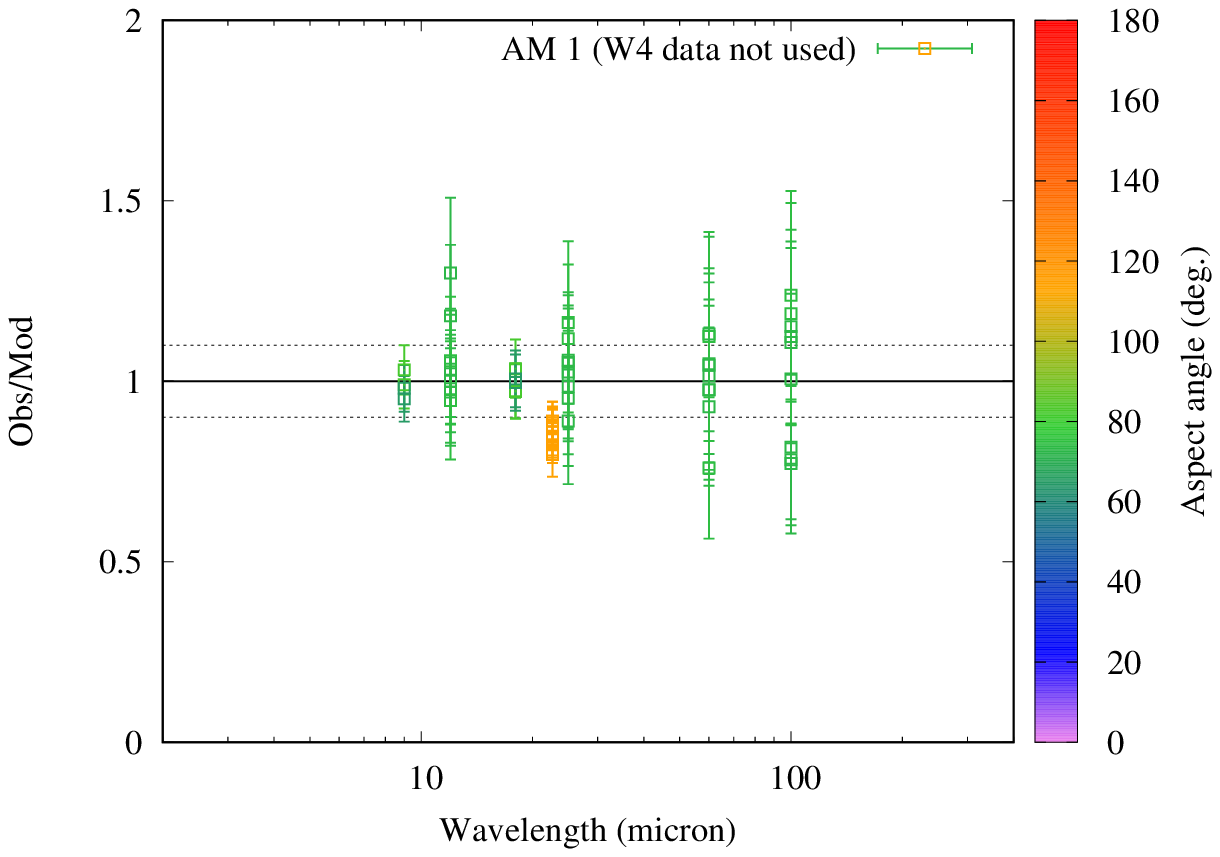}  
\\
  \includegraphics[width=0.90\linewidth]{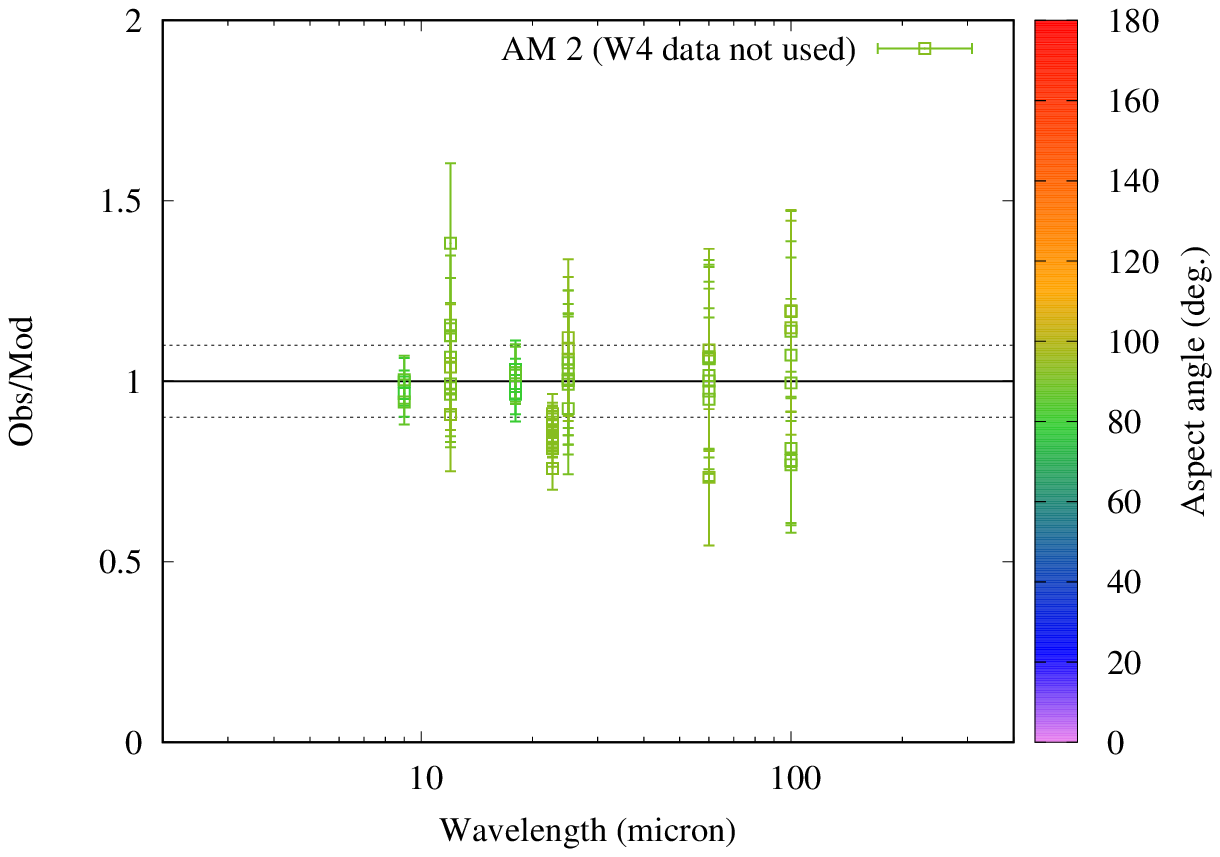}
\captionof{figure}{Asteroid (195) Eurykleia, left: observation to model ratios versus wavelength for
    models AM 1 (first and third plot) and AM 2 (second and fourth) using all data to minimise the $\chi^2$
    (top) and excluding the W4 data (bottom). The colour indicates the aspect
    angle at which the observations were taken. The AM 1 model's pole
    orientation is such that the WISE data would have been taken at a
    sub-observer latitude of $\sim$25$^\circ$ south, and the rest of the data at
    $\sim$25$^\circ$ north. On the other hand, if  pole 2 is correct, then
    all data would have been taken at similar ``equator-on'' aspect angles
    between 80$^\circ$ and 100$^\circ$. 
    Right: $\chi^2$ versus thermal inertia curves for model AM1 (top) and AM2 (bottom) using all thermal data. 
  }    \label{fig:195_OMR}
\\
\end{tabularx}    
\end{table*}  

\clearpage

\begin{figure}
  \centering

  \includegraphics[width=0.90\linewidth]{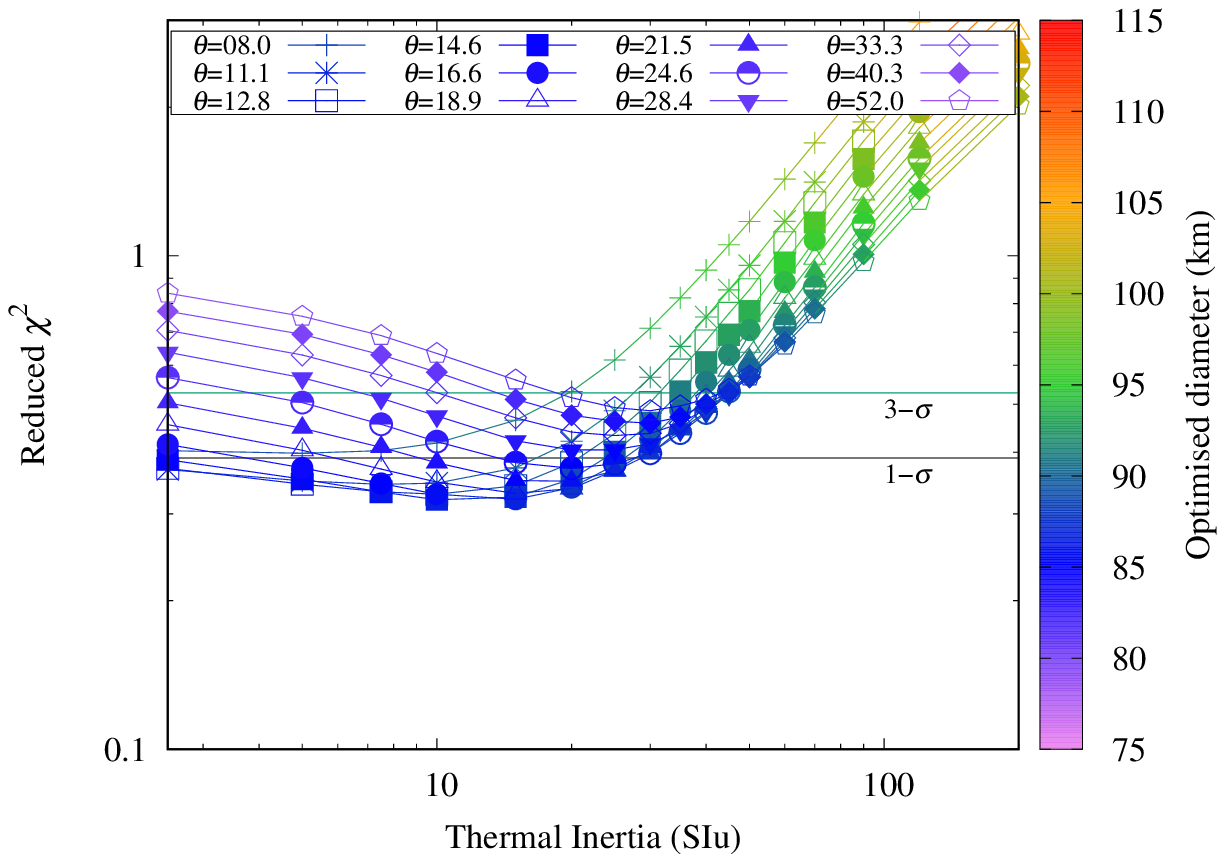}

  \includegraphics[width=0.90\linewidth]{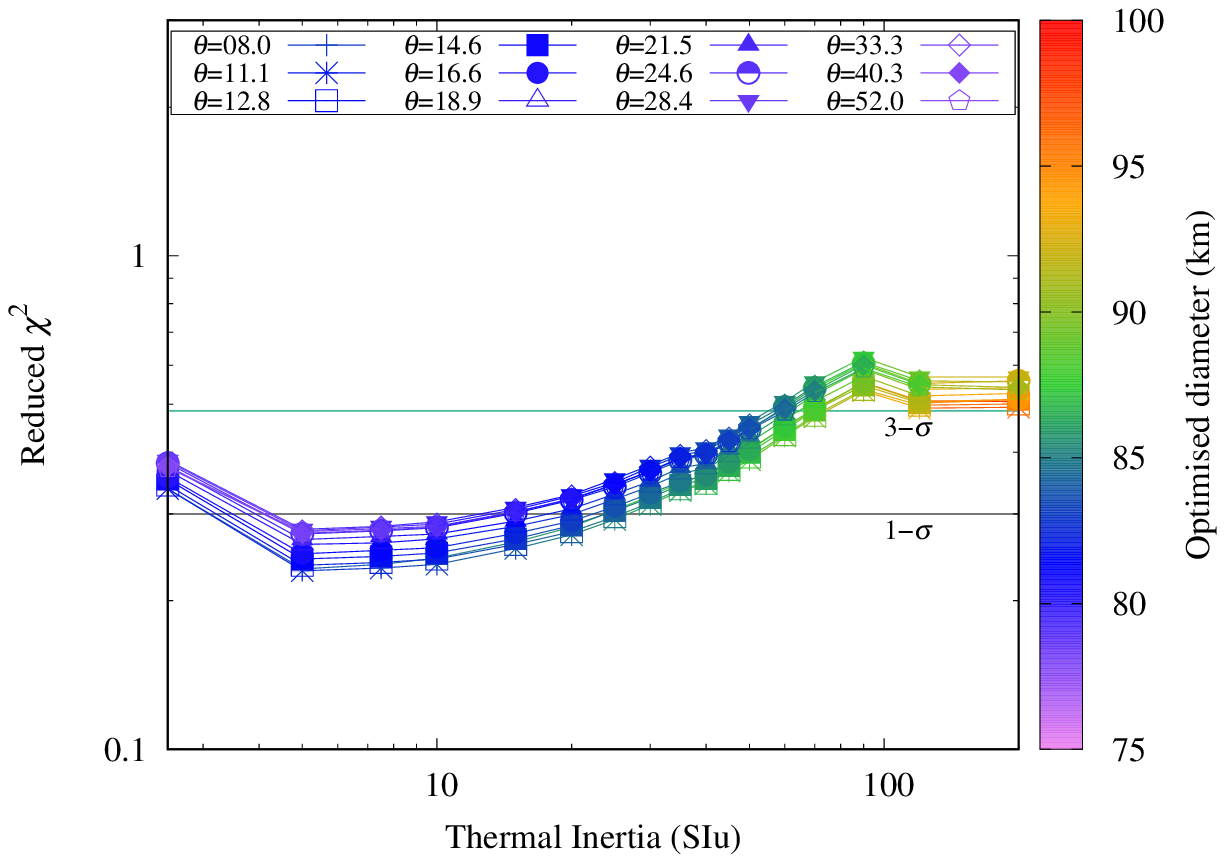}

  \includegraphics[width=0.90\linewidth]{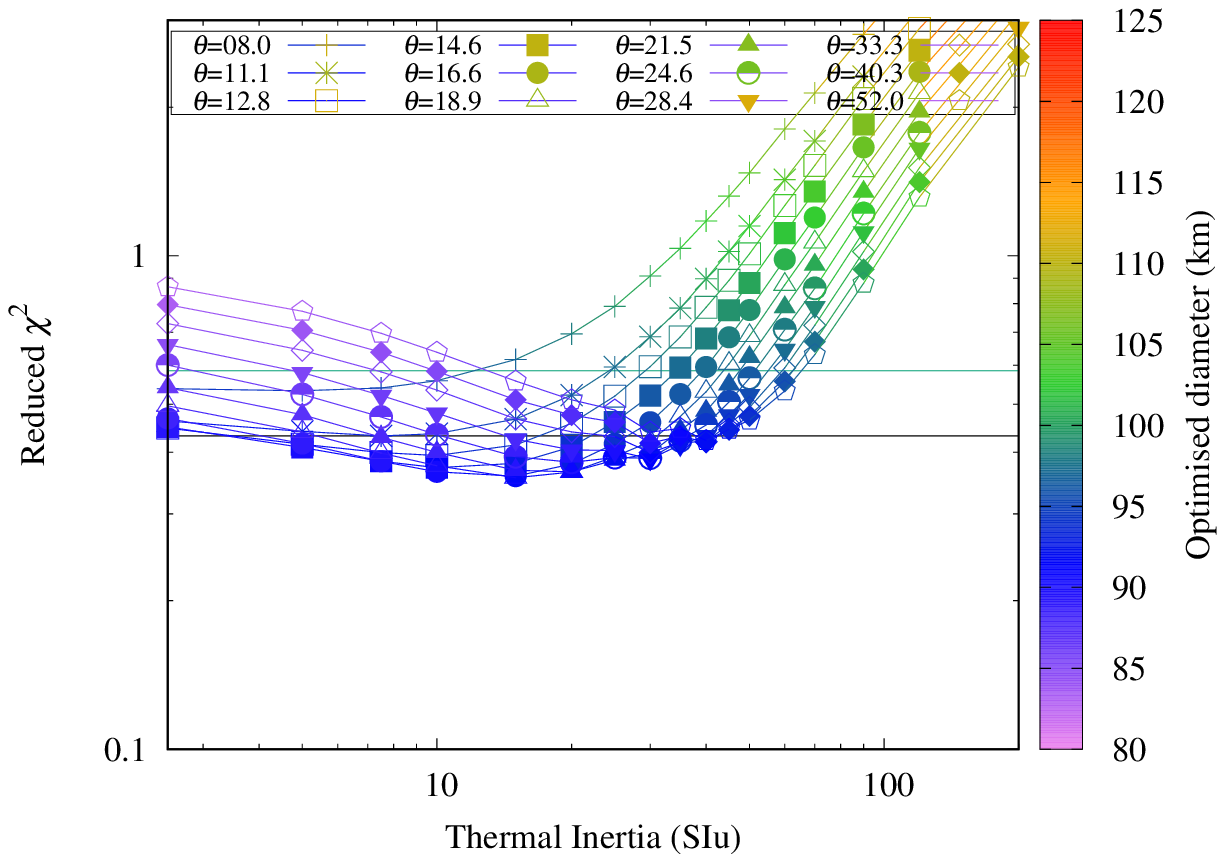}

  \includegraphics[width=0.90\linewidth]{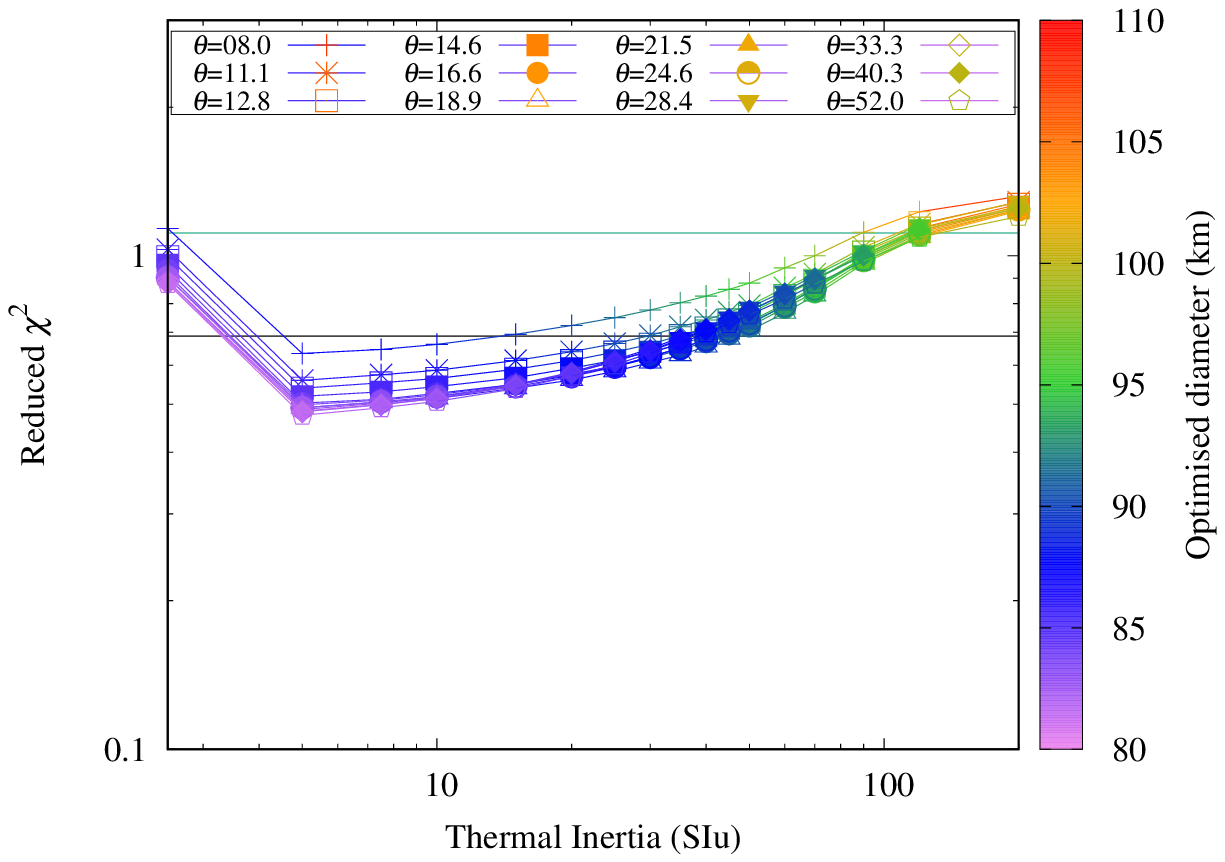}
 
  \caption{Additional $\chi^2$ versus thermal inertia plots for (195) Eurykleia: 
  Top to bottom: Model AM1 fitted to data excluding WISE W4 data, AM1 with W4 data only, 
  model AM2 fitted to data excluding WISE W4 data, AM2 with W4 data only. 
  }\label{chi2_195}
\end{figure}

\clearpage

    \begin{table*}[ht]
    \centering
\begin{tabularx}{\linewidth}{XX}
  \includegraphics[width=0.90\linewidth]{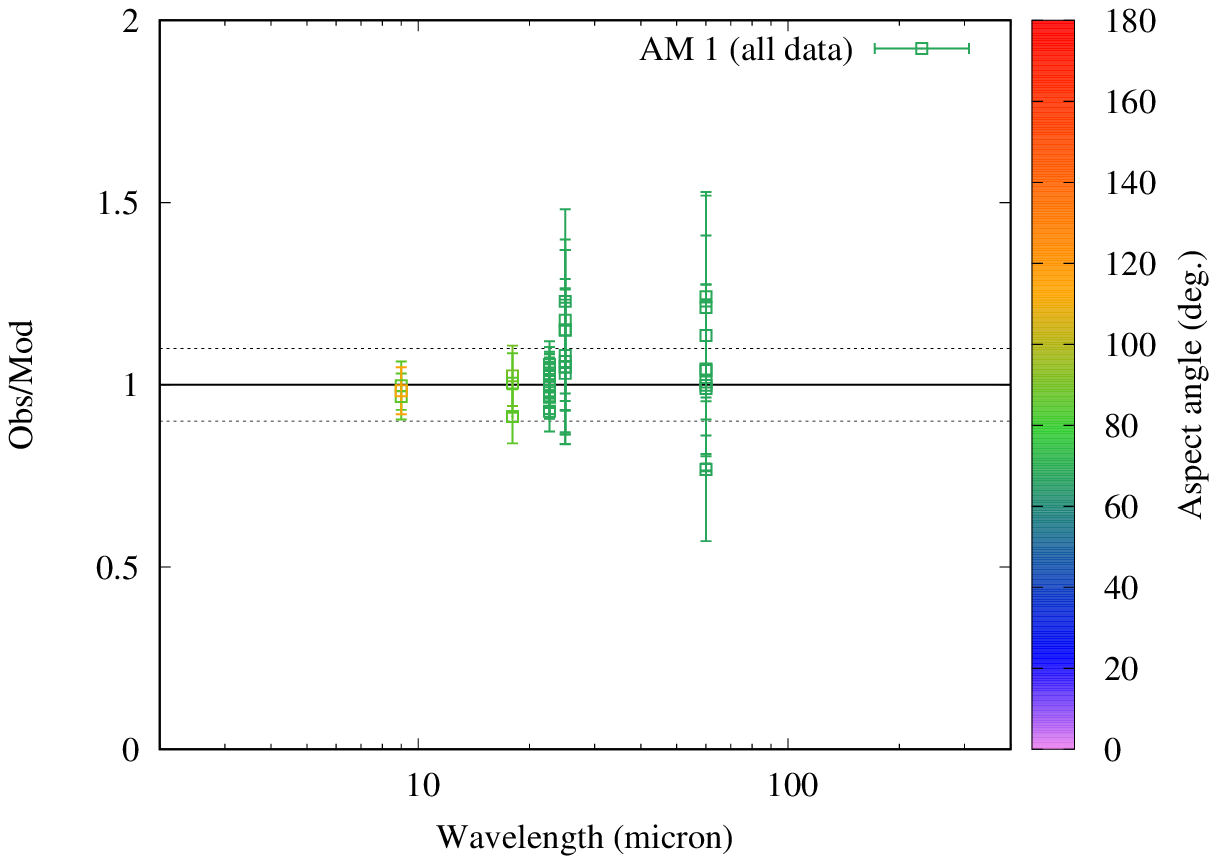}
&
  \includegraphics[width=0.90\linewidth]{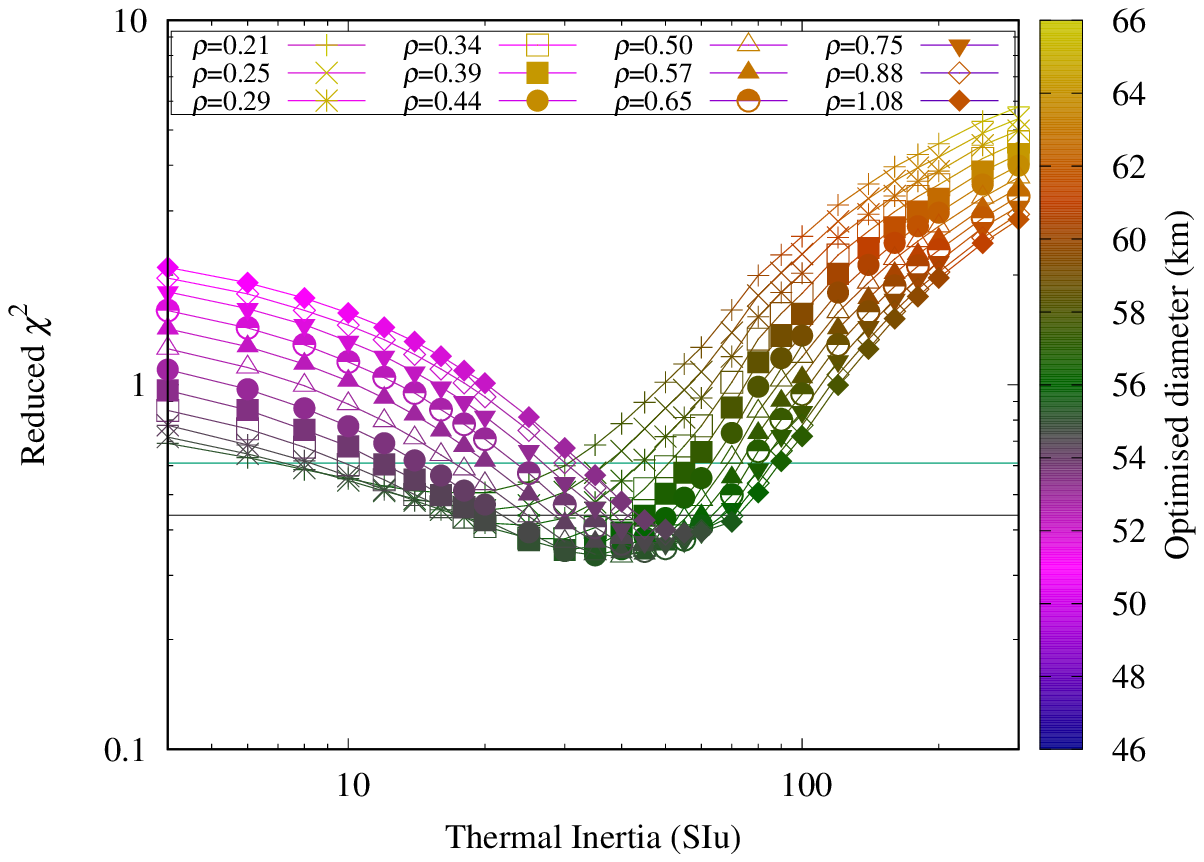}
\\
  \includegraphics[width=0.90\linewidth]{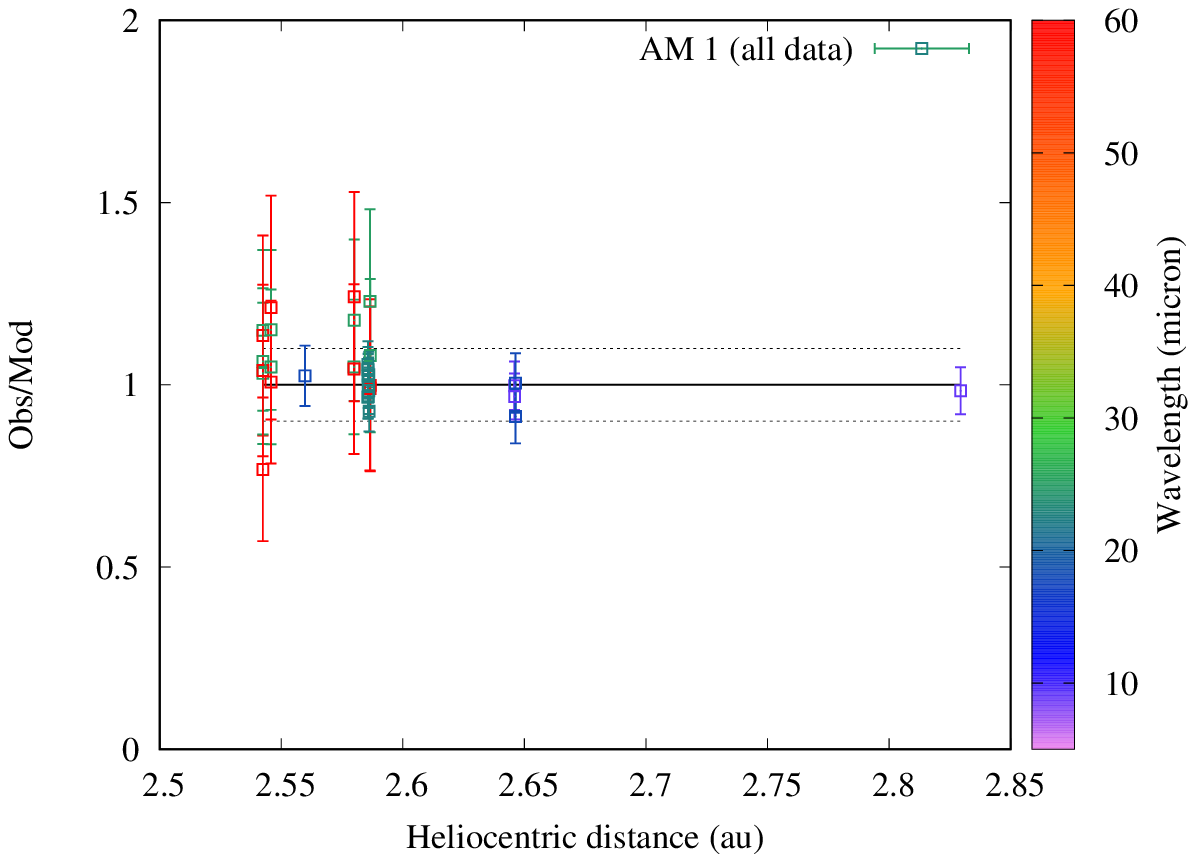}
&
  \includegraphics[width=0.90\linewidth]{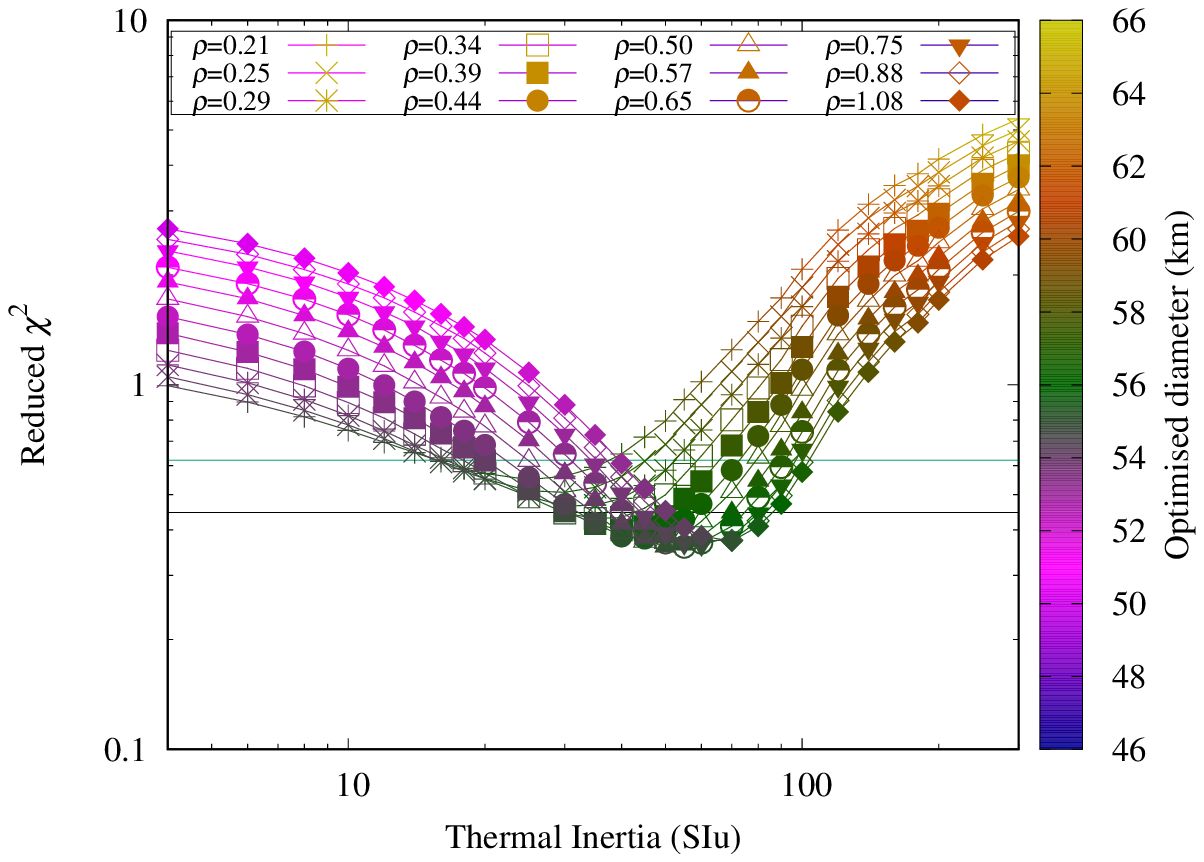}
\\
  \includegraphics[width=0.90\linewidth]{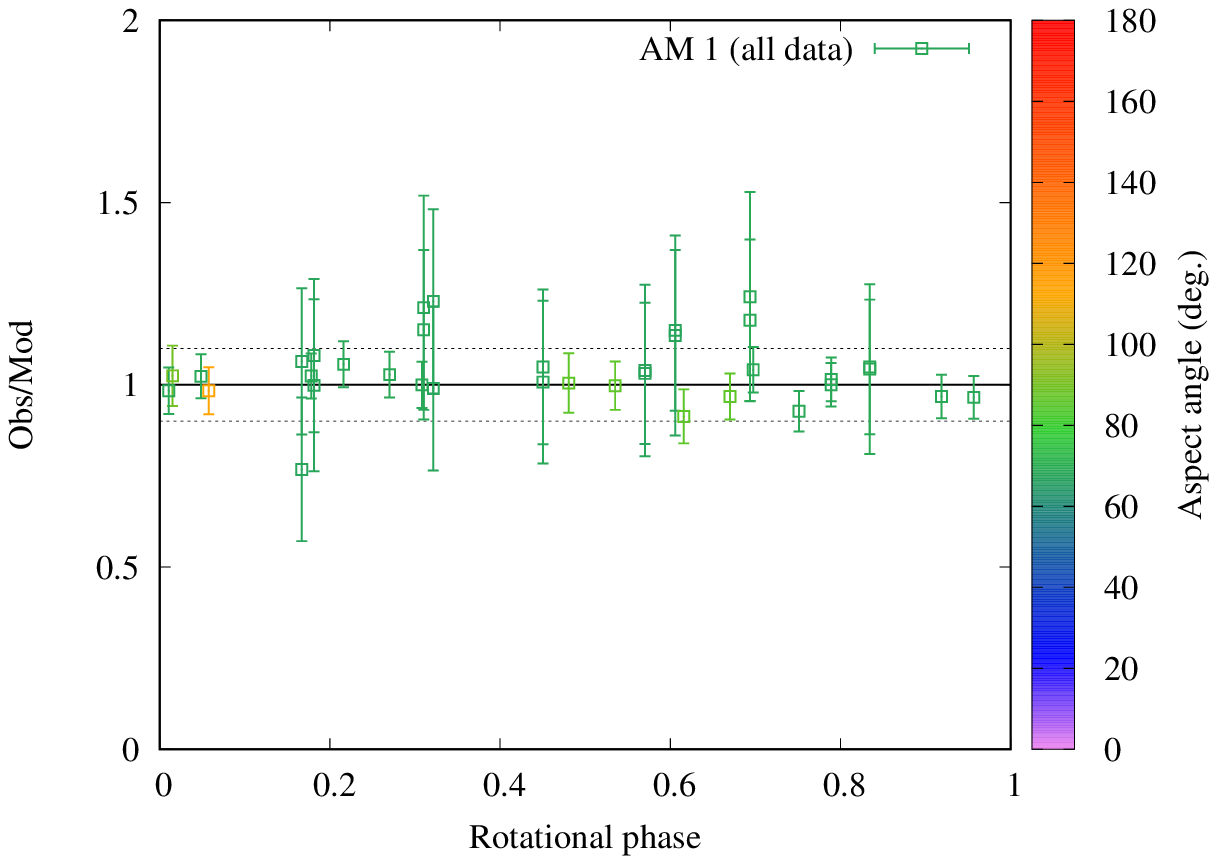}
\\
  \includegraphics[width=0.90\linewidth]{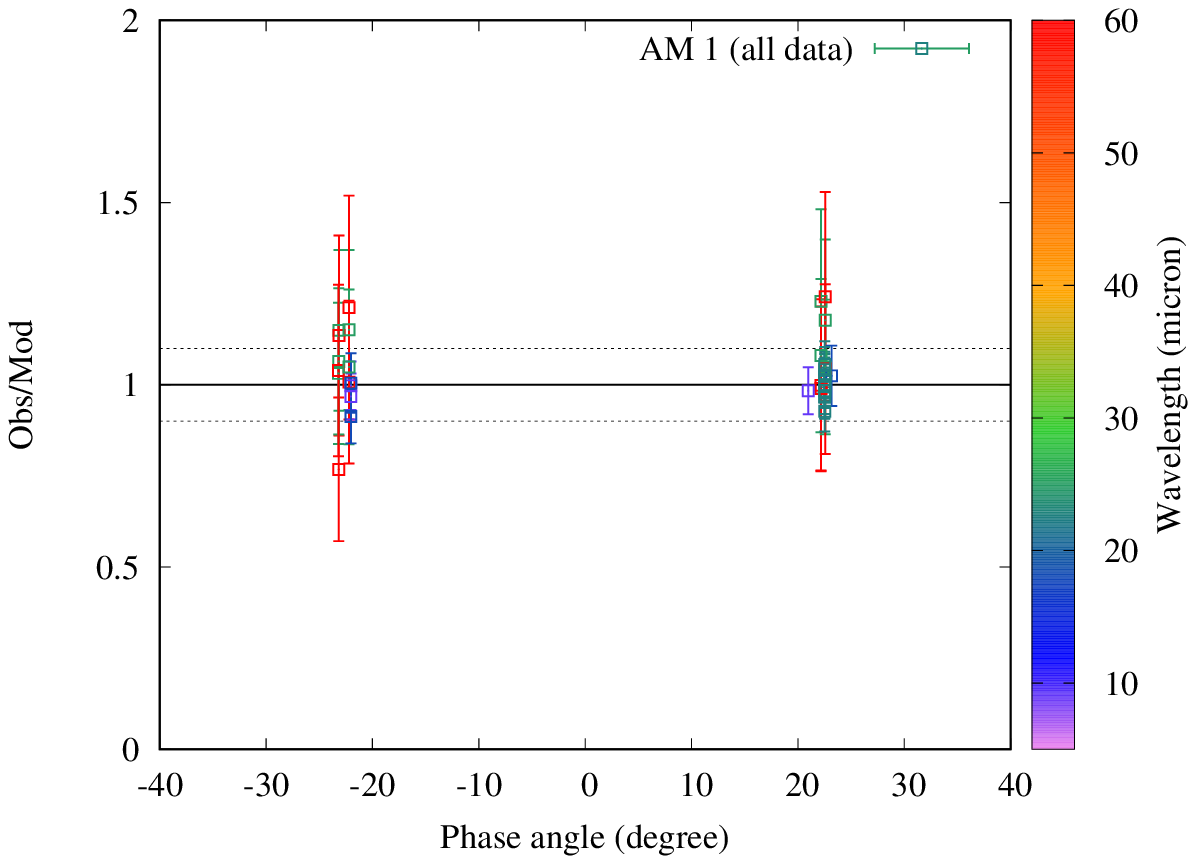}
  \captionof{figure}{Asteroid (301) Bavaria, left, from top to bottom: observation to model ratios
    versus wavelength, heliocentric distance, rotational phase, and phase
    angle for model AM 1 (the same plots for AM 2 looked similar). 
    Right: $\chi^2$ versus thermal inertia curves for model AM1 (top), and AM2 (bottom). 
  }\label{fig:301_OMR}
\\  
\end{tabularx}  
\end{table*}

\clearpage

    \begin{table*}[ht]
    \centering
\begin{tabularx}{\linewidth}{XX}
  \includegraphics[width=0.90\linewidth]{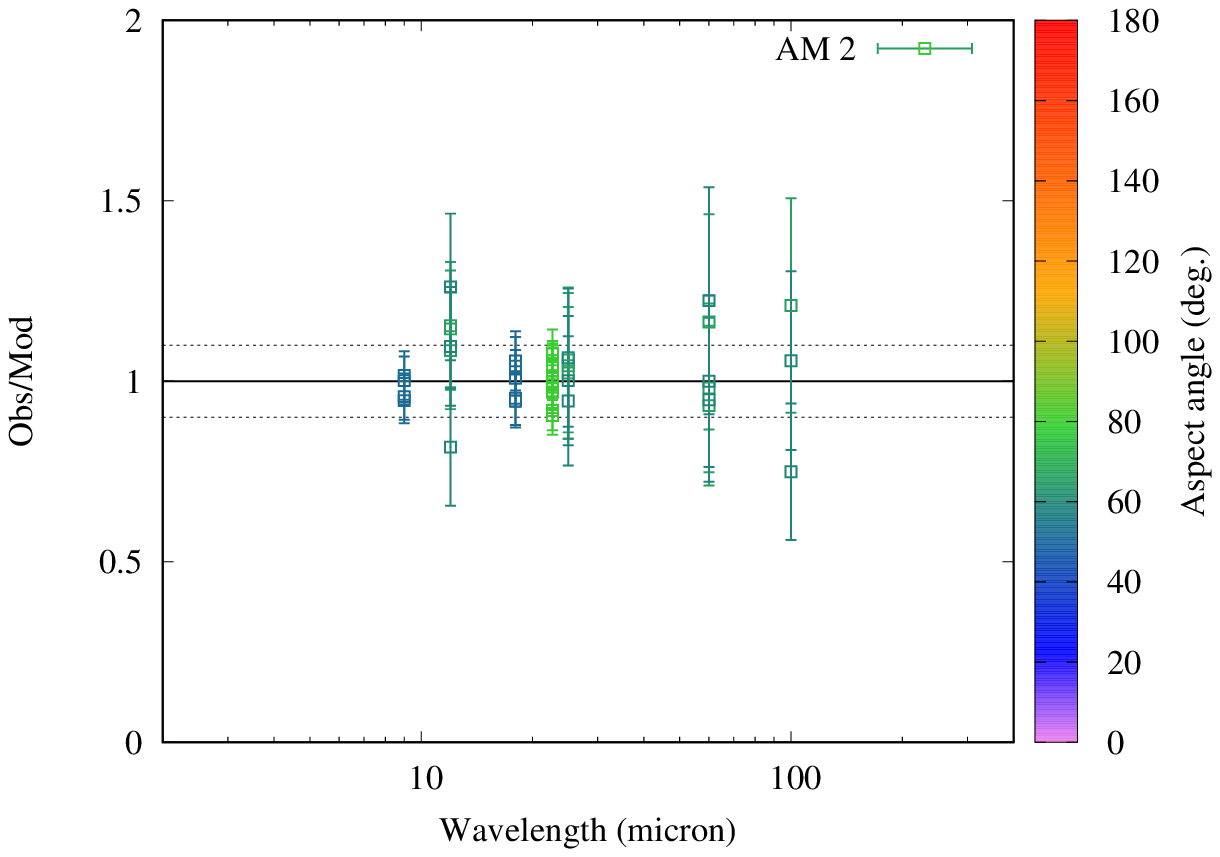}  
&
  \includegraphics[width=0.90\linewidth]{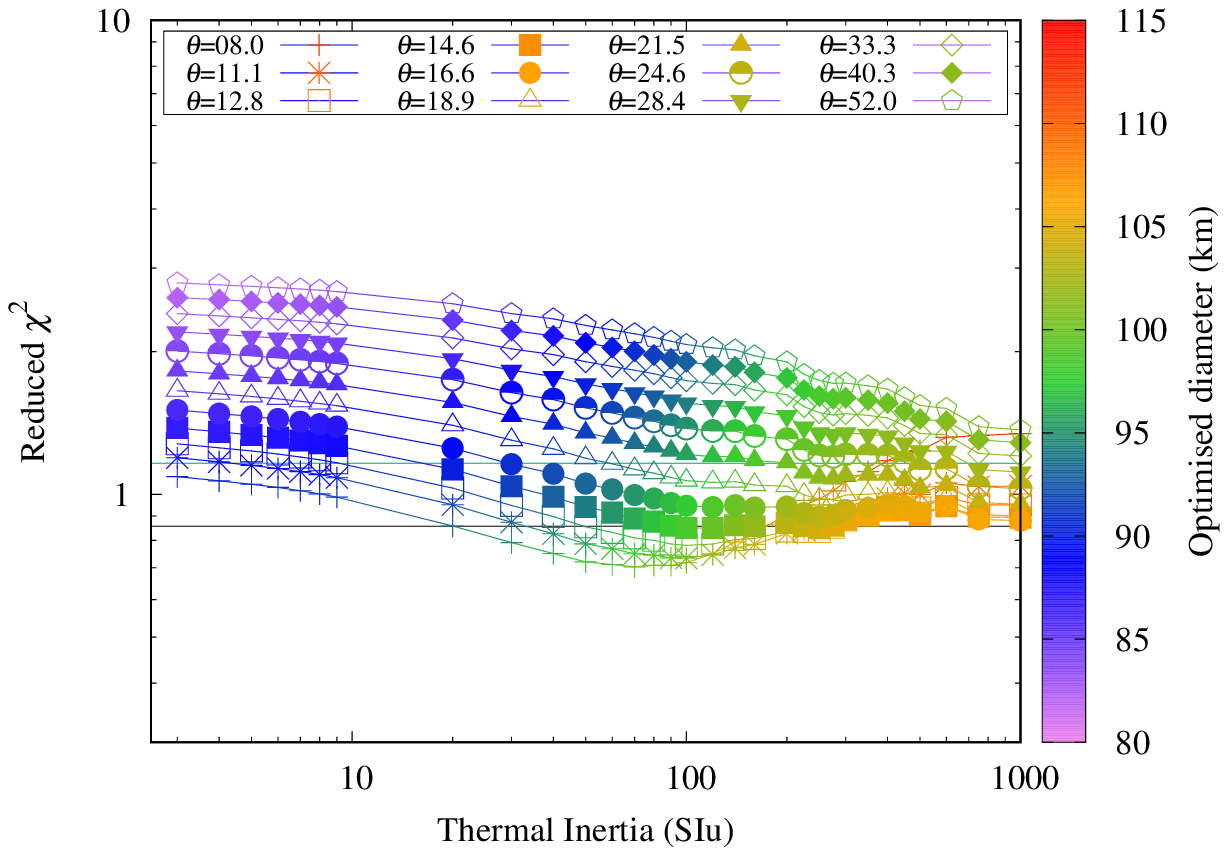}
\\
  \includegraphics[width=0.90\linewidth]{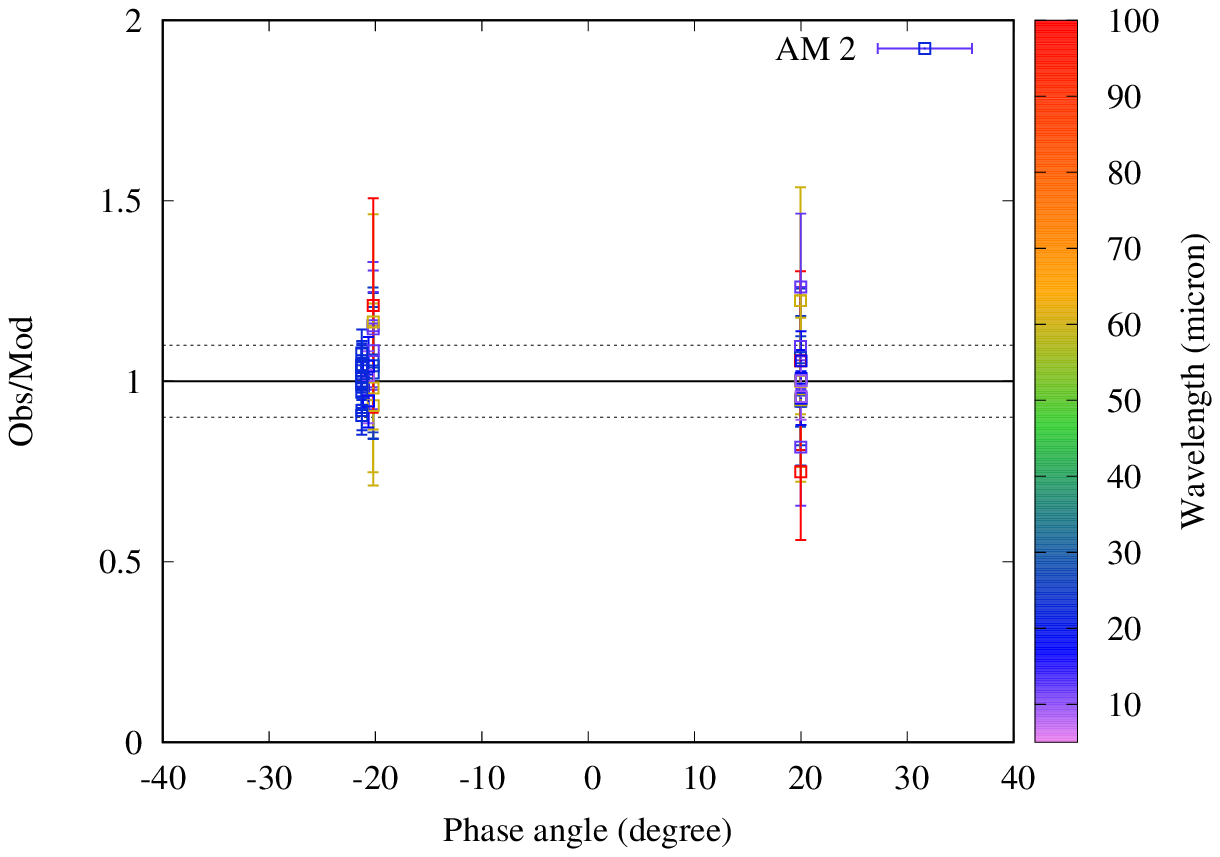}
&
  \includegraphics[width=0.90\linewidth]{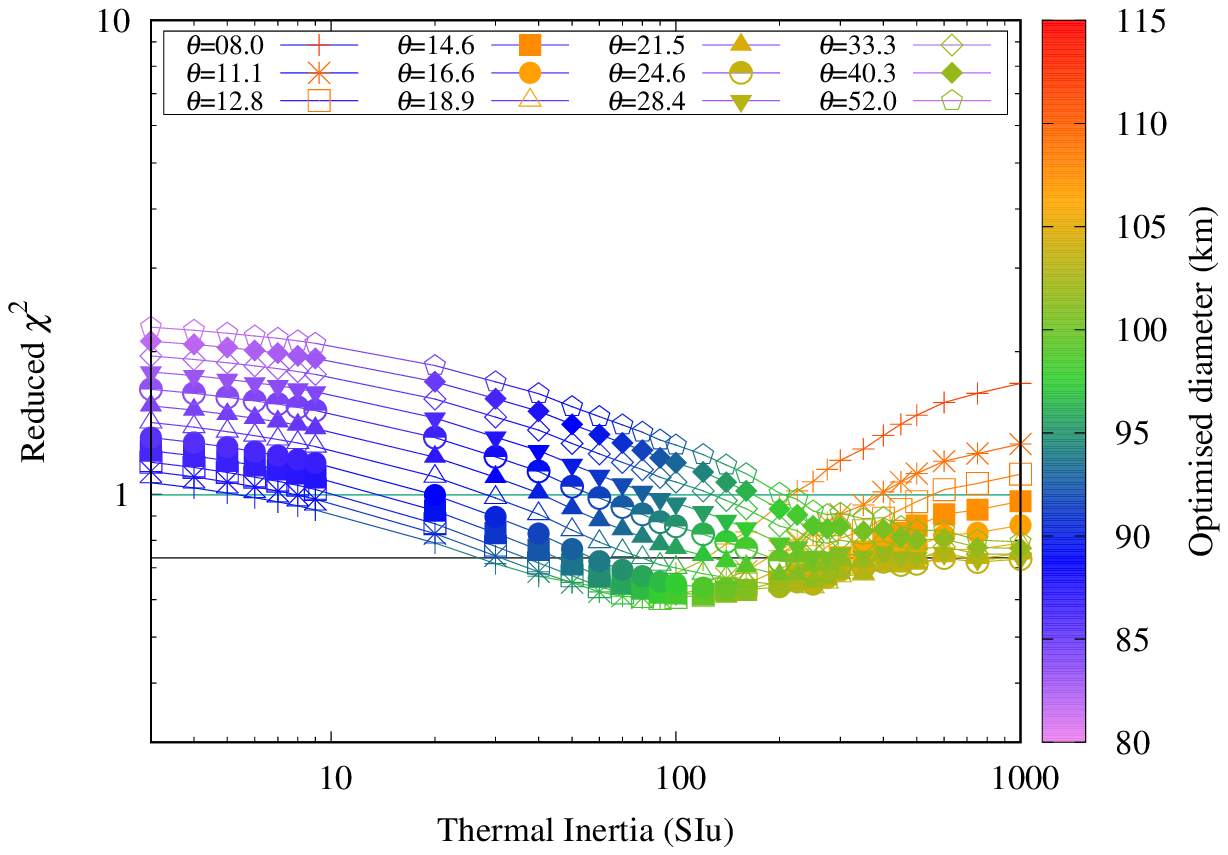}
\\
  \includegraphics[width=0.90\linewidth]{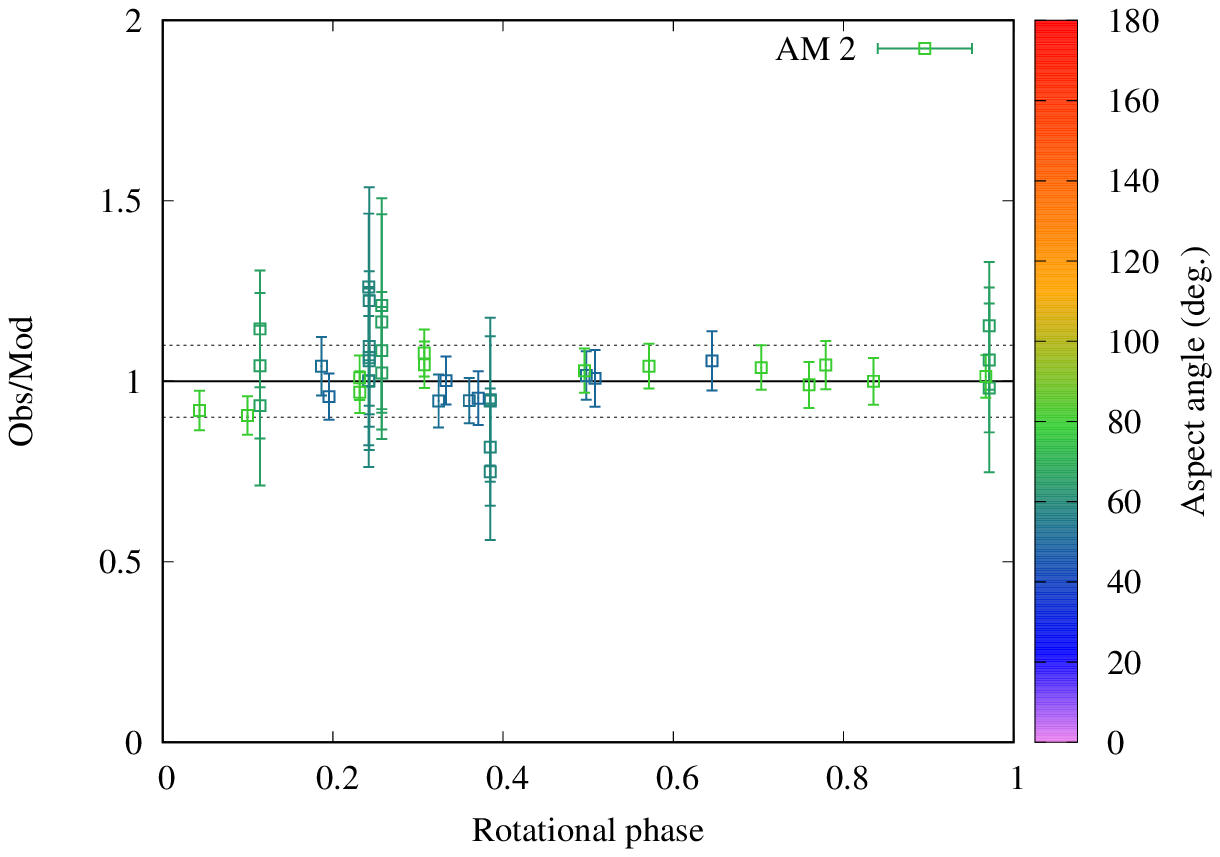}  
\\
  \includegraphics[width=0.90\linewidth]{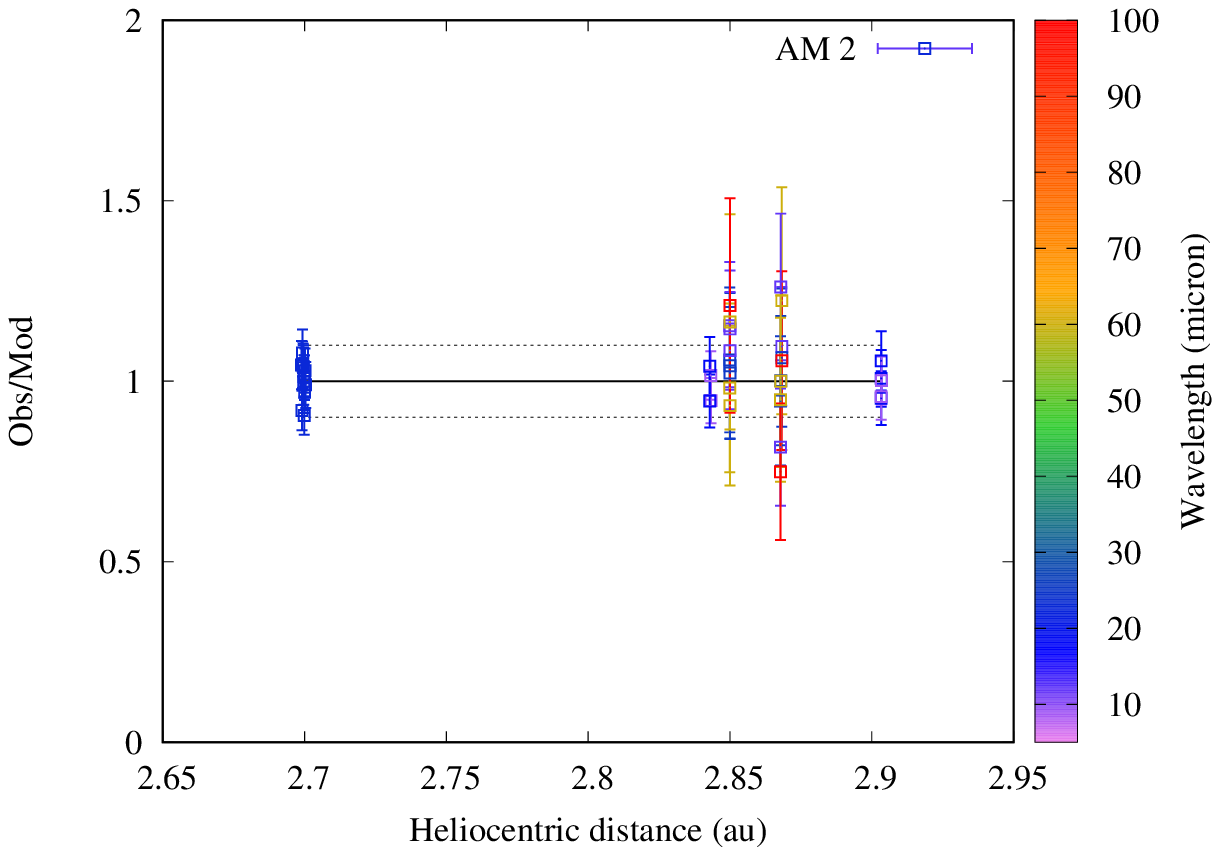}
  \captionof{figure}{Asteroid (335) Roberta, left: observation to model ratios versus wavelength,
  phase angle, rotational phase, and heliocentric distance for model AM 2.
  The plots for AM 1 looked similar. 
  Right: $\chi^2$ versus thermal inertia curves for model AM1 (top), and AM2 (bottom). 
  }    \label{fig:335_OMR}
\\  
\end{tabularx}  
\end{table*}  

\clearpage

    \begin{table*}[ht]
    \centering
\begin{tabularx}{\linewidth}{XX}
  \includegraphics[width=0.90\linewidth]{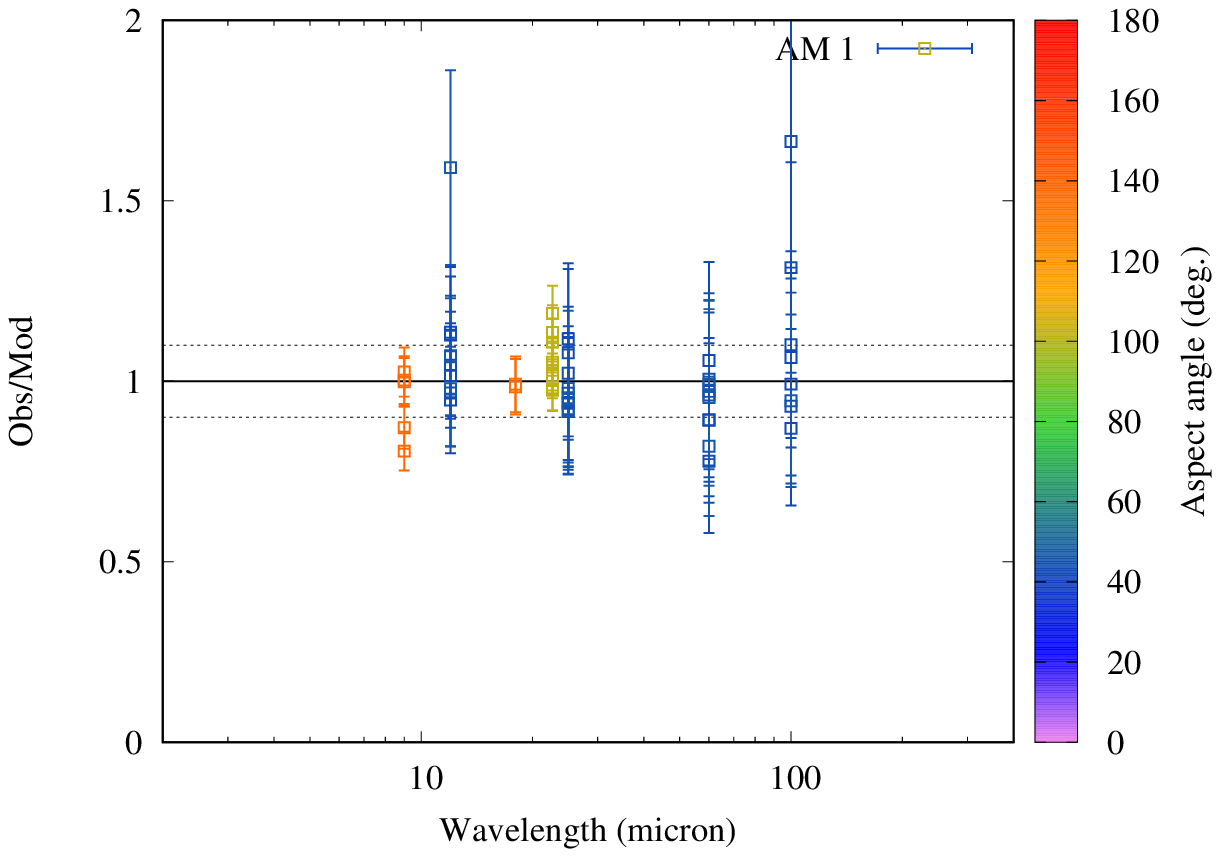}  
&
  \includegraphics[width=0.90\linewidth]{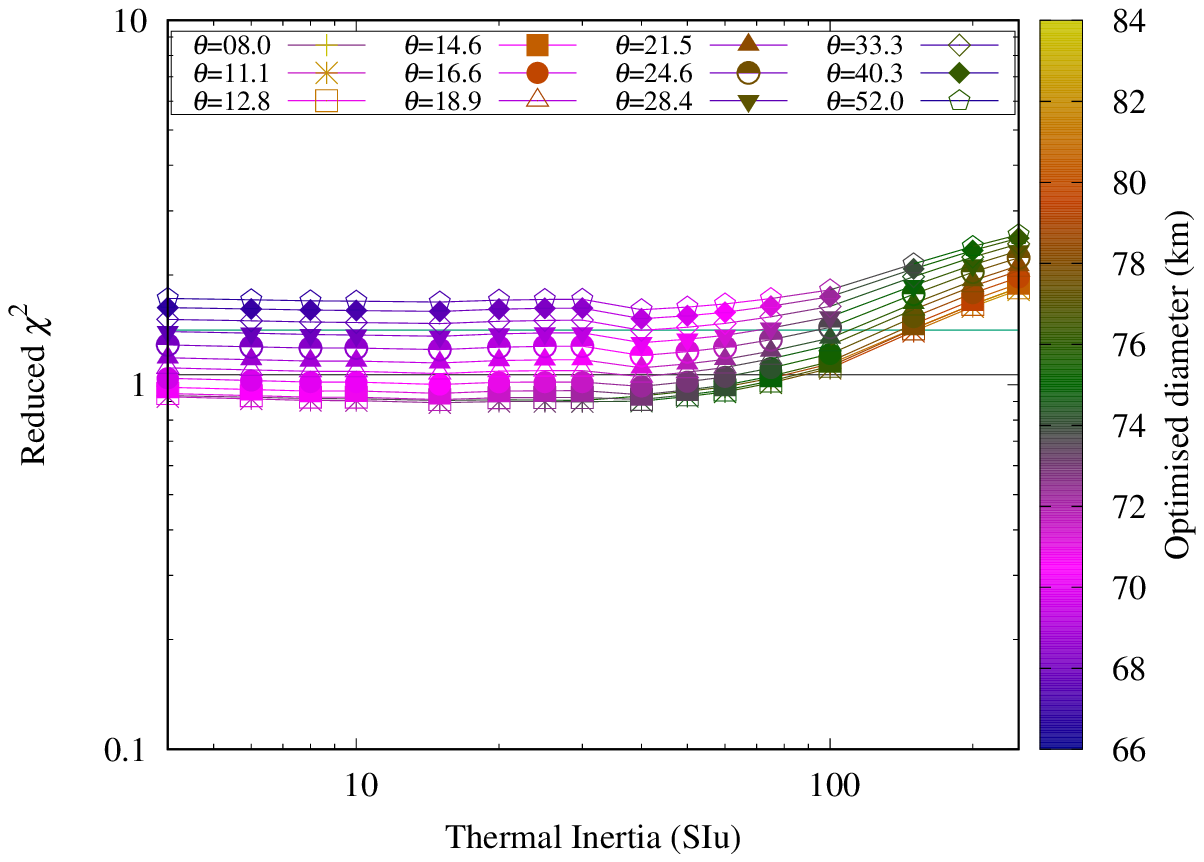}
\\
  \includegraphics[width=0.90\linewidth]{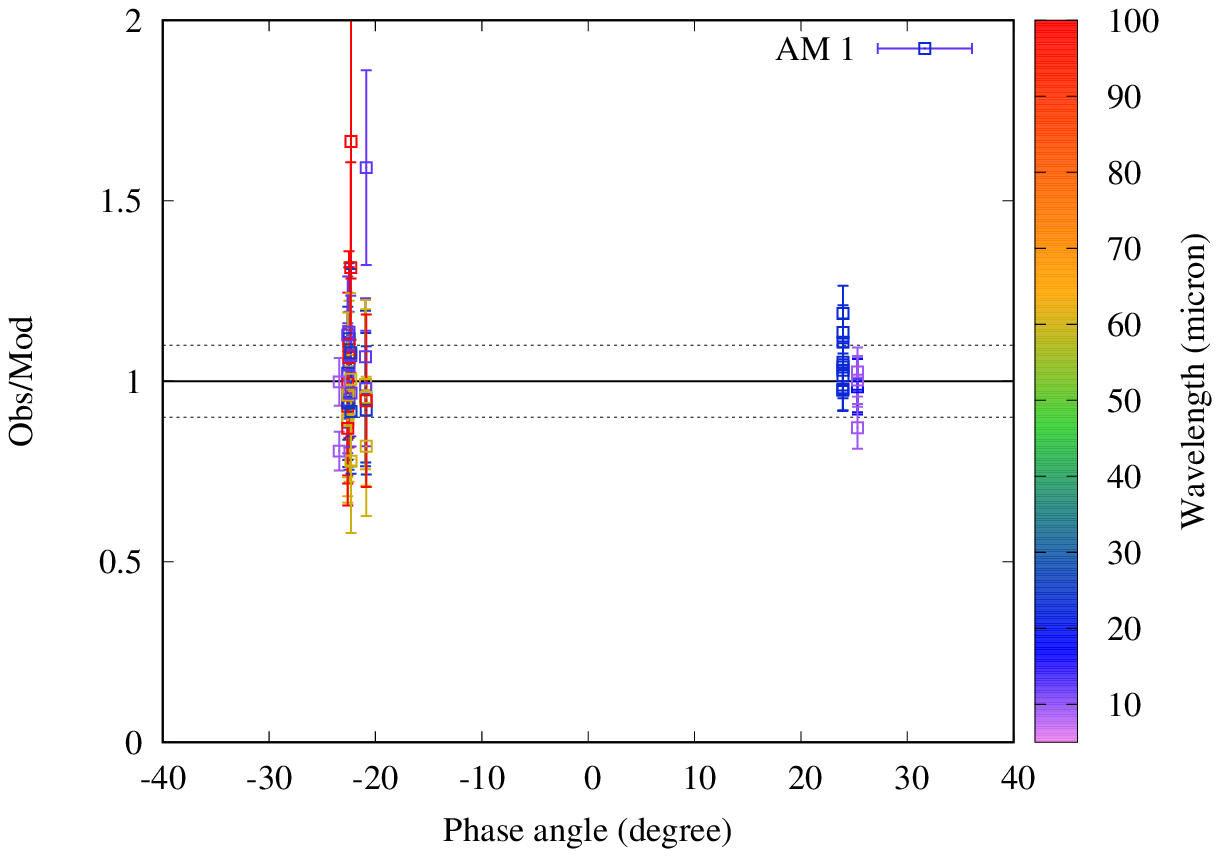}
&
  \includegraphics[width=0.90\linewidth]{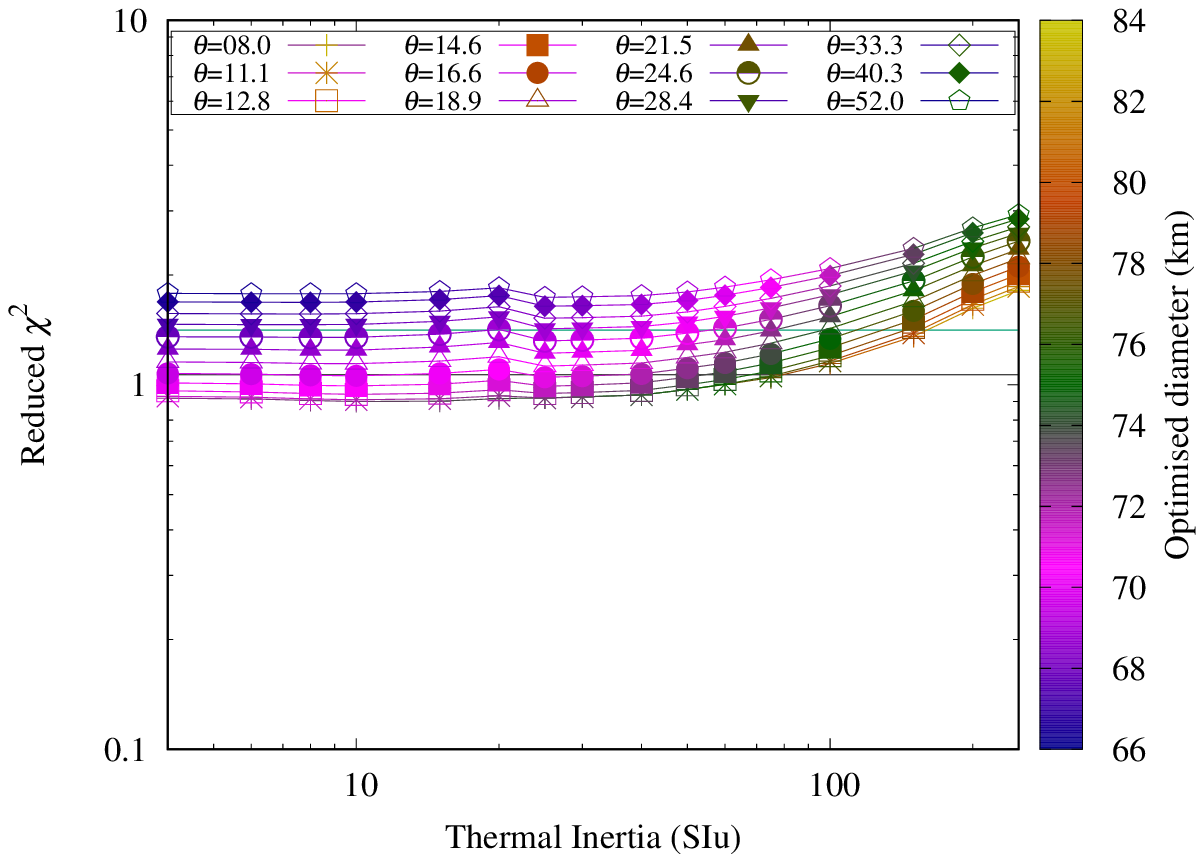}
\\
  \includegraphics[width=0.90\linewidth]{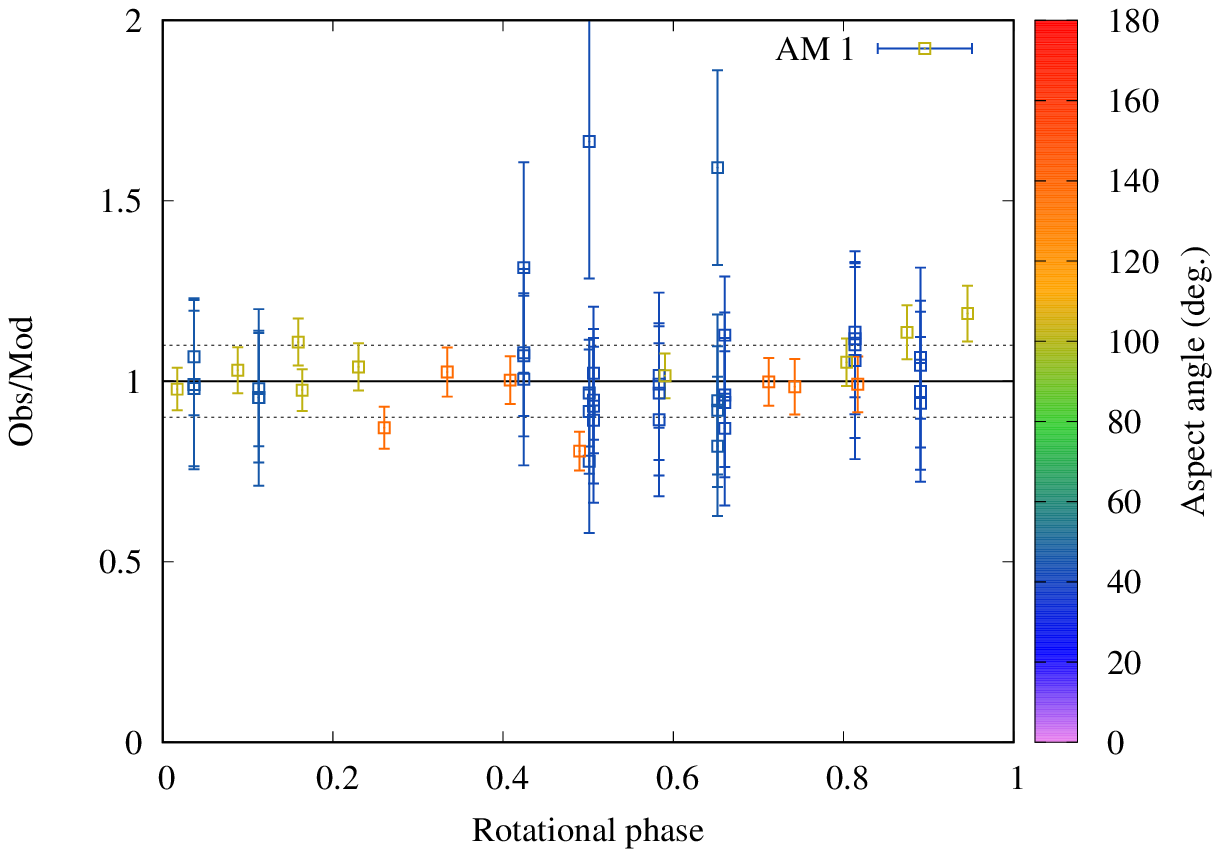}  
\\
  \includegraphics[width=0.90\linewidth]{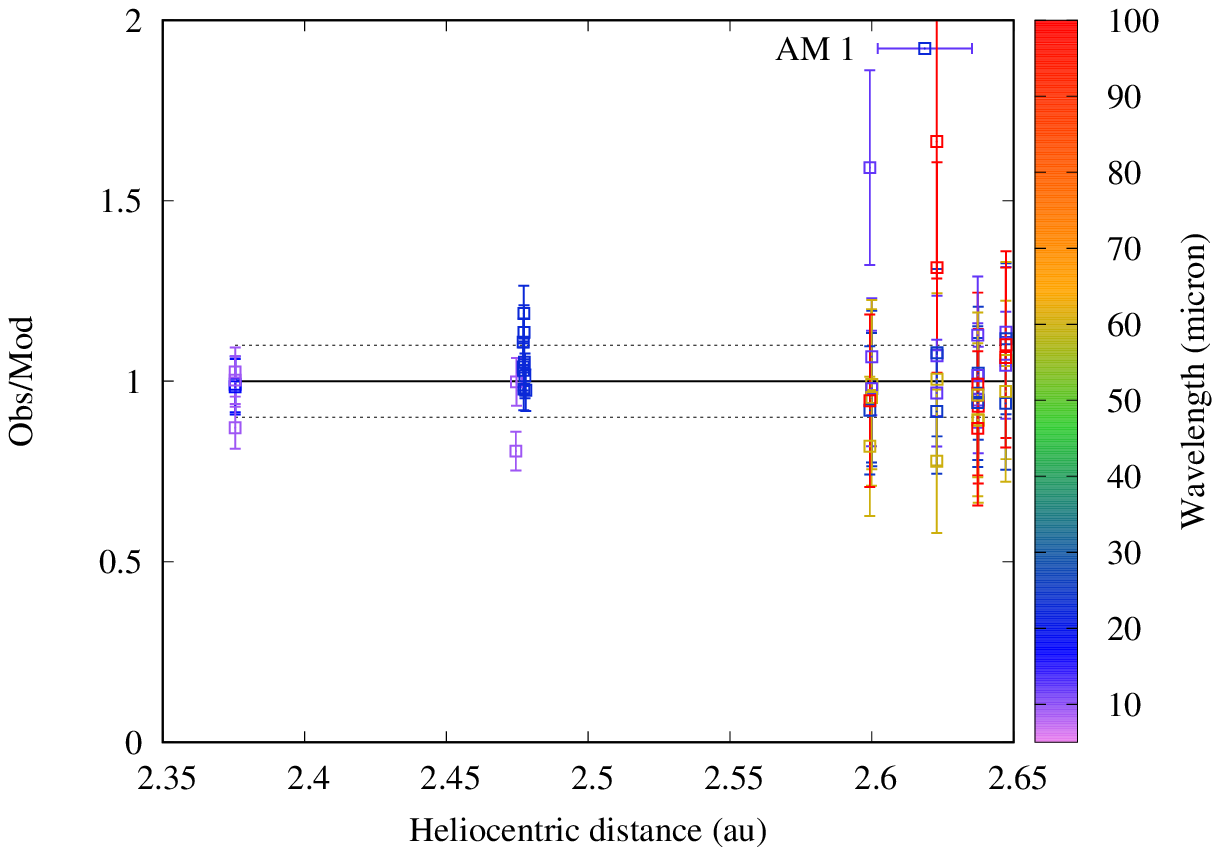}
  \captionof{figure}{Asteroid (380) Fiducia, left: observation to model ratios versus wavelength, 
    phase angle, rotational phase, and heliocentric distance for model AM 1.
    The plots for AM 2 looked similar. 
    Right: $\chi^2$ versus thermal inertia curves for model AM1 (top), and AM2 (bottom).  
  }    \label{fig:380_OMR}
\\  
\end{tabularx}  
\end{table*}  

\clearpage

    \begin{table*}[ht]
    \centering
\begin{tabularx}{\linewidth}{XX}
  \includegraphics[width=0.90\linewidth]{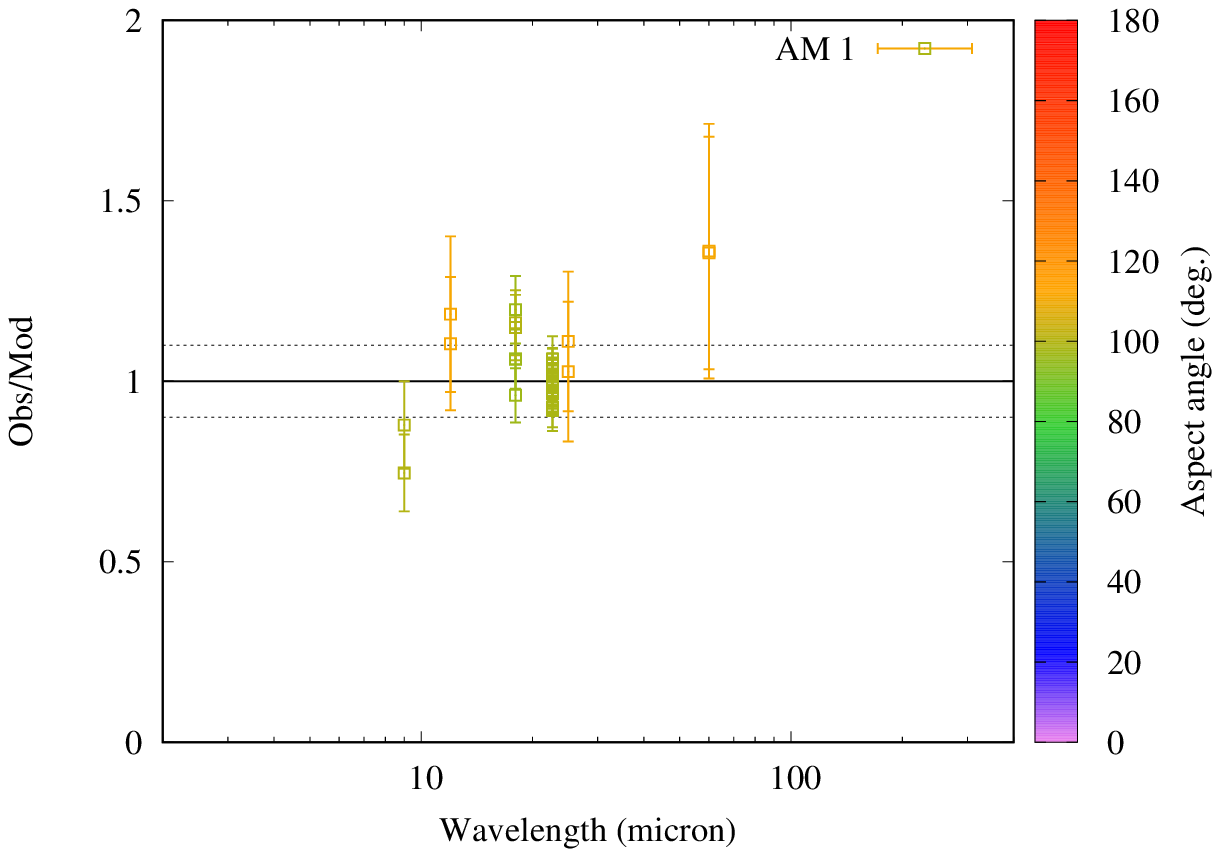}  
&
  \includegraphics[width=0.90\linewidth]{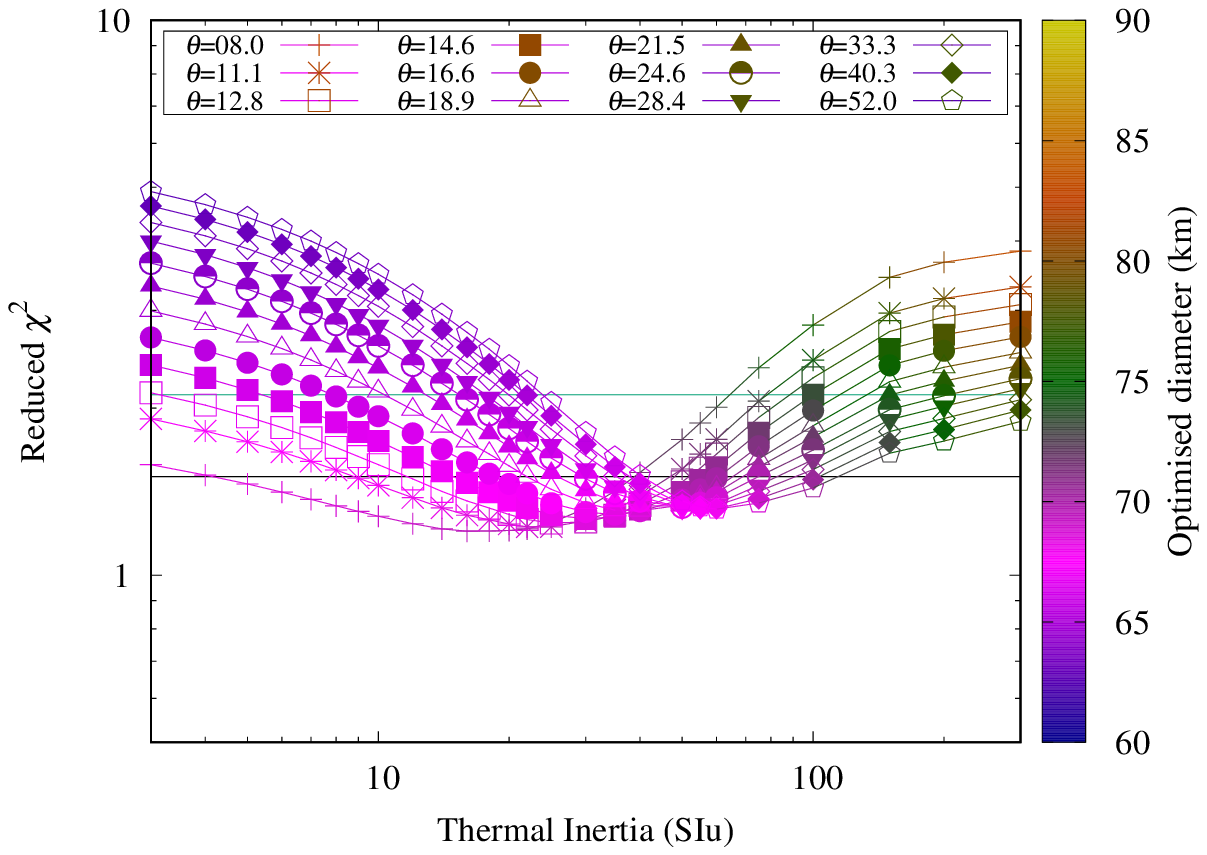}
\\
  \includegraphics[width=0.90\linewidth]{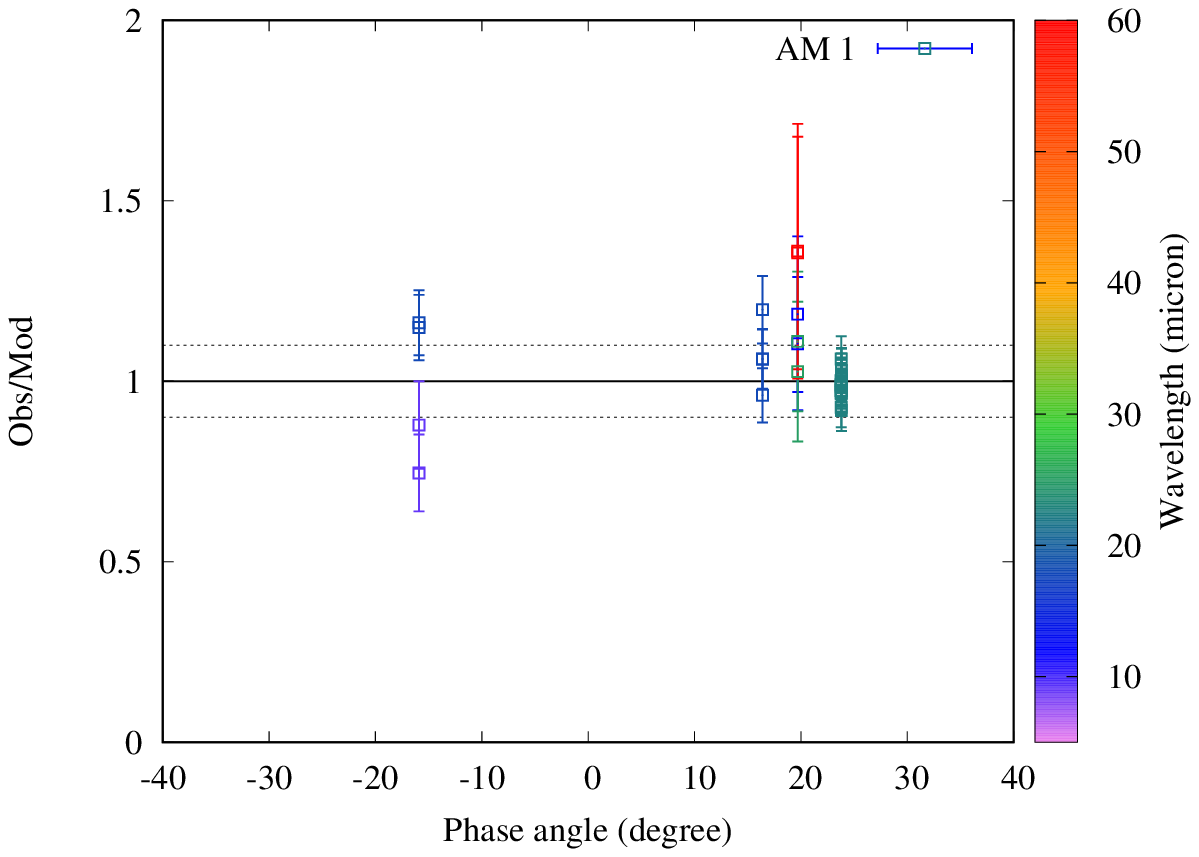}
&
  \includegraphics[width=0.90\linewidth]{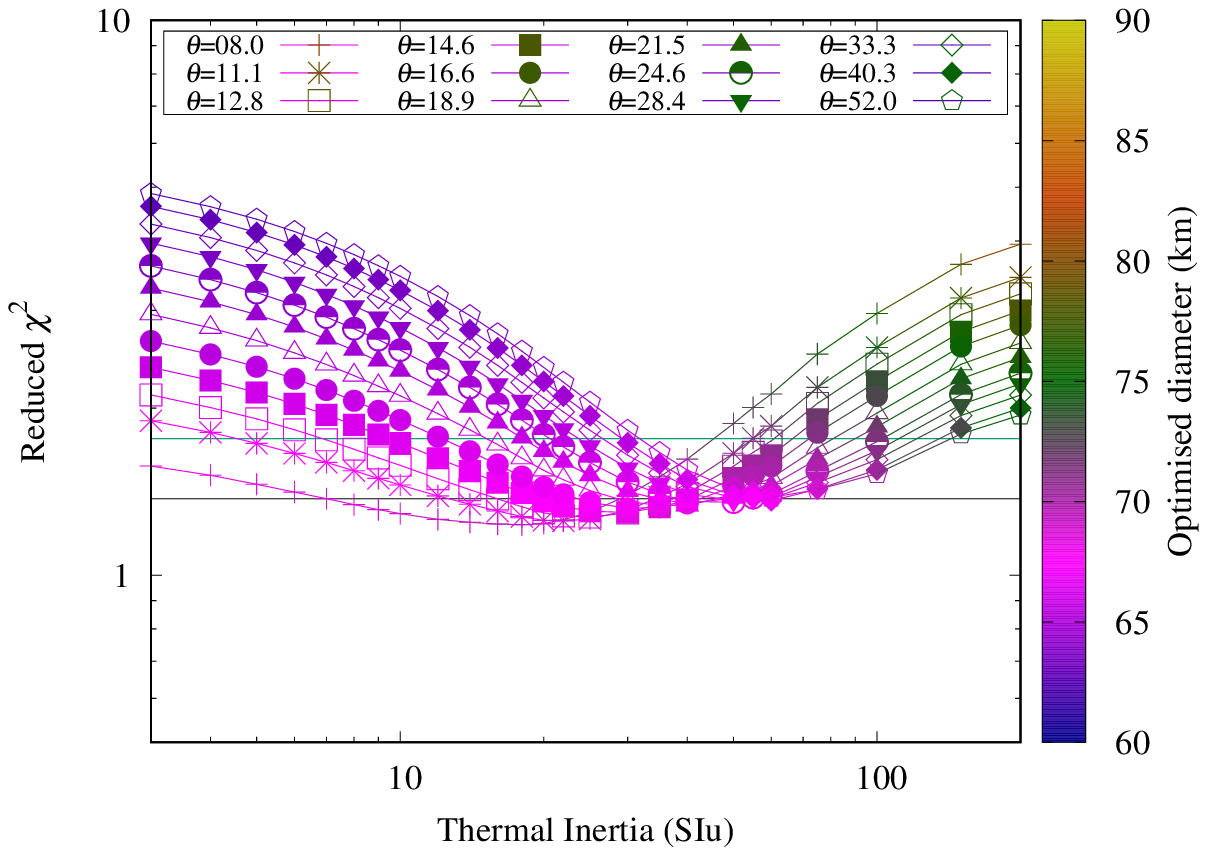}
\\
  \includegraphics[width=0.90\linewidth]{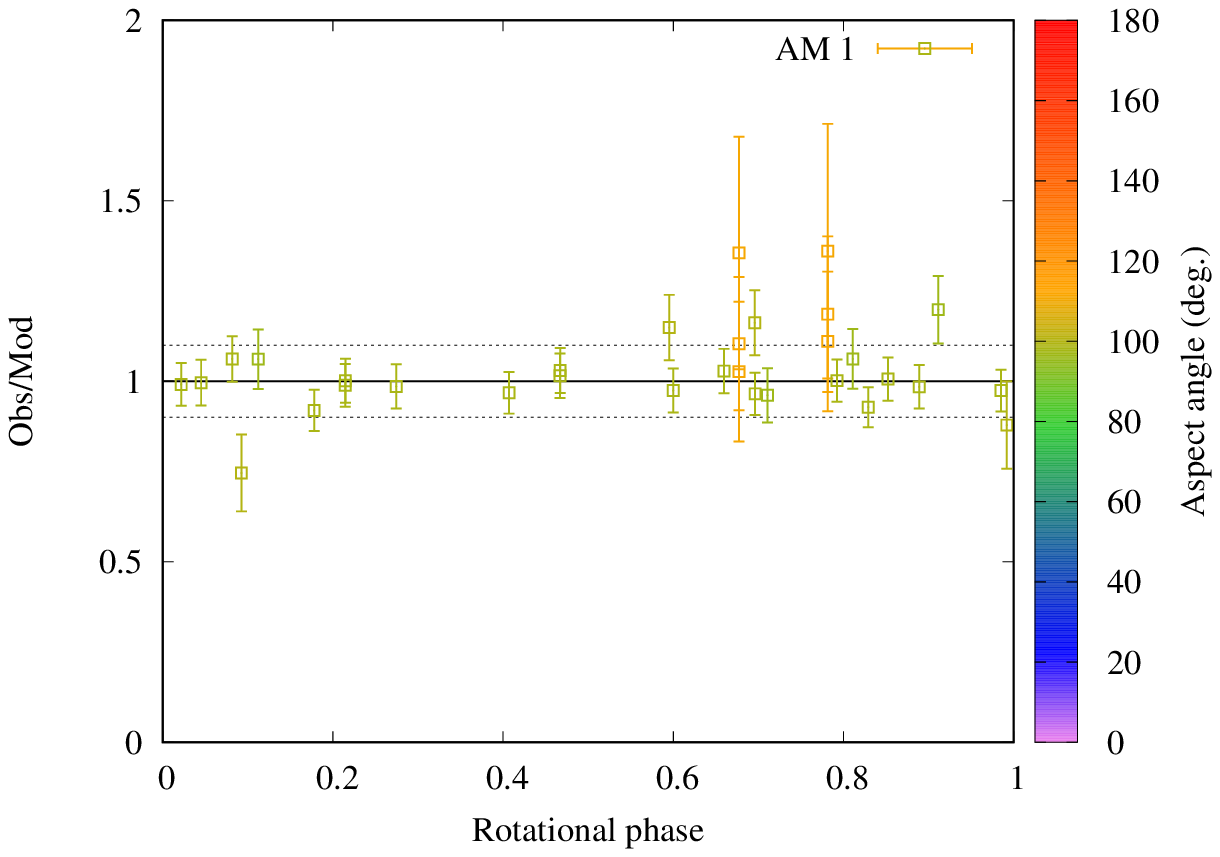}  
\\
  \includegraphics[width=0.90\linewidth]{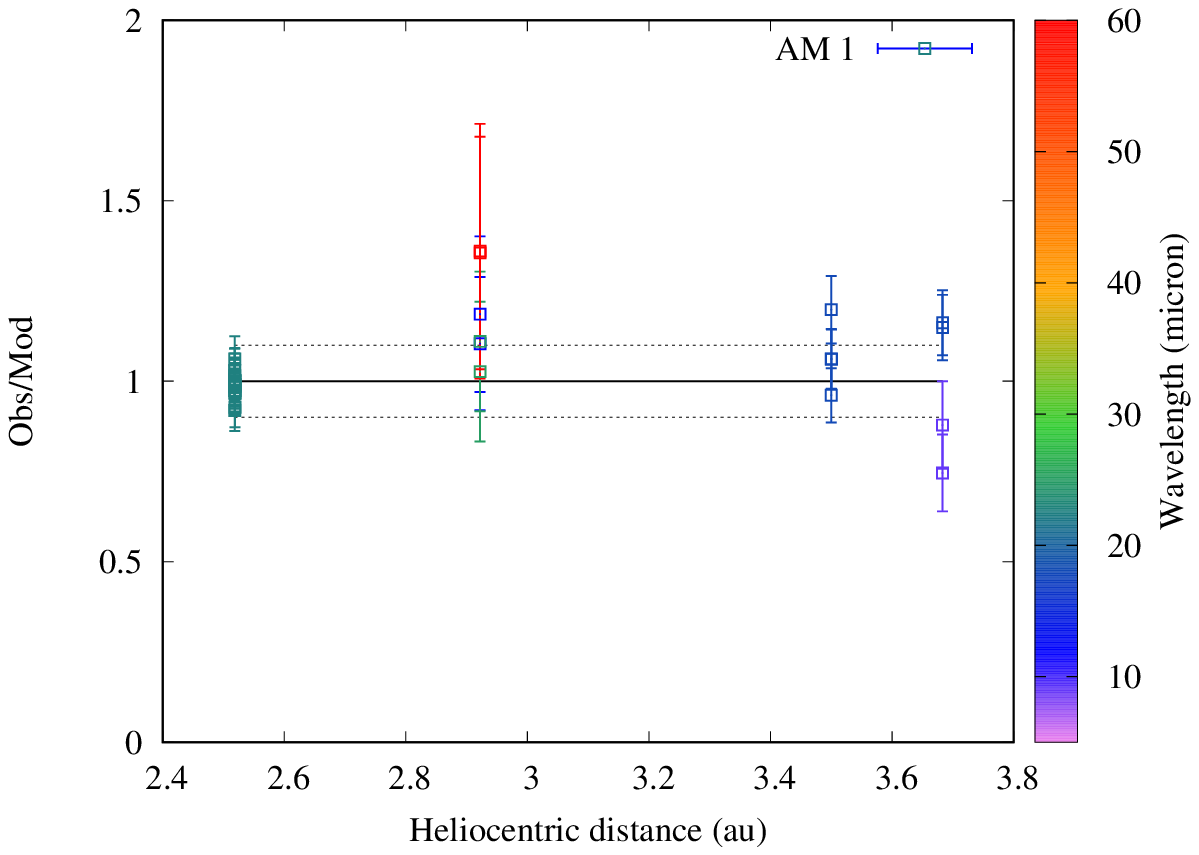}
  \captionof{figure}{Asteroid (468) Lina, left: observation to model ratios versus wavelength,
  phase angle, rotational phase, and heliocentric distance for model AM 1.
    The plots for AM 2 looked similar. 
   Right: $\chi^2$ versus thermal inertia curves for model AM1 (top), and AM2 (bottom).  
  }    \label{fig:468_OMR}
\\  
\end{tabularx}  
\end{table*}  

\clearpage

    \begin{table*}[ht]
    \centering
\begin{tabularx}{\linewidth}{XX}
  \includegraphics[width=0.90\linewidth]{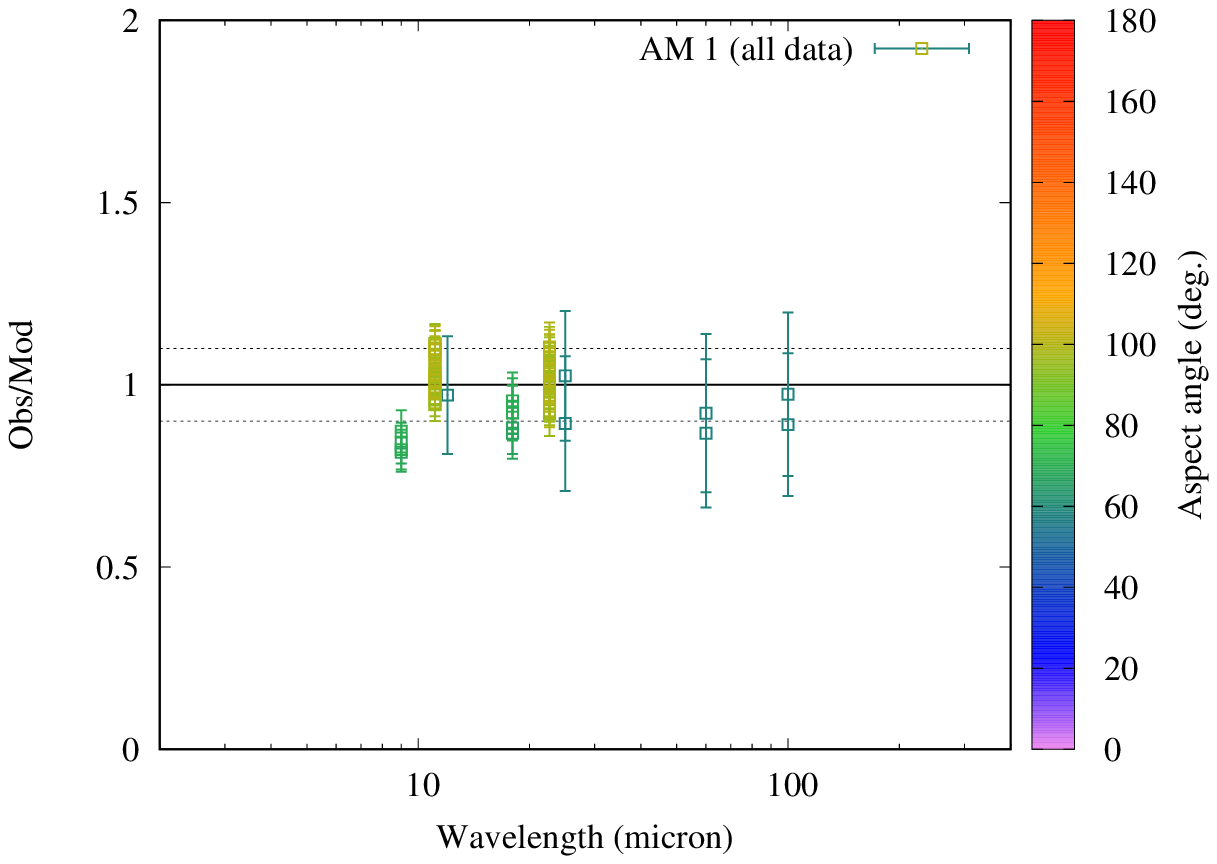}
&
  \includegraphics[width=0.90\linewidth]{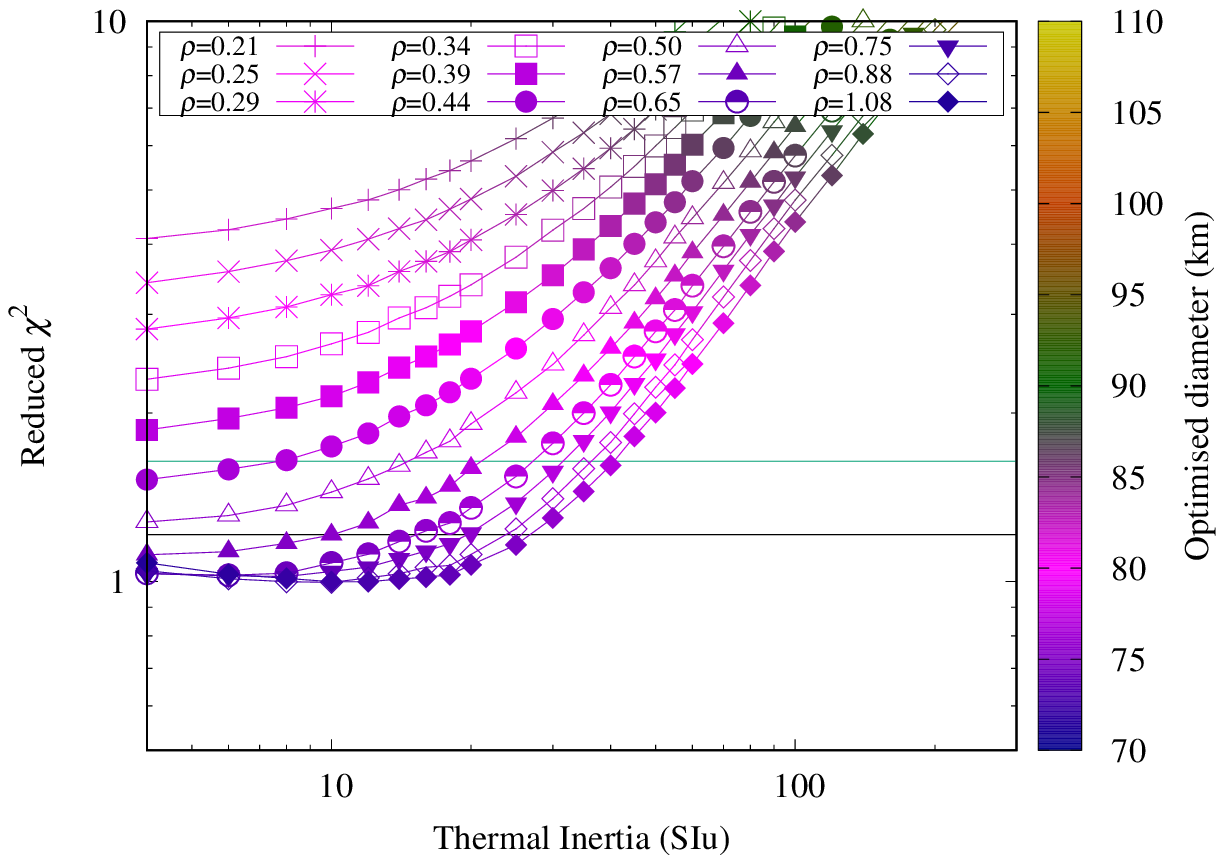}
\\
  \includegraphics[width=0.90\linewidth]{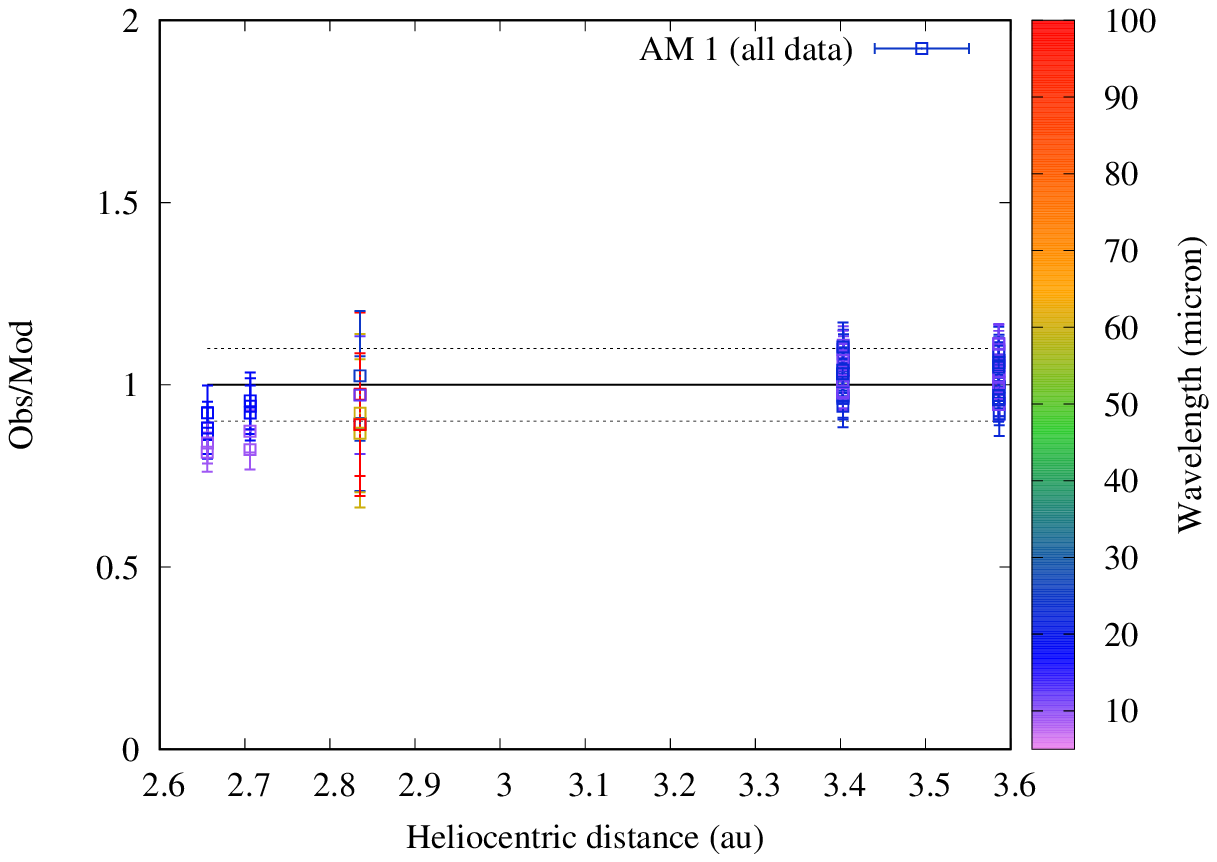}
&
  \includegraphics[width=0.90\linewidth]{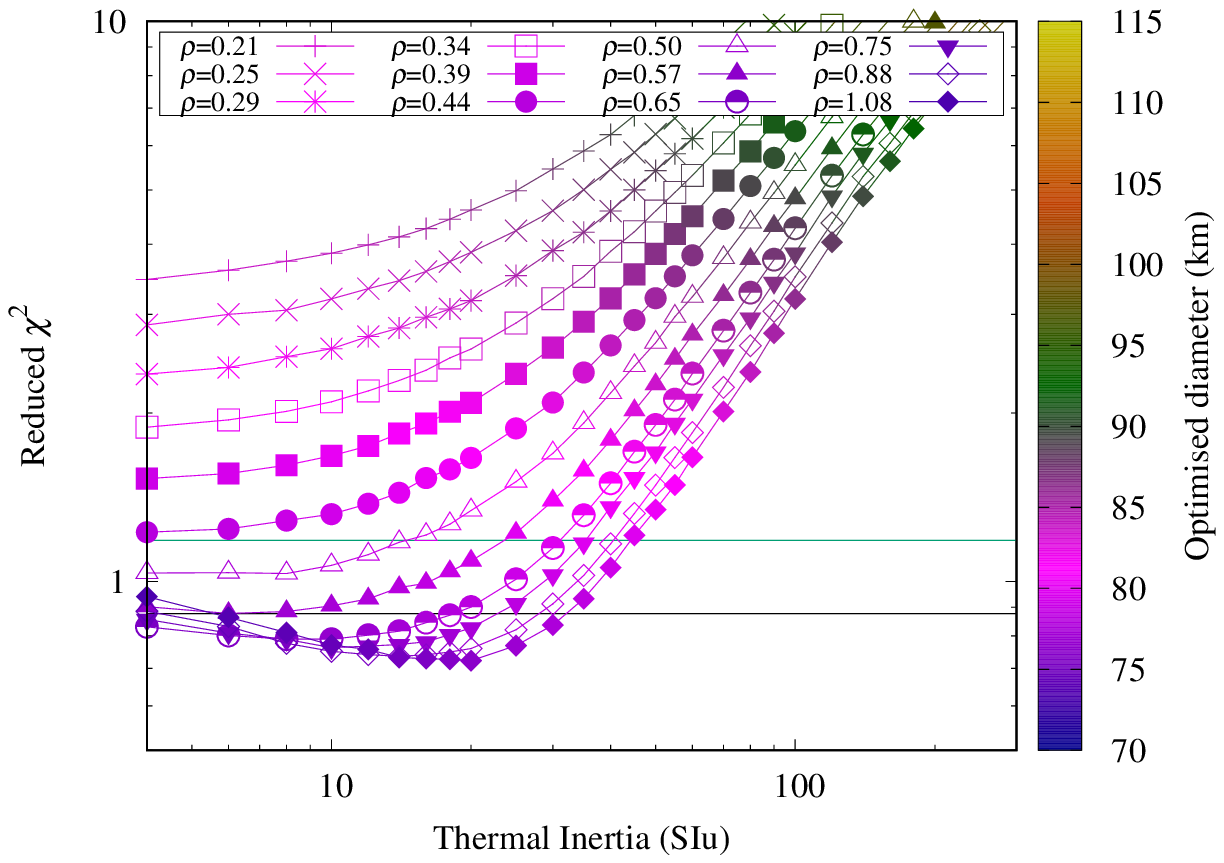}
\\
  \includegraphics[width=0.90\linewidth]{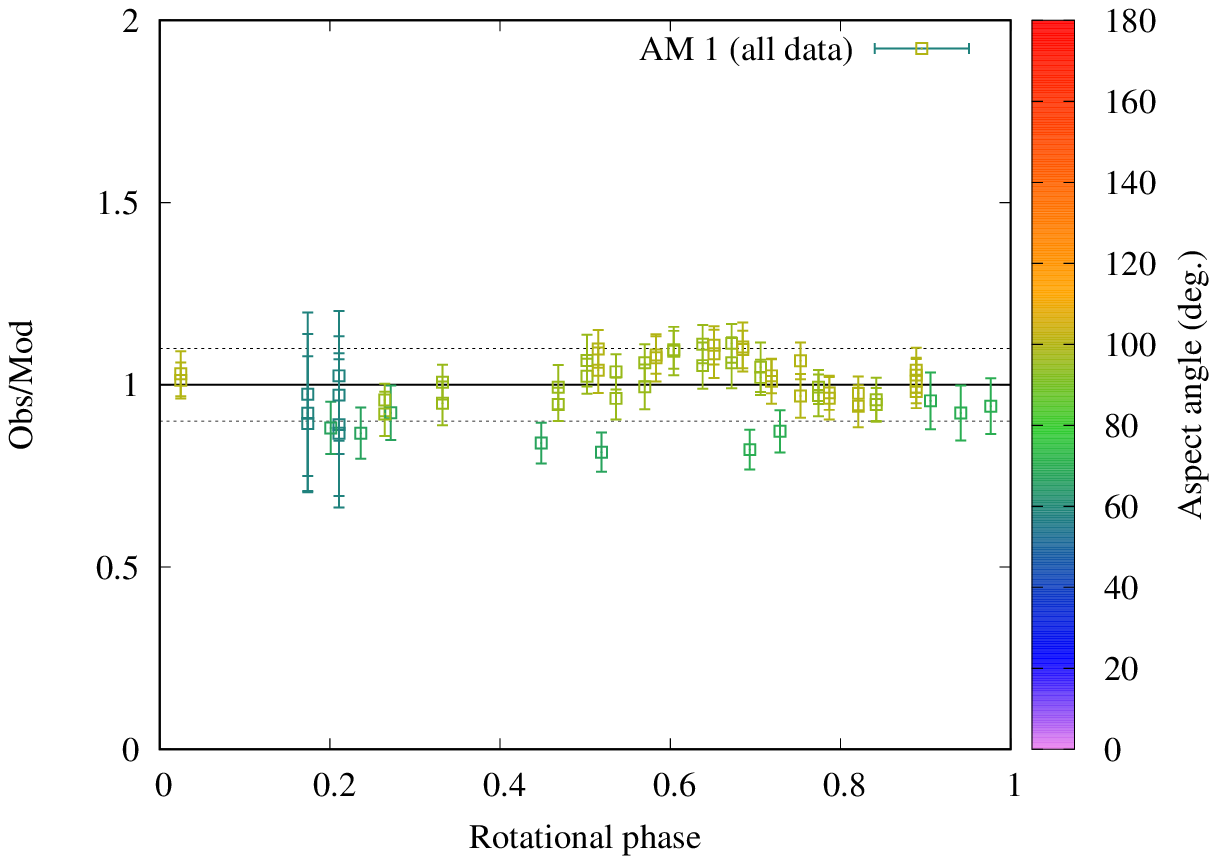}
\\
  \includegraphics[width=0.90\linewidth]{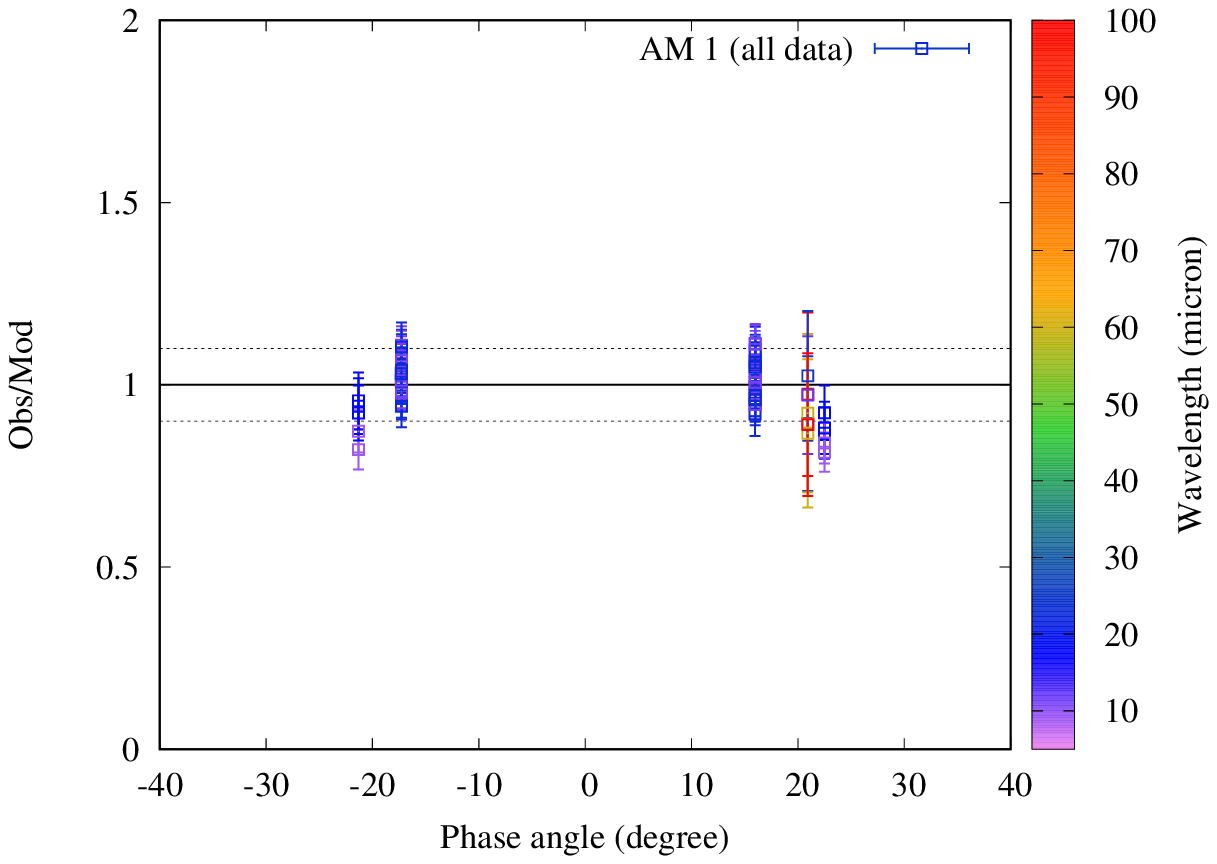}
  \captionof{figure}{Asteroid (538) Friederike, from top to bottom: observation to model ratios
    versus wavelength, heliocentric distance, rotational phase, and phase
    angle for model AM 1 (the AM 2 model on all data did not provide an acceptable fit and
    was rejected).  
   Right: $\chi^2$ versus thermal inertia curves for model AM1 (top), and AM2 (bottom) using WISE W3 and W4 data only.  
  }\label{fig:538_OMR}
\\  
\end{tabularx}  
\end{table*}

\clearpage

    \begin{table*}[ht]
    \centering
\begin{tabularx}{\linewidth}{XX}
  \includegraphics[width=0.90\linewidth]{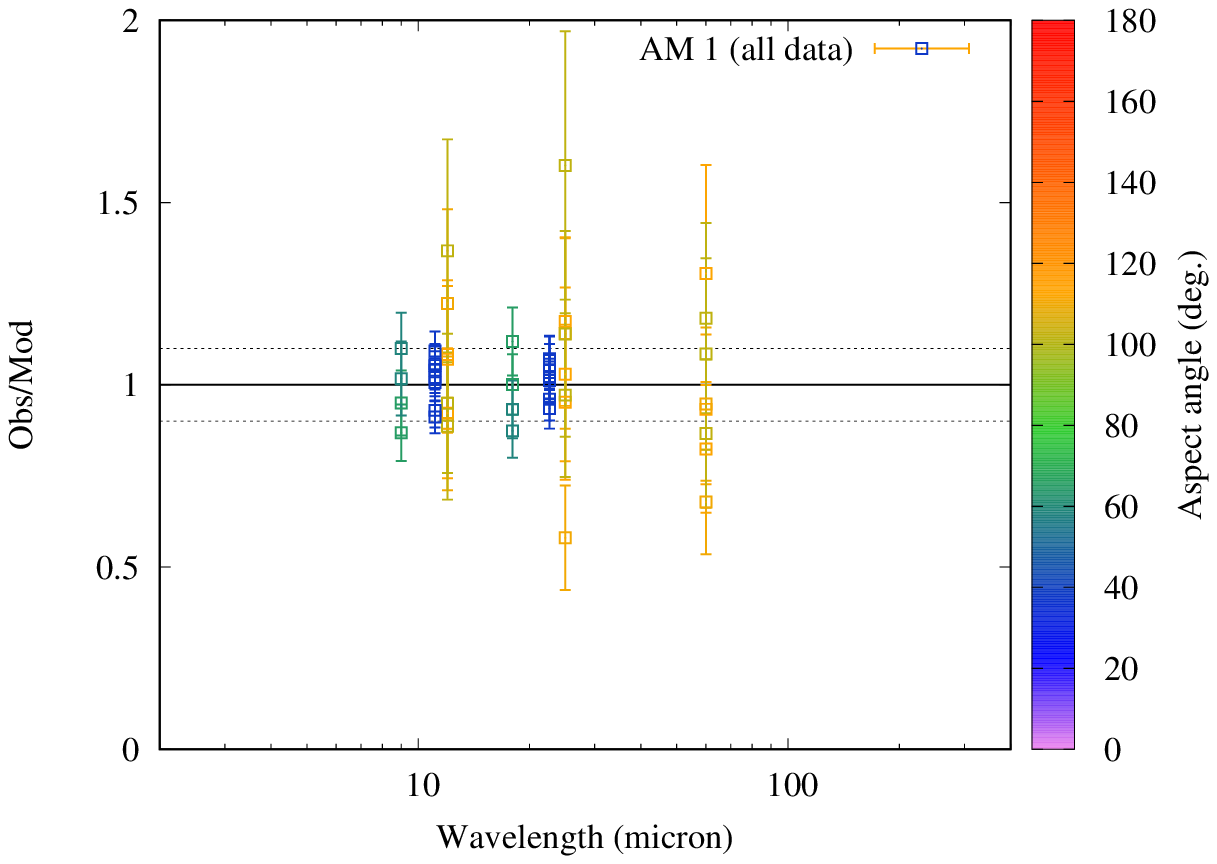}
&
  \includegraphics[width=0.90\linewidth]{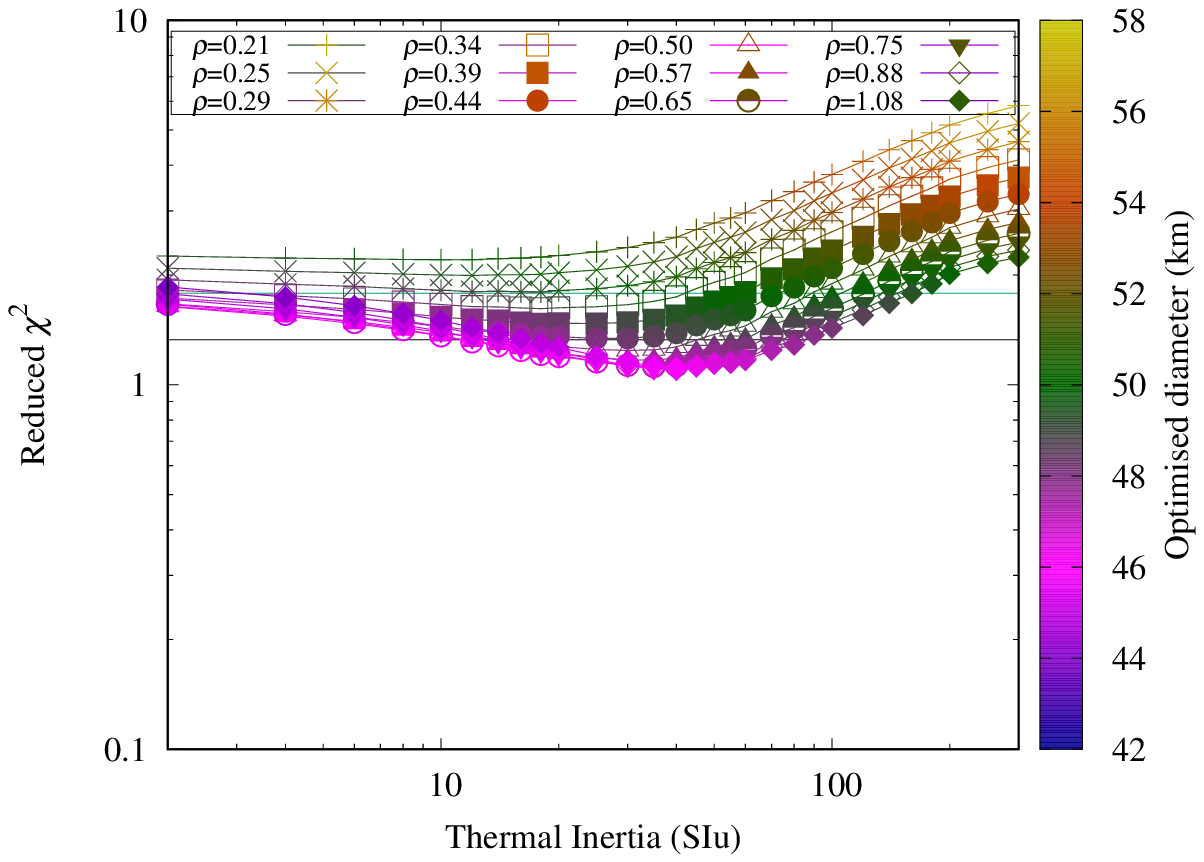}
\\
  \includegraphics[width=0.90\linewidth]{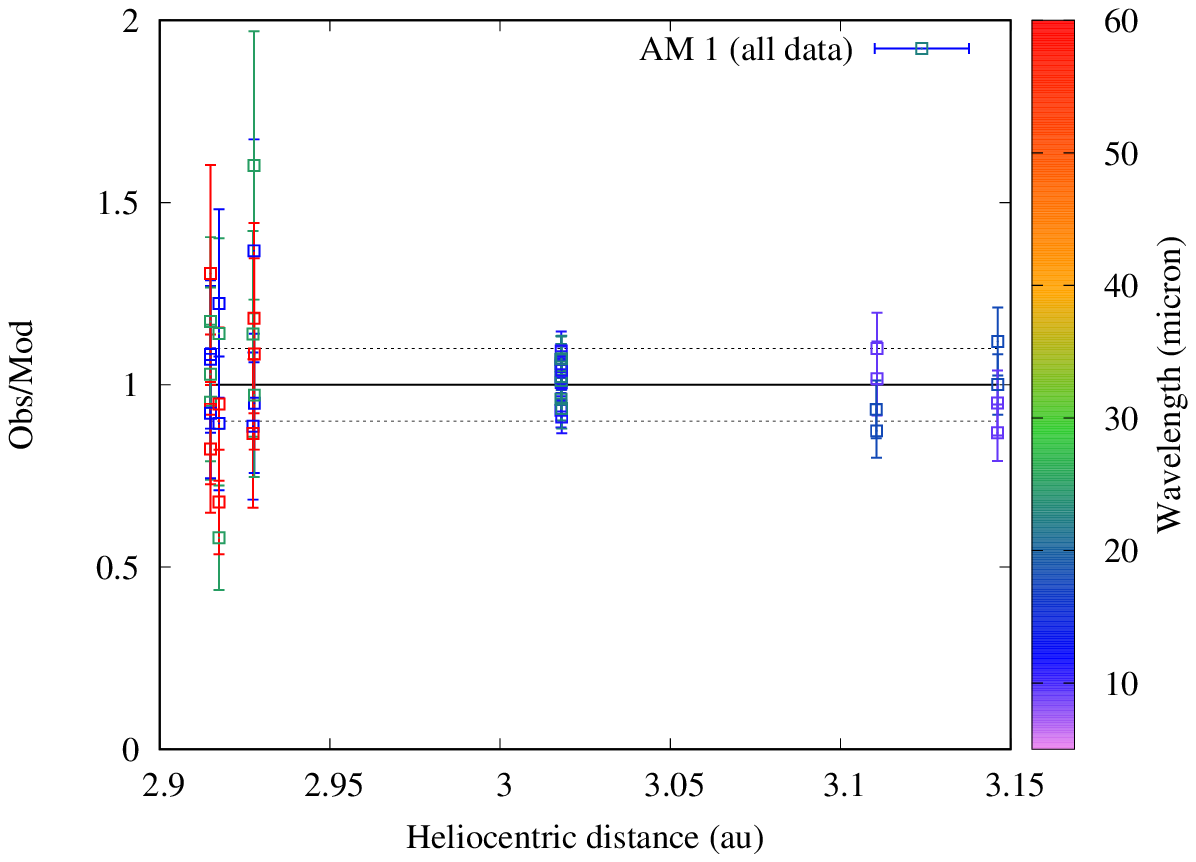}
\\
  \includegraphics[width=0.90\linewidth]{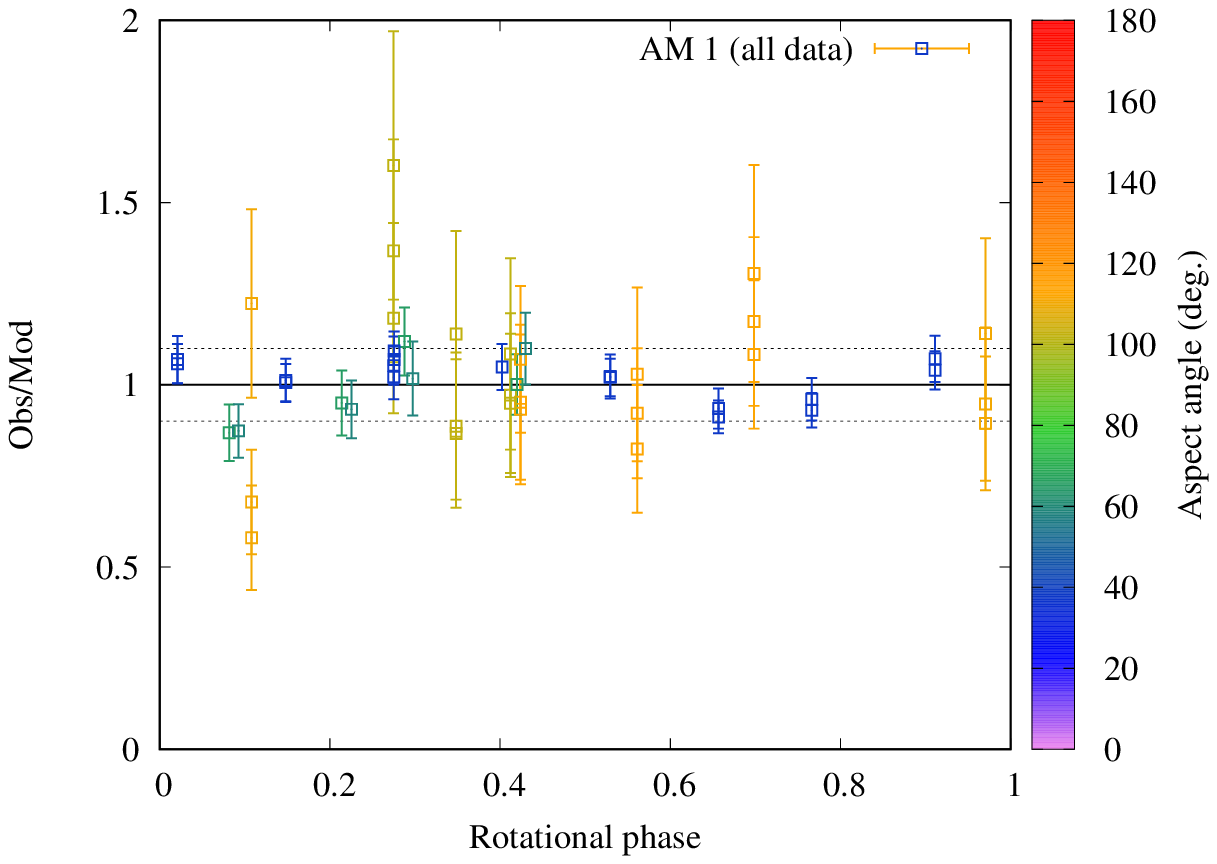}
\\
  \includegraphics[width=0.90\linewidth]{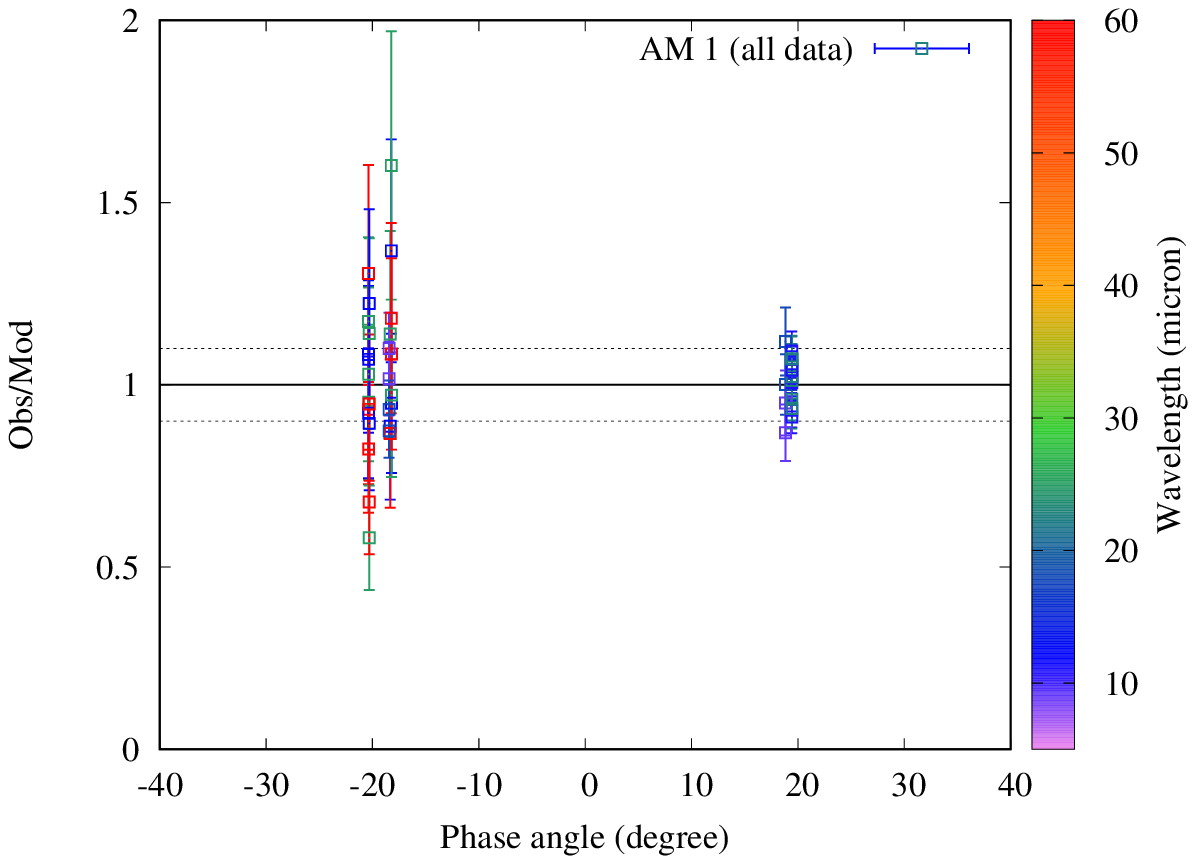}
  \captionof{figure}{Asteroid (653) Berenike, left, from top to bottom: observation to model ratios
    versus wavelength, heliocentric distance, rotational phase, and phase
    angle for model AM 1 (the AM 2 model did not provide an acceptable fit and
    was rejected). 
    Right: $\chi^2$ versus thermal inertia curves for model AM1.  
  }\label{fig:653_OMR}
\\  
\end{tabularx}  
\end{table*}  

\clearpage

    \begin{table*}[ht]
    \centering
\begin{tabularx}{\linewidth}{XX}
  \includegraphics[width=0.90\linewidth]{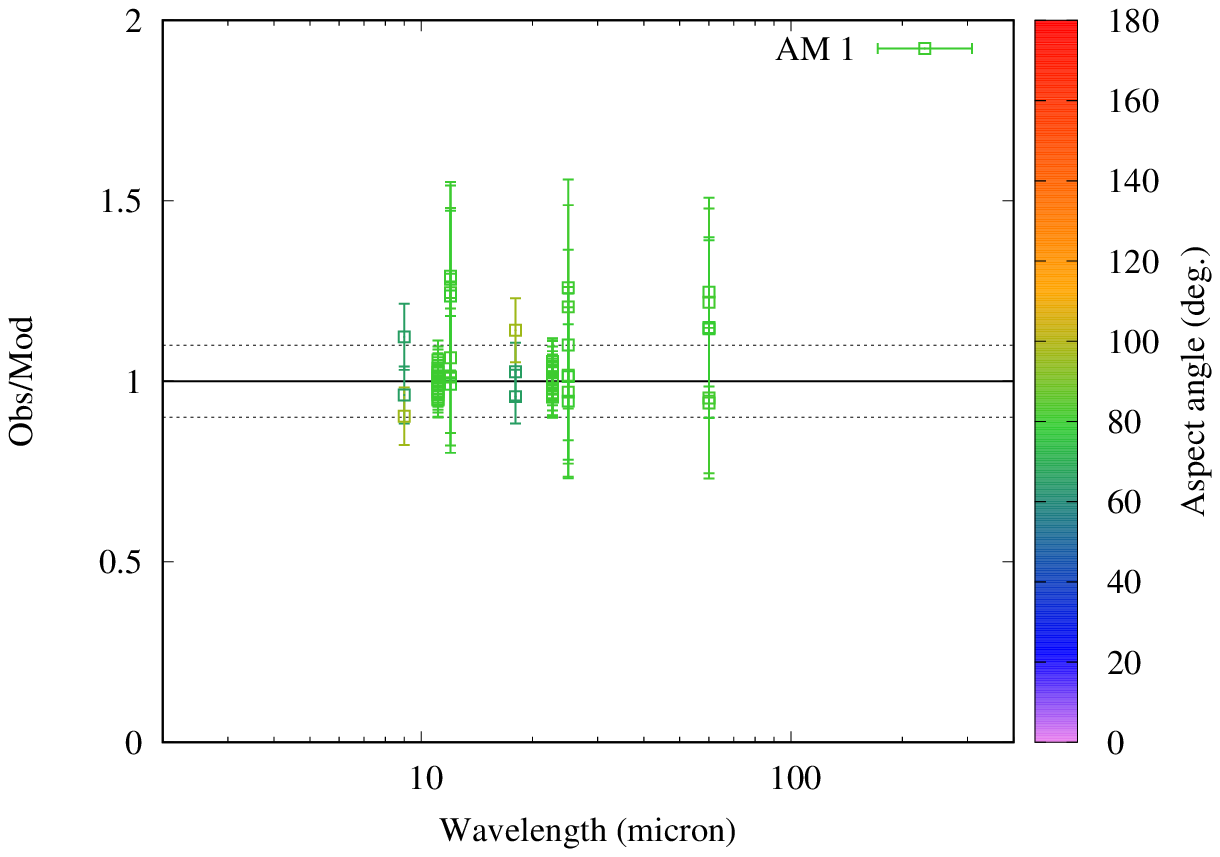}  
&
  \includegraphics[width=0.90\linewidth]{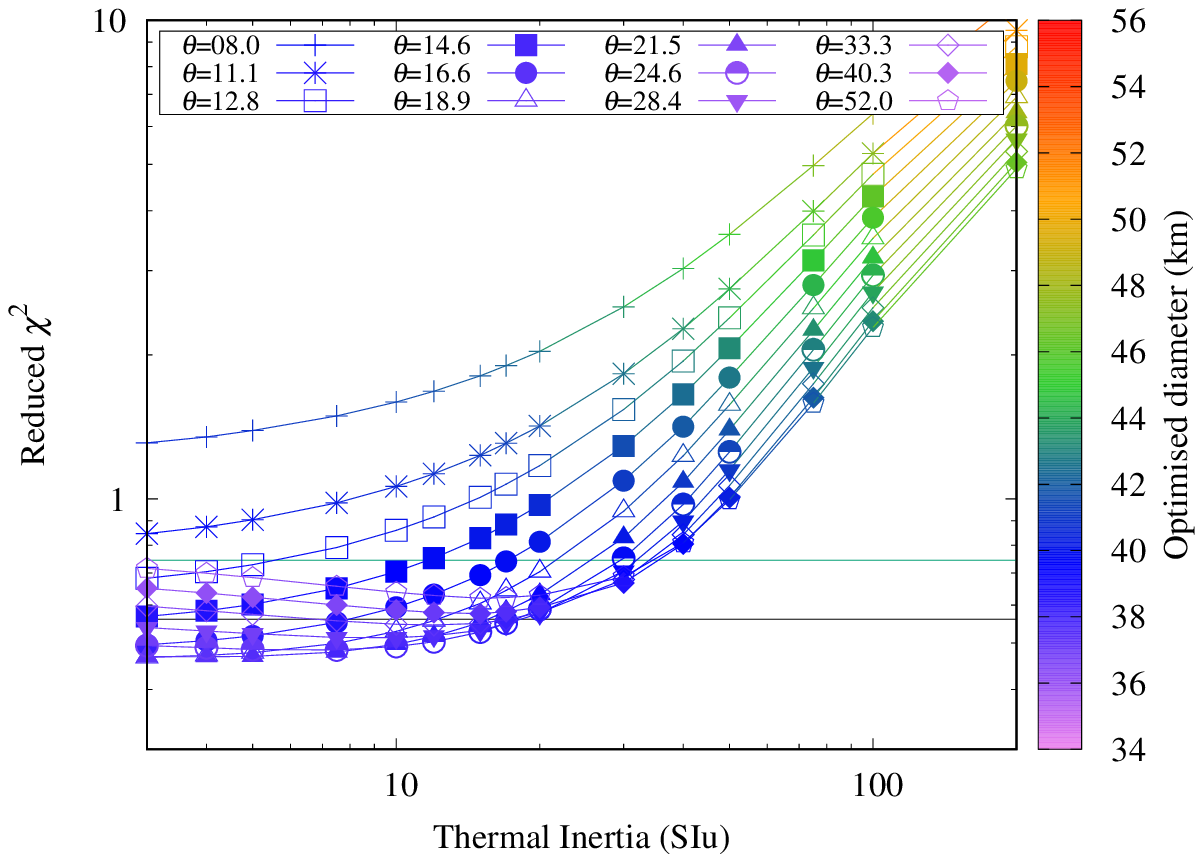}
\\
  \includegraphics[width=0.90\linewidth]{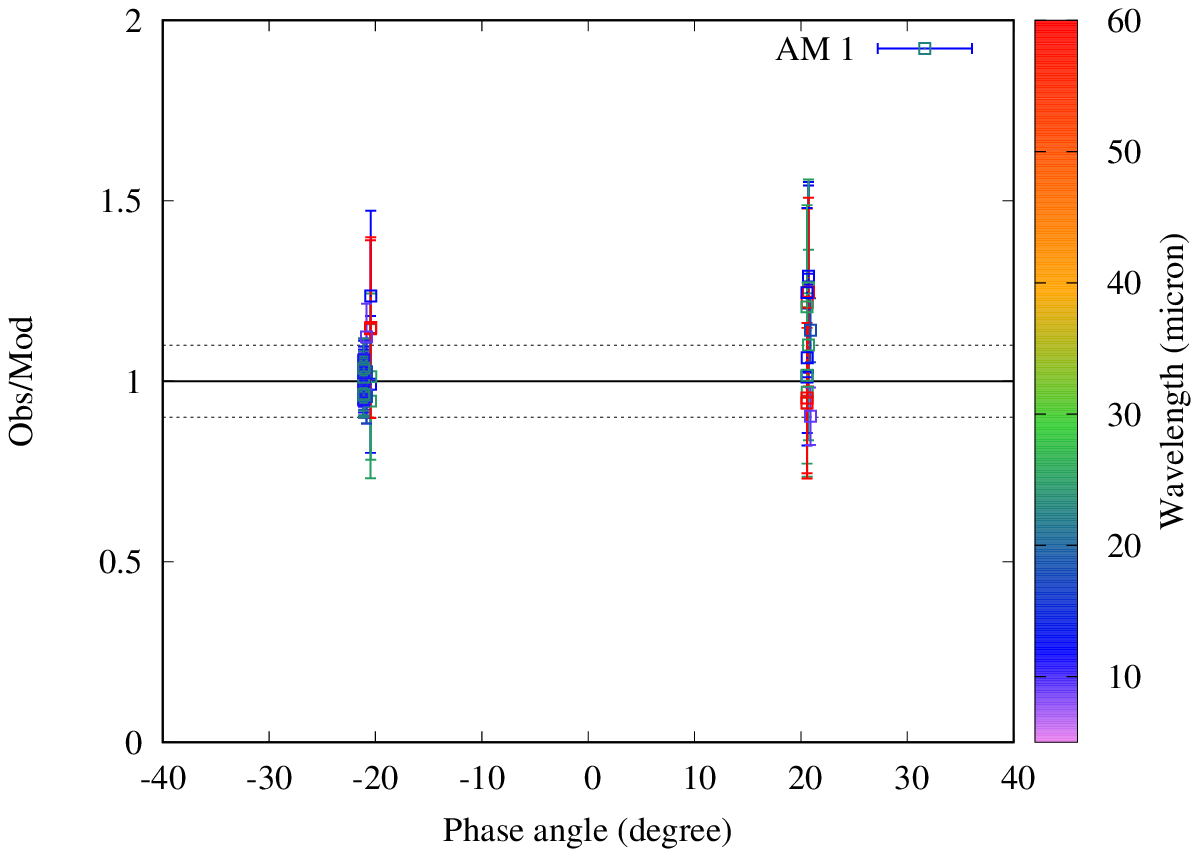}
&
  \includegraphics[width=0.90\linewidth]{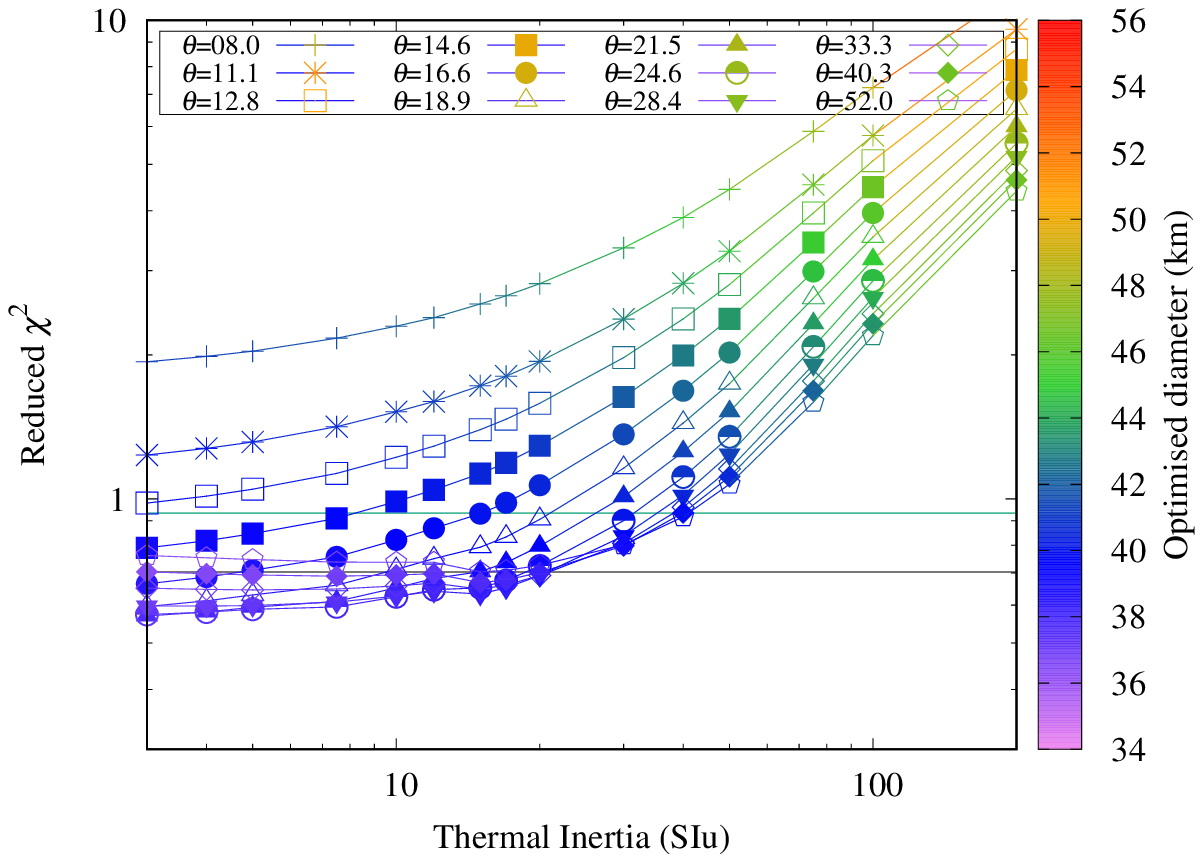}
\\
  \includegraphics[width=0.90\linewidth]{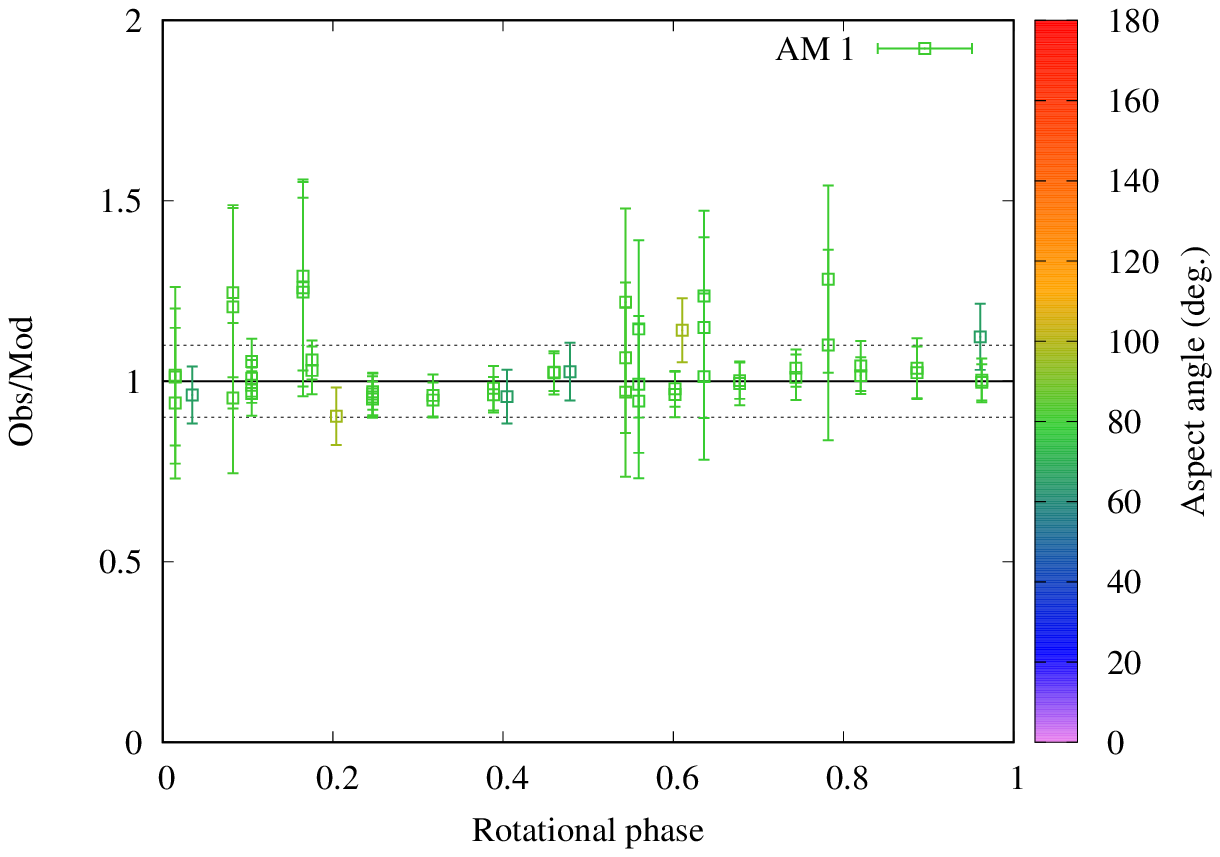}  
\\
  \includegraphics[width=0.90\linewidth]{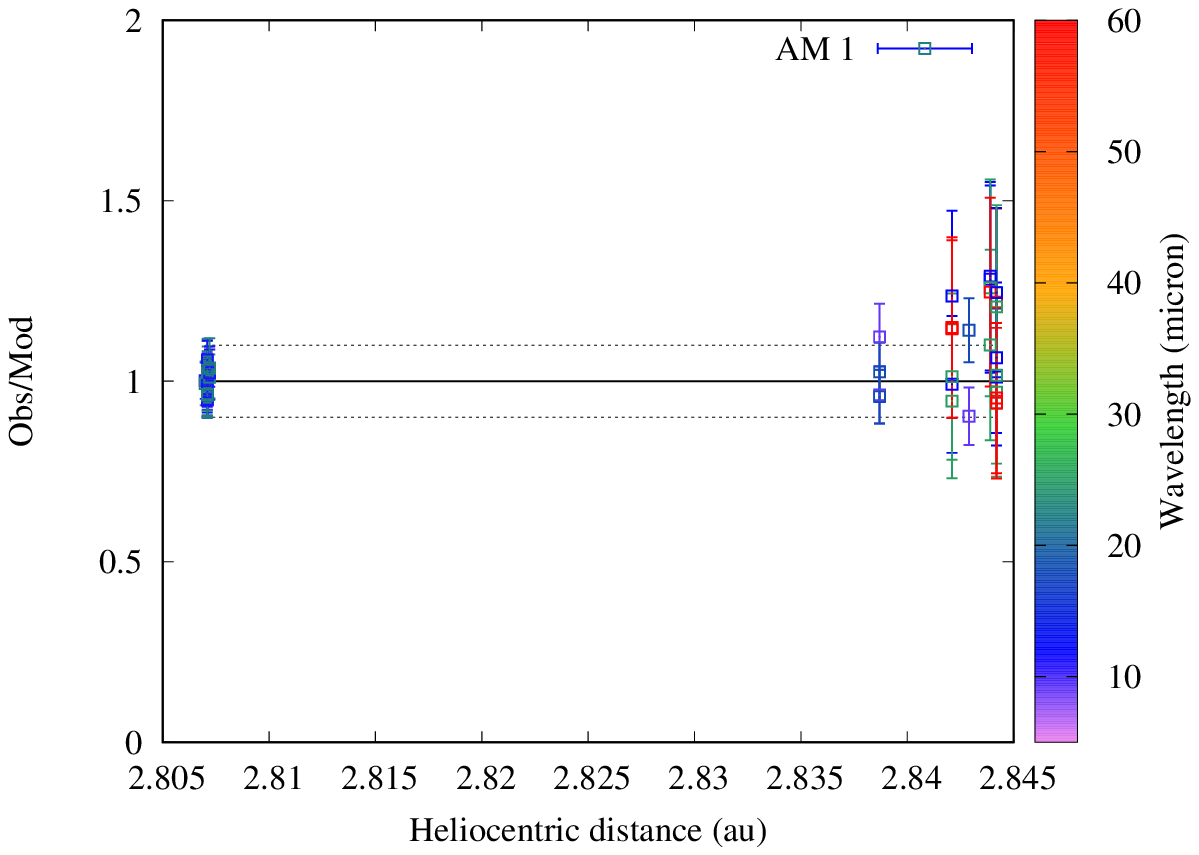}
  \captionof{figure}{Asteroid (673) Edda, left: observation to model ratios versus wavelength,
  phase angle, rotational phase, and heliocentric distance for model AM 1.
    The plots for AM 2 looked similar. 
    Right: $\chi^2$ versus thermal inertia curves for model AM1 (top) and AM2 (bottom).  
  }    \label{fig:673_OMR}
\\  
\end{tabularx}  
\end{table*}  

\clearpage

    \begin{table*}[ht]
    \centering
\begin{tabularx}{\linewidth}{XX}
  \includegraphics[width=0.90\linewidth]{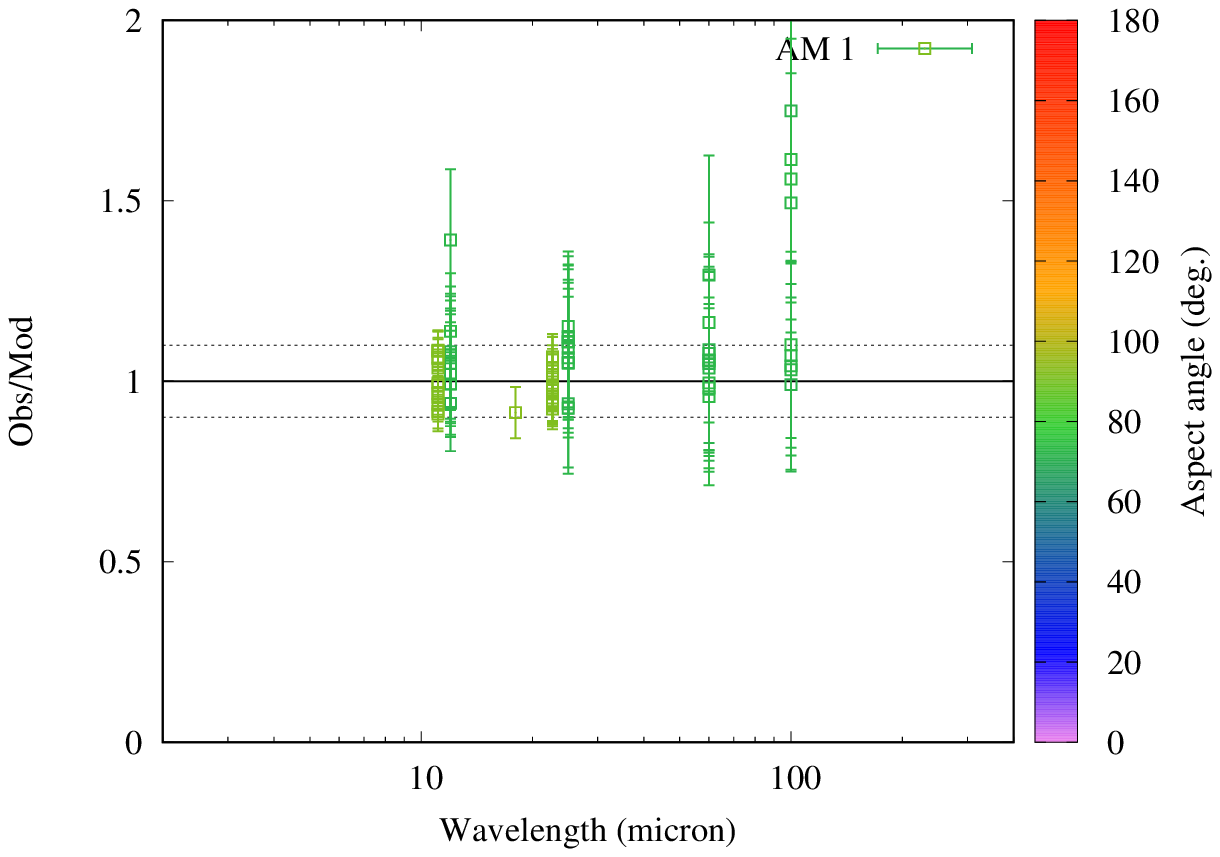}  
&
  \includegraphics[width=0.90\linewidth]{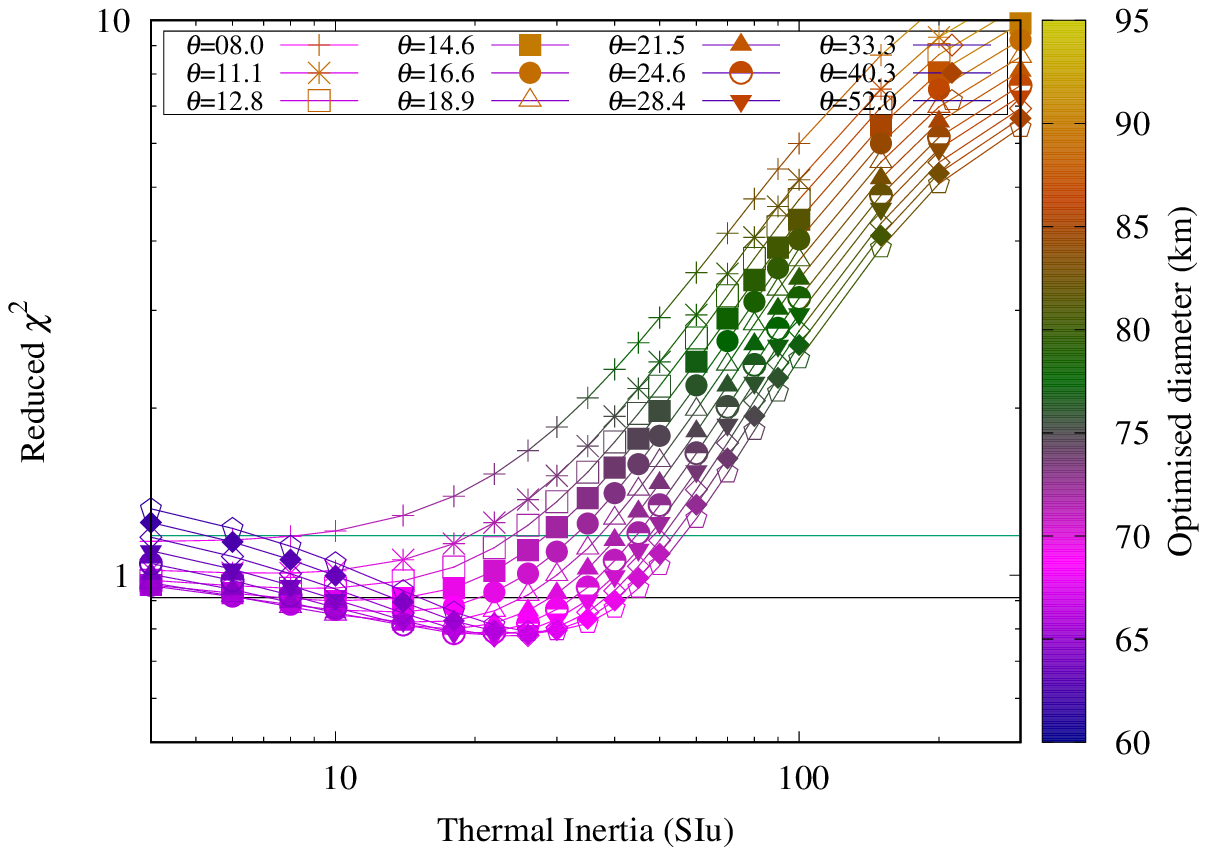}
\\
  \includegraphics[width=0.90\linewidth]{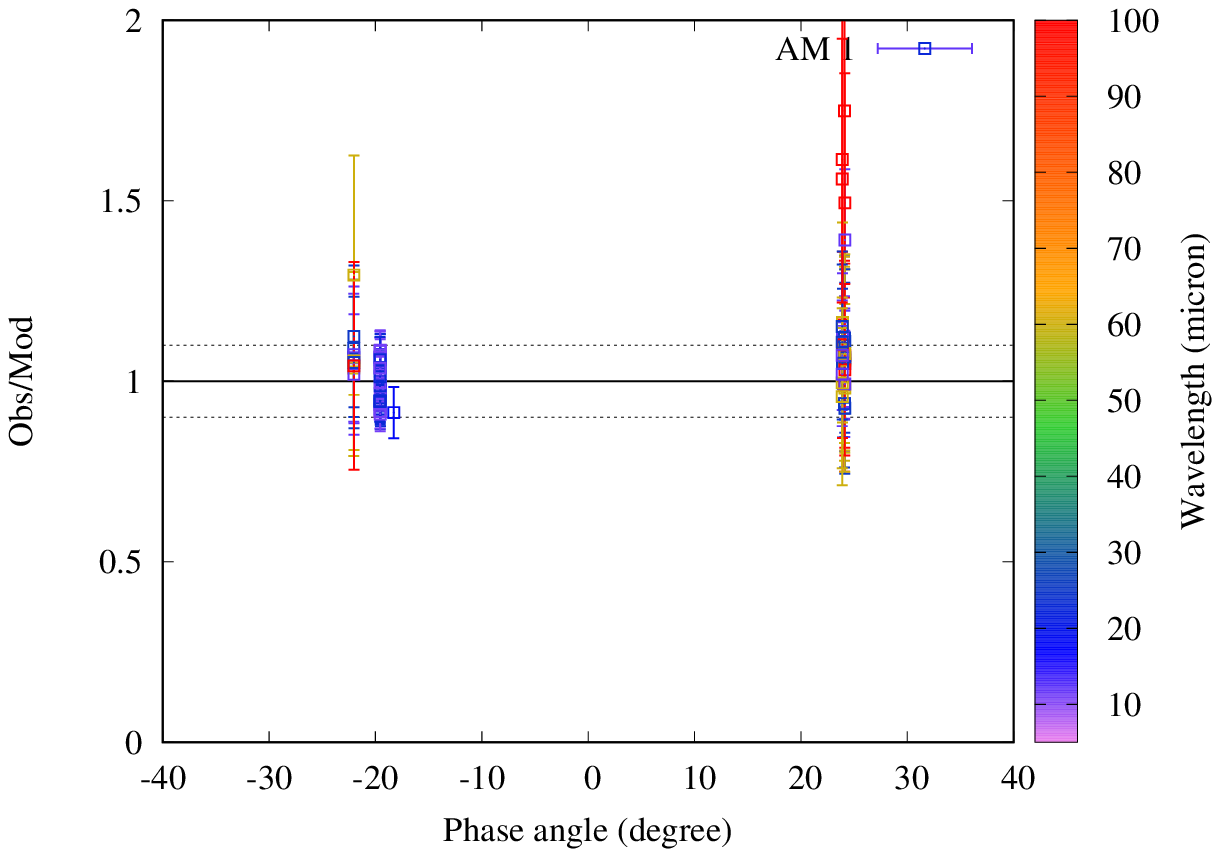}
&
  \includegraphics[width=0.90\linewidth]{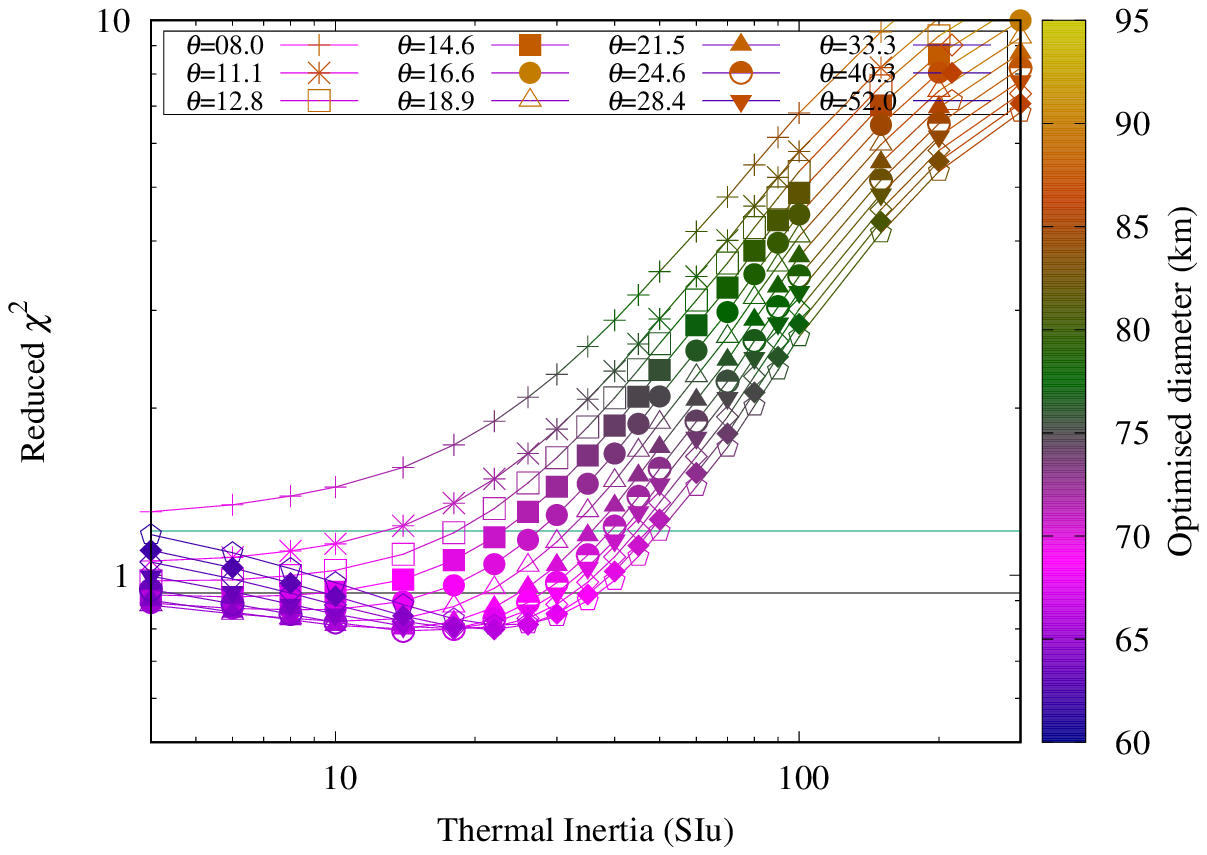}
\\
  \includegraphics[width=0.90\linewidth]{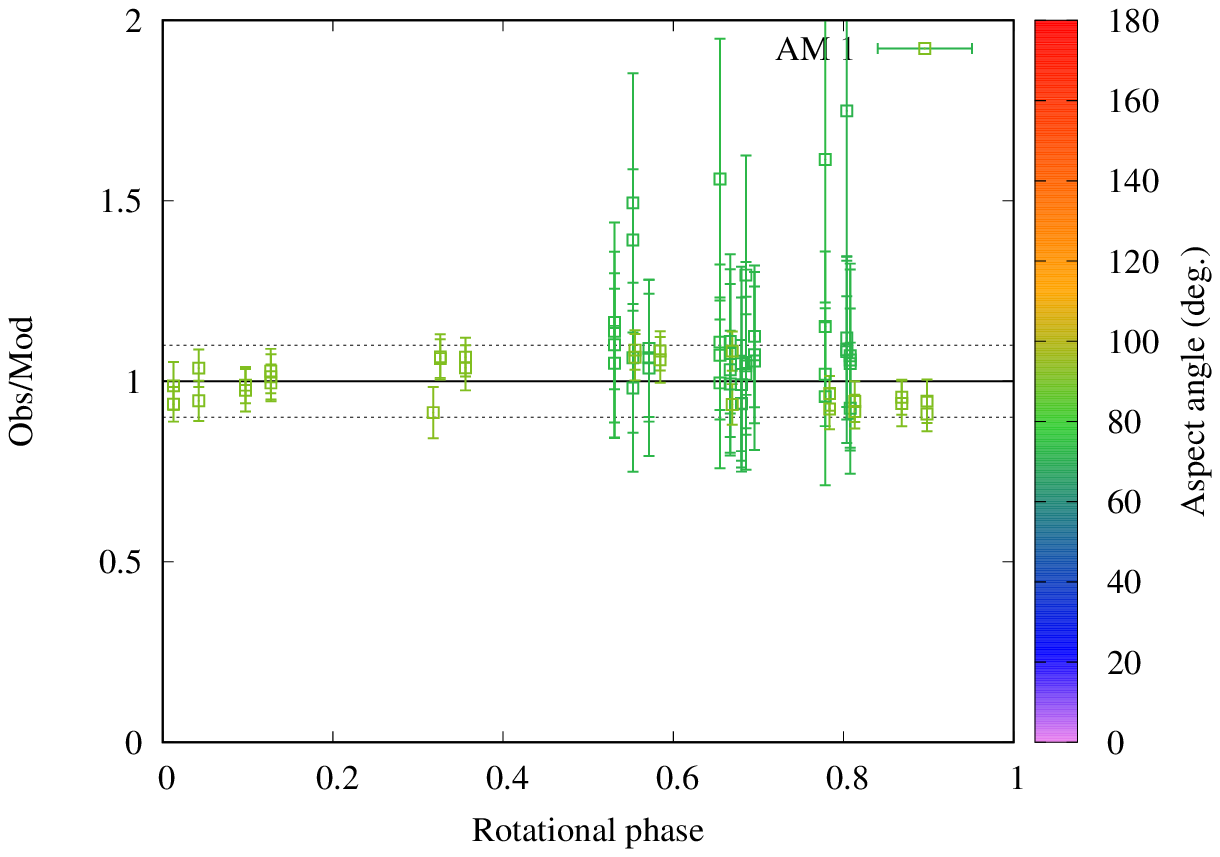}  
\\
  \includegraphics[width=0.90\linewidth]{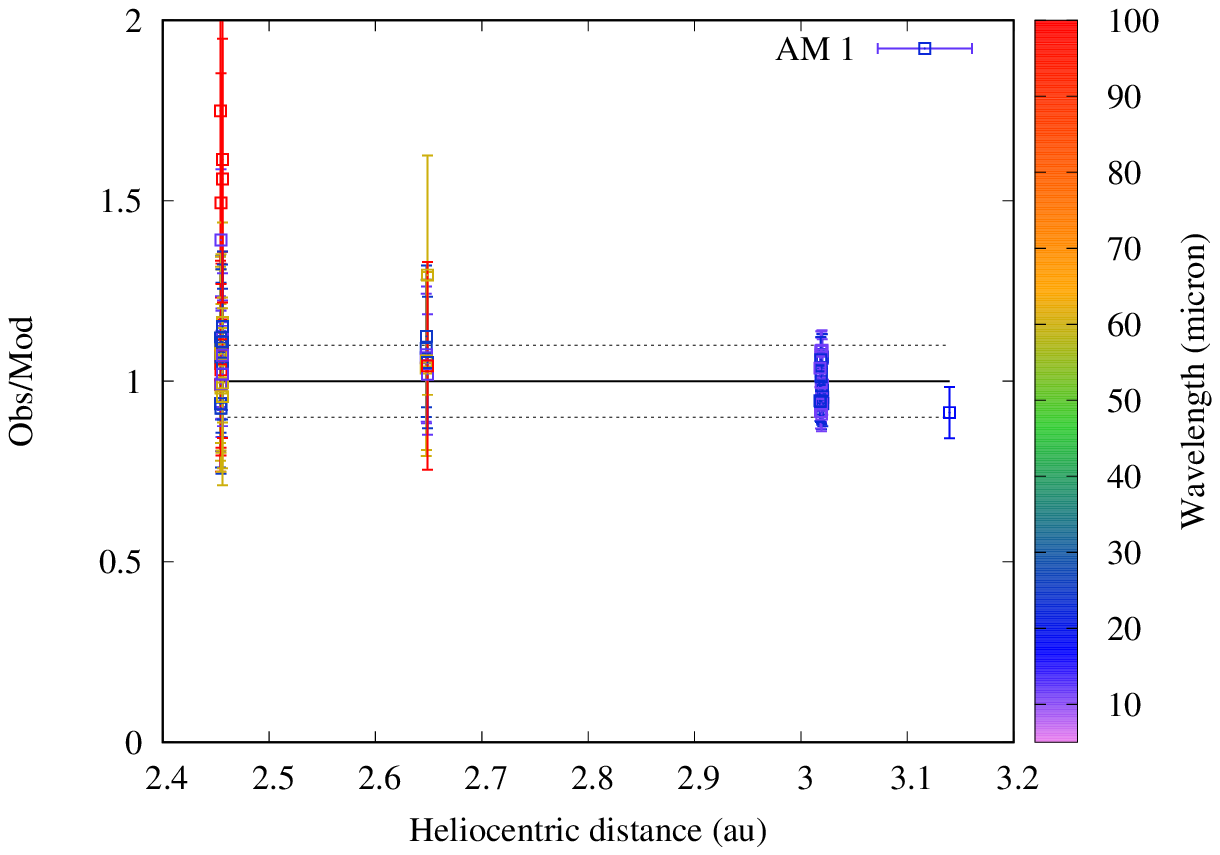}
  \captionof{figure}{Asteroid (834) Burnhamia, left: observation to model ratios versus wavelength,
  phase angle, rotational phase, and heliocentric distance for model AM 1.
    The plots for AM 2 looked similar. 
        Right: $\chi^2$ versus thermal inertia curves for model AM1 (top) and AM2 (bottom).  
  }    \label{fig:834_OMR}
\\  
\end{tabularx}  
\end{table*}
\end{appendix}

\clearpage

\end{document}